\documentclass[journal]{vgtc} 
\usepackage{float}

\usepackage{tabularx}
\usepackage{multirow}

\onlineid{0}

\vgtccategory{Research}

\vgtcpapertype{please specify}

\usepackage{graphicx}
\newcommand{\papertitle}{Using Tactile Charts to Support Comprehension and Learning of Complex Visualizations for Blind and Low-Vision Individuals}
\title{\papertitle}

\author{%
  \authororcid{Tingying He}{0000-0002-9670-5587},
  \authororcid{Maggie McCracken}{0009-0006-5280-0546.},
  \authororcid{Daniel Hajas}{0000-0002-2811-1197},
  \authororcid{Sarah Creem-Regehr}{0000-0001-7740-1118}, and
  \authororcid{Alexander Lex}{0000-0001-6930-5468}
}

\authorfooter{
    \item T. He (\raisebox{-.5pt}{\includegraphics[height=6pt]{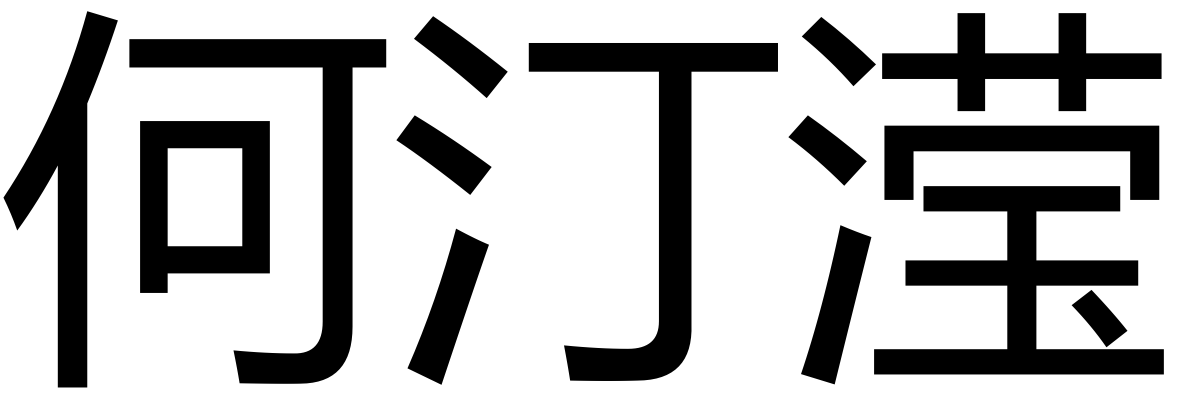}}), M. McCracken, S. Creem-Regehr, and A. Lex are with the University of Utah, USA. E-mails: hetingying.hty@gmail.com, \{maggie.mccracken\,$|$\,sarah.creem\}@psych.utah.edu, alex@sci.utah.edu.
    \item D. Hajas is with the Global Disability Innovation Hub, UK. E-mail: d.hajas@ucl.ac.uk.
}

\abstract{%
We investigate whether tactile charts support comprehension and learning of complex visualizations for blind and low-vision (BLV) individuals and contribute four tactile chart designs and an interview study.
Visualizations are powerful tools for conveying data, yet BLV individuals typically can rely only on assistive technologies---primarily alternative texts---to access this information. Prior research shows the importance of mental models of chart types for interpreting these descriptions, yet BLV individuals have no means to build such a mental model based on images of visualizations. Tactile charts show promise to fill this gap in supporting the process of building mental models. Yet studies on tactile data representations mostly focus on simple chart types, and it is unclear whether they are also appropriate for more complex charts as would be found in scientific publications.
Working with two BLV researchers, we designed 3D-printed tactile template charts with exploration instructions for four advanced chart types: UpSet plots, violin plots, clustered heatmaps, and faceted line charts. 
We then conducted an interview study with 12 BLV participants comparing whether using our tactile templates improves mental models and understanding of charts and whether this understanding translates to novel datasets experienced through alt texts. Thematic analysis shows that tactile models support chart type understanding and are the preferred learning method by BLV individuals. We also report participants' opinions on tactile chart design and their role in BLV education.
}


\teaser{
    \centering
    \includegraphics[width=\columnwidth]{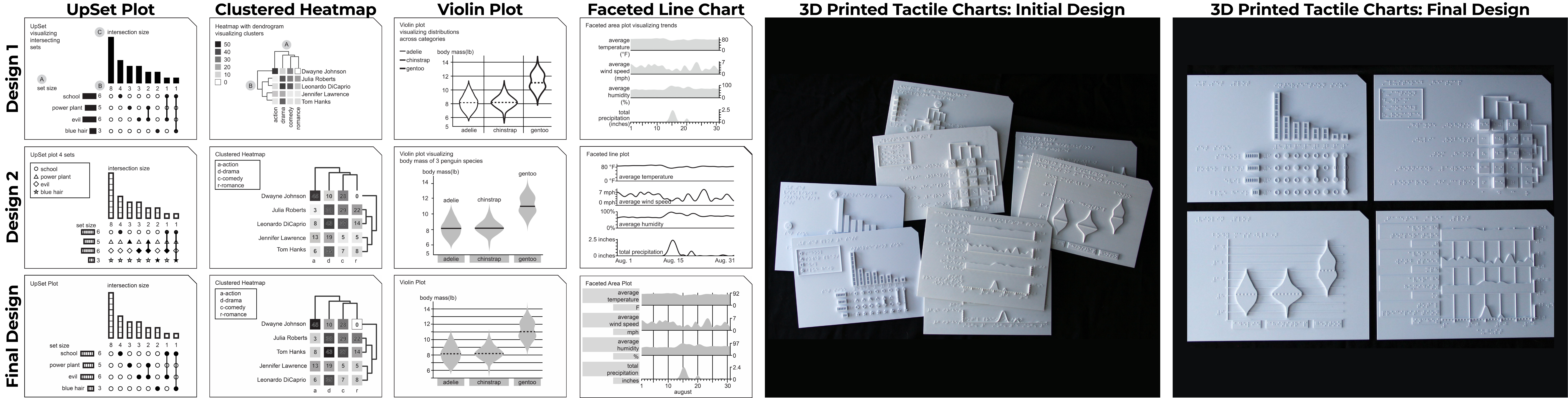}
    \caption{\rev{Design iterations and final designs for tactile template charts for four visualization techniques: UpSet plots, clustered heatmaps, violin plots, and faceted line charts. Left: Sighted versions; larger in \autoref{sec:template-chart-design-variations} and \autoref{sec:template-chart-final-design}. Right: 3D-printed tactile charts, with final designs on the far right; larger in \autoref{sec:template-chart-model-photos-design-variations} and \autoref{sec:template-chart-model-photos-final-design}. Models and additional materials are also available on \osfrepo\ and our accessible website: \companionweb.}}
    \label{fig:teaser}
}

\graphicspath{{figs/}{figures/}{pictures/}{images/}{./}} % where to search for the images

\usepackage{tabu}                      % only used for the table example
\usepackage{booktabs}                  % only used for the table example
\usepackage{lipsum}                    % used to generate placeholder text
\usepackage{mwe}                       % used to generate placeholder figures

\usepackage{mathptmx}                  % use matching math font

\usepackage{cuted} % appendix
\usepackage{ccicons} %copyright icons

\newcommand{\hquote}[1]{``\textit{#1}''} %highlight quotes, only italics

\newcommand{\inlinevis}[3]{\raisebox{#1}[0pt][0pt]{\includegraphics[height=#2]{#3}}}

\newcommand{\eg}{e.\,g.}
\newcommand{\ie}{i.\,e.}

\newcommand{\osfrepo}{{\href{https://osf.io/9dwgq/}{\texttt{osf\discretionary{}{.}{.}io\discretionary{/}{}{/}9dwgq}}}}
\newcommand{\companionweb}{\href{https://vdl.sci.utah.edu/tactile-charts/}{\texttt{vdl\discretionary{}{.}{.}sci\discretionary{}{.}{.}utah\discretionary{}{.}{.}edu\discretionary{/}{}{/}tactile\discretionary{}{-}{-}charts\discretionary{/}{}{/}}}}

\newcommand{\rev}[1]{\textcolor{Crimson}{#1}} %reivision for VIS25
\renewcommand{\rev}[1]{\textcolor{Black}{#1}} %reivision for VIS25

\newcommand{\parhead}[1]{\noindent\textbf{#1}.}
\DeclareRobustCommand{\gobblefour}[5]{}
\newcommand*{\SkipTocEntry}{\addtocontents{toc}{\gobblefour}}

\begin{document}

%%%%%%%%%%%%%%%%%%%%%%%%%%%%%%%%%%%%%%%%%%%%%%%%%%%%%%%%%%%%%%%%
%%%%%%%%%%%%%%%%%%%%%% START OF THE PAPER %%%%%%%%%%%%%%%%%%%%%%
%%%%%%%%%%%%%%%%%%%%%%%%%%%%%%%%%%%%%%%%%%%%%%%%%%%%%%%%%%%%%%%%

\firstsection{Introduction}
\label{sec:introduction}
{\renewcommand{\addtocontents}[2]{}\maketitle}
% \maketitle

%% \section{Introduction} %for journal use above \firstsection{..} instead

% \SkipTocEntry\section{Introduction}
% \label{sec:introduction}

Visualization is an important tool for communicating complex information in professional contexts, making visualization literacy essential across many careers. However, blind and low vision (BLV) individuals---as audiences of visualizations \cite{Lee:2020:Reaching, Siu:2021:Covid, Joyner:2022:Visualization}---often face barriers to access them~\cite{Chundury:2022:Towards, Marriott:2021:Inclusive, lyi:2023:comprehensive}.
Most commonly, BLV individuals have to access the information contained in visualizations through alternative text (alt text), which provides descriptive text for the content of charts \cite{Joyner:2022:Visualization, Lundgard:2022:Accessible}. 
% Although useful alt text is still not widespread, even in scientific journals~\cite{McNutt:2025:Accessible}, 
Alt text, in principle, works well for simple charts, yet it can be difficult to both read and write when applied to complex visualizations \cite{Joyner:2022:Visualization}. The challenge is amplified when BLV individuals lack a corresponding mental model of these chart types, thus hindering them in effectively interpreting its alt text \cite{Kim:2023:Exploring, Jung:2022:Communicating, Kim:2023:Explain}. 

Tactile charts---charts that can be experienced by touch---are another modality that makes visualization accessible for BLV individuals. They are beneficial in helping users understand spatial relationships and data patterns \cite{Watanabe:2012:Development, Marriott:2021:Inclusive, BANA:2022:Guidelines, Yang:2020:Tactile}. Prior work on tactile charts has focused on simple chart types and direct data representations rather than as educational tools (\eg, \cite{Holloway:2019:3D, Chundury:2022:Towards, Chundury:2022:Towards, Engel:2018:User, khalaila:2024:They}). However, as data representations, tactile charts are slower and more costly to produce, and cannot be updated \cite{chen:2025:Tactile, Phutane:2022:Tactile}, compared to alt texts or other digital methods such as interactive data explorers~\cite{Zong:2022:Rich}, or  sonifications~\cite{Hoque:2023:Accessible}. There are prototypes of refreshable tactile displays (\eg, \cite{Reinders:2025:When, Holloway:2024:Refreshable}), but they are expensive, low resolution, and not widely available.

Considering the strengths and limitations of both tactile charts and alt texts, \rev{this work explores the use of tactile charts as tools for learning chart types rather than solely for data representation.} 
We thus propose that tactile charts can function as ``template charts'' to support BLV individuals in developing mental models when learning about chart types. \rev{We adopt the concept of a mental model from cognitive psychology \cite{johnson:1983:mental} to refer to the internal representation of a chart type's structure, aligning with related work in our field (\eg, \cite{Holloway:2018:Accessible, Yang:2020:Tactile, Phutane:2022:Tactile, Jung:2022:Communicating}).} Once individuals form mental models of specific chart types, we hypothesize that they will be better equipped to interpret digital representations, such as alt texts, that utilize those chart types. 
\rev{We note that our approach is visual-first, \ie, only minimally adapts chart layouts to better suit tactile representation, as we aim to support BLV individuals when consuming extant visualizations and to facilitate collaboration between sighted and blind readers of visualizations.}
Tactile charts are widely used in BLV education \cite{Sheppard:2001:Tactile, Fan:2024:Tangible} \rev{and to explain interactive systems \cite{Zhao:2024:TADA}}, but their use for teaching complex chart types remains underexplored.
Given their infrequent need for updates, creating custom 3D-printed charts is a practical approach for educational purposes.
% since concerns like production cost and speed are less relevant.

% When we use tactile charts as an educational tool, it is worthwhile to design bespoke high resolution 3D-printed tactile charts, and disadvantages such as production cost and slowness are less relevant.

To evaluate this hypothesis, we designed tactile template charts for four chart types: UpSet plots \cite{Lex:2014:UpSet}, clustered heatmaps \cite{Eisen:1998:Cluster}, violin plots, and faceted line charts. We chose these charts because they are complex, yet commonly used, \eg, in scientific publications and professional contexts. 
\rev{Since our goal is to use tactile charts as a learning tool, we designed them with simple, familiar datasets to help people focus on the chart types rather than the content.}
% To help people focus on chart types rather than chart contents, we designed template tactile charts using simple, familiar datasets. 
We followed existing tactile chart design guidelines (\eg, \cite{BANA:2022:Guidelines, Schuffelen:2002:Editing, Prescher:2016:Richtlinien, TactileGraphicsAHowToGuide}) where available and sensible, yet we found that many guidelines do not capture essential aspects of the plots we intended to design. 

To address gaps in existing guidelines, we leveraged our expertise in visualization and collaborated closely with our blind coauthor, and also consulted another BLV psychology researcher. Following an iterative design process, we initially created two design variations for each chart type, developed accompanying exploration instructions, and then refined these designs based on feedback from our BLV collaborators to one tactile model per chart type. 

For logistical reasons (recruiting, distribution, length of interviews), we selected two of our four tactile charts---the clustered heatmap and the violin plot---to conduct an interview study with 12 BLV participants to evaluate their utility for learning transferable knowledge about the chart type. Participants first learned about the chart using either a tactile model paired with textual instructions, or just textual instructions. We then provided them with an alt-text of a different dataset visualized with the chart type, and asked questions about the new dataset. 
We compared BLV participants' learning experiences, outcomes, and preferences between these two training modalities and elicited experiences and feedback on the tactile charts.

In summary, this work contributes: 
(1) The design of tactile template charts for four complex chart types with exploration instructions;
(2) Results from an interview study with 12 BLV participants, showing that tactile charts with exploration instructions support comprehension and learning of complex visualizations and are a preferred learning method; 
(3) Insights into BLV individuals' learning strategies and experience with unfamiliar chart types, their perspectives on tactile chart design, and the potential of tactile charts in BLV education;
and (4) Refined design recommendations for creating tactile charts for complex visualizations for \rev{learning purposes}.

\vspace{3mm}
\SkipTocEntry\section{Related Work}
\label{sec:relatedwork}
In this section, we first discuss visualization accessibility, and then focus on chart type comprehension challenges and tactile charts.

\SkipTocEntry\subsection{Visualization Accessibility}
In our increasingly data-driven society, visualizations are prevalent, both in a professional and personal context. However, because visualizations require the visual system for interpretation \cite{Choi:2019:Visualizing}, they inherently bring challenges for BLV individuals. Despite these challenges, BLV individuals need to access to the rich information conveyed through visualizations and need to reason about data \cite{Marriott:2021:Inclusive}. Unfortunately, they frequently encounter inaccessible visualizations across various contexts \cite{Lundgard:2022:Accessible}, which can create substantial difficulties in their daily lives \cite{Siu:2021:Covid} and limit access to employment opportunities in the knowledge economy. Ensuring information accessibility is therefore a matter of social justice. In recognition of this, researchers in the visualization community have increasingly called for greater attention to accessibility \cite{Marriott:2021:Inclusive, Chundury:2022:Towards} and to provide evaluation guidelines \cite{Elavsky:2022:How, lyi:2023:comprehensive}.

To improve visualization accessibility, researchers have explored various nonvisual approaches to information representation \cite{Kim:2021:Accessible}, including speech \cite{Zong:2022:Rich, McNutt:2025:Accessible, Lundgard:2022:Accessible}, sonification (\eg, \cite{Hoque:2023:Accessible, Daunys:2008:Sonification, Franklin:2003:Pie}), and haptic or tactile representations (\eg, \cite{Yang:2020:Tactile, Engel:2019:User, Goncu:2011:GraVVITAS}). Researchers also recommend for multimodal approaches 
% (\eg, \cite{Seo:2024:MAIDR,Wall:2006:Feeling, Mackowski:2023:Multimodal, Thevin:2019:Creating})
(\eg, \cite{Seo:2024:MAIDR,Wall:2006:Feeling, Gotzelmann:2018:Visually, Mackowski:2023:Multimodal, Thevin:2019:Creating}) to reduce the limitations of individual modalities \rev{and enhance data access and comprehension.}
These approaches, however, often come with higher economic and learning costs \cite{Chundury:2022:Towards, khalaila:2024:They, Goncu:2010:Tactile}. 
Other work has focused on methods, such as extracting original data from chart images \cite{Choi:2019:Visualizing}, developing question-answering systems \cite{Kim:2023:Exploring}, interactive exploration systems \cite{Zong:2022:Rich, Elavsky:2024:Data} or using LLM for low-level visual analysis tasks \cite{Xu:2024:Exploring}. These methods are helpful; however, to fully benefit from them, users need to understand the used chart types.

Among these alternative modalities, alternative text (also alt text or text descriptions) is the most common and practical, particularly in digital environments such as web pages \cite{Joyner:2022:Visualization, Lazar:2007:What}. Alt text consists of textual descriptions of visual content that BLV individuals can access, e.g., through screen readers \cite{Joyner:2022:Visualization}. Researchers have studied the generation of alt text and evaluated its effectiveness (\eg, \cite{Lundgard:2022:Accessible, Jung:2022:Communicating, Ault:2002:Evaluation}). Recent efforts to develop text descriptions for complex charts \cite{Smits:2024:AltGosling, McNutt:2025:Accessible} are an attempt to remedy the dismal situation of accessibility of scientific articles and data resources~\cite{lyi:2023:comprehensive}. Given the widespread use and practicality of alt text, we adopt it as a representative accessibility method in our evaluation study to assess the learning outcomes of BLV participants.

% Although complex visualizations are common today, and new chart types are continues emerging, 
% research on advanced chart type also gain interests.

% using llms for low level visual analysis tasks \cite{10.1109/VIS55277.2024.00033}

\SkipTocEntry\subsection{Chart Type Understanding}
Understanding chart types is crucial for BLV individuals to access visualizations, as it might allow them to form a mental model for interpreting the data and the encoding mechanisms used in the visualization \cite{Kim:2023:Explain, Jung:2022:Communicating}. Visualizations are often more than a raw representation of the data. A histogram, for example, communicates the distribution of a dataset. It hence makes sense to describe visualizations instead of describing the underlying data directly.
Prior studies show that BLV individuals express a need for understanding chart types when using different assistive technologies. 
For example, Jung et al. \cite{Jung:2022:Communicating} investigate how to better write alt text and recommend that alt text should first describe the chart type for uncommon charts.
%, providing a brief explanation if the chart type is uncommon, because they suggest that the chart type is the starting point for BLV individuals to understand visualizations' alt texts. 
Kim et al. \cite{Kim:2023:Exploring} explored the use of question-answering systems to help BLV individuals interpret visualizations and found that participants frequently asked about visualization layouts and tried to understand chart types. 

BLV individuals are generally familiar with basic chart types (\eg, bar, pie, and line charts); however, they often struggle with more advanced chart types. Engel et al.  \cite{Engel:2018:User, Engel:2017:Improve} find BLV individuals are less familiar with stacked bar charts, scatter plots, and area charts. Wang et al.~\cite{Wang:2022:Seeing} report that BLV individuals are relatively unfamiliar with violin plots, and also unfamiliar with donut charts and area charts. 
These findings align with the fact that BLV individuals are typically exposed only to basic chart types in school \cite{Sheppard:2001:Tactile}. In addition, most existing visualization accessibility technologies and research focus primarily on simple chart types \cite{Kim:2021:Accessible, Wimer:2024:Beyond}, whereas studies on accessible representations of complex chart types remain under-represented \cite{Wimer:2024:Beyond}. 

Therefore, to address the challenges of making complex visualization accessible \cite{Marriott:2021:Inclusive}, it is essential to study how to teach BLV individuals about advanced chart types---an area that remains largely unexplored. Kim et al.~\cite{Kim:2023:Explain} investigate different strategies for explaining unfamiliar chart types to BLV individuals, including referencing familiar charts, using declarative versus procedural knowledge, and providing abstract versus concrete explanations. They also developed a prototype system for automatically generating explanations for 50 chart types. Similarly, Smits et al.~\cite{Smits:2024:Explaining} found using a gradual explanation method can help BLV individuals to comprehend unfamiliar genomics data visualizations (sequence logos and Circos plots). These work both focus only on textual explanation; our work shares a similar goal but extends the scope by incorporating tactile charts and covering different chart types.

\SkipTocEntry\subsection{Tactile Charts}
\label{sec:tactile-charts}
Tactile charts are a subset of tactile graphics and a form of data physicalization \cite{Huron:2022:Making, Jansen:2015:Opportunities}. In this work, we focus on their use for accessibility---enabling BLV individuals to access data representations through touch \cite{Engel:2018:User}. 
Prior research points out the benefits of tactile representations in supporting the comprehension of visualizations, particularly for obtaining an overview of charts \cite{Chundury:2022:Towards, Engel:2021:Heatmaps, Goncu:2010:Tactile,Fan:2024:Tangible,Fan:2022:Accessibility} and conducting data analysis tasks \cite{Watanabe:2019:Usefulness, Engel:2021:Heatmaps, khalaila:2024:They, Yang:2020:Tactile}. Moreover, studies identify tactile charts as a preferred method for BLV individuals to interpret and analyze data \cite{Watanabe:2019:Usefulness, Engel:2021:Heatmaps}. In addition, blind participants have expressed a desire for tactile representations for understanding specific chart types, such as UpSet plots, when hearing alt texts \cite{McNutt:2025:Accessible}.
% Despite these benefits, researchers have also pointed out that tactile materials are difficult to create, obtain, and update, therefore, we should only produce tactile charts when needed \cite{Phutane:2022:Tactile}.

Tactile charts can be static or refreshable. Static tactile charts are physical charts that are hard or impossible to change once produced. The most common methods for creating these charts include embossing and the use of swell paper, which is cost-effective and efficient \cite{Rowell:2003:World}. However, these techniques are limited in terms of embossed height and resolution. In contrast, 3D printing enables the production of more detailed and durable tactile charts \cite{Lundgard:2019:Sociotechnical}. Prior research has shown that BLV individuals tend to prefer 3D-printed tactile charts over embossed ones \cite{Lundgard:2019:Sociotechnical, Engel:2021:Heatmaps}. Refreshable tactile displays, on the other hand, are dynamic devices that show tactile graphics on an updatable surface (\eg, \cite{Holloway:2024:Refreshable, Reinders:2025:When}). Although they offer the advantage of data flexibility, they are generally expensive and provide lower resolution compared to static alternatives. Given our aim of using tactile charts as an educational tool, affordability and resolution are key considerations. Since our charts serve as templates and do not need updates with new datasets, we focus on static tactile charts. Moreover, considering the complexity of the chart types explored in our study and our emphasis on small-scale production for \rev{learning}, rather than rapid manufacturing for novel datasets, we use 3D printing to create charts.
% \al{I'd also mention data physicalization and cite the book; but say that you're not going into it if it's not specifically designed for accessibility }

The design of tactile charts influence their readability \cite{Engel:2019:User}. 
There exist established guidelines for tactile graphics (\eg, \cite{BANA:2022:Guidelines, Schuffelen:2002:Editing, Prescher:2016:Richtlinien, TactileGraphicsAHowToGuide}), and we can also find suggestions and recommendation articles online, such as from educational websites.\footnote{Including \href{https://accessiblegraphics.org/}{\texttt{accessiblegraphics\discretionary{}{.}{.}org}}, 
\href{https://www.aph.org/}{\texttt{aph\discretionary{}{.}{.}org}}, 
\href{https://Brailleaustralia.org/}{\texttt{Brailleaustralia\discretionary{}{.}{.}org}}, 
\href{https://www.pathstoliteracy.org/}{\texttt{pathstoliteracy\discretionary{}{.}{.}org}}, \href{https://www.perkins.org/}{\texttt{perkins\discretionary{}{.}{.}org}}, etc.
}
% (\eg, \cite{accessible_graphics, american_printing_house, australian_Braille_authority, paths_to_literacy, perkins_school_for_the_blind}).
%like Accessible Graphics \cite{accessible_graphics}, American Printing House \cite{american_printing_house}, Australian Braille Authority \cite{australian_Braille_authority}, Paths to Literacy \cite{paths_to_literacy}, Perkins School For Blind \cite{perkins_school_for_the_blind}, etc. 
However, these resources are scattered, and most focus on tactile graphics in general. They offer broad design principles (\eg, ensuring clarity) and practical recommendations (\eg, on Braille size, element dimensions, spacing for discriminability, and pattern filling). They also cover only basic chart types and provide limited guidance on chart-specific elements \cite{Engel:2019:User}. 

Beyond general guidelines, researchers have explored specific design considerations for tactile charts, including texture-based encoding \cite{Watanabe:2018:Textures, Prescher:2017:Consistency}, grid lines \cite{Barth:1984:Incised, Aldrich:1987:Tangible, Lederman:1982:Tangible}, Braille labels \cite{Puerta:2024:Effect}, and tactile line styles \cite{Chau:2021:Composite}. Some studies also propose additional design considerations based on user feedback and empirical evaluations \cite{Engel:2019:User, Engel:2017:Improve, Engel:2018:User, Engel:2021:Heatmaps}. In our work, we incorporate relevant recommendations from these sources (discussed in detail in \autoref{sec:transcribe-to-tactile-charts}) and also supplement existing tactile chart design guidelines with insights gained from our design process.

To facilitate tactile chart creation, researchers explored different automated and semiautomated tactile chart creation approaches driven by images or data \cite{chen:2025:Tactile}. Image-driven systems 
% (\eg, \cite{Goncu:2014:Generation, Gonzalez:2019:Tactiled, Mech:2014:Edutactile, Moured:2024:Chart4Blind})
\cite{Crombie:2004:Bigger, Mech:2014:Edutactile, Moured:2024:Chart4Blind} 
typically use image-processing techniques to convert charts from visual format into tactile formats. These systems have the risk of data loss for complex charts, since they rely on image input. In contrast, data-driven systems generate tactile charts directly from structured data inputs (\eg, \cite{Goncu:2009:Generation, Araki:2014:Development, Watanabe:2012:Development, Engel:2019:SVGPlott}). 
% Researchers have developed interfaces that allow designers to input raw data and adjust parameters to create tactile chart representations (\eg, \cite{Goncu:2009:Generation, Araki:2014:Development, Watanabe:2012:Development, Engel:2019:SVGPlott}). 
Chen et al. \cite{chen:2025:Tactile} introduced Tactile Vega-Lite and simplified the programmatic creation of accessible tactile charts. These data-driven tools, however, all focus only on basic chart types (\ie, bar charts, pie charts, line charts, and scatter plots). Since our study focuses on complex charts that they do not cover, we adopt a manual design process to ensure optimal tactile representation.

\SkipTocEntry\section{Template Chart Design}
\label{sec:template-chart-design}

% To investigate whether tactile charts can effectively support BLV individuals to learn complex visualizations, we needed to design tactile chart stimuli, which we describe in this section.
\rev{To effectively teach BLV users about different chart types, we need well-designed example tactile charts,} which we describe in this section.

\SkipTocEntry\subsection{Chart Types and Datasets}

We chose chart types that are common in scientific publications but that were not covered by previous tactile designs. We also chose charts that have different visual encodings to improve the generalizability of our approach. 
We ultimately selected four advanced chart types: UpSet plots, which are used to visualize set intersections; clustered heatmaps, which are used to visualize large numerical tables; violin plots, which are used to visualize (multiple) distributions; and faceted line charts, which are common \eg, in genome browsers.

To help people focus on understanding chart types, we used real-world datasets on familiar topics to create the charts.
We refined these datasets by selecting subsets and modifying data to balance the simplicity with sufficient data features for showing each chart type’s unique characteristics. 
For example, to create a bi-modal distribution in the violin plot, we selectively removed entries from the original dataset. 
\autoref{sec:template-chart-datasets} provides details on the dataset selection and modifications. %Using these refined datasets, we generated the template charts with Python.

% \begin{figure*}[t]
%     \centering
%         % \includegraphics[width=1\columnwidth]{figures/model-photos/violin_final_back.pdf}
%         \includegraphics[width=1\linewidth]{figures/final-designs}
%     \vspace{-4mm}
%     \caption{The final designs for the four chart types, showing English labels and the 3D printed versions with Braille beneath it.}
%     \vspace{-6mm}
%     \label{fig:final-designs}
% \end{figure*}  

\SkipTocEntry\subsection{Design Process and Design Variations}
\label{sec:transcribe-to-tactile-charts}
We began by generating charts with Python (see \autoref{sec:python-generated-charts-simple}), and ``transcribed'' these charts into 2D designs for tactile charts, carefully considering our design choices.  Following existing tactile chart design guidelines (discussed in \autoref{sec:tactile-charts}), we prioritized simplicity while ensuring their meaningfulness for BLV readers. In addition, to facilitate communication between BLV and sighted individuals, we kept our design as similar as possible to the commonly used sighted version. For aspects not covered by extant design recommendations, we drew on our design expertise and feedback from two BLV collaborators. We also developed exploration instructions alongside the charts.

% \subsubsection{Design Variations}
% , which we then refined to a final, third design after an extensive feedback process with two BLV collaborators. 

% \sarah{I think Gordon identifies as someone with low vision, not blind, so I might change to BLV instead of blind when you refer to the collaborators}

%For those points that are not clear in the guideline and we are concerned about the advanced chart type specific we provided two design variations of each chart type to explore different approaches to tactile representation. Below we provide our chart specific considerations and the design variations. 

We followed an iterative design process. To explore different options for designs, we created two initial variations of each chart (see \autoref{fig:teaser}). We provide details on our design considerations regarding chart size, orientation cues, title and legend, spacing and styling of chart elements, and the use of raised labels in \autoref{sec:detailed-design-considerations}. 

To compare design variations, gather feedback, and refine our prototypes and instructions, we consulted our blind coauthor and a second BLV researcher. We shipped our initial designs of four pairs of chart types to them and consulted them via videoconferencing. We asked them to explore the charts one by one and also review the accompanying instructions, while ``thinking aloud'' during the process. We then asked for their feedback on the design and instructions, and refined each chart into a third, final design. \autoref{fig:teaser} shows the two initial versions and the final designs (more figures see \autoref{sec:template-chart-design-variations}--\ref{sec:template-chart-model-photos-final-design}). 
Next we present our design variations and the rationale going into each design. 

\parhead{UpSet Plots}
We experimented with two main aspects for UpSet plots: bar styles and direct labeling vs.\ legends. 
For bar styles, we compared solid bars (Design 1)  to bars with ``notches'' (Design 2), as recommended by Schuffelen~\cite{Schuffelen:2002:Editing}. The idea of notches is to make the bar size ``countable''.  
We also tried different shapes for each set (circles, triangles, diamonds and stars, as shown in Design 2) and a legend, as opposed to direct labels and uniform shapes (Design 1). Our collaborators appreciated the countable bars, but preferred the direct labeling to support easier reading, resulting in the final design.

\parhead{Clustered Heatmaps}
For the heat map, we experimented with direct labeling vs.\ a legend for the bar heights (darkness), dendrogram placement, and vertical labels vs.\ abbreviations with a legend. Design 1 used smaller, unlabeled cells; readers inferred values based on cell height and by referring to a legend. The second design used larger squares with embedded Braille numbers, avoiding the need for a legend. Our collaborators again preferred direct labeling. For the dendrogram placement, Design 1 followed conventional layouts seen in sighted versions, positioning the dendrogram on the left and movie actor names on the right. Design 2 reversed this arrangement, with the names on the left and the dendrogram on the right. Feedback from our collaborators indicated that reading order is critical, especially as tactile perception lacks the ability to see an overview. Hence, our final design places the labels on the left. Finally, we again experimented with direct labeling vs.\ legend, but this time for vertical labels below the heat map. 
It is generally recommended that Braille text should be written horizontally, and our collaborators agreed. Also, foregoing direct labels allowed us to print the heat map significantly larger, resulting in the only instance where our final design uses a legend instead of direct labeling. 

\parhead{Violin Plots}
For violin plots, we explored filled shapes vs.\ outlines, the use of grids, stippled vs.\ solid median lines, whether to use raised labels, and again whether to use direct labeling vs.\ legends. Design 2 also used redundant labels (above and below the violin). 
Our collaborators preferred filled violins and the horizontal grid lines. For vertical distinction of the categories, they preferred the raised labels over the vertical grid lines. They found redundant labels unnecessary. %These decisions are reflected in our final design.

%To differentiate between species in violin plots, we provide the first design varied contour thicknesses for each violin and adjusted the thickness of the median bars, with a legend beneath the title explaining these distinctions. The second design used uniformly filled violins with solid median bars of the same thickness. Since this design did not differentiate between species, no legend was provided.

%Regarding the grid, the first design incorporated a tactile grid to help users align violin shapes with Y-axis values. In contrast, the second design omitted the grid.

%For species name labeling, the first design placed names along the X-axis, separated by extended vertical grid lines. The second design used background blocks to differentiate species names and additionally labeled them near the top of each violin.

%For the violin plot, one participant prefered Design 2 without redundant labels, and another participant also disliked the redundant labels but expressed no strong preference on the remaining design features. One participant recommended that we label tick marks evenly to reduce confusion.

\begin{figure*}[t]
    \centering
        \includegraphics[width=1\linewidth]{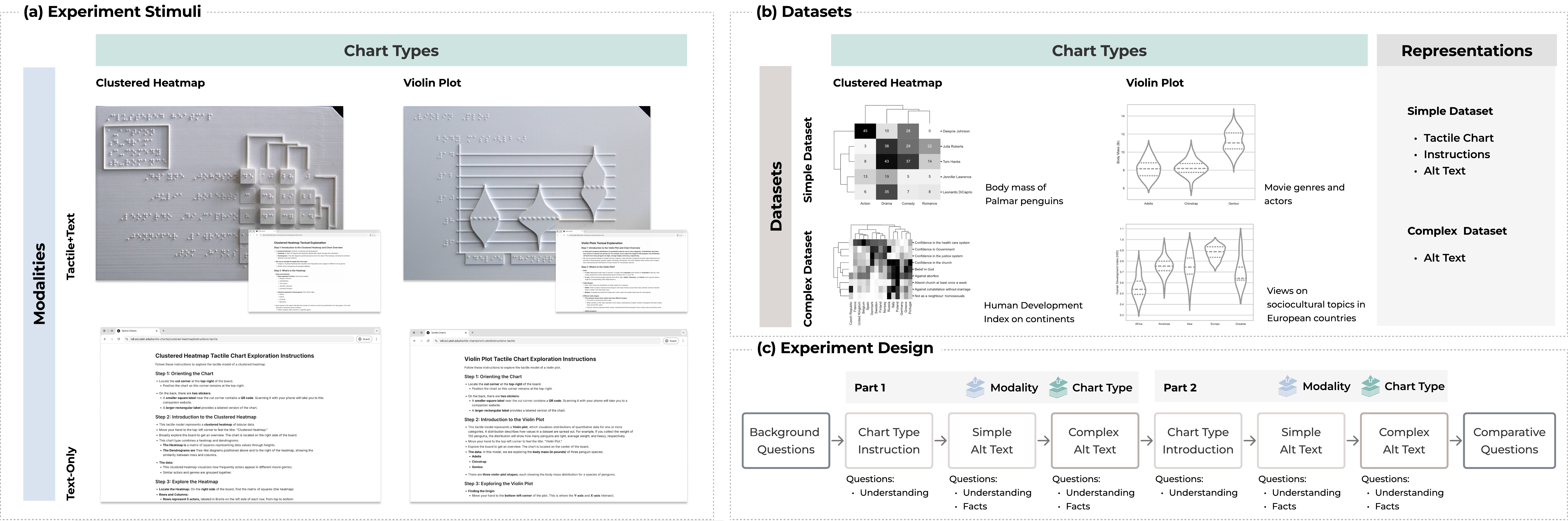}
        \vspace{-5mm}
    \caption{Overview of the experiment. (a) Participants were assigned one combination of chart type (violin plot, clustered heatmap) and modality (tactile chart with instruction, textual introduction), for one part of the experiment, and the opposite combination for the other part. (b) We used four different datasets in the experiments. The simple datasets are shown in the tactile charts, described in the instructions and in the alt text. For the complex datasets, we provided only alt texts. (c) After background questions, the experiment was conducted in two parts, with different modality / chart type combinations. Each experiment part first introduced the chart type, then provided alt-text for the simple dataset and the complex dataset. These stages was accompanied by factual and understanding questions. The interview ended with comparative questions.
    }
    \label{fig:experiment-overview}
    \vspace{-3mm}
\end{figure*}

\parhead{Faceted Line Charts}
For faceted line charts, we explored the use of filled area charts vs.\ line charts, and the amount of ticks to use. The line charts require a lot of labels, communicating not only what a particular line encodes, but also the units and the scale. We did not consider a legend, but instead attempted to place labels within the chart (Design 1). Our collaborators found this hard to read, as the lines interfered with the labels. Hence, our final design uses labels on the left, adopts the raised labels from the violin plot, and moves the axis and scales to the right. We again opted for a solid representation over the outline, and used horizontal but not vertical ticks. One collaborator also suggested reducing the density of tick marks on the Y-axis. Another suggested adding the name of the month to the X-axis to improve interpretability, and adding a vertical grid line for reference. 

% For encoding trends, we tested two designs. The first design used solid areas to represent trends, while the second design depicted trends solely with curved lines. We considered these because in fact trends are encoded with the lines (even we use solid areas for them, they in fact the top edge of the area), but if we uses lines there might be too many lines on the charts (trend lines and axes)

% For the X-axis, the first design included 31 tactile tick marks, one for each day of the month, with numeric labels at 1, 10, 20, and 30. The second design marked only key dates (Aug. 1, Aug. 15, and Aug. 31) without tick marks.

% Label placement for variable names also varied. In the first design, we aligned the sub-chart's name and units to the left side of each section consistenly. In this case, we place the y-axis on the right. In the second design, they were positioned in the bottom-left corner, sometimes overlapping with the trend lines. The units are near the y-axis.  In this case, we place the y-axis on the left.

% Regarding the faceted line plots, participants generally preferred Design 1. In addition, to be consistent with the design on violin plot, we added background blocks behind the Y-axis labels to improve their distinguishability.
\parhead{\rev{Adapting Visual Encodings for Tactile Perception}}
\rev{Our design is based on existing visualizations; however, we adapted them to support tactile reading. When a visual channel is spatial and thus compatible with touch (\eg, position), we preserved it. For purely visual channels, we identified alternative spatial attributes to replace them. For example, in heatmaps, we translated color intensity into height.}

\parhead{General Design Feedback} 
% We designed different titles for each chart, 
% In Design 1 we also attempted raised labels (A, B, C) in \autoref{fig:teaser}, which were referenced in the textual exploration instruction, but abandoned those based on feedback in all charts. One collaborator was worried that these are confusing, and only make sense when also using the exploration instructions. Similarly, 
We tested raised labels (A, B, C) on one design of UpSet plot and clustered heatmap, and referenced them in the instructions. However, we removed them because a collaborator found them potentially confusing without the instructions.
We experimented with long and short titles for each chart, and adopted a collaborator’s suggestion to directly use chart type names as titles.

Regarding the Braille, participants noted that the placement of Braille labels was sometimes too close to raised chart elements, and thus hard to read. Although we initially followed spacing guidelines from existing tactile design literature, those guidelines might be developed primarily for embossed swell paper, which has limited height variation. In contrast, our 3D-printed charts include higher elements. As a result, we increased the distance between Braille labels and raised elements to improve tactile readability. Participants also reported that the Braille dots felt too sharp and might be too high. We experimented with dome-shaped Braille to reduce sharpness, but they did not print well, given the 3D printer’s resolution. Ultimately, we retained the cylindrical dot shape but reduced the Braille height to 0.6mm (the minimum recommended value) to improve comfort.

\SkipTocEntry\subsection{Exploration Instructions}
Our instructions use four steps: (1) orienting the chart, (2) an introduction to the chart type and the data shown in the chart, followed by (3) detailed exploration instructions for the chart. The exploration instructions consistently use spatial guidance (\eg, on the top left corner) and combine it with specific information about the data. The instructions conclude with (4) a brief recap, summarizing key points that are essential to remember. We elicited feedback from our blind collaborators on our initial instructions. Both experts appreciated the overall structure and clarity of our instructions, as well as the content used to explain each chart. One collaborator suggested that we should improve our spatial references. Based on this feedback, we revised the instructions for the updated chart designs. We provide the initial and final versions of instructions in \autoref{sec:exploration-instruction-initial-version} and \autoref{sec:exploration-instruction-final-version}. We also host the final version on our accessible website: \companionweb.

\SkipTocEntry\subsection{3D Model Creation} 
3D printing was a suitable modality that could produce charts with the level of detail we wanted to include. Based on the 2D designs, we created digital 3D models (available in \osfrepo\ and on our companion website) and printed them. \rev{We provide details on the model creation procedure and the associated costs in \autoref{sec:3D-model-ceation-procedure-and-cost}.}

\SkipTocEntry\subsection{Chart Backside} 
We attached two stickers to the back: a small one that links to the exploration instructions, and a large one with a labeled version of the chart. The labeled version is intended to facilitate collaboration with sighted people, \eg, when sighted educators teach blind students, but also as a reference for those with residual vision.
%We attached two adhesive label stickers on the back
% a round label showing a QR code that links to the exploration instructions, and a large rectangular sticker with a labeled version of the chart. The labeled version is intended to facilitate collaboration with sighted people, \eg, when sighted educators teach blind students, but also as a reference for those with residual vision. 

% \subsubsection{modalitiy selection}
% We will use 3D printing. While swell paper is cheap and commonly used, it is in fact 2.5D. 3D printing achieves true 3D representations with higher resolution, which is preferred by BLV users. A previous study (Lundgard et al.m 2019) found students from the Perkins School prefer 3D-printed graphics due to their higher resolution. Similarly, Engel et al. compared embossed and 3D-printed heatmaps and found that BLV participants clearly favored the 3D-printed format. Since the charts in our study are complex, resolution is important for us.
%  Laser cutting: It does allow for depth that is interesting, reminds me that some guidelines suggest using dented grid lines. However, for creating embossed elements, 3D printing might be easier?

% \subsection{Model Backside Design}

\SkipTocEntry\section{Evaluation Study}
 \label{sec:evaluation-study}

% Using our finalized tactile charts, we evaluated their utility with BLV individuals in an interview study. 
% Recall that the goal of template charts is not only to introduce BLV individuals to these types of charts, but also to ensure that they lead to improved comprehension for BLV individuals interacting with digital, easy to create (compared to 3D printed charts) representations for arbitrary datasets such as text descriptions.

Using our finalized tactile charts, we conducted an interview study with BLV individuals to evaluate their effectiveness. We preregistered our study on OSF (\href{https://osf.io/uhq68}{\texttt{osf\discretionary{}{.}{.}io\discretionary{/}{}{/}uhq68}}) and received IRB approval from the University of Utah (\textnumero\,IRB\_00180924).

Our goal of using tactile charts is not only to introduce BLV individuals to these chart types, but also to improve their comprehension when engaging with digital, easily generated representations (\eg, alt text) of arbitrary datasets presented using these chart types. Based on our goal and the benefits of tactile charts discussed in the \autoref{sec:relatedwork}, we developed the following research questions:\footnote{Our preregistration also contains hypothesis statements associated with these research questions. However, we later realized that our predominantly qualitative analysis does not fit well with falsifiable hypothesis, and hence omit explicit hypothesis statements.} 

\noindent\textbf{RQ1:} Does a tactile model help BLV individuals develop a better understanding of new advanced chart types compared to a textual explanation alone?

\noindent\textbf{RQ2:} Does a tactile model enhance BLV individuals’ comprehension of alt-text for an advanced chart? Specifically, we expect tactile charts support high-level comprehension but not memorizing individual facts.

\noindent\textbf{RQ3:} Do BLV individuals prefer learning new advanced chart types with a tactile model rather than textual explanations alone?

\noindent\textbf{RQ4:} Do BLV individuals believe that tactile models can teach them transferable knowledge about a chart type.

\begin{table*}[t]
\centering
\small
%\vspace{2mm}
\caption{Participant demographics. IP = degree in progress; $\bar{X}$ is mean; $Y$ is mode.}
\vspace{-3mm}
\begin{tabular}{ccccccccc}
\toprule
\textbf{pID} & \textbf{Age} & \textbf{Gender} & \textbf{First Interview} & \textbf{Data Intensive} & \textbf{Braille} & \textbf{Vision Loss} & \textbf{Onset of} & \textbf{Highest Level}\\ 
 &  & & \textbf{Condition (modality)} & \textbf{Career} & \textbf{Years Exp.} & \textbf{Level} & \textbf{Vision Loss} & \textbf{of Education}  \\ 
 \toprule
P1 & 37 & Female & Heatmap (tactile) & Yes & 25 & Severe low vision & Lost vision gradually & Master's \\ 
P2 & 32 & Male & Heatmap (text) & Yes & 29 & No residual vision & Blind since birth & Master's  \\ 
P3 & 44 & Male & Violin (text) & Yes & 7 & No residual vision & Lost vision gradually & Master's  \\ 
P4 & 24 & Female & Violin (tactile) & No & 18 & Light perception & Lost vision suddenly & Bachelor's (IP) \\ 
P5 & 29 & Male & Violin (tactile) & No & 15 & No residual vision & Blind since birth & Bachelor's  \\ 
P6 & 57 & Female & Heatmap (tactile) & No & 52 & Light perception  & Blind since birth & Associate's  \\ 
P7 & 40 & Female & Heatmap (tactile) & Yes & 37 & Severe low vision & Blind since birth & Master's  \\ 
P8 & 23 & Male & Violin (text) & No & 11 & Severe low vision & Lost vision gradually & Associate's (IP)  \\ 
P9 & 42 & Female & Heatmap (text) & No & 38 & No residual vision & Blind since birth & Master's  \\ 
P10 & 26 & Male & Heatmap (text) & Yes & 17 & Light perception & Lost vision suddenly & High School  \\ 
P11 & 27 & Female & Violin (tactile) & No & 26 & No residual vision & Blind since birth & High School   \\ 
P12 & 28 & Female & Violin (text) & No & 15 & No residual vision & Lost vision suddenly & Master's  \\ 

\midrule
& $\bar{X}=34 \pm 10$ 
& 7/12 F  
&
& 5/12 Yes  
%& $\bar{X}=7.5 \pm 3$ 
& $\bar{X}=24 \pm 13$ 
%&$Y$=
& $Y$ = No residual vision (6)
& $Y$ = Blind since birth (6)
& $Y$ = Master's (6) \\
\bottomrule
\end{tabular}
\label{tab:demographics}
\vspace{-5mm}
\end{table*}

\SkipTocEntry\subsection{Study Design}
We followed methodology we previously successfully employed with the target population~\cite{McNutt:2025:Accessible} to conduct semistructured interviews over Zoom. We used a paired interview technique with one researcher asking questions and another researcher asking follow-ups \cite{akbaba2023two}.  

Considering the expected length of the interview, we deemed it infeasible to test all four chart types with individual participants. Challenges in recruiting from our desired population made a four-condition between-subject design that also counteracts order effects equally infeasible. To winnow down our charts, we ranked them by difficulty (from most to least difficult): UpSet plot, clustered heatmap, violin plot, and faceted line chart. To balance the need for evaluating complex charts while avoiding overwhelming participants, we then selected the clustered heatmap and violin plot for testing. 

We used a mixed-design approach with two factors, as illustrated in \autoref{fig:experiment-overview}: (1) chart type: clustered heatmap or violin plot and (2) teaching modality: tactile model with exploration instruction (\texttt{Tactile+Text}) or textual instruction alone (\texttt{Text-Only}). 
% All participants completed the evaluation procedure described in the following section with both chart types. 
Each participant experienced one chart type with \texttt{Tactile+Text} and the other chart type with \texttt{Text-Only}. The specific pairing and order of chart type and teaching method were counterbalanced across participants to control for order effects. 
Participants completed the interview on their personal computer or phone. 
The length of interviews averaged 110 min $\pm$ 25 min (SD). 
% The length of interviews averaged 1h 50m $\pm$ 25 minutes (SD). 
Each participant received a \$100 Amazon gift card as compensation.

\SkipTocEntry\subsection{Study Material Preparation}
For each chart type, participants experienced two datasets in the study, which we refer to as the simple dataset and the complex dataset (see \autoref{fig:experiment-overview}).
We used the simple datasets to create the tactile charts and developed the corresponding exploration instructions for the \texttt{Tactile+Text} condition (see \autoref{sec:exploration-instruction-final-version}), as well as the textual instructions for the \texttt{Text-Only} condition (see \autoref{sec:textual-explanation}).
% The simple datasets are the ones used for the \texttt{Tactile+Text} condition (see \autoref{sec:exploration-instruction-final-version}). For the \texttt{Text-Only} condition, participants read textual instructions (see \autoref{sec:textual-explanation}). 
To ensure comparability between conditions, we prepared textual instructions by removing tactile chart-specific content from the tactile chart exploration instructions and adapting the text accordingly. We also wrote the alt text for the simple dataset, which we refer to as \textit{simple alt texts} (see \autoref{sec:simple-alt-text}). The complex datasets we selected are larger and address more specialized topics---Human Development Index (HDI) across continents for the violin plot, or sociocultural values across European countries for the clustered heatmap 
(see \autoref{sec:complex-datasets}). %Since the participants come from different domains, we avoided using niche scientific topics. 
We created charts with Python (see \autoref{fig:experiment-overview} and \autoref{sec:complex-charts}) and wrote corresponding alt texts, which we refer to as \textit{complex alt texts} (see \autoref{sec:complex-alt-texts}). 
We followed established guidance~\cite{Lundgard:2022:Accessible, McNutt:2025:Accessible} and refined them with our blind coauthor.
We also prepared interview questions (see 
\autoref{sec:evaluation-study-scripts}). \rev{The questions assess general understanding of the charts and datasets, factual questions targeting specific information in the alt texts, and comparative questions evaluating the two modalities.}

% We also prepared interview questions (see \autoref{sec:evaluation-study-scripts}).

% Each participant will complete two trials, experiencing both chart types and both teaching methods, resulting in two conditions: (1) violin plot with tactile modeling + clustered heatmap with textual explanation, (2) violin plot with textual explanation + clustered heatmap with tactile modeling. To control for order effects, we counterbalanced the sequence of chart types and teaching method combinations across participants.

\SkipTocEntry\subsection{Participants}
We recruited 12 participants through social media and by re-engaging participants who had previously expressed interest in participating in further studies~\cite{McNutt:2025:Accessible} . 
%We referred to other similar studies (\eg, 11 participants in \cite{Engel:2021:Heatmaps}, 8 participants in \cite{Yang:2020:Tactile}) related to tactile visualizations in our field to decide this sample size. 
We required participants to be (1) 18 years or older, (2) proficient in English and Braille, (3) legally blind (\eg, a visual acuity of 20/200 or worse), (4) have a valid U.S.\ mailing address, and (5) be employed or attending school at least part-time. We required student or employment status, because the focus of our study is a professional data analysis context. 
All participants provided informed consent. \autoref{tab:demographics} shows participants' demographic information.
% to the procedures approved by the University of Utah's IRB and to have the audio of the interview recorded. 

\SkipTocEntry\subsection{Procedure}
\parhead{Pre-Interview} After expressing interest in participating, individuals completed a screening phone call to assess eligibility. Eligible participants were then provided with information about the study procedures and consent process, and scheduled for the remote interview. Participants subsequently read and signed the consent form and provided their shipping address.
% We then shipped one tactile chart in a well-protected package \cite{deGreef:2021:Interdependent}, along with its corresponding exploration instructions , to each participant. 
We shipped each participant \rev{a well-protected tactile chart \cite{deGreef:2021:Interdependent}} with exploration instructions, and a additional link to text instructions for a second chart type (see \autoref{sec:evaluation-study-package-examples}).

We then shipped one tactile chart in well protect package  \cite{deGreef:2021:Interdependent} along with its corresponding exploration instructions to each participant. We also included links to text instructions for the second chart type (example packages see \autoref{sec:evaluation-study-package-examples}). 
All instructions were accessible via QR code and were also linked in an email. 
Participants were permitted, but not required, to review the instructions for both conditions and explore the tactile chart prior to the interview.

%Before the experiment begins, participants sign an informed consent form. Based on their assigned condition, we ship the corresponding tactile model to them along with QR codes pointing to both tactile model instructions and textual explanation. They are permitted to explore the tactile model (and the associated instructions) and the textual explanation before the study. 

\parhead{Background Questions (13 $\pm$ 3 min)} Each interview began with background questions about demographics, vision loss, screen reader and Braille experience, and familiarity with tactile graphics, violin plots, and clustered heatmaps. 
% Background questions lasted 13 $\pm$ 3 min. 
The researchers then directed the participants to a series of pages via Zoom chat or email. Throughout the interview, participants were permitted to read the current page as many times as needed to feel comfortable with the information. However, they were not allowed to return to a previous page. %They go through the following three sessions two rounds with two chart types.

\parhead{Chart Type Introductions (13 $\pm$ 5 min)} Participants first read a chart type instruction page. Depending on their modality condition, they explore the tactile chart or not. 
We then asked three \rev{understanding questions} about the chart type (\eg, ``What types of data are best suited for visualization using a violin plot?'') 
% The training phase lasted a 13 $\pm$ 5 min.

% \maggie{Not sure the best to list the four links here. Anyone, feel free to change it!}(\asLink{https://vdl.sci.utah.edu/tactile-charts/clustered-heatmap/instructions-tactile}{/vdl.sci.utah.edu/tactile-charts/clustered-heatmap/instructions-tactile}, \asLink{https://vdl.sci.utah.edu/tactile-charts/clustered-heatmap/instructions-text}{vdl.sci.utah.edu/tactile-charts/clustered-heatmap/instructions-text}, \asLink{https://vdl.sci.utah.edu/tactile-charts/violin-plot/instructions-tactile}{vdl.sci.utah.edu/tactile-charts/violin-plot/instructions-tactile}, \asLink{https://vdl.sci.utah.edu/tactile-charts/violin-plot/instructions-text}{vdl.sci.utah.edu/tactile-charts/violin-plot/instructions-text}). 
% This page was designed to familiarize participants with the assigned chart type, either by reading the text explanation alone or by reading instructions while exploring the tactile model. 

%Screen reader settings, such as reading speed, could be adjusted at any time during the interview. 

\parhead{Simple Alt Text (9 $\pm$ 3 min)} Next, participants read the simple alt text, and answered three questions about it, \rev{\eg, ``What is the dataset about?'' (understanding) or ``Which species had the least variation in body mass, and why?'' (factual)}. 

\parhead{Complex Text Description (18 $\pm$ 6 min)} Next, participants read the complex alt text and again answered seven questions about it, \rev{\eg, ``What is the dataset about?'' (understanding) or ``How many clusters were the countries divided into, and which is an outlier?'' (factual)}.
Then we asked participants to rate their understanding of the dataset, reflect on any difficulties or surprising aspects, and provide additional comments about their experience with the chart.  

\parhead{Comparative Questions (18 $\pm$ 7 min)} After completing the above procedure for both chart types, we asked participants to compare the two training formats (\texttt{Tactile+Text} or \texttt{Text-Only}) and indicate their preference, and elaborate on their reasoning. They then rated various aspects of the tactile model, including its helpfulness, ease of use, and potential for building transferable knowledge. Finally, participants provided their opinions on the usefulness of each training format for educational contexts and on the tactile model in general. 

\SkipTocEntry\subsection{Quantitative Results}
In this section, we report our quantitative results: correctness of \rev{understanding and factual questions}, and the subjective ratings. \rev{These results suggest some trends, but should be interpreted with caution, as our small sample size precluded meaningful statistical testing.}

\parhead{Understanding and Factual Questions}
Two researchers independently coded the correctness of the answers to these questions and reached agreement through discussion. %We relate questions asked after the chart type instructions to RQ1, and those asked after reading alt texts to RQ2. We further split these questions to those  about general understanding of the datasets, and questions targeting specific factual recall.
As shown in \autoref{tab:factual-questions-correctness}, participants did not perform better on questions related to chart types after \texttt{Tactile+Text} training.
%It is important to note that the differences between conditions are small---the 2.77\% difference at N=6 is basic a noise. 
Participants performed slightly better in the \texttt{Tactile+Text} condition related to general understanding of new datasets (91.67\% correct vs 83.33\% for the \texttt{Text-Only} condition). As expected, there was no improvement in factual recall.

\begin{table}[t!]
\centering
\small
%\vspace{-5mm}

% \renewcommand{\arraystretch}{1.2} % Slightly increase row spacing
\caption{Rate of correct, partially correct, and wrong answers for understanding and factual questions (N = 6).}
\vspace{-3mm}
\begin{tabularx}{\linewidth}{>{\raggedright\arraybackslash}X l c c c}
\toprule
  & \textbf{Training} & \textbf{Correct} & \textbf{Partially Correct} & \textbf{Wrong} \\
\midrule
\multirow{2}{\linewidth}{Understanding questions after instructions} 
& \textbf{Tactile} & 30.56\% & 50.00\% & 19.44\% \\
& \textbf{Text}    & 33.33\% & 50.00\% & 16.67\% \\
\midrule
\multirow{2}{\linewidth}{Understanding questions after alt text} 
& \textbf{Tactile} & 91.67\% & 8.33\% & 0.00\% \\
& \textbf{Text }   & 83.33\% & 12.50\% & 4.17\% \\
\midrule
\multirow{2}{\linewidth}{Factual questions after alt text} 
& \textbf{Tactile} & 44.05\% & 28.57\% & 27.38\% \\
& \textbf{Text}    & 50.00\% & 23.81\% & 26.19\% \\
\bottomrule
\end{tabularx}
\label{tab:factual-questions-correctness}
\vspace{-5mm}
\end{table}

\begin{figure}[t]
    \centering
        \includegraphics[width=1\linewidth]{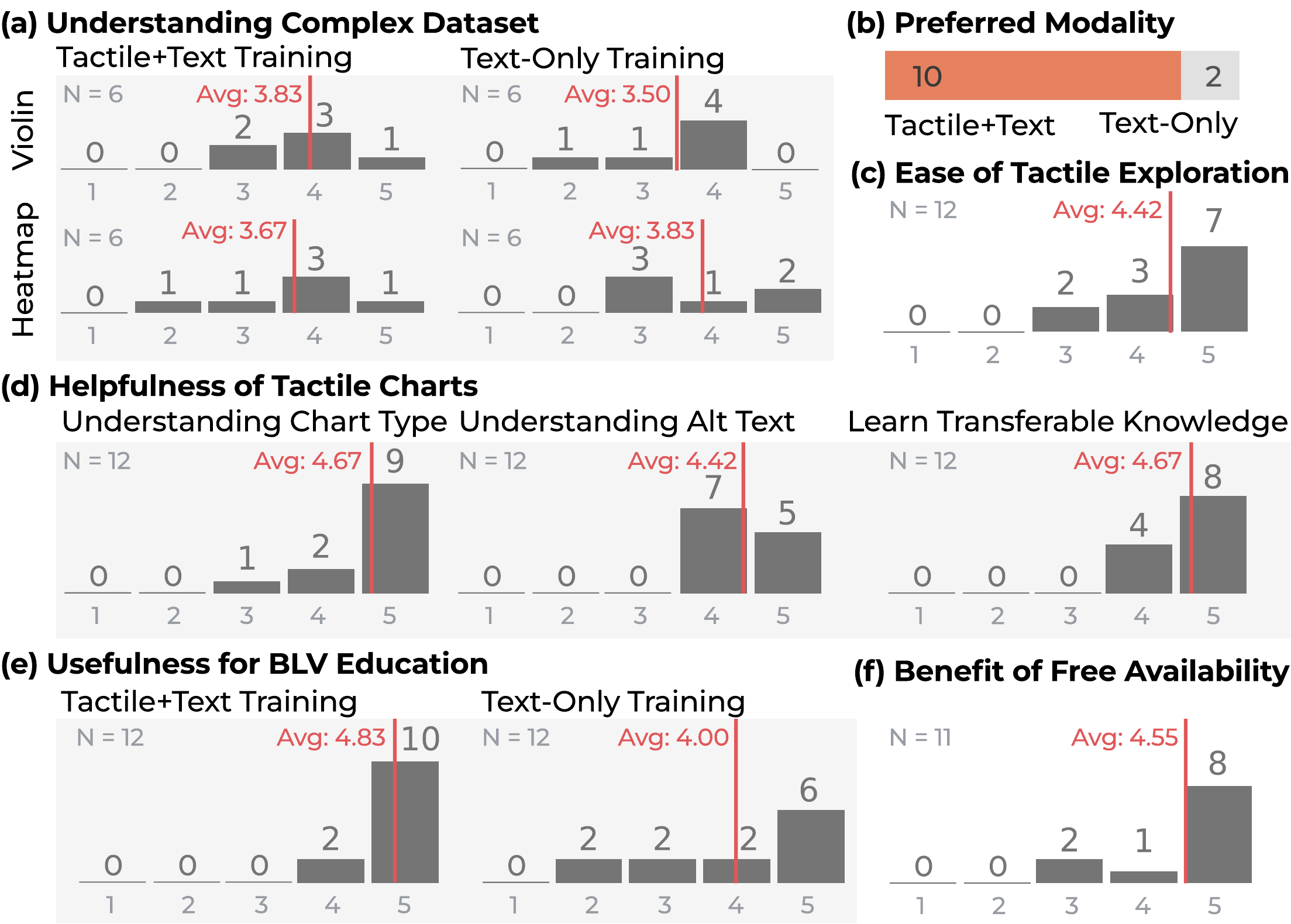}
    \vspace{-4mm}
    \caption{(a), (b), (d), (e), (f): Histograms of participant ratings on various questions on a 5-point Likert scale (1 = strongly disagree; 5 = strongly agree). (c): Participant preferences between the two learning modalities.}
    \vspace{-6mm}
    \label{fig:ratings}
\end{figure}

\parhead{Subjective Ratings}
% \al{TODO: reference figure with (a) (b) (c) ...}
We summarize the results of the subjective ratings in \autoref{fig:ratings}. Participants' perceived understanding of complex datasets across the two training modalities, and the datasets are overall similar, ranging from an average of 3.5 to 3.83 on a 5-point Likert scale (\autoref{fig:ratings}(a)).
% \marginpar{\tiny \sarah{I would emphasize that the subjective ratings test/support hypotheses 3 and 4. This gets buried in this paragraph}} 
% When asked which learning method they prefer (\texttt{Tactile+Text} or \texttt{Text-Only}), 
Regarding the two learning methods, 10 participants preferred \texttt{Tactile+Text}, and 2 preferred \texttt{Text-Only} (\autoref{fig:ratings}(b)). 
Participants also rated the tactile charts as easy to explore (\autoref{fig:ratings}(c), $M = 4.42$) and expressed positive perceptions regarding their helpfulness in chart learning (\autoref{fig:ratings}(d)). Specifically, participants see tactile charts as helpful in supporting understanding chart types ($M = 4.67$), interpreting alt texts ($M = 4.42$), and learning transferable knowledge about chart types ($M = 4.67$). As shown in \autoref{fig:ratings}(e), participants rated the perceived utility as high for \texttt{Tactile+Text} ($M = 4.83$), and \rev{somewhat lower} for \texttt{Text-Only} ($M = 4.00$). They also find value in tactile charts being freely available (\autoref{fig:ratings}(f), $M = 4.55$).\footnote{P5 did not rate for (f), but he expressed a negative opinion on this question.}

\SkipTocEntry\section{Themes}
\label{sec:themes}
\rev{We analyzed our corrected transcriptions using thematic analysis \cite{Lazar:2014:Research, Braun:2006:Using, Braun:2023:Toward}. The first author began by familiarizing herself with the data, generating initial codes, and grouping them into themes based on our research questions. All authors then reviewed and refined the codes and theme names through an iterative discussion process. During this process, the first author continued to actively construct additional codes through deeper engagement with the data. We finally extracted five themes, including how tactile charts support the construction of mental models, the transferability of chart-related knowledge across datasets, the role of tactile models in education, the use of other modalities, and design recommendations for educational tactile charts (see \autoref{sec:codebook} for the codebook). The first author drafted the initial report of themes, and all authors collaboratively refined it and contributed to the version reported in this section.} 

% Our themes include how tactile charts support building mental models, what knowledge about charts gained with tactile models is transferrable between datasets, the role of tactile models in education, the relationship of tactile chart to other modalities such as raw data or the use of AI tools, and the design recommendations for educational tactile charts.

% For our qualitative analysis, \rev{one researcher} open-coded the corrected transcriptions and conducted a thematic analysis \cite{Lazar:2014:Research}, and developed an initial codebook based on our research questions.
% and then iteratively refined the codebook by adding emerging codes identified during the coding process. 
% After the first round of coding, 
% we revisited the transcriptions to add the emerging codes, and iterated for several rounds. 
% \rev{we revisited the transcripts to actively construct additional codes through further engagement with the data.}
% In this section, we summarize the interviews based on the themes that emerged in our thematic analysis, covering how tactile charts support building mental models, how knowledge about charts gained with tactile models is transferrable between data types, the use of tactile models in education, the relationship of tactile chart to other modalities such as raw data or the use of AI tools, and the design recommendations for tactile charts.

\SkipTocEntry\subsection{Building Mental Models of Chart Types}

In this section, we discuss how tactile charts support facilitate learning and comprehension of new chart types for BLV individuals.
% \marginpar{\tiny \sarah{refer to H1} \hty{TH: We only use factual questions to support H1, as we consider H1 is subjective}}
Tactile charts offer direct access to spatial information and enable hands-on data exploration through touch. These benefits enhance the learning experiences of BLV individuals compared to other visual substitution modalities, such as auditory descriptions.

\paragraph{Supporting Mental Model Creation} 
Tactile charts serve as direct substitutes for visual charts for BLV individuals. 
As P5 described, tactile representations are \hquote{our equivalent} compared to how sighted people see pictures, and P8 similarly noted exploring the tactile charts as \hquote{like seeing it}. 
P7 emphasized that tactile charts provide \hquote{an overall understanding of what [the chart] would look like.} 
This information can help with the mental model-building process. As P9 described: \hquote{The tactile model helps put it into a picture that you can use, you can create a mental model, to picture things in your head.} 

Even without tactile charts, participants attempted to visualize charts mentally, indicating that creating mental images aligns with their existing learning strategies. However, \rev{many participants (P1, P2, P3, P5, P6, P8, P9, P11, P12)} noted the difficulty of doing so with only textual instructions.
P1 explained, \hquote{sometimes even though I’m listening to the instructions, I still don’t get the bigger picture because I can’t see what I’m looking at.} 
P6 added that even well-written textual instructions are not as effective as tactile models in helping BLV individuals create mental images.
%:  \hquote{[It is] so much harder to build a concept without very, very clear and concise descriptions, and some people will still not be able to visualize it.}
P12 was in the \texttt{Text-Only} condition for the violin plot, commented:
%in a condition where she heard the textual explanation, but she really wants to say the tactile version, and ask us to send her one: 
\hquote{I just wish I can see the graph itself, because that would make so much more sense for me, because I think right now, I have a general image. But I'm still not really sure how it looks like visually.} She then asked us to send her the tactile model after the study. 

Participants highlighted that tactile charts are particularly beneficial for those who have never seen a chart before, such as people who were blind from birth, a point raised by both those who are blind since birth (P2, P6, P7) and not (P4, P10). 
%For example, P6 remarked \hquote{It's hard to create something on your own that you've never seen.} 
%P2 shared a similar sentiment: \hquote{I don't speak for everyone, but especially as someone who's never has never had vision.}

\textbf{Understanding chart layout.} 
Participants shared the difficulty of understanding the spatial layout of chart elements using textual explanation alone. 
As P9 pointed out, \hquote{not having the tactile representation, you have to draw it in your own mind, so that makes it harder.} 
We noticed this difficulty from the reactions of participants in the \texttt{Text-Only} condition in our study.
For example, P2 struggled with grasping the multidimensional structure of the clustered heatmap: \hquote{When I'm reading things top to bottom, I'm trying to picture them in my mind. Because I have no visual context of what this could even look like, I find myself thinking, ``Wait, up and to the left? Up and down?'' I'm just trying to imagine all these things, [..] which doesn't always work.} 
Reflecting on the learning experience with only textual explanations, P2 noted: \hquote{It's hard to visualize. When you have all those clusters, and this going this way and the other things going on this way on top of that. You lose some of the ability to really analyze it and answer questions about it, and retain the information and really gain insights about it.}  
Participants also had difficulty in visualizing violin plots. For example, P8 misunderstood violin plots based on textual information, imagining overlapping violins. P12 tried to visualize a violin plot but was uncertain: \hquote{I still don't really understand how the violin plot looks like, I just end up creating my own visual graph.} 
\rev{Participants’ experiences suggest that even though text can convey spatial cues like direction or overlap, it is hard to mentally form a coherent structure from text alone. Thus, text may be insufficient for building mental models.}

\textbf{Understanding shapes.} 
Tactile charts also help participants understand the shapes of chart elements, which is especially important when the chart has a unique shape encoding like the violin plot.
Five participants (P6, P8, P9, P10, P12) expressed confusion when they first heard ``violin shapes.''
% Five out of six participants (P6, P8, P9, P10, P12) expressed confusion about ``violin shapes'' in the text condition. 
P9 noted that \hquote{sighted people name it in ways that make perfect sense visually, but might not make sense to someone who wouldn't make the visual connection.} 
After interacting with the tactile chart, however, participants expressed understanding. As P10 noted, \hquote{I understand now why you call it a violin plot. Makes sense!} 
Beyond our study, P9 offered a real-life example, recalling the early pandemic: \hquote{When COVID hit, everybody was talking about flattening the curve. [...] What the hell does that mean? And somebody came up with a tactile representation of flatten the curve because we didn't realize, it's a bell curve and flattening the curve actually meant stopping the increase in COVID cases.} 
%This example shows how tactile charts can help BLV individuals understand the shape of chart elements and help them access to important public information.

\paragraph{Learning by Touch}

Tactile charts, as a form of data physicalization, enable learning through material interaction \cite{Jansen:2015:Opportunities}.
\rev{Eight} participants preferred this hands-on learning approach, which supports exploratory and multisensory learning.
P10 stated, \hquote{I like the hands-on stuff.} 
P8 mentioned that \hquote{some people do learn better visually or tactilely than just verbally} and noted \hquote{My brain would just like to see one.}
P3 describes how this hands-on learning approach aligns with BLV individuals' natural tendencies toward tactile learning: \hquote{Blind people are like children. I want to touch everything.}

Several participants (P1, P4, P6, P9, P12) identified themselves as ``visual, tactile, or kinesthetic learners''. Although these labels are debated by cognitive psychologists, they reflect these participants' learning experience and preferences. These participants emphasized that they obtain information better through physical interaction than through auditory input alone. 
P4 describes her preference for tactile charts over alt texts, stating \hquote{I'm blind, but I do consider myself a visual or a tactile learner in the sense that I'm able to feel what I'm being told. I did fine with the alt text, but I do love the nicely added touch with the tactile}. 
P6 also explained, \hquote{I'm a visual learner, but not truly visually. I touch things with my hands, and that's how I learned best. [..] When I'm listening, I don't get as much information or don't process it as well.} 
% Similarly, P9 stated, \hquote{I'm more of a kinesthetic learner...I learned by doing.} 

\textbf{Exploratory learning.} 
Tactile models enable BLV individuals to freely explore tactile charts, as opposed to the prescribed sequence and contents of textual instructions.
We found that this autonomy in exploration fosters reflection and insights that might be overlooked in textual or auditory descriptions.
For example, P2 described a moment of tactile discovery:
\hquote{There was that huge thing in the middle. That thing must represent something in terms of the actual visual data that the text isn't going to give me.}
%For P8, who has severe low vision, tactile exploration made it easier to focus on the most meaningful parts of the chart. He explained, \hquote{it's telling of what the more prominent aspect is. On the visual side [..] the denser parts are darker. But that can still be overwhelming for someone. The tactile part gives you something more useful. [..] You’re just focusing on the number, the height, and the connection.} 

\textbf{Multisensory learning.} 
Touching tactile charts while also hearing instructions engages multiple sensory channels, and thus enriches participant's understanding of the data. 
P7 articulated that \hquote{it just gives you more of a reference point, another way to remember the data. Auditorily is great, but I also like using another sense. [..] Reading it, gives you one perspective, but touching and understanding the different relationships.}
In addition, tactile charts can reinforce and complement textual explanations. P1 emphasized the value of combining touch with audio for building a mental model: 
\hquote{Being able to feel the map helps me better understand the information I am hearing. When you can't see the visual map, it’s almost like you're missing pieces.} P1 also noted the reassurance that comes from the physical representation: \hquote{You wouldn't have to worry about missing anything, because it's right there in front of you.}
Similarly, P9 shared, \hquote{the tactile diagram really, really reinforces the text, especially when you know what the hell you're looking at. [...] Having something in front of you to look at while JAWS is yammering away, it helps.} 
% \marginpar{\tiny \sarah{I like this section 5.2 and it supports H1, but the length makes it difficult to keep track of the high-level heading, "building mental models" so would be good to cut some}}

\SkipTocEntry\subsection{Developing Transferable Knowledge}
\rev{Participants believe the mental models of chart types are reusable and transferable across different datasets.} These mental models help them in interpreting new data and alt texts. 

Participants reported that tactile chart exploration helped them develop a structural understanding of chart types independent of the dataset, which could be generalized to new contexts. P5 noted this process as \hquote{like getting a mental model of how this particular chart type functions.} 
This understanding allowed participants to mentally reconstruct the chart with different data. As P12 explained, \hquote{It helps me understand the chart itself, then it allows me to recreate it mentally with a different data set.} 

This structural mental model also supported participants in understanding alt texts.
P1 shared, \hquote{It was different information, but it helped you understand how it could be placed and how you could read it.}  
%She described her interpretive strategy: \hquote{I'm just trying to look at it all, and remember what I read, and picture it all, and put it all together.} 
P2 made a comparison to math learning: \hquote{You can use the same type of chart to transfer the information you know about it to be able to understand a different set of information, because it's the same chart. [...]
%It's just the information's different. 
When you learn math, it's the same process to solve a problem. It's just that the numbers are different.} 

Participants emphasized having a reference for comparison helps them understand the new datasets. P4 explained, \hquote{Having one example to feel helped me understand the descriptions of other ones, and understand what they would potentially look like. It definitely helped a lot to have something to compare it to.} 
Similarly, P9 explained, \hquote{It gave me a frame of reference to picture the data that I was given. It gave me a model to use, like a mental template. Once you have a concept of a diagram, you can insert other data into that same concept. You have a framework in which to make pictures in your head of the data. If you've never seen the diagram before, you don't have that framework.}

This idea aligns with our observation that some participants relied on tactile reference points, both physically and mentally, when they encounter new datasets. 
When interpreting the complex alt text for chart types previously encountered through tactile exploration, participants often referenced their prior tactile experience. 
%As P7 noted, \hquote{I'm just comparing it to this because it's what's in front of me,} and 
For example, P11  said, \hquote{I had that reference point to go back to, because I kept the chart in front of me the entire time.} 
We also observed that when asked to imagine the corresponding tactile chart for a new dataset, some participants (P2, P3, P9, P11) used tactile charts they had just explored as a reference in their descriptions. For instance, when P2 was asked to describe a new violin plot about continents after they explored the tactile violin plot about penguins, he said, \hquote{I'm comparing Asia, which has the widest HDI range to the Gentoo penguin on this tactile map because it also had the widest range of body mass.} 

\SkipTocEntry\subsection{BLV Visualization Education}
\label{sec:blv-visualization-education}
Participants discussed barriers that BLV individuals face in learning about data visualizations, their interest in gaining equal access to information, and how tactile charts can support more effective and inclusive visualization education. 
%\hty{These believe in teaching potential of tactile charts support H4}. 
% \marginpar{\tiny \sarah{should we have a separate (or more explicit) hypothesis relating to perceived value for teaching/education? Currently there is text that comes before H3 and H4 but those hypotheses are not really specific to education} \hty{should have, but we already preregistered our hypotheses so cannot update}}
%Participants reported feeling left out when learning about data visualizations and reflected on what caused these barriers.

\paragraph{Lack of Institutional Support} 
Participants reported on challenges in their education, where instructors were either unaware of the needs of BLV students or lacked the tools and strategies to support them, \rev{consistent with findings by Butler et al. \cite{Butler:2017:Understanding}}.
As a result, BLV learners \rev{could be} excluded from lessons involving data visualizations. P2 recalled feeling completely lost in a graduate-level visualization course where the instructor made no effort on inclusion: \hquote{He was just writing a bunch of data on the board and drawing it. I had no clue what's going on.} 
Even when instructors tried to be inclusive, they \rev{could lack} effective teaching strategies or resources. As P2 recalled when encountering visualizations in a statistics course:  \hquote{They didn't know how to help me. So we just skipped that part of the curriculum for me.}
%If there's a visual thing, we'll just skip it.}.  

\paragraph{Self-Exclusion}
Some participants noted that BLV students themselves might give up on engaging with visual content. 
P1 reflected, \hquote{As a blind student, it is so easy for you to say, oh, I'm blind, I can't use this map, and they will let you bypass it.} 
However, she also observed positive changes over time: \hquote{Back when I was in school, blind students weren’t really pushed. We could opt out of situations that required reading data tables, graphs, and similar materials. But now, with advancements in technology, things are much better in that regard. Blind students are encouraged not to take that cop-out because there are ways to access the information.}  

\paragraph{Interest in Learning Visualizations}
Participants expressed interest in learning new visualizations and emphasized the importance of equal access to information. 
%As P2 stated, for tactile charts, \hquote{maybe I couldn't have produced one on my own. But at least having the ability to participate and understand it, and be able to understand what it was that was going on, and what I was looking at would have been nice.} 
For P3, the violin plot offered a clearer alternative to raw data: \hquote{[I think it's a simple way of looking at it rather than having lines and rows of thousands of numbers.} After exploring the tactile violin plot, P4 shared her enthusiasm: \hquote{This is a really cool format. I've never seen this type of plot for any type of data or information. I really like the concept.} 
She also appreciated the clustered heatmap, noting, \hquote{all of this in terms of heatmap, clusters, and dendrograms were new to me, but it was really interesting to learn and see this sort of information be plotted in this way.}  P11 shared similar enthusiasm after interacting with both chart types: \hquote{It is an eye opening experience to see different ways that data are visualized.}
Beyond interest, learning visualizations is important to expend future opportunities. P1 explained: \hquote{Information, like how to read a map or access data, is so important, especially when you get a job. [..]
%And just in general, knowing the placement of things is crucial. 
You feel left out when you don’t have that information while your peers do.}

\paragraph{Empowering BLV Education With Tactile Charts}
Participants \rev{discussed} the educational value of tactile charts and discussed their availability and potential. P3, an educator herself, emphasized that tactile materials are \hquote{the cornerstone} and \hquote{the bare baseline,} stating, \hquote{I think this is extremely valuable.} 
P3 pointed out the limitation of relying only on the verbal instructions: \hquote{If you're just teaching by talking to them, for a blind person, they can't get the stimulus, you’re missing out on a whole lot of information. It's really difficult to grasp the whole concept. That missing piece causes disengagement. So, the tactile part is absolutely essential.} 
Participants also emphasized the potential of tactile charts to support learning and expand future opportunities. 
P2 noted that tactile charts \hquote{would really change our ability to study a lot of different things,} and P1 added, \hquote{it is extremely important for students to be able to have access to that, because it just changes the game for you.} She added that such tools could enable blind students to be \hquote{on the same playing field as their sighted peers.}  
%After interacting with the tactile charts in our study, P6 noted it's helpfulness, \hquote{I think that it's a really great idea and it'd be so helpful and probably help many people learn it once they catch on how to use it.} 

%These reflections reinforces with our earlier discussions on the advantages of tactile charts in supporting mental model-building.

Despite their value, tactile charts are often inaccessible. 
P6 shared, \hquote{People who are blind don't get exposed to this information, so if I would have had something like this in college, I probably would have done a lot better.} 
P9 added, \hquote{It's something that blind people don't often get access to, because they're so expensive to produce.}
%Participants expressed a desire for integrating tactile charts more broadly into education. P6 described them as \hquote{something every teacher with a blind student in their class should be exposed to.} 
Similarly, P1 pointed out the ongoing demand within the BLV community: \hquote{We're always looking for things like this.} She explained: \hquote{I know a lot of individuals who are blind have requested things like this be put into schools, their workplaces, and their communities---anywhere a sighted person can go and read a map or access data.}% Individuals in the blindness community are asking: we want the same thing, how can we put it in a way that we can get the same or similar information?} 

\SkipTocEntry\subsection{Other Modalities}

Participants also reflected on alternative modalities for accessing data visualizations, including textual descriptions, raw data, and AI tools. In this section, we discuss participants’ experiences with these modalities and highlight how they can complement---but not replace---the need for chart type learning, particularly through tactile charts.

% Of the two participants who preferred the \texttt{Text-Only} condition, P3 explained that text alone can convey all the necessary information: \hquote{[Text] will 100\% give you the information.} 

\rev{Two participants (P3, P5) preferred the \texttt{Text-Only} condition. 
P3 explained that text can convey all the necessary information and presents a lower barrier to access, serving as a minimal form of support when other modalities are unavailable. 
For example, P3 noted that not everyone can read Braille: 
\hquote{Only some very strong Braille users would say that they want to look at the Braille 100\%. [..] If you have a tactile piece with [text]---the more the merrier---but I think [text] meets the threshold. [Text] will 100\% give you the information.} P3 also emphasized that Seeing AI, an accessibility tool he uses, cannot read Braille. He pointed out: \hquote{It's a game of availability. [..] Alternate text is the bare minimum.} 
P12 shared a similar view:
\hquote{[Text] is better than nothing. But it is really hard to digest and comprehend. It's doable, just takes a lot of energy.} 
These perspectives align with our earlier discussions: while textual explanations often fall short in supporting the development of mental models, they offer broad accessibility. 
P5 preferred AI tools and was not fully satisfied with either training format, but acknowledged the practicality of text: \hquote{Neither of these [modalities] was super effective, but I'd say [I prefer] the text because it summarizes the info I need to know in one easy package.}}

% Other participants also highlighted that textual explanations are often easier to access using mainstream technologies. As P10 noted, \hquote{If you have a QR code, you can scan it, or even have different like apps that I can scan the text and my phone would read it.} However, he also had concerns, adding, \hquote{I don't know how accurate it is, though it's not 100\% accurate. Just scanning technology.} \al{Maybe kill this paragraph? I don't know what I'm getting from it. Or are they advocating that printed text and graphics is better than tactile charts? }

% Participants acknowledged that textual descriptions can serve as a minimal form of support, especially when other modalities are unavailable. As P3 stated, \hquote{It's a game of availability. [...] Alternate text is the bare minimum.} P12 also said: \hquote{[Text] is better than nothing. But it is really hard to digest and comprehend. It's doable, just takes a lot of energy.} These perspectives align with our earlier discussions: while textual explanations often fall short in supporting the development of mental models compared to tactile charts, they offer broad accessibility.
%To address this, we plan to develop a website where BLV individuals and educators can freely order tactile charts to support the learning of data visualizations.

A few participants expressed a preference for raw data in tables over charts. P9 stated that a table might be more accessible: 
 \hquote{As a blind person, you'd probably be more pragmatic just to have a table with the actors and the movie genres, and then the numbers.}  
 %Similarly, P11 noted, \hquote{If I'm looking at a but if you just flush out the data that makes it so much easier for me to look at and read, because I don't do well with graphs. } 
P5, who prefers using AI tools, advocated for direct access to raw data and the use of AI summaries and question answering: \hquote{When you're trying to take in a data visualization, if I have time I want to take it in as a table, and if I don't have time I need a summary.} 
\rev{We argue that these workflows are sensible, and our tactile charts are meant to support workflows for arbitrary datasets. In addition, raw datasets are often not available or become impractical when dealing with complex data or layered structures \cite{Joyner:2022:Visualization}---as seen in many scientific visualizations.}

\SkipTocEntry\subsection{Design Better Tactile Charts \rev{For Learning Chart Type}}

% \sarah{the recommendations embedded in this theme are confusing to me. Seems like we need a better way to transition from 5.1-5.3 where it is just communication of themes to 5.4 where there are recommendations, and then back to 5.5 where it is other reflections (no recommendations)} 

Participants appreciated our design decisions. For instance, P3, who had prior experience with poorly made charts, noted, \hquote{I'm very, very impressed, because I've seen some bad ones.} Positive feedback reinforced the importance of thoughtful chart design and construction process. 
% To enhance the effectiveness of tactile charts for educational purposes, 
In this section, we identify key design considerations based on participant feedback and provide guidance on how to create more accessible and pedagogically effective tactile charts for BLV learners. \rev{We note that in the future, these design suggestions should be validated in real learning scenarios.}

\paragraph{Topic Familiarity}
We observed that familiar content helped participants use prior knowledge to build a mental model of the chart, even when the data was complex. 
For example, P9, who was more interested in European countries (our complex dataset) than in movies, stated that she could imagine the heatmap for countries but probably not for movies.
% We observed that topic familiarity and interest influence how easily participants understand tactile charts. 
% Familiar content helped them use prior knowledge to build a mental model, even when the data was complex. 
% P9, for example, was more interested in data about European countries (our complex dataset) than movies, stated that she could imagine the heat map for countries, but probably not for movies. 
%Similarly, P12 said, 
%\hquote{this second data set on the continents and wellness index was easier to make sense of than the one on the penguins.} 
Hence, we recommend \textbf{using topics that are relatable or familiar to the targeted audience \rev{in educational tactile charts}.}
% Hence, we recommend \textbf{selecting chart topics that are relatable or familiar to the targeted audience.}

% \parhead{Chart Size}
% We used an A4-sized board for all tactile charts, and participants found this size intuitive and appropriate. As P5 noted, \hquote{I think this is fine. It's the size of a paper.}
% Participants emphasized that if a chart is too large, there is too much information they need to remember, due to the constrain of tactile reading. As P5 explained, \hquote{you can't scan with your fingers as easily as you can scan with your eyes. So there's a limited amount of context window. And so if it gets bigger, you still have to scan more, it just takes more time, and you forget what you've scanned at the beginning by the time you've hit the end.} 
% We thus recommend using \textbf{A4/letter sized charts and carefully balancing chart size with tactile readability.}

\paragraph{Clarity of Chart Elements}
Clear layout and spacing helps reading tactile charts, \rev{as suggested by guidelines (\eg, \cite{BANA:2022:Guidelines})}. Participants appreciated the clarity of our chart design. %P1 commented that \hquote{the way it was laid out was very nicely done.} 
For example, P4 noted that the chart \hquote{has a nice and neat touch} 
% For example, P4 noted that the chart \hquote{has a nice and neat touch. It's very easy to read.} 
and added, \hquote{It's clear, legible, and everything is spaced out perfectly. The lines, and all of the different shapes on the graph are easy to identify, and just all of the details were absolutely fantastic.} These responses highlight the the value of clear organization and visual separation in tactile formats.

Participants also emphasized the value of clear labels. P4 noted that \hquote{it is especially helpful to have the labels along the Y and X-axis} for the violin plot. P8 also thought the labels of the heatmap columns and rows are very helpful, and further suggested adding \hquote{a more distinct outline, whether between the names or the y-axis columns} to enhance their separation.
% While we included background blocks behind the violin plot labels, feedback suggests that this should be applied consistently across all chart types. 
We thus recommend designing tactile charts with \textbf{clear layout, well-spaced elements, and providing clear labels.}

\paragraph{Braille}
All participants who commented on the 3D-printed Braille described it as legible and well-formed.
P1 and P4 both described it as \hquote{easy to read} and P3 confirmed that \hquote{Braille dots are not dead. You can read it really well}. 
P1 further explains that \hquote{the Braille is very sharp and crisp, too. That's a good thing.} 
These positive responses were reassuring, as we had initially been concerned that 3D-printed Braille might be unpleasant to touch compared with embossed paper. %However, none of the participants reported issues or provide negative feedback on the touching feeling. 

Beyond readability, participants also discussed the choice between Grade 1 (uncontracted) and Grade 2 (contracted) Braille. We used Grade 1 across all charts in the study. Participants commented that Grade 1 is more accessible (P4, P7, P9), particularly for beginners. 
%As P4 explained, \hquote{I did notice that it's an uncontracted Braille, which is also really great, for maybe beginner Braille users. Not everyone learns Braille as a child, especially for people who lose their vision later in life, and we all start with uncontracted Braille, no matter how old we are.} 
However, they also expressed concerns that Grade 1 takes up more space (P7, P9) and may reduce reading speed (P9). P7 further noted that people in advanced classes should be comfortable with Grade 2. 
These perspectives reflect a trade-off between accessibility and efficiency. 
In addition, P7 also noted the absence of numeric indicators for the Braille numbers in the heatmap and suggested adding a sentence to indicate that these Braille characters represent numbers to avoid confusion. We thus recommend \textbf{selecting the Braille grade based on the target audience’s proficiency, and ensuring that all indicators are used correctly or that deviations are clearly explained.}

% \hquote{[P07L568] P07 thinks Grade 1 Braille takes up more space but make content more widely accessible. People in advanced class are comfortable with Grade 2. }

% Participants who discussed our Braille think they are easy to read,but discussed Grade 1 or Grade 2 Braille. They think grade 1 is more accessible but grade 2 take less space and can be read quicker.

% \hquote{[P04L688] The Braille is very easy to read. I did notice that it's an uncontracted Braille, which is also really great, for maybe beginner Braille users. Not. Everyone learns Braille as a child, especially for people who lose their vision later in life, and we all start with uncontracted Braille, no matter how old we are.}

% \hquote{[P07L568] P07 thinks Grade 1 Braille takes up more space but make content more widely accessible. People in advanced class are comfortable with Grade 2. }

% \hquote{[P09L868] It's more accessible. But those of us who read grade 2 read it faster.}

% \hquote{[P09L65] (in background questions) I learned regular Grade 2 Braille when they shoved ueb down our throats. I'm sticking to my grade 2, because that's what I know. I don't understand why make Braille take up more space.}

% \hquote{[P01L676] The Braille was easy to read.}

% \textbf{the use of the indicators}

\paragraph{Production Method}
Participants expressed appreciation for the 3D-printed tactile charts and identified several advantages of 3D printing. One commonly praised benefit was the sturdiness of the 3D-printed models.
P2 noted, \hquote{Regular paper smushes and bends, not helpful at all.}
%I like the surface. It's better. It's sturdy.} 
%P04 remarked, \hquote{I think it was really cool that you guys made it like a placard. It's nice and solid and sturdy.}
P4 also explained that the sturdy surface enhanced readability: \hquote{A lot of a lot of tactile stuff that I've seen is flimsy.}
Participants also highlighted the resolution of 3D printing, which allows for more realistic, accurate, and detailed representations, \rev{which aligns with prior discussions \eg, \cite{Lundgard:2019:Sociotechnical, Holloway:2018:Accessible}}.
P11 praised our 3D-printed charts:
\hquote{It's very realistic looking. [..] That might draw people in. I'm amazed that this can be done.}
P2 contrasted it with embossed paper: \hquote{The problem I always had with them was that you couldn't create highly realistic 3D representations, at least not when I was in school. With 3D printing, you can.}
%, but on paper, it's really hard to make an accurate 3D tactile graph. If you were looking at a sphere, you'd literally just see a circle with a hole in it, which isn't really a sphere.} 
%P2 expressed his concern about the accuracy of paper-based tactile charts, noting \hquote{I always was concerned about the accuracy of a lot of 3D charts or graphs on paper} He also added that \hquote{the more complex the the graph got, the more frustrated and confused I got.} 
% Although embossed paper remains more common, participants were drawn to the potential of 3D printing.
% P11 is fascinated by the technology, saying \hquote{This is just this stuff is kind of fascinating to me in its own way, just to see where technology is.}
% P07, who was also new to 3D printing technology, shared a simliar interest: \hquote{I've never seen a 3D printer do this kind of thing. So it's interesting to see how detailed it can be.} 
P9 expressed optimism about improvements in tactile chart quality:
\hquote{I'm hopeful that with 3D printing technology [..] that's changing. Back in my day...you rarely found a good tactile representation of something.} 
Participants also showed interest in the process behind 3D printing itself. P10 commented, 
\hquote{I was interested in, surprised that it was printed with a 3D printer. Is that just something you'd have to type in code, and it prints out? [..] Can anyone buy a 3D printer?}
We recommend \textbf{using 3D printing to produce tactile charts for complex visualizations for educational purpose.} 

% We recommend \textbf{using 3D printing to produce tactile charts for complex visualizations, especially those intended as reusable educational templates.}

%\subsection{Provide support for education and learning}

\paragraph{Providing a Sighted Version}  
We included a sighted version on the back of the chart to facilitate communication between blind and sighted individuals, which proved helpful. 
Before noticing it, P3, an educator, asked, \hquote{If a professor is teaching a blind student [..] Would they be able to guide that person based on the tactile model?} 
% \hquote{If a professor is teaching a blind student, and he has this to look at it. Would they be able to guide that person based on the tactile model?} 
% Considering the educational purpose of our tactile charts, we made some design considerations to support education, and these were appreciated by participants. 
% Participants noted that 

%Upon learning that we labeled the back of the tactile chart with a sighted-version reference, he commented, \hquote{That label would help in multiple ways, not the tech, not only the tactile model, but also seeing AI would also read.}

% \parhead{Orientation cues.} Orientation is essential for independent tactile exploration. Our use of a cut corner on the top right was consistently described as helpful. \hty{P8 remarked, \hquote{once I got the orientation it was pretty simple to understand the names and the kind of the layout, and then the kind of the values.} }
% %P01 explained, \hquote{you put a little cut out then a person would know this is the right way up... You could look at the Braille and check, but that was also automatic, you know, this is the way it should look.}
% P2 also appreciated that the corner guided them to the sticker containing a QR code for exploration instructions: \hquote{there is only one cut corner the stickers right there, that stuff's pretty straightforward.}

\paragraph{Exploration Instructions} 
% We designed our tactile charts to support independent learning, which is also preferred by participants, as P1 said: \hquote{Someone can explain things to you all day, but being able to look at it yourself opens up a whole new window of information and opportunity.} 
\rev{To support independent learning, we provided instructions for tactile charts, which participants appreciated.} 
As P1 noted: \hquote{Someone can explain things to you all day, but being able to look at it yourself opens up a whole new window of information and opportunity.} 
\rev{While we did not directly compare conditions with and without instructions, many participants (P2, P3, P4, P6, P7, P9, P11) emphasized that guidance was crucial for effective interpretation.}
Participants often struggled to make sense of the tactile charts before reading the instructions but understood them more clearly afterward. For example, P4 recalled, \hquote{When I first opened this package and I saw what this was, I was like: ``What is this? This doesn't make any sense!'' But with the instructions, it clicked right away.} \rev{Similarly, P2 said, \hquote{having the instructions on how to read it let this makes obviously more sense, and I know what I'am looking at.} P6 also mentioned that the instructions made the charts \hquote{easy to follow, especially since you guys sent that little instruction thing.} This impact was reflected in ratings: P3 and P7 rated ease of exploration as four without instructions and five with them.}
% P9 also mentions \hquote{Once I knew what I was looking at. It was easy.} 
% This benefit was also reflected in participant ratings. When asked about the ease of exploring the tactile model, two participants (P3 and P7) rated it a 4 without instructions but a 5 with instructions.
We recommend \textbf{pairing tactile charts with exploration instructions and a label referencing the sighted version to support independent learning and communication with sighted individuals.}

\paragraph{Suggestions on Instructions}

% Participants discussed what they thought made good accompanying instructions. First, they recommended to give a high-level summary before diving into details (\eg, P2, P3, P8). At the same time, participants also recommended immediately explaining attention-grabbing features of the tactile chart, such as the 3D bars in the heat map (P3).
% P2 suggested that authors of alt texts should not be afraid of describing color in the equivalent tactile chart because these help them understand things, which also applies to instructions. He explained, 

Participants shared what makes effective exploration instructions. They recommended giving a high-level summary before diving into details (P2, P3, P8) and immediately explaining attention-grabbing features of the tactile chart, such as the 3D cells in the heatmap (P3). P2 also emphasized that describing visual features like color is helpful: 
\hquote{You might not think color is important because we can't see it, but it is still good to know. [..] It would help us understand.} We thus recommend \textbf{structuring exploration instructions to begin with an overview of the dataset and chart structure, followed by a sequence that aligns with the tactile exploration flow. Authors should not avoid describing visual features.}

\SkipTocEntry\section{Discussion and Future Work}
Our initial quantitative results suggest that \texttt{Tactile+Text} and \texttt{Text-Only} training support similar levels of understanding of advanced chart types (RQ1). \texttt{Tactile+Text}, however, resulted in slightly higher accuracy when interpreting complex alt texts (RQ2). This pattern suggests that \rev{although tactile input may not significantly improve immediate basic comprehension, it can support retaining and applying chart type knowledge in more complex or long-term scenarios.}
% while tactile input may not significantly improve basic chart comprehension, it could offer additional support for retaining and applying chart type knowledge in more complex contexts.

%Our limited quantitative results do not indicate that tactile models help BLV individuals develop a better understanding of advanced chart type (RQ1), but slightly higher correctness on alt text understanding questions may show that tactile models lead to a better understanding of complex alt texts (RQ2). These results are somewhat contradictory with each other, which may point to the low power of our study, and may indicate that our questions did not accurately test understanding.

The subjective ratings and qualitative data show a more consistently positive picture.  Participants felt that tactile charts helped them develop a better understanding of chart types, emphasizing the benefits of building mental models, understanding chart layouts, learning about shapes, and appreciating the advantages of hands-on exploration. These findings \rev{align with the benefits of tactile charts shown in prior work (\eg, \cite{Chundury:2022:Towards, Engel:2021:Heatmaps, Goncu:2010:Tactile,Fan:2024:Tangible,Fan:2022:Accessibility, Watanabe:2019:Usefulness, Engel:2021:Heatmaps, khalaila:2024:They, Yang:2020:Tactile})} and further indicate that tactile models do lead to a better understanding of advanced chart types (RQ1).  
Furthermore, participants reported that tactile models helped them transfer knowledge to new datasets (RQ4) and even referenced analogous patterns in prior charts, yet self-reported understanding of specific complex datasets did not vary much between conditions, even though self-assessment on the helpfulness of tactile charts for learning transferable knowledge was high ($M=4.67$ out of 5). 
Preference is clear: most participants (10/12) prefer learning with tactile models (RQ3), which is evident both from the quantitative and from the qualitative data. 
\rev{Only two participants were hesitant about tactile models: one preferred alt text for its lower barrier to access, and the other preferred querying ChatGPT for information.}
% Only two participants were hesitant about tactile models, and stated that digital means, such as LLMs, which they can query are more useful to them. 
As for the utility in learning, participants strongly believed that tactile template charts are highly useful in education, and reported many personal situations where they could have benefited from them, \rev{aligning with previous findings \cite{Sheppard:2001:Tactile}}.
Taken together, our design and evaluation shows the potential of tactile template charts to promote more inclusive access to data. \rev{To our knowledge, this is the first empirical study to explore how tactile charts can be leveraged to teach BLV individuals to interpret complex visualizations. We thus argue that we address a BLV education gap.}

Several limitations should be noted in interpreting these results. First, we evaluated only two chart types, limiting the generalizability of our findings. Whereas violin plots and heatmaps allow us to examine the comprehension of different structures, like distributions versus clustering, they do not fully capture the complexity or variation found across all data visualizations. 
In addition, the interview format creates a structured setting differing from real-world use, where exploration could be self-directed without expert guidance.
% where exploration is self-directed and expert guidance is not always available. 
Third, our study focused on short-term comprehension and  did not assess long-term retention, learning outcomes, or educational impact. Finally, participants may have self-selected to participate based on an interest in tactile charts.

Reflecting on our process, conducting online interviews has several advantages with BLV participants. Remote participation made it easier to recruit eligible individuals without geographic restrictions. Participants join the interview using their personal computers, which are already configured with their preferred accessibility tools, such as specific screen readers, settings, and additional software or hardware (\eg, screen magnification, Braille display). 
\rev{In addition, future researchers should note the complexity of remote studies involving tactile charts \cite{deGreef:2021:Interdependent}. In our experience, this adds extra logistics, such as an asynchronous consent processes as well as shipping costs and accounting for delivery timelines.}
% In addition, future researchers should note the complexity of remote collaboration with tactile charts \cite{deGreef:2021:Interdependent}, we note that that printing and mailing materials require extra logistics on the researchers' behalf, such as an asynchronous consent process, shipping costs, and delivery times. 
% and ensuring a timely delivery before the scheduled interview. 
% We had initially planned to send all participants three tactile charts, but the limited throughput of the printer limited us to a single chart per participant.  

\rev{Future work could explore the educational potential of tactile charts by extending them to additional chart types or modalities (\eg, comparing to or integrating with LLM question-answering systems \cite{Kim:2023:Exploring, Seo:2024:MAIDRAI}) and by supporting BLV individuals to create them independently \cite{Fan:2020:Constructive}.}
%While we used single-material 3D printing for scalability and generalizability, using multiple materials (\eg, \cite{Ebermann:2024:From}) may enhance engagement and easily show different visual encodings. 
It would also be interesting to explore embedding tactile cues for common interactive features (\eg, on-demand tooltips). 
% In addition, our current focus was on static charts as a first step. As interactive visualizations become more common, future work could explore ways to encode hints about typical interactions into tactile charts. For example, tactile cues could indicate that scatterplots often provide on-demand tooltips that readers can interact with.
Finally, given the importance participants placed on exploration instructions, an automated pipeline that generates them from existing chart specifications could streamline the creation of educational tactile charts.
% \section{Future work}

\SkipTocEntry\section{Conclusion}
We show that tactile charts, when thoughtfully designed and with clear exploration instructions, can be a helpful and preferred educational tool for BLV individuals to learn complex visualizations. 
% Our evaluation with 12 BLV participants show that brief exposure to tactile charts can help build transferable mental models, and support better understanding new alt texts of the same chart types. 
These findings suggest a promising research direction toward improving data visualization literacy among BLV individuals with tactile charts and contribute to the broader goal of making information more accessible and inclusive. To this end, we make our models and instructions freely available and hope to build out a library of charts in the future.

\acknowledgments{
We thank Gordon Legge for his valuable feedback on our tactile chart design. We appreciate Jinol Shah for digital modeling, Omar Shami and Ryan Manwill for 3D printing, Nathan Galli for chart photography, and Iara Delgado for transcription corrections. 
% We thank Gordon Legge for his valuable feedback on our tactile chart design. We appreciate Jinol Shah for creating the digital models, Omar Shami and Ryan Manwill for 3D printing, and Nathan Galli for photographing the charts. We also thank Iara Delgado for help with correcting our interview transcriptions. 
We gratefully acknowledge Funding by the Chan Zuckerberg Initiative.}

\section*{Supplemental Material Pointers}
The preregistration for our study can be found at \href{https://osf.io/uhq68}{\texttt{osf.io/uhq68}}. 
Supplementary materials are available at {\osfrepo}, {\companionweb}, and 
\href{https://github.com/visdesignlab/tactile-charts}{
\texttt{github\discretionary{}{.}{.}com\discretionary{/}{}{/}visdesignlab\discretionary{/}{}{/}tactile\discretionary{}{-}{-}charts}}.

\section*{Images/graphs/plots/tables/data license/copyright}
We as authors state that all of our own figures, graphs, plots, and data tables in this article are, and remain, under our own personal copyright, with the permission to be used here. We also make them available under the \href{https://creativecommons.org/licenses/by/4.0/}{Creative Commons At\-tri\-bu\-tion 4.0 International (\ccLogo\,\ccAttribution\ \mbox{CC BY 4.0})} license and share them at \osfrepo.

\bibliographystyle{abbrv-doi-hyperref-narrow}
\bibliography{abbreviations,processed-output}

\appendix % You can use the `hideappendix` class option to skip everything after \appendix

\clearpage

\begin{strip} % requires \usepackage{cuted}
\noindent\begin{minipage}{\textwidth}
\makeatletter
\centering%
\sffamily\bfseries\fontsize{15}{16.5}\selectfont
\papertitle \\[.5em]
\large Appendix\\[.75em]
\makeatother
\normalfont\rmfamily\normalsize\noindent\raggedright 
In this appendix, we provide detailed discussion and additional figures that we could include in the main paper due to space limitations or because it was not essential for explaining our approach. .%\vspace{-.5em}
\end{minipage}
\end{strip}
% \begin{strip} % requires \usepackage{cuted}
% \noindent\begin{minipage}{\textwidth}
% \makeatletter
% \centering%
% \sffamily\bfseries\fontsize{15}{16.5}\selectfont
% \papertitle\\[.5em]
% \large Appendix\\[.75em]
% \makeatother
% \normalfont\rmfamily\normalsize\noindent\raggedright In this appendix we provide additional explanations, tables, plots, and charts that show data beyond the material that we could include in the main paper due to space limitations or because it was not essential for explaining our approach.%\vspace{-.5em}
% \end{minipage}
% \end{strip}

% so that we get page numbers on the appendix
\makeatletter
\renewcommand{\@oddfoot}{\hfil\textrm{\thepage}\hfil}
\renewcommand{\@evenfoot}{\@oddfoot}
\makeatother

\tableofcontents
% toc only for appendix
%\parttoc % Insert the appendix TOC

% \raggedbottom 
% \flushbottom
\section*{Images/graphs/plots/tables/data license/copyright}
We as authors state that all of our own figures, graphs, and plots in this appendix are, and remain, under our own personal copyright, with the permission to be used here. We also make them available under the \href{https://creativecommons.org/licenses/by/4.0/}{Creative Commons At\-tri\-bu\-tion 4.0 International (\ccLogo\,\ccAttribution\ \mbox{CC BY 4.0})} license and share them at \osfrepo. 

\clearpage
\section{Datasets for Template Charts}
\label{sec:template-chart-datasets}

To help people focus on understanding chart types, we used real-world datasets on familiar topics to create the template charts.
We refined these datasets by selecting subsets and modifying data to balance the simplicity with sufficient data features for showing each chart type’s unique characteristics. 
% For example, to create a bi-modal distribution in the violin plot, we selectively removed entries from the original dataset. 
In this section, we provide details on the dataset selection and modifications.
% In this section, we present how we selected and created the datasets for the four chart types.

\subsection{UpSet Plot: The Simpsons dataset}
For the UpSet plot, we used a dataset representing attribute overlaps among Simpsons characters. We selected this dataset because it shows various intersection patterns including empty intersections, individual character attributes, and multiple overlapping attributes.
We downloaded the original dataset from the UpSet GitHub repository (\href{https://github.com/VCG/upset/tree/master/data/simpsons}{\texttt{github\discretionary{}{.}{.}com\discretionary{/}{}{/}VCG\discretionary{/}{}{/}upset\discretionary{/}{}{/}tree\discretionary{/}{}{/}master\discretionary{/}{}{/}data\discretionary{/}{}{/}simpsons}}). This dataset was curated by Alexander Lex and originally sourced from The Simpsons character database (\href{http://www.thesimpsons.com/}{\texttt{thesimpsons\discretionary{}{.}{.}com}}).

\subsection{Clustered Heatmap: Movies by actors and genres}
For the clustered heatmap, we used a dataset representing the number of movies performed by five well-known actors across four common movie genres. We selected this dataset because it shows different distribution and clustering patterns. 
We derived this dataset from the IMDb Non-Commercial Datasets (\href{https://developer.imdb.com/non-commercial-datasets/}{\texttt{developer\discretionary{}{.}{.}imdb\discretionary{}{.}{.}com\discretionary{/}{}{/}non\discretionary{}{-}{-}commercial\discretionary{}{-}{-}datasets\discretionary{/}{}{/}}}). 
To create this dataset, we processed three IMDb datasets: \texttt{title\discretionary{}{.}{.}basics\discretionary{}{.}{.}tsv}, \texttt{title\discretionary{}{.}{.}principals\discretionary{}{.}{.}tsv}, and \texttt{name\discretionary{}{.}{.}basics\discretionary{}{.}{.}tsv}. These datasets refresh daily.We downloaded these original datasets from \href{https://datasets.imdbws.com/}{\texttt{datasets\discretionary{}{.}{.}imdbws\discretionary{}{.}{.}com}} on November 2, 2024.

\subsection{Violin Plot: Penguin dataset}
For the violin plot, we used a modified version of the Palmer Penguins dataset, which contains body mass values for three penguin species. We selected and modified this dataset to demonstrate three different distribution patterns. We obtained the original dataset from the Palmer Penguins R package (\href{https://allisonhorst.github.io/palmerpenguins/}{\texttt{allisonhorst\discretionary{}{.}{.}github\discretionary{}{.}{.}io\discretionary{/}{}{/}palmerpenguins}}). To show a clearer bimodal distribution, we selectively removed 21 records from the original dataset.

\subsection{Faceted Line Chart: Weather dataset}
For the faceted line chart, we used a dataset representing daily weather measurements (average temperature, average humidity, average wind speed, and total precipitation) in Austin, Texas during August 2016. We selected this dataset because it shows different types of trend patterns across multiple weather variables. We downloaded the original dataset from Kaggle (\href{https://www.kaggle.com/datasets/grubenm/austin-weather}{\texttt{kaggle.com\discretionary{/}{}{/}datasets\discretionary{/}{}{/}grubenm\discretionary{/}{}{/}austin\discretionary{}{-}{-}weather}}). 

% \clearpage
\section{Python Generated Plots for Template Charts}
\label{sec:python-generated-charts-simple}

To create our tactile charts, we first generated visualizations using Python, Matplotlib, and Seaborn. In this section, we present these Python-generated plots.

% UpSet plot
\begin{figure}[t]
    \centering
        \includegraphics[width=1\columnwidth]{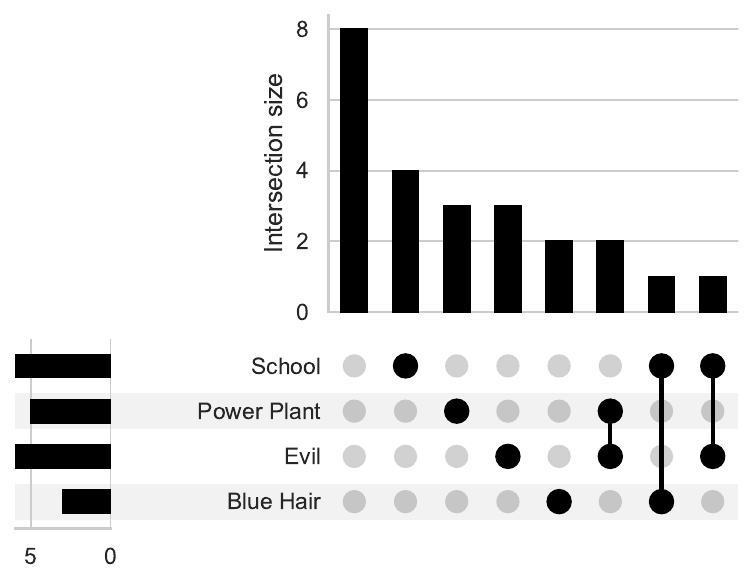}
    \caption{Python-generated UpSet plot used to create the tactile chart.}
    \label{fig:python-generated-upset-template-chart}
\end{figure}

% Clustered heatmap
\begin{figure}[t]
    \centering
        \includegraphics[width=1\columnwidth]{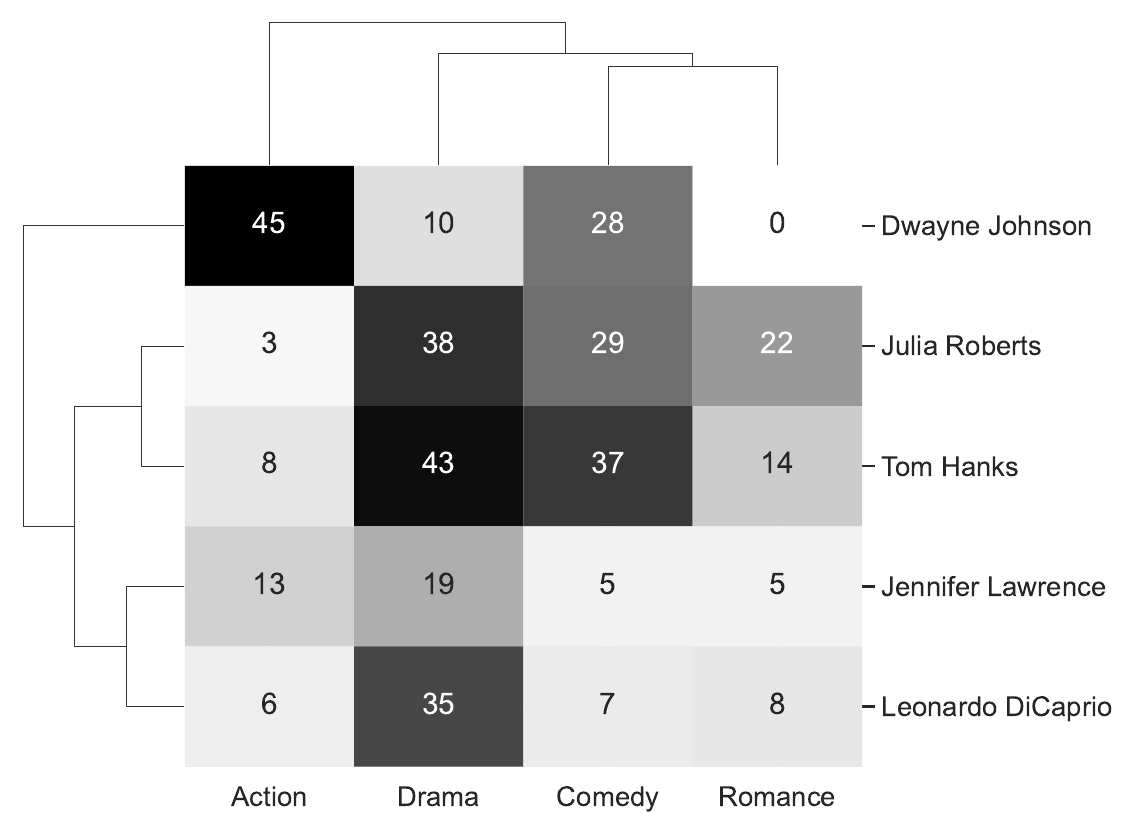}
    \caption{Python-generated clustered heatmap used to create the tactile chart.}
    \label{fig:python-generated-heatmap-template-chart}
\end{figure}

% Violin plot
\begin{figure}[t]
    \centering
        \includegraphics[width=1\columnwidth]{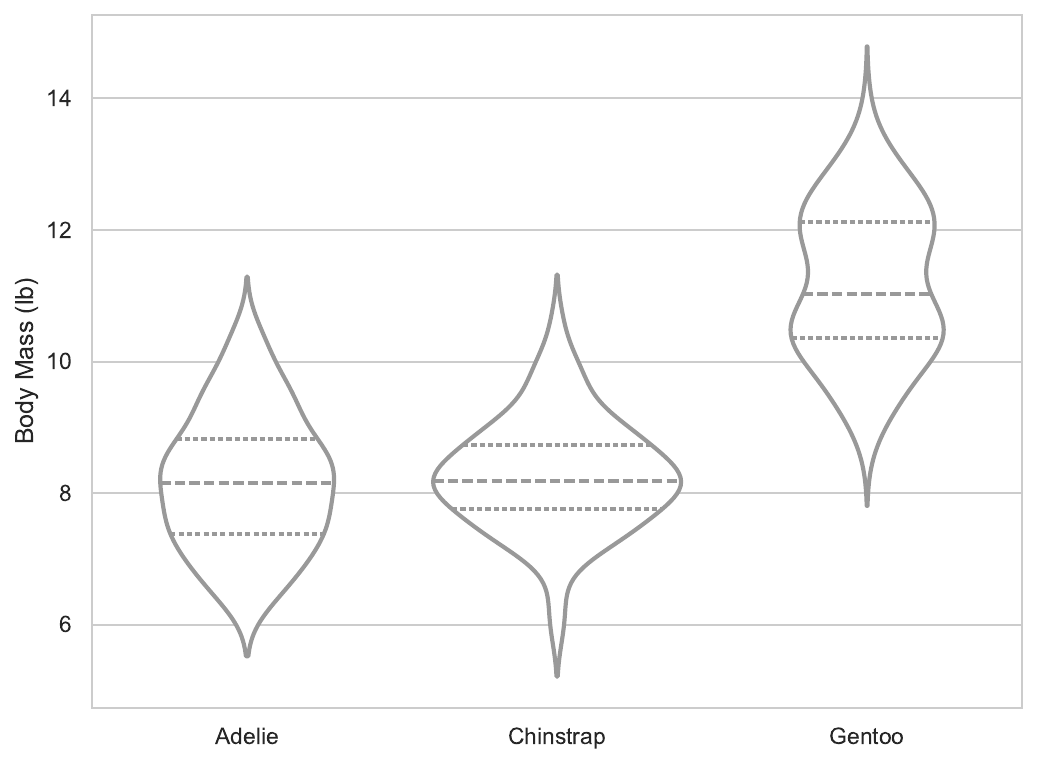}
    \caption{Python-generated violin plot used to create the tactile chart.}
    \label{fig:python-generated-violin-template-chart}
\end{figure}

% Faceted plot
\begin{figure}[!t]
    \centering
        \includegraphics[width=1\columnwidth]{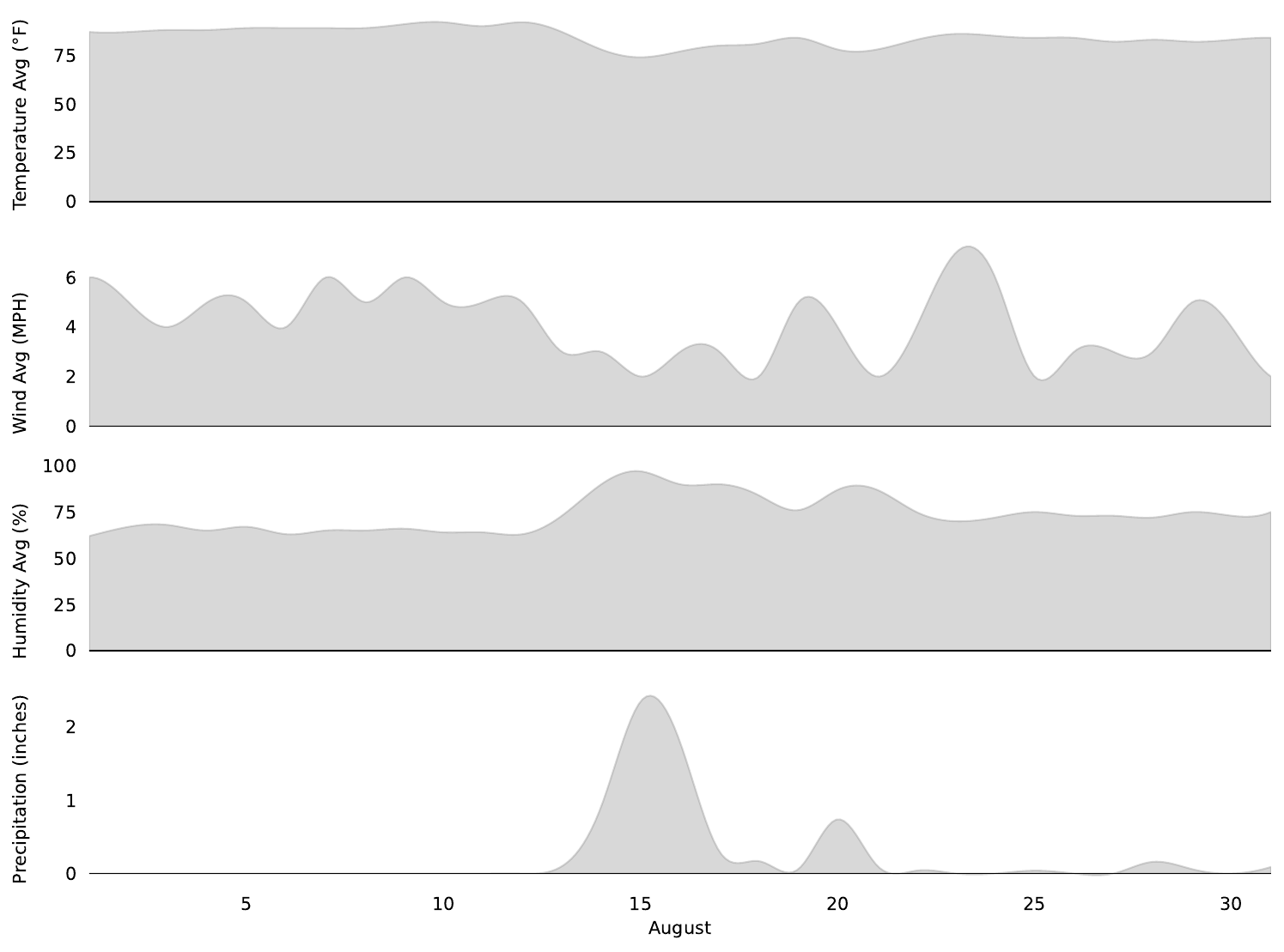}
    \caption{Python-generated faceted line chart used to create the tactile chart.}
    \label{fig:python-generated-faceted-template-chart}
\end{figure}    

% \section{Two variations for each chart type}
% We designed two variations for each of our four chart types (UpSet plot, clustered heatmap, violin plot, and faceted plot). 

% \clearpage
\section{Detailed Design Considerations}
\label{sec:detailed-design-considerations}

In this section, we describe our design choices based on tactile chart design guidelines and related research (as we discussed in \autoref{sec:tactile-charts}). These considerations primarily concern the spacing and styling of chart elements, and the using of the tactile markers. 

\noindent \paragraph{Chart Size}
We chose A4-sized boards, maximizing the size of the chart while keeping it manageable in total size \cite{BANA:2022:Guidelines}. We consistently placed chart titles and legends (if needed) at the top left of each chart. To indicate orientation, there are two common conventions. 

\noindent \paragraph{Orientation Cues} 
To indicate the orientation of tactile graphics, guidelines suggest two conventions (\cite{Race:2019:Designing} and \href{https://www.colorado.edu/project/bbb/creating-tactile-graphics}{\texttt{www\discretionary{}{.}{.}colorado\discretionary{}{.}{.}edu\discretionary{/}{}{/}project\discretionary{/}{}{/}bbb\discretionary{/}{}{/}creating\discretionary{}{-}{-}tactile\discretionary{}{-}{-}graphics}}: one is marking the top-right corner with a triple slash \inlinevis{-1pt}{1em}{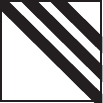}, and the other is physically cutting the top-right corner. We adopted the latter for its simplicity.

\noindent \paragraph{Chart Elements}
Following the guidelines (\eg, \cite{Prescher:2016:Richtlinien, Schuffelen:2002:Editing, BANA:2022:Guidelines}) and related research (\eg, \cite{Engel:2019:User, Engel:2017:Analysis, Engel:2018:User}), we separated elements by 3-6 mm and also positioned labels 3–6 mm from their referenced components. We differentiated line widths by at least 25\%. For dashed lines, we maintained a minimum unit length of 2.5 cm with at least three repetitions of the pattern to ensure clear tactile recognition. For the two bar charts in the UpSet plots, we followed the recommendation for bar charts, and used a bar width of 10 mm. We included both x- and y-axes for plots for violin plots and faceted line plots. We provided the Y-axis labels at the top of the axis (rather than vertically along the axis), while the x-axis labels were placed at the bottom. We avoided arrows at axis endpoints and clearly marked all starting points. We used tick marks only on numerical scales, not on categorical scales. We placed tick marks outside the axis line. 
% UpSet plots and matrices do not have an axis, as they are a tabular representation. 
Given the complexity of UpSet plots and their matrix-based structure, we omitted the axes. Instead, we directly annotated each bar with its corresponding value to enhance interpretability.

\noindent \paragraph{Raised Labels}
We also introduced raised labels to the UpSet plots and clustered heatmaps to guide readers in exploring different parts of the visualization. For example, in the UpSet plots, we labeled the set bar chart, the intersection matrix, and the intersection bar chart as A, B, and C, respectively. We consulted with our blind coauthor on this design, and he suggested that for complex charts, when designers wish to highlight or emphasize specific regions or elements, using raised labels can be an effective approach.

\noindent \paragraph{Braille}
We used Unified English Braille (UEB) Grade 1 consistently for all text labels.
Grade 1 Braille spells words letter by letter without contractions (as Grade 2 does). Although Grade 1 results in longer text compared to Grade 2, it is easier to understand by Braille reading beginners.
Braille uses indicators to denote special symbols that follow. These indicators are special Braille characters. We used them as recommended. Specifically, we used the number indicator (\inlinevis{-2pt}{1em}{inline-indicator-number}) to indicate numeric values, except in contexts where the meaning was unambiguous and space constraints, such as when labeling individual cells in the heatmap. We generally limited the use of capital letters, using it only for cases such as names, where we also included the appropriate capitalization indicators (\inlinevis{-2pt}{1em}{inline-indicator-capital}).
For size, we followed standard Braille sizing guidelines (\eg, \cite{BANA:2022:Guidelines}).

% \clearpage

\section{\rev{3D Model Creation Procedure and Cost}}
\label{sec:3D-model-ceation-procedure-and-cost}

\rev{Based on the 2D designs, we created digital 3D models in SolidWorks (available in \osfrepo\ and on our companion website \companionweb). These models were then fabricated as 3D-printed tactile charts. }

\rev{In the first round of model creation, we designed two variations for each of the four chart types, resulting in a total of eight unique designs. We printed five copies of each design, producing 40 models in total.
Two UpSet designs were printed at the Merrill Engineering Building (MEB) of the University of Utah with a high-quality large-format printer (Prusa XL), yielding 10 models at a total cost of USD 84.15. Given the substantial number of charts required, we also experimented with commercial 3D printing services. We ordered additional models from a provider listed on Craftcloud (\href{https://craftcloud3d.com/}{\texttt{craftcloud3d.com}}), chosen for its fast turnaround time. The remaining 30 models were printed via this provider at a total cost of USD 311.07 (USD 244.85 for production and USD 66.22 for shipping). All prints used white standard PLA with 20\% infill.}

\rev{To reduce both printing time and cost, we initially used a 5mm base board. Two models printed at the university have a 5mm base (one for each UpSet design). Finding this base thickness stable enough, we transitioned to a 3mm base for the remaining 38 models.}

\rev{From the first round of printing, we found the quality of the online printing to be inconsistent. For example, some charts required extensive polishing with sandpaper to make them usable. Consequently, for the second round of printing---which included four final designs for each chart type---we only use the printing services from our university. Given that the 3 mm base had proven sufficiently stable, we further reduced the base thickness to 2 mm. In this second round, we printed eight violin plots, eight clustered heatmaps, two UpSet plots, and two faceted line charts, at a total cost of USD 160.95 (approximately USD 8 per model).}

% We had initially planned to send all participants three tactile charts, but the limited throughput of the printer limited us to a single chart per participant.  

\clearpage
\section{Tactile Chart Exploration Instructions (Initial Version)}
\label{sec:exploration-instruction-initial-version}

We provide instructions on how to explore the four tactile models.
We first drafted an initial version of the instructions and then elicited feedback from our BLV collaborators. Based on their input, we revised the instructions accordingly. In this section, we present the initial version. The initial version corresponds to Design 1 of the tactile chart for each chart type (see \autoref{fig:teaser}). 

\subsection{Tactile Chart Exploration Instructions (Initial Version): UpSet Plot }

Follow these instructions to explore the \textbf{UpSet Plot} tactile model.   
Feel free to stop the exploration at any time or take breaks as needed.

\subsubsection*{Step 1: Orienting the Chart}

\begin{itemize}
    \item Begin by locating the \textbf{cut corner} at the top-right of the page.
    \item Position the chart so that this corner is at the top-right. This ensures the chart is correctly oriented for your exploration.
    \item The tactile model is at the front of the chart. There are two stickers on the back, one with a labeled version of the chart, and a round-shaped one with a QR code that can bring you to a companion website, if you scan it with your phone.
\end{itemize}

\subsubsection*{Step 2: Introduction to the UpSet Plot}

This tactile model represents an UpSet plot, which visualizes how different sets intersect. 

\textbf{Key Concepts:}

\begin{itemize}
    \item \textbf{Sets:} Groups of \textbf{elements} sharing a specific attribute.
    \item \textbf{Intersection:} The overlap of \textbf{elements} between these sets.
\end{itemize}

\textbf{Example Context: } 
This example plot uses attributes of Simpson characters:

\begin{itemize}
    \item Each character represents an element.
    \item Their attributes (e.g., "School," "Blue Hair") form sets. Characters with a specific attribute belong to that set.
    \item Intersections represent overlaps of attributes; for example, a character can belong to multiple sets.
\end{itemize}

\subsubsection*{Step 3: Overview}

To get started, move your hand to the \textbf{top-left corner} to feel the \textbf{title} of the chart, which reads:  ``UpSet: Visualizing Intersecting Sets.'' The \textbf{UpSet plot} has three main sections:

\begin{enumerate}
    \item \textbf{Bottom-Left:} A horizontal bar chart representing the size of each set (i.e., how many elements are in each set).  
    \item \textbf{Bottom-Center:} A matrix of circles showing different types of intersections between sets.  
    \item \textbf{Top-Right:} A vertical bar chart showing the size of each intersection (i.e., how many elements belong to each intersection).
\end{enumerate}

\subsubsection*{Step 4: Exploring the Sections}

\noindent \paragraph{A-Set Size Bars (Bottom-Left)}

\begin{enumerate}
    \item Move your hand to the \textbf{left edge of the chart} and locate \textbf{Marker A}.
    \item Below this marker are \textbf{four horizontal bars}. Each bar represents the size of a set.
\end{enumerate}

\textbf{How to Explore:}

\begin{itemize}
    \item Labels: On the left end of each bar, feel the Braille labels for set names.
    \item Bar Lengths: Longer bars represent larger sets.
    \item Numbers: On the right of each bar, feel the Braille numbers indicating set sizes.
\end{itemize}

From top to bottom, the sets and their size are:

\begin{itemize}
    \item School, 6  
    \item Power Plant, 5  
    \item Evil, 6  
    \item Blue Hair, 3
\end{itemize}

\noindent \paragraph{B-Intersection Matrix (Bottom-Center)}

\begin{enumerate}
    \item Move your hand to the \textbf{center} and locate \textbf{Marker B}.  
    \item Below and to the right of this marker is the \textbf{intersection matrix}, a grid of circles.  
\end{enumerate}

\textbf{Grid Overview:}

\begin{itemize}
    \item \textbf{Rows:} Represent sets, aligned horizontally with the set size bars.  
    \item \textbf{Columns:} Represent different intersections.
\end{itemize}

\textbf{Symbols in the Grid:}

\begin{itemize}
    \item \textbf{Filled Circles:} The set is included in the intersection.  
    \item \textbf{Empty Circles:} The set is not included in the intersection.  
    \item \textbf{Vertical Lines:} Connect filled circles to help trace the columns.
\end{itemize}

\textbf{Types of Intersections:}

\begin{itemize}
\item \textbf{Empty Intersection:} A column with no filled circles.  \textbf{Example:} The first column (leftmost) represents an empty intersection. Trace along the column to feel that all circles are empty. The intersection size is labeled above as \textbf{8}, meaning 8 characters have none of the four attributes.  
   \item \textbf{Individual-Set Intersection:} A column with only one filled circle.  \textbf{Example:} The second column from the left represents an individual-set intersection. It has one filled circle (School). The intersection size is labeled above as \textbf{4}, meaning 4 characters have only the "School" attribute.  
   \item \textbf{Multi-Set Intersection:} A column with multiple filled circles.  \textbf{Example:} The last column (rightmost) has filled circles in the rows for "School" and "Blue Hair" and empty circles for the other two attributes. The intersection size is labeled above as \textbf{1}, meaning 1 character has both attributes but not others.  
   \item \textbf{All-Set Intersection:} A column with all circles filled.  \textbf{Example:} In this plot, no column has all circles filled, meaning no character has all four attributes.
\end{itemize}

\textbf{Intersection Degrees:}

\begin{itemize}
    \item The number of sets included in the intersection increases from \textbf{empty intersections} to \textbf{all-set intersections}, 
    \item \textbf{Low-Degree Intersection:} Few sets are included.  
    \item \textbf{High-Degree Intersection:} Many sets are included.
\end{itemize}

\noindent \paragraph{C-Intersection Sizes (Top-Right)}

\begin{enumerate}
    \item Move your hand to the \textbf{top-right} and locate \textbf{Marker C}.  
    \item Below this marker are eight vertical bars, each corresponding to a possible intersection.
\end{enumerate}

\textbf{How to Explore:}

\begin{itemize}
    \item \textbf{Height:} Taller bars indicate larger intersection sizes.  
    \item \textbf{Labels:} Below each bar, the intersection size is written in Braille.  
    \item \textbf{Correspondence:} The filled-in cells in the column directly below the bar show which sets are part of that intersection.
\end{itemize}

\textbf{Features of Intersections in This Plot:}

\begin{itemize}
    \item The tallest bar (leftmost) represents the \textbf{empty set}, with a size of \textbf{8}.  
    \item Bars are sorted by size in descending order. Moving from left to right, the bar heights decrease, representing intersection sizes down to \textbf{1}.
\end{itemize}

\noindent \paragraph{Step 5: Recap}

\begin{itemize}
    \item Left Section: Horizontal bars show set sizes.  
    \item Center Grid: Circles and lines represent intersections.  
    \item Top-Right Section: Vertical bars show the size of these intersections.
\end{itemize}

% Take your time exploring, and remember that the key is to understand how the three sections relate to each other.

\clearpage
\subsection{Tactile Chart Exploration Instructions (Initial Version): Clustered Heatmap}

Follow these instructions to explore the tactile model of a \textbf{Clustered Heatmap}.  
Feel free to stop at any time or take breaks as needed.

\subsubsection*{Step 1: Orienting the Chart}

\begin{enumerate}
    \item Locate the \textbf{cut corner} at the \textbf{top-right of the page}.  
    \item Position the chart so this corner remains at the \textbf{top-right}. This ensures the chart is properly aligned for your exploration.  
    \item The tactile model is at the front of the chart. There are two stickers on the back, one with a labeled version of the chart, and a round-shaped one with a QR code that can bring you to a companion website, if you scan it with your phone. 
\end{enumerate}

\subsubsection*{Step 2: Introduction to the Clustered Heatmap}

\begin{itemize}
    \item This tactile model represents a \textbf{clustered heatmap}, a chart type that shows patterns and clusters within data.  
    \item The chart combines a \textbf{heatmap} (a grid of squares representing data values) with a \textbf{dendrogram} (a tree-like structure showing clusters).
\end{itemize}

\textbf{Example Context:}

\begin{itemize}
    \item This heatmap represents the number of movies actors have acted in, categorized by genre.
    \begin{itemize}
    \item \textbf{Rows} represent actors.  
    \item \textbf{Columns} represent genres.  \end{itemize}
    \item Each square in the grid represents the number of movies:  
    \begin{itemize}
    \item \textbf{Higher squares} mean more movies.  
    \item \textbf{Lower squares} mean fewer or no movies.
    \end{itemize}
\end{itemize}

\subsubsection*{Step 3: Overview}

Move your hand to the \textbf{top-left corner} to feel the title: ``Heatmap with dendrogram: visualizing clusters.'' Below the title, explore the \textbf{legend}:

\begin{itemize}
    \item A vertical line of \textbf{5 tactile squares} represents different values (from 0 to 50).  
    \item Each square's \textbf{height} corresponds to a data value and is labeled in Braille.
\end{itemize}

You can feel the chart at the right of the legend. 

The chart consists of \textbf{three main sections:}

\begin{enumerate}
    \item \textbf{Heatmap (Center):}  
        A grid of squares showing the intensity of values.  
    \item \textbf{Row Dendrogram (Left):}  
   A branching structure grouping similar rows (actors).  
    \item \textbf{Column Dendrogram (Top):}  
        A similar branching structure grouping similar columns (genres).
\end{enumerate}

\subsubsection*{Step 4: Exploring the Sections}

\noindent \paragraph{Explore the Legend (Top-Left)}

\begin{itemize}
    \item Locate the vertical line of \textbf{5 tactile squares}, representing values:  
    \begin{itemize}
        \item \textbf{Highest} square=50.  
        \item \textbf{Lowest} square=0.
    \end{itemize}
\end{itemize}

\noindent \paragraph{Explore the Heatmap Matrix (Center)}

\begin{itemize}
    \item Find the grid of squares at the center.  
    \item Rows represent \textbf{actors}: Braille labels are along the \textbf{right side} of each row.  
    \item Columns represent \textbf{genres:} Braille labels are along the \textbf{bottom} of each column. Please note that these labels are written vertically. To read them, you will need to rotate your hand 90 degrees counterclockwise or board orientation by 90 degrees clockwise.
\end{itemize}

\textbf{Explore each square:}

\begin{itemize}
    \item Feel its height to determine the value it represents. 
    \begin{itemize}
    \item Higher squares \= more movies in that genre.  
    \item Lower or flat squares \= fewer or no movies.
    \end{itemize}
\end{itemize}

\textbf{Example:}

\begin{itemize}
    \item \textbf{Top row (Dwayne Johnson):}
    \begin{itemize}
        \item \textbf{First square (Action):} Very high (many movies).  
        \item \textbf{Second square (Drama):} Low (few movies).  
        \item \textbf{Third square (Comedy):} Medium-high (moderate number of movies).  
        \item \textbf{Fourth square (Romance):} Flat (no movies).
        \end{itemize}
\end{itemize}

\noindent \paragraph{Explore the Dendrograms}

\textbf{Column Dendrogram (Top)}

\begin{itemize}
    \item Move your hand to the \textbf{top edge} of the heatmap.  
    \item Locate \textbf{Marker A} at the top of the dendrogram.  
    \item Trace the lines downward toward the heatmap grid.  
    \item Feel how the branches connect, grouping similar genres.  \textbf{Example:} The \textbf{rightmost columns} (Romance and Comedy) group together.  Drama joins next, followed by Action, which is the most distinct.
\end{itemize}

\textbf{Row Dendrogram (Left)}

\begin{enumerate}
    \item Move your hand to the \textbf{left edge} of the heatmap.  
    \item Locate \textbf{Marker B} at the top of the dendrogram.  
    \item Trace the lines to the right toward the heatmap grid.  
    \item Feel how branches group similar actors.  \textbf{Example:}  
    \begin{itemize}
       \item  \textbf{Dwayne Johnson} is distinct (specializes in Action).  
       \item  \textbf{Julia Roberts and Leonardo DiCaprio} cluster together (focus on Drama and Comedy).  
       \item  \textbf{Jennifer Lawrence and Tom Hanks} cluster next (focus mainly on Drama).
       \end{itemize}
\end{enumerate}

\subsubsection*{Step 5: Recap}

\begin{itemize}
    \item The \textbf{Heatmap} displays values using a matrix of raised squares.  
    \item The \textbf{Dendrograms} show clustering relationships between rows (actors) and columns (genres).  
    \item Together, they illustrate patterns in the data.
\end{itemize}

\clearpage
\subsection{Tactile Chart Exploration Instructions (Initial Version): Violin Plot}

Follow these step-by-step instructions to explore the tactile model of the \textbf{Violin plot}. Take breaks as needed to ensure a thorough understanding.

\subsubsection*{Step 1: Orienting the Chart}

\begin{enumerate}
    \item Begin by locating the \textbf{cut corner} at the \textbf{top-right} of the tactile model.  
    \item Position the model so that this cut corner is at the \textbf{top-right}, ensuring correct orientation for exploration.  
    \item The tactile model is at the front of the chart. There are two stickers on the back, one with a labeled version of the chart, and a round-shaped one with a QR code that can bring you to a companion website, if you scan it with your phone. 
\end{enumerate}

\subsubsection*{Step 2: Introduction to the Violin Plot}

This tactile model represents a \textbf{Violin plot}, which visualizes distributions of quantitative data for one or more categories.

\textbf{Example Context: }

\begin{itemize}
    \item In this model, we are exploring the \textbf{body mass (in pounds)} of three penguin species:

\begin{itemize}
    \item \textbf{Adelie}  
    \item \textbf{Chinstrap}  
    \item \textbf{Gentoo}
\end{itemize}
\item Each \textbf{violin shape} shows how body mass varies within each species.
\end{itemize}
\subsubsection*{Step 3: Overview}

\begin{itemize}
    \item Move your hand to the \textbf{top-left corner} to locate the title, which reads: ``Violin plot visualizing distributions across categories.''
    \item Below the title, explore the \textbf{legend}. From top to bottom, there are three lines with different thicknesses:
    \begin{enumerate}
        \item \textbf{Thin line}: Represents \textbf{Adelie} penguins.
        \item \textbf{Middle line}: Represents \textbf{Chinstrap} penguins.
        \item \textbf{Thick line}: Represents \textbf{Gentoo} penguins.
    \end{enumerate}
    \item The species names are in Braille next to each line in the legend.
    \item To the right of the legend, you will find the \textbf{Violin plot}. It has three main areas:
    \begin{enumerate}
        \item \textbf{Y-axis} (Left): A vertical line with Braille tick marks, representing body mass in pounds.
        \item \textbf{X-axis} (Bottom): A horizontal line with evenly spaced Braille labels for each species: \textbf{Adelie}, \textbf{Chinstrap}, and \textbf{Gentoo}.
        \item \textbf{Violin Shapes} (Center): Three violin shapes, each positioned above a species label, showing the distribution of body masses for that species.
    \end{enumerate}
\end{itemize}

\subsubsection*{Step 4: Detailed Exploration}

\textbf{Finding the Origin:  } 
   \begin{itemize}
       \item Move your hand to the \textbf{bottom-left corner} of the plot. This is where the \textbf{Y-axis} and \textbf{X-axis} intersect.
   \end{itemize}

\textbf{Y-Axis and Scale:  }
   \begin{itemize}
       \item Start at the \textbf{bottom-left corner} and trace the \textbf{Y-axis} upward.
       \item Feel the evenly spaced tick marks and their Braille labels, which represent body mass measurements:
         \begin{itemize}
               \item The bottom tick (at the corner) is \textbf{5 pounds}.
               \item The next tick above is \textbf{6 pounds}, then \textbf{8 pounds}, \textbf{10 pounds}, and so on, increasing evenly to the top tick, which is \textbf{14 pounds}.
        \end{itemize}
   \end{itemize}

\textbf{X-Axis and Species Labels:  }
   \begin{itemize}
       \item Return your hand to the \textbf{bottom-left corner} and trace the \textbf{X-axis} horizontally.
       \item From left to right, you will feel three Braille labels identifying the penguin species:  \textbf{Adelie}, \textbf{Chinstrap}, and \textbf{Gentoo}.
       \item These labels correspond to the violins directly above them.
   \end{itemize}

\textbf{Violin Shapes:  }
   \begin{itemize}
       \item Locate the \textbf{first violin} above \textbf{Adelie} on the X-axis. This represents the distribution of body masses for Adelie penguins.
       \item Trace the \textbf{outer contour} of the violin with your fingers:
       \begin{itemize}
           \item The violin is symmetric.
           \item Wider parts (more curved sections) represent more penguins with body masses at those values.
           \item Narrower parts represent fewer penguins with those body masses.
       \end{itemize}
       \item Refer to the \textbf{Y-axis} or tactile \textbf{grid lines} to identify the corresponding body mass values.
       \item Inside each violin, you will feel a \textbf{dashed horizontal line}, which marks the \textbf{median body mass} for that species.
       \item Move your hand to the right and repeat this process for the \textbf{Chinstrap} and \textbf{Gentoo} violins.
       \item Touch the three violins. Compare the shapes and medians to understand differences in distribution.
       \begin{itemize}
           \item \textbf{Left violin (Adelie penguins):} The distribution is unimodal, and it has one peak in the middle. The median body mass is about 8.2 pounds, with most Adelie penguins' body masses clustering close to this value.
           \item \textbf{Middle violin (Chinstrap penguins):} The distribution is also unimodal, with a single peak in the middle. The median body mass is about 8.2 pounds. Most Chinstrap penguins' body masses are concentrated around 8.2 pounds, and their weights are more tightly clustered near the middle compared to Adelie penguins.
           \item \textbf{Right violin (Gentoo penguins):} The distribution is bimodal. It has two peaks, one near each end. The median body mass of Chinstrip penguins is roughly 11 pounds, which is larger than Adelie and Chinstrap penguins. Gentoo penguins' body masses are concentrated around 10.5 pounds and 12 pounds, which form the two peaks.
       \end{itemize}
   \end{itemize}

\subsubsection*{Step 5: Recap}

\begin{itemize}
    \item  \textbf{Y-Axis}: Represents body mass in pounds, increasing from 5 to 14 (vertical scale).
    \item  \textbf{X-Axis}: Identifies penguin species (horizontal labels).
    \item  \textbf{Violin Shapes}: Show the distribution of body masses, with wider sections indicating higher density and dashed horizontal lines marking the medians.
\end{itemize}

\clearpage
\subsection{Tactile Chart Exploration Instructions (Initial Version): Faceted Line Chart}

Follow these step-by-step instructions to explore the tactile model of the \textbf{Faceted Area Plot}. 
Take breaks as needed to ensure a thorough understanding.

\subsubsection*{Step 1: Orienting the Chart}

\begin{enumerate}
    \item Begin by locating the \textbf{cut corner} at the \textbf{top-right} of the tactile model.  
    \item Position the model so that this cut corner is at the \textbf{top-right}, ensuring correct orientation for exploration. 
    \item The tactile model is at the front of the chart. There are two stickers on the back, one with a labeled version of the chart, and a round-shaped one with a QR code that can bring you to a companion website, if you scan it with your phone. 
\end{enumerate}

\subsubsection*{Step 2: Introduction to the Faceted Area Plot}

\begin{itemize}
\item This tactile model represents a \textbf{Faceted Area Plot}, which visualizes trends in multiple related datasets.  
\item \textbf{Example Context:}  
  The model displays weather trends in Austin, Texas, using four variables:  
  \begin{itemize}
    \item \textbf{Average Temperature} (°F)  
    \item \textbf{Average Wind Speed} (mph)  
    \item \textbf{Average Humidity} (\%)  
    \item \textbf{Total Precipitation} (inches)
  \end{itemize}
\end{itemize}

\subsubsection*{Step 3: Overview}

\textbf{Title:}  
Move your hand to the \textbf{top-left corner} to locate the Braille title:  ``Faceted area plot visualizing trends.''
     
\textbf{Chart Structure:}  
   \begin{itemize}
       \item The plot consists of \textbf{four area charts}, stacked vertically. Each chart represents one variable.
       \item All charts \textbf{share the same X-axis} (at the bottom) but have their \textbf{own Y-axis} (on the right).
       \item The layout can be divided into three vertical sections:
       \begin{itemize}
           \item \textbf{Left:} Braille labels indicating variable names and units.
           \item \textbf{Middle:} The embossed area chart for each variable.
           \item \textbf{Right:} The Y-axis with tick marks showing the scale for that variable.
        \end{itemize}
   \end{itemize}

\subsubsection*{Step 4: Detailed Exploration}

\noindent \paragraph{Variable Names and Units}

\begin{itemize}
\item Place your hand on the \textbf{left side} of the board and move \textbf{vertically}.  
\item You will feel the Braille labels for each variable, listed from \textbf{top to bottom}:  
    \begin{enumerate}
        \item \textbf{Average Temperature} (°F)  
        \item \textbf{Average Wind Speed} (mph)  
        \item \textbf{Average Humidity} (\%)  
        \item \textbf{Total Precipitation} (inches)
    \end{enumerate}
\end{itemize}

\noindent \paragraph{Exploring the X-Axis}

\begin{itemize}
\item Locate the \textbf{bottom of the plot} to find the shared X-axis. 
\item Move your hand \textbf{horizontally} along this axis, which represents the days of the month.
    \begin{itemize}
      \item Tick marks are \textbf{evenly spaced}.  
      \item \textbf{Longer tick marks} indicate intervals of 5 or 10 days.
    \end{itemize}
\end{itemize}

\noindent \paragraph{Exploring an Example Section} 

\textbf{Total Precipitation (Bottom Section):}  
   \begin{itemize}
       \item Start at the \textbf{bottom section}, which represents \textbf{Total Precipitation (inches)}.
       \item Move to the \textbf{rightmost corner}, where the X-axis and Y-axis intersect.
       \item Trace the \textbf{Y-axis vertically upwards}; the scale ranges from \textbf{0 to 2.5 inches}.
       \item Return to the X-axis and \textbf{move your hand to the left} to start tracing the curve.
       \item Horizontally trace the area curve from \textbf{left to right}:
       \begin{itemize}
           \item Initially, there is no curve (no rainfall for the first half of the month).
           \item Around \textbf{August 15}, you will feel a \textbf{large peak}, indicating the maximum rainfall.
           \item Beyond the peak, smaller bumps represent \textbf{minor rainfall} later in the month.
           \end{itemize}
       \item While tracing the curve, refer to the \textbf{Y-axis} for rainfall amounts and the \textbf{X-axis} for the corresponding dates.
   \end{itemize}

\noindent \paragraph{Exploring the Other Variables} 

\textbf{Average Humidity\%): }
   \begin{itemize}
       \item Move your hand \textbf{one section up} to find the humidity plot.
       \item Locate the \textbf{Braille label} on the left: ``Average Humidity (\%).''
       \item Trace the curve of the area:
       \begin{itemize}
           \item Humidity is generally \textbf{high and stable}.
           \item A \textbf{slight peak} occurs around the middle of the month.
       \end{itemize}
       \item Compare this peak with the precipitation section below; you will notice a rainfall event corresponds to increased humidity.
       \item 
   \end{itemize}
   
\textbf{Average Wind Speed (mph):}  
   \begin{itemize}
       \item Move your hand \textbf{one section up} to the wind speed plot.
       \item Locate the \textbf{Braille label} on the left: *"Average Wind Speed (mph)."*
       \item Trace the curve:  You will feel \textbf{periodic peaks and valleys}, indicating fluctuations in wind speed.
   \end{itemize}

\textbf{Average Temperature (°F):}  
   \begin{itemize}
       \item Move to the \textbf{top section} of the plot.
       \item Locate the \textbf{Braille label} on the left: ``Average Temperature (°F).''
       \item Trace the curve:
       \begin{itemize}
           \item Temperature remains \textbf{stable throughout the month}, hovering around \textbf{85°F}.
           \item Even during the mid-month rainfall, temperature changes only slightly, indicating \textbf{minimal impact}.
       \end{itemize}
   \end{itemize}

\noindent \paragraph{Comparison Between Variables} 

To compare variables at a specific time, trace \textbf{vertically} across the plots.  For example, during the mid-month rainfall, compare the \textbf{humidity and precipitation peaks} while noting the stable temperature.

\subsubsection*{Step 5: Recap}

\begin{itemize}
    \item \textbf{X-Axis:} Represents days of the month (at the bottom), shared by all area plots.
    \item \textbf{Y-Axis:} Represents the scale for each variable (on the right side of each plot).
    \item \textbf{Faceted Areas:} Show trends for four weather variables:
    \begin{itemize}
        \item \textbf{Top:} Average Temperature.
        \item \textbf{Second:} Average Wind Speed.
        \item \textbf{Third:} Average Humidity.
        \item \textbf{Bottom:} Total Precipitation.
    \end{itemize}
\end{itemize}

\textbf{Exploration Tips:}

\begin{itemize}
    \item Trace \textbf{horizontally} to observe trends for a single variable.  
    \item Trace \textbf{vertically} to compare different variables at the same time.
\end{itemize}

\clearpage
\section{Tactile Chart Exploration Instructions (Final Version)}
\label{sec:exploration-instruction-final-version}
We provide instructions on how to explore the four tactile models.
We first drafted an initial version of the instructions and then elicited feedback from our BLV collaborators. Based on their input, we revised the instructions accordingly. The initial version is provided in \autoref{sec:exploration-instruction-initial-version}. 
In this section, we present the final versions, which correspond to the tactile chart design for each chart type (see \autoref{fig:teaser}).
These final version instructions are also hosted on our accessible website: \companionweb.

\subsection{Tactile Chart Exploration Instructions (Final Version): UpSet Plot}

Below is the final version of the tactile UpSet plot exploration instructions, which can also be accessed at \href{https://vdl.sci.utah.edu/tactile-charts/upset-plot/instructions-tactile}{\texttt{vdl\discretionary{}{.}{.}sci\discretionary{}{.}{.}utah\discretionary{}{.}{.}edu\discretionary{/}{}{/}tactile\discretionary{}{-}{-}charts\discretionary{/}{}{/}upset\discretionary{}{-}{-}plot\discretionary{/}{}{/}instructions\discretionary{}{-}{-}tactile}}.

\vspace{5mm}

Follow these instructions to explore the tactile model of a \textbf{UpSet Plot}.

\subsubsection*{Step 1: Orienting the Chart}

\begin{itemize}
    \item Locate the \textbf{cut corner} at the \textbf{top-right} of the board.  Position the chart so this corner remains at the top-right.
    \item On the back, there are \textbf{two stickers}:
    \begin{itemize}
        \item A \textbf{smaller square label} near the cut corner contains a \textbf{QR code}. Scanning it with your phone will take you to a companion website.
        \item A \textbf{larger rectangular label} provides a labeled version of the chart.
    \end{itemize}
\end{itemize}

\subsubsection*{Step 2: Introduction to the UpSet Plot}

\begin{itemize}
    \item This tactile model represents an UpSet plot, which visualizes how different sets intersect.
    \begin{itemize}
        \item \textbf{Sets:} Groups of \textbf{elements} sharing a specific attribute.
        \item \textbf{Intersection:} The overlap of \textbf{elements} between these sets.
    \end{itemize}
    \item Move your hand to the top-left corner to feel the title: ``UpSet Plot.''
    \item Explore the board to get an overview. The chart is located on the right side of the board.
    \item The \textbf{UpSet plot} has three main sections:
    \begin{itemize}
        \item \textbf{Bottom-Left} of the chart: A horizontal bar chart representing the size of each set (i.e., how many elements are in each set).
        \item \textbf{Bottom-Center} of the chart: A matrix of circles showing different types of intersections between sets.
        \item \textbf{Top-Right} of the chart: A vertical bar chart showing the size of each intersection (i.e., how many elements belong to each intersection).
    \end{itemize}
    \item \textbf{The data: }
    \begin{itemize}
        \item In this model, we are exploring the \textbf{Simpson characters} and their attributes: School, Blue Hair, Evil, Power Plant.
        \item Each character represents an element.
        \item Their attributes (e.g., "School," "Blue Hair") form sets. Characters with a specific attribute belong to that set.
        \item Intersections represent overlaps of attributes; for example, a character can belong to multiple sets.
    \end{itemize}
\end{itemize}

\subsubsection*{Step 3: Exploring the Sections}

\noindent \paragraph{Set Size Bars}

\begin{itemize}
    \item Move your hand to the \textbf{left edge of the chart}
    \item There are \textbf{four horizontal bars}. Each bar represents the size of a set.
    \item \textbf{Labels:} On the left end of each bar, feel the Braille labels for set names.
    \item \textbf{Bar Lengths:} Longer bars represent larger sets.
    \item \textbf{Reference Lines:} Feel the \textbf{lines and holes} within each bar:
    \begin{itemize}
        \item Each line represents \textbf{one unit}
        \item Every \textbf{fifth unit} is marked with a slightly thicker reference line
        \item These tactile markers help you count the total value of each bar
    \end{itemize}
    \item Numbers: On the right of each bar, feel the Braille numbers indicating set sizes.
    \item From top to bottom, the sets and their size are:
    \begin{itemize}
        \item School, 6
        \item Power Plant, 5
        \item Evil, 6
        \item Blue Hair, 3
    \end{itemize}
\end{itemize}

\noindent \paragraph{Intersection Matrix} 

Move your hand to the \textbf{center} of the chart  
Below and to the right of this marker is the \textbf{intersection matrix}, a grid of circles.

\textbf{Grid Overview:}
   \begin{itemize}
       \item \textbf{Rows:} Represent sets, aligned horizontally with the set size bars.
       \item \textbf{Columns:} Represent different intersections.
   \end{itemize}
   
\textbf{Symbols in the Grid:}
\begin{itemize}
    \item \textbf{Filled Circles:} The set is included in the intersection.
    \item \textbf{Empty Circles:} The set is not included in the intersection.
    \item \textbf{Vertical Lines:} Connect filled circles to help trace the columns.
\end{itemize}

\textbf{Types of Intersections:}
\begin{itemize}
    \item \textbf{Empty Intersection:} A column with no filled circles.  This is a special intersection with elements that have none of the attributes. \textbf{Example:} The first column (leftmost) represents an empty intersection. Trace along the column to feel that all circles are empty. The intersection size is labeled above as \textbf{8}, meaning 8 characters have none of the four attributes.
    \item \textbf{Individual-Set Intersection:} A column with only one filled circle. \textbf{Example:} The second column from the left represents an individual-set intersection. It has one filled circle (School). The intersection size is labeled above as \textbf{4}, meaning 4 characters have only the "School" attribute.
    \item \textbf{Multi-Set Intersection:} A column with multiple filled circles. \textbf{Example:} The last column (rightmost) has filled circles in the rows for "School" and "Blue Hair" and empty circles for the other two attributes. The intersection size is labeled above as \textbf{1}, meaning 1 character has both attributes but not others.
    \item \textbf{All-Set Intersection:} A column with all circles filled. \textbf{Example:} In this plot, no column has all circles filled, meaning no character has all four attributes.
\end{itemize}

\textbf{Intersection Degrees:}
\begin{itemize}
    \item The number of sets included in the intersection increases from \textbf{empty intersections} to \textbf{all-set intersections}.
    \item \textbf{Low-Degree Intersection:} Few sets are included.
    \item \textbf{High-Degree Intersection:} Many sets are included.
\end{itemize}

\noindent \paragraph{Intersection Size Bars}
\begin{itemize}
    \item Move your hand to the \textbf{top-right} of the chart
    \item There are \textbf{eight vertical bars}, each corresponding to a possible intersection.
    \item \textbf{Height:} Taller bars indicate larger intersection sizes.
    \item \textbf{Labels:} Below each bar, the intersection size is written in Braille.
    \item \textbf{Reference Lines:} Same as the horizontal bars.
    \begin{itemize}
        \item Each line represents \textbf{one unit}
        \item Every \textbf{fifth unit} is marked with a slightly thicker reference line
        \item These tactile markers help you count the total value of each bar
    \end{itemize}
    \item \textbf{Correspondence:} The filled-in cells in the column directly below the bar show which sets are part of that intersection.
    \item \textbf{Features of Intersections in This Plot:}
    \begin{itemize}
        \item The tallest bar (leftmost) represents the \textbf{empty set}, with a size of \textbf{8}.
        \item Bars are sorted by size in descending order. Moving from left to right, the bar heights decrease, representing intersection sizes down to \textbf{1}.
    \end{itemize}
\end{itemize}

\subsubsection*{Recap}
\begin{itemize}
    \item \textbf{Left Section:} Horizontal bars show set sizes.
    \item \textbf{Center Grid:} Circles and lines represent intersections.
    \item \textbf{Top-Right Section:} Vertical bars show the size of these intersections.
\end{itemize}

\clearpage
\subsection{Tactile Chart Exploration Instructions (Final Version): Clustered Heatmap}
\label{sec:tactile-instructions-final-heatmap}
Below is the final version of the tactile clustered heatmap exploration instructions, which can also be accessed at \href{https://vdl.sci.utah.edu/tactile-charts/clustered-heatmap/instructions-tactile}{\texttt{vdl\discretionary{}{.}{.}sci\discretionary{}{.}{.}utah\discretionary{}{.}{.}edu/tactile\discretionary{}{-}{-}charts\discretionary{/}{}{/}clustered\discretionary{}{-}{-}heatmap\discretionary{/}{}{/}instructions\discretionary{}{-}{-}tactile}}.

\vspace{5mm}

Follow these instructions to explore the tactile model of a clustered heatmap.

\subsubsection*{Step 1: Orienting the Chart}

\begin{itemize}
    \item Locate the \textbf{cut corner} at the \textbf{top-right} of the board.  Position the chart so this corner remains at the top-right.
    \item On the back, there are \textbf{two stickers}:
    \begin{itemize}
        \item A \textbf{smaller square label} near the cut corner contains a \textbf{QR code}. Scanning it with your phone will take you to a companion website.
        \item A \textbf{larger rectangular label} provides a labeled version of the chart.
    \end{itemize}
\end{itemize}

\subsubsection*{Step 2: Introduction to the Clustered Heatmap}

\begin{itemize}
    \item This tactile model represents a \textbf{clustered heatmap} of tabular data.   
    \item Move your hand to the top-left corner to feel the title: "Clustered Heatmap."  
    \item Broadly explore the board to get an overview. The chart is located on the right side of the board.  
    \item This chart type combines a heatmap and dendrograms:
    \begin{itemize}
      \item \textbf{The Heatmap is} a matrix of squares representing data values through heights.  
      \item \textbf{The Dendrograms are} Tree-like diagrams positioned above and to the right of the heatmap, showing the similarity between rows and columns.  
      \end{itemize}
    \item \textbf{The data:} 
     \begin{itemize}
      \item This clustered heatmap visualizes how frequently actors appear in different movie genres.   
      \item Similar actors and genres are grouped together. 
      \end{itemize}
\end{itemize}

\subsubsection*{Step 3: Explore the Heatmap}

\textbf{Locate the Heatmap:} On the \textbf{right side} of the board, find the matrix of squares (the heatmap).

\textbf{Rows and Columns:}  
\begin{itemize}
    \item \textbf{Rows represent 5 actors, } labeled in Braille on the left side of each row, from top to bottom:
    \begin{itemize}
        \item Dwayne Johnson
        \item Julia Roberts
        \item Tom Hanks
        \item Jennifer Lawrence
        \item Leonardo DiCaprio
    \end{itemize}
    \item \textbf{Columns represent 4 movie genres, } labeled in Braille at the bottom of each column using abbreviations, from left to right:
    \begin{itemize}
        \item a: Action
        \item d: Drama
        \item c: Comedy
        \item r: Romance
    \end{itemize}
\end{itemize}

\textbf{Legend:} On the top-left of the board, below the chart title, find the legend enclosed in a rectangular frame. It explains the column abbreviations.  

\textbf{Squares:}  
\begin{itemize}
    \item \textbf{Height:} The height of the squares encodes the number  of movies an actor has performed in in a specific genre. High squares indicate many moves, whereas lower or flat squares indicate fewer or no movies.
    \item \textbf{Braille Numbers:} Each square has a Braille number representing its value.
    \item \textbf{Example Exploration:}   Top row (Dwayne Johnson), the four squares from left to right:
    \begin{itemize}
        \item Action: Very high (45 movies). This is the highest square in the whole heatmap, so it is the biggest value.
        \item Drama: Low (10 movies).
        \item Comedy: Medium-high (28 movies).
        \item Romance: Flat (0 movies). This is the smallest value in the heatmap.
    \end{itemize}
\end{itemize}

\subsubsection*{Step 4: Exploring the Dendrograms}

Dendrograms show hierarchical grouping based on similarity.

\noindent \paragraph{Row Dendrogram (Actor Clusters)}   

Locate the Row Dendrogram: On the right edge of the heatmap, find the tree-like structure.  From the right end of each row of the heatmap, you'll feel a line extending to the right.  These lines connect rows to each other and group together hierarchically. The more quickly the lines group together, the more similar the two rows are.  

Clusters are groups of similar rows with close connections.  This chart shows two main clusters among actors.

\textbf{Cluster 1: Julia Roberts and Tom Hanks} (rows 2 and 3 from top):  
    \begin{itemize}
      \item Trace their lines to see how they connect to each other quickly.  
      \item Confirm in the heatmap: Feel the similar height patterns of squares in their rows:  
      \begin{itemize}
        \item Both are low in Action (column 1 from left).  
        \item Both are high in Drama and Comedy (columns 2 and 3).  
        \item Both are moderately high in Romance (column 4).
    \end{itemize}
    \end{itemize}

\textbf{Cluster 2: Jennifer Lawrence and Leonardo DiCaprio} (rows 4 and 5):  
\begin{itemize}
    \item Similarly, trace their lines to see how they group quickly.
    \item Confirm in the heatmap: Feel the similar height patterns of squares in their rows:
    \begin{itemize}
        \item Both are high in Drama.
        \item Both are low in other genres.
    \end{itemize}
\end{itemize}

\textbf{Outlier:} Rows that deviate from the main clusters are outliers.  
\begin{itemize}
    \item This chart shows one outlier among actors.
    \item Dwayne Johnson (row 1):
    \begin{itemize}
        \item Trace his line to see how it remains separate for a long time.
        \item Confirm in the heatmap: The pattern in row 1 differs significantly from other rows. For example, row 1 shows a very high value in Action (column 1), while other rows show low values in this column.
    \end{itemize}
\end{itemize}

\noindent \paragraph{Column Dendrogram (Movie Genre Clusters)}  

Locate the Column Dendrogram at the top of the heatmap.Lines extend from each column upwards, clustering genres hierarchically. Its principle is the same as the row dendrogram, but this time it applies to the genres in the columns.
    
\textbf{Cluster and Outlier:}
    \begin{itemize}
    \item The movie genres are clustered incrementally in a step-by-step manner, starting with closely related pairs and gradually incorporating more distinct genres. The tall tree indicates that they are generally not very similar.
    \begin{itemize}
        \item \textbf{Romance and Comedy (columns 3 and 4 from left):} These two genres group together first, indicating they are the most similar movie genres among the four genres in terms of actor involvement patterns.
        \item \textbf{Drama (column 2):} Joins next.
        \item \textbf{Action (column 1):} Joins last, being the most distinct genre (outlier).
    \end{itemize}
    \end{itemize}

\subsubsection*{Recap}

\begin{itemize}
    \item The heatmap displays values using a matrix of raised squares.  
    \item The dendrograms reveal similarity between rows and columns.
\end{itemize}

\clearpage
\subsection{Tactile Chart Exploration Instructions (Final Version): Violin Plot}
\label{sec:tactile-instructions-final-violin}

Below is the final version of the tactile violin plot exploration instructions, which can also be accessed at \href{https://vdl.sci.utah.edu/tactile-charts/violin-plot/instructions-tactile}{\texttt{vdl\discretionary{}{.}{.}sci\discretionary{}{.}{.}utah\discretionary{}{.}{.}edu/tactile\discretionary{}{-}{-}charts\discretionary{/}{}{/}violin\discretionary{}{-}{-}plot\discretionary{/}{}{/}instructions\discretionary{}{-}{-}tactile}}.

\vspace{5mm}

Follow these instructions to explore the tactile model of a violin plot.
\subsubsection*{Step 1: Orienting the Chart}

\begin{itemize}
    \item Locate the \textbf{cut corner} at the \textbf{top-right} of the board.  Position the chart so this corner remains at the top-right.
    \item On the back, there are \textbf{two stickers}:
    \begin{itemize}
        \item A \textbf{smaller square label} near the cut corner contains a \textbf{QR code}. Scanning it with your phone will take you to a companion website.
        \item A \textbf{larger rectangular label} provides a labeled version of the chart.
    \end{itemize}
\end{itemize}

\subsubsection*{Step 2: Introduction to the Violin Plot}

\begin{itemize}
\item This tactile model represents a \textbf{Violin plot}, which visualizes distributions of quantitative data for one or more categories. A distribution describes how values in a dataset are spread out. For example, if you collect the weight of 100 penguins, the distribution will show how many penguins are light, average weight, and heavy, respectively.   
\item Move your hand to the top-left corner to feel the title: "Violin Plot."  
\item Explore the board to get an overview. The chart is located on the center of the board.  
\item \textbf{The data:} In this model, we are exploring the \textbf{body mass (in pounds)} of three penguin species:  
\begin{itemize}
    \item \textbf{Adelie}  
    \item \textbf{Chinstrap}  
    \item \textbf{Gentoo}  
\end{itemize}
\item There are \textbf{three violin-plot shapes}, each showing the body mass distribution for a species of penguins.
\end{itemize}

\subsubsection*{Step 3: Exploring the Violin Plot}

\noindent \paragraph{Finding the Origin}  

Move your hand to the \textbf{bottom-left corner} of the plot. This is where the \textbf{Y-axis} and \textbf{X-axis} intersect.  

\noindent \paragraph{Y-Axis and Scale}
\begin{itemize}
    \item Start at the \textbf{bottom-left corner} and trace the \textbf{Y-axis} upward.
    \item The Y-axis is a vertical line with tick marks and Braille labels, encoding body mass in pounds.
    \item At the top of Y-axis is the y-axis label in Braille: body mass (lb)
    \item Feel the evenly spaced tick marks and their Braille labels, which represent body mass measurements:
    \begin{itemize}
        \item The bottom tick (at the corner) is \textbf{5 pounds}, which means the Y-axis starts from 5 pounds.
        \item The next tick above is \textbf{6 pounds}, then \textbf{7 pounds}, \textbf{8 pounds}, and so on, increasing evenly to the top tick, which is \textbf{14 pounds}.
        \item The even numbers are marked out with Braille numbers.
    \end{itemize}
\end{itemize}

\noindent \paragraph{X-Axis and Species Labels}  
  \begin{itemize}
      \item Return your hand to the \textbf{bottom-left corner} and trace the \textbf{X-axis} horizontally.
      \item The X-axis is a horizontal line with evenly spaced Braille labels for each species, from left to right: \textbf{Adelie}, \textbf{Chinstrap}, and \textbf{Gentoo}.
      \item These labels correspond to the violin plots directly above them.
  \end{itemize}

\noindent \paragraph{Violin Shapes}  
Three violin shapes, each positioned above a species label show the distribution of body masses for that species.  
\begin{itemize}
    \item Locate the \textbf{first violin} from the left. This represents the distribution of body masses for \textbf{Adelie} penguins.
    \item Trace the \textbf{outer contour} of the violin with your fingers:
    \begin{itemize}
        \item The violin is symmetrical left to right.
        \item Wider sections of the violin represent more values, meaning that a higher number of penguins have a body mass around this value.
        \item Skinnier sections represent fewer values, meaning that few  penguins have a body mass around this value.
    \end{itemize}
    \item Refer to the \textbf{Y-axis} or tactile \textbf{grid lines} to identify the corresponding body mass values.
    \item Inside each violin, you will feel a \textbf{dashed horizontal line}, which marks the \textbf{median body mass} for that species.
    \item Move your hand to the right and repeat this process for the \textbf{Chinstrap} and \textbf{Gentoo} violins.
\end{itemize}

Touch the three violins to compare the shapes and medians to understand differences in distribution.   

\textbf{For the left violin representing Adelie penguins note that:}
    \begin{itemize}
        \item The distribution has one peak, meaning that it is unimodal, and is \textbf{symmetrical}.
        \item The widest section is centered at the \textbf{median (8.16 pounds)}, where most data points are concentrated, indicating that most Adelie penguins have body masses around this value.
        \item The density tapers off symmetrically on both sides, showing a relatively balanced distribution with fewer penguins at the lower and upper extremes.
    \end{itemize}

\textbf{For the middle violin, representing Chinstrap penguins:}
    \begin{itemize}
        \item The distribution also has one peak and is \textbf{symmetrical}. The peak near the median body mass is about 8.18 pounds.
        \item The peak is more distinct compared to Adelie penguins. The violin is wider, indicating that Chinstrap penguins' body masses are more tightly clustered around the median. The tails at both ends are skinnier than that of Adelie penguins. This suggests lower variability in body mass compared to Adelie penguins.
    \end{itemize}
\textbf{For the right violin, representing Gentoo penguins:}
    
    \begin{itemize}
    \item The distribution is \textbf{bimodal}, meaning there are two distinct peaks in body mass. The first peak appears around \textbf{10.5 pounds}, and the second peak near \textbf{12 pounds}, forming two areas of high density. This suggests that Gentoo penguins exhibit greater variation in body mass, possibly due to differences in sex with males having distinctly higher body mass than females.
    \item The median body mass of \textbf{11.02 pounds} is \textbf{higher} than that of Adelie and Chinstrap penguins, and about as high as the heaviest penguins of the other species.
    \begin{itemize}
    \item The median is closer to the lower peak (10.5 pounds) rather than centered between the two peaks, indicating that more Gentoo penguins have body masses around the lower peak than the higher peak. This suggests an asymmetry in distribution, where the higher peak represents fewer but more widely spread heavier penguins.
    \item The violin also has a more elongated distribution compared to the other two species, showing Gentoo penguins have a larger body mass range.
    \end{itemize}
\end{itemize}

\subsubsection*{Step 4: Recap}

\begin{itemize}
    \item \textbf{Y-Axis}: Represents body mass in pounds, increasing from 5 to 14 (vertical scale).  
    \item \textbf{X-Axis}: Identifies penguin species (horizontal labels).  
    \item \textbf{Violin Shapes}: Show the distribution of body masses, with wider sections indicating higher density and horizontal lines marking the medians.
\end{itemize}
\clearpage
\subsection{Tactile Chart Exploration Instructions (Final Version): Facted Line Chart}

Below is the final version of the tactile faceted line chart exploration instructions, which can also be accessed at \href{https://vdl.sci.utah.edu/tactile-charts/faceted-plot/instructions-tactile}{\texttt{vdl\discretionary{}{.}{.}sci\discretionary{}{.}{.}utah\discretionary{}{.}{.}edu/tactile\discretionary{}{-}{-}charts\discretionary{/}{}{/}faceted\discretionary{}{-}{-}plot\discretionary{/}{}{/}instructions\discretionary{}{-}{-}tactile}}.

\vspace{5mm}

Follow these instructions to explore the tactile model of a violin plot.

\subsubsection*{Step 1: Orienting the Chart}

\begin{itemize}
    \item Locate the \textbf{cut corner} at the \textbf{top-right} of the board.  Position the chart so this corner remains at the top-right.
    \item On the back, there are \textbf{two stickers}:
    \begin{itemize}
        \item A \textbf{smaller square label} near the cut corner contains a \textbf{QR code}. Scanning it with your phone will take you to a companion website.
        \item A \textbf{larger rectangular label} provides a labeled version of the chart.
    \end{itemize}
\end{itemize}

\subsubsection*{Step 2: Introduction to the Faceted Line Chart}

\begin{itemize}
    \item This tactile model represents a \textbf{Faceted Line Chart}, which visualizes trends in multiple related datasets.
    \item Move your hand to the top-left corner to feel the title: ``Faceted Line Chart.''
    \item Explore the board to get an overview. The chart is located on the center of the board.
    \item \textbf{The data:} This faceted line chart visualizes weather trends in Austin, Texas, focusing on four variables:
    \begin{itemize}
        \item Average Temperature (°F)
        \item Average Wind Speed (mph)
        \item Average Humidity (\%)
        \item Total Precipitation (inches)
    \end{itemize}
    \item \textbf{Chart Structure:}
    \begin{itemize}
        \item The plot consists of \textbf{four line charts}, stacked vertically. Each chart represents one variable.
        \item All charts \textbf{share the same X-axis} (at the bottom) but have their \textbf{own Y-axis} (on the right).
        \item The layout can be divided into three vertical sections:  \textbf{Left:} Braille labels indicating variable names and units.  \textbf{Middle:} The embossed area chart for each variable. \textbf{Right:} The Y-axis with tick marks showing the scale for that variable.
    \end{itemize}
\end{itemize}

\subsubsection*{Step 3: Exploring the Faceted Line Chart}
\noindent \paragraph{Variable Names and Units}
\begin{itemize}
    \item Place your hand on the \textbf{left side} of the board and move \textbf{vertically}.
    \item You will feel the Braille labels for each variable. Each variable has two parts:
    \begin{itemize}
        \item A larger rectangular block containing the variable name
        \item A smaller rectangular block containing the units
    \end{itemize}
    \item The variables listed from \textbf{top to bottom} are:
    \begin{itemize}
        \item Average Temperature (F)
        \item Average Wind Speed (mph)
        \item Average Humidity (\%)
        \item Total Precipitation (inches)
    \end{itemize}
\end{itemize}

\noindent \paragraph{Exploring the X-Axis}
\begin{itemize}
    \item Locate the \textbf{bottom of the plot} to find the shared X-axis.
    \item At the bottom of the plot, you will feel the Braille labels for the X-axis: "August".
    \item Move your hand \textbf{horizontally} along this axis, which represents the days of the month.
    \begin{itemize}
        \item Tick marks are \textbf{evenly spaced}.
        \item \textbf{Longer tick marks} indicate intervals of 5 or 10 days.
    \end{itemize}
\end{itemize}

\noindent \paragraph{Exploring an Example Section} Total Precipitation (Bottom Section)
\begin{itemize}
    \item Start at the \textbf{bottom section}, which represents \textbf{Total Precipitation (inches)}.
    \item Move to the \textbf{rightmost corner}, where the X-axis and Y-axis intersect.
    \item Trace the \textbf{Y-axis vertically upwards}; the scale ranges from \textbf{0 to 2.4 inches}.
    \item Return to the X-axis and \textbf{move your hand to the left} to start tracing the curve.
    \item Horizontally trace the area curve from \textbf{left to right:}
    \begin{itemize}
        \item Initially, there is no curve (no rainfall for the first half of the month).
        \item Around the middle of the month, you will feel a \textbf{large peak}, indicating the maximum rainfall.
        \item Beyond the peak, smaller bumps represent \textbf{minor rainfall} later in the month.
        \item While tracing the curve, refer to the \textbf{Y-axis} for rainfall amounts and the \textbf{X-axis} for the corresponding dates.
    \end{itemize}
\end{itemize}
   
\noindent \paragraph{Exploring the Other Variables}
\textbf{Average Humidity (\%):}
\begin{itemize}
    \item Move your hand \textbf{one section up} to find the humidity plot.
    \item Locate the \textbf{Braille label} on the left: ``Average Humidity (\%).''
    \item Trace the curve of the area:
    \begin{itemize}
        \item Humidity is generally \textbf{high and stable}.
        \item A \textbf{slight peak} occurs around the middle of the month.
    \end{itemize}
    \item Compare this peak with the precipitation section below; you will notice a rainfall event corresponds to increased humidity.
\end{itemize}

\textbf{Average Wind Speed (mph):}
\begin{itemize}
    \item Move your hand \textbf{one section up} to the wind speed plot.
    \item Locate the \textbf{Braille label} on the left: ``Average Wind Speed (mph).''
    \item Trace the curve:  You will feel \textbf{periodic peaks and valleys}, indicating fluctuations in wind speed.
\end{itemize}

\textbf{Average Temperature (°F): }
\begin{itemize}
    \item Move to the \textbf{top section} of the plot.
    \item Locate the \textbf{Braille label} on the left: "Average Temperature (F)."
    \item Trace the curve:
    \begin{itemize}
    \item Temperature remains \textbf{stable throughout the month}, hovering around \textbf{85°F}.
    \item Even during the mid-month rainfall, temperature changes only slightly, indicating \textbf{minimal impact}.
    \end{itemize}
\end{itemize}

\noindent \paragraph{Comparison Between Variables}
  \begin{itemize}
      \item To compare variables at a specific time, trace \textbf{vertically} across the plots. The vertical grid lines can help you align the plots.
      \item For example, during the mid-month rainfall, compare the \textbf{humidity and precipitation peaks} while noting the stable temperature.
  \end{itemize}

\subsubsection*{Recap}
\begin{itemize}
    \item \textbf{X-Axis:} Represents days of the month (at the bottom), shared by all line plots.
    \item \textbf{Y-Axis:} Represents the scale for each variable (on the right side of each plot).
    \item \textbf{Faceted Areas:} Show trends for four weather variables:
    \begin{itemize}
        \item \textbf{Top:} Average Temperature.
        \item \textbf{Second:} Average Wind Speed.
        \item \textbf{Third:} Average Humidity.
        \item \textbf{Bottom:} Total Precipitation.
    \end{itemize}
    \item \textbf{Exploration Tips:}
    \begin{itemize}
        \item Trace \textbf{horizontally} to observe trends for a single variable.
        \item Trace \textbf{vertically} to compare different variables at the same time.
    \end{itemize}
\end{itemize}

\clearpage
\section{Example Packages}
\label{sec:evaluation-study-package-examples}

\begin{figure}[ht]
    \centering   \includegraphics[width=1\columnwidth]{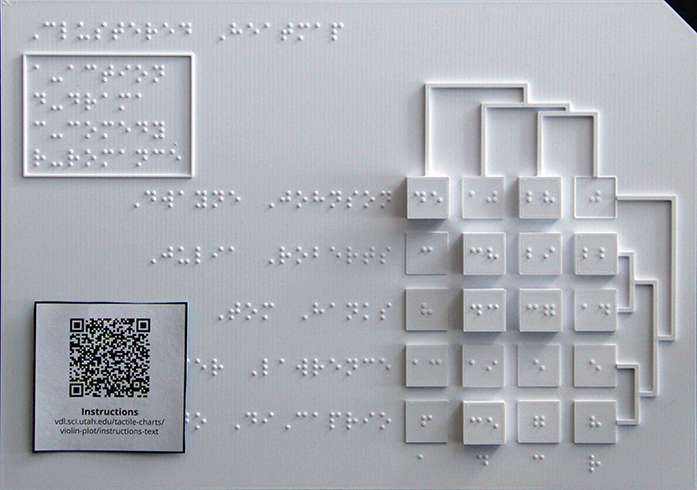}
    \caption{A package we shipped to a participant in our evaluation study under the tactile clustered heatmap condition.}
    \label{fig:package-example-heatmap}
\end{figure}

\begin{figure}[ht]
    \centering    \includegraphics[width=1\columnwidth]{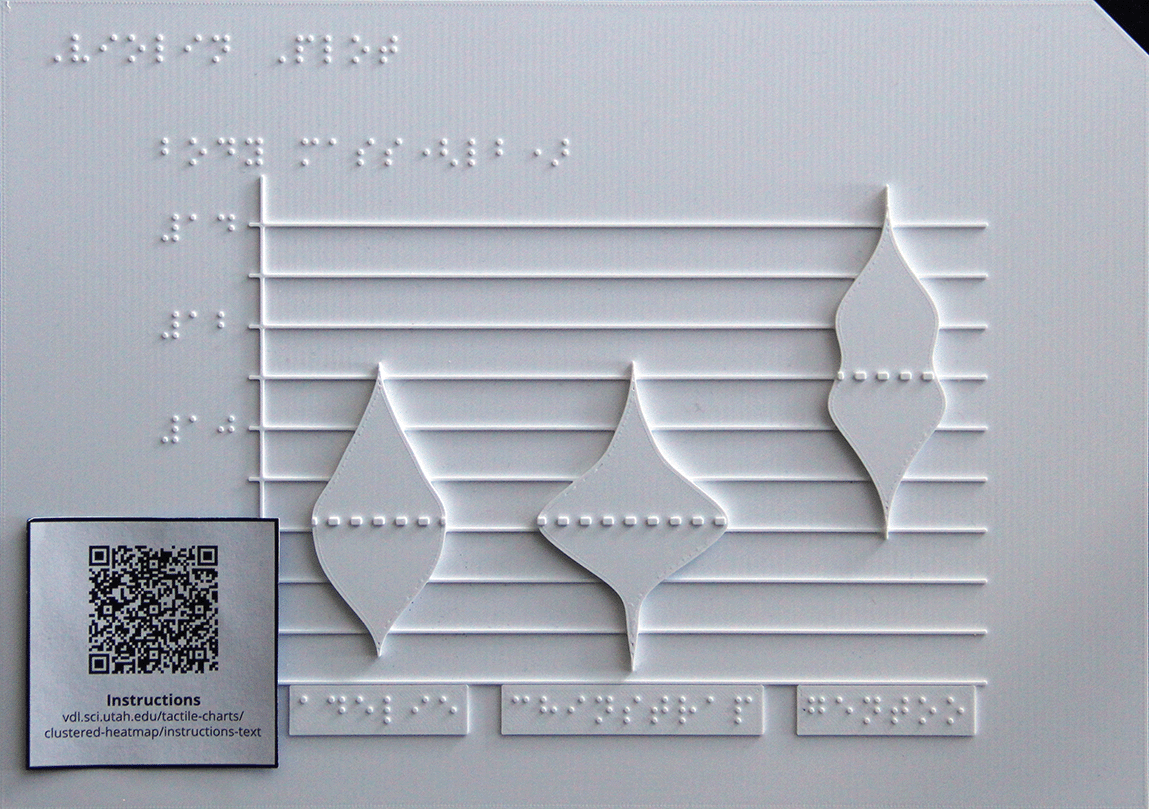}
    \caption{A package we shipped to a participant in our evaluation study under the tactile violin plot condition.}
    \label{fig:package-example-violin}
\end{figure}

\autoref{fig:package-example-heatmap} and \autoref{fig:package-example-violin} show the example packages we shipped to participants in our evaluation study (\autoref{sec:evaluation-study}).

\clearpage
\section{Textual Instructions}
\label{sec:textual-explanation}

In our evaluation study (\autoref{sec:evaluation-study}), we used textual instructions of the clustered heatmap and the violin plot for the \texttt{Text-Only} condition. 
To ensure comparability between conditions, we prepared textual instructions by removing tactile chart-specific content from the tactile chart exploration instructions and adapting the text accordingly.
In this section, we present these textual instructions used in our evaluation study.

\subsection{Textual Instructions: Clustered Heatmap}
\label{sec:textual-instructions-heatmap}
\subsubsection*{Step 1: Introduction to the Clustered Heatmap and Chart Overview}

\begin{itemize}
\item A \textbf{clustered heatmap} combines a heatmap and dendrograms:  
\begin{itemize}
  \item  \textbf{Heatmap:} A matrix of squares that represent tabular data values through color saturation.  
  \item  \textbf{Dendrograms:} Tree-like diagrams positioned above and to the right of the heatmap, showing the similarity between rows and columns.  
  \end{itemize}
\item \textbf{We use an example to explain this chart type.}
\begin{itemize}
  \item  Imagine a  clustered heatmap that visualizes how frequently actors appear in different movie genres.   
  \item  Similar actors and genres are grouped together. 
    \end{itemize}
\end{itemize}
 
\subsubsection*{Step 2: What's in the Heatmap}

\textbf{Rows and Columns:}  
  \begin{itemize}
      \item \textbf{Rows represent 5 actors, } from top to bottom:
      \begin{itemize}
          \item Dwayne Johnson
          \item Julia Roberts
          \item Tom Hanks
          \item Jennifer Lawrence
          \item Leonardo DiCaprio
          \end{itemize}
      \item \textbf{Columns represent 4 movie genres,} from left to right:
      \begin{itemize}
          \item Action
          \item Drama
          \item Comedy
          \item Romance
      \end{itemize}
  \end{itemize}
    
Each square in the matrix indicates the number of movies an actor has performed in for each genre. The color saturation represents these numbers:  
  \begin{itemize}
      \item Darker squares: More movies in a specific genre.
      \item Lighter or blank squares: Fewer or no movies.
  \end{itemize}

\textbf{Example:} For Dwayne Johnson (his row):  
  \begin{itemize}
      \item \textbf{Action}: Very dark (45 movies). This is the darkest square in the heatmap, representing the highest value.
      \item \textbf{Drama}: Light (10 movies).
      \item \textbf{Comedy}: Medium-dark (28 movies).
      \item \textbf{Romance}: Blank (0 movies). This is the lowest value in the heatmap.
  \end{itemize}

\subsubsection*{Step 3: What do Dendrograms Show}

Dendrograms show hierarchical grouping based on similarity.  
\noindent \paragraph{Row Dendrogram}

The row dendrogram shows similarities between actors. If rows are connected directly with a short branch it indicates that these actors are similar in terms of the movie genres they typically perform in.  
Clusters are groups of similar rows with close connections.  
This chart has two main actor clusters.

\textbf{Cluster 1}: Julia Roberts and Tom Hanks  
    \begin{itemize}
      \item Both have few Action movies.  
      \item Both have many Drama and Comedy movies.  
      \item Both are some Romance movies.  
  \end{itemize}

\textbf{Cluster 2}: Jennifer Lawrence and Leonardo DiCaprio  
    \begin{itemize}
        \item Both have many Drama movies.  
        \item Both have few movies in other genres.  
  \end{itemize}

\textbf{Outlier:} The elements deviating significantly from the main clusters are outliers.  
    \begin{itemize}          
        \item This chart has one outlier among actors. 
        \item \textbf{Outlier:} Dwayne Johnson  
              * His movie genre pattern is different from other actors. For example, he is heavily involved in Action movies, but the other four actors have very low numbers in this genre. 
    \end{itemize}

\noindent \paragraph{Column Dendrogram}  

The column dendrogram shows similarities of movie genres.   
 The principle is the same as the row dendrogram but applied to the columns. If columns group closer, it means these genres are more similar in terms of actor involvement patterns.  
Again, groups of similar columns can form clusters.   

\textbf{Cluster and Outlier:} 
\begin{itemize}
\item   In this chart, the movie genres are clustered incrementally in a step-by-step manner, starting with closely related pairs and gradually incorporating more distinct genres. The tall tree indicates that they are generally not very similar.   
\item  \textbf{Romance and Comedy} group together first, indicating they are the most similar genres.  
\item  \textbf{Drama} joins next.  
\item  \textbf{Action} joins last, being the most distinct genre (outlier).
\end{itemize}

\subsubsection*{Recap}

\begin{itemize}
    \item The heatmap displays values using a matrix of color-saturation coded squares.  
    \item The dendrograms reveal similarity between rows and columns.
\end{itemize}

\clearpage
\subsection{Textual Instructions: Violin Plot}
\label{sec:textual-instructions-violin}
 \subsubsection*{Step 1: Introduction to the Violin Plot and Chart Overview}

\begin{itemize}
\item \textbf{A Violin plot visualizes distributions of quantitative data for one or more categories. A distribution describes how values in a dataset are spread out. For example, if you collect the weight of 100 penguins, the distribution will show how many penguins are light, average weight, and heavy, respectively.} 
\item  We use an example dataset to explain the plot. Imagine a violin plot that compares the body mass distributions (in pounds) of three penguin species: Adelie, Chinstrap, and Gentoo. The chart displays three vertical violin shapes, each representing the distribution of body masses for one penguin species.
\end{itemize}

\subsubsection*{Step 2: What's in the Violin Plot}

\noindent \paragraph{Axes}  
  \begin{itemize}
      \item \textbf{Y-axis:} Represents body mass in pounds. It ranges from \textbf{5 pounds} at the bottom to \textbf{14 pounds} at the top, with evenly spaced tick marks representing whole numbers (5, 6, 7, up to 14).
      \item \textbf{X-axis:} Lists the three penguin species from left to right: \textbf{Adelie}, \textbf{Chinstrap}, and \textbf{Gentoo}. Each species label is under its corresponding violin shape above it.
  \end{itemize}
  
\noindent \paragraph{Violin Shapes}:  
\begin{itemize}
    \item Each violin shows the distribution of body masses for a species:
    \item \textbf{Width:} Wider sections indicate more penguins with body masses around that value; skinnier sections indicate fewer numbers with that body mass.
    \item \textbf{Median:} A dashed horizontal line inside each violin marks the median body mass for that species.
\end{itemize}

\noindent \paragraph{Different violin shapes:}  
\textbf{This datasets shows three violins that have different shapes}  
    \begin{itemize}
        \item The violin is symmetrical left to right.
        \item Wider sections of the violin represent more values, meaning that a higher number of penguins will have a body mass around this value.
        \item Skinnier sections represent fewer values, meaning that  that few penguins have a body mass around this value
    \end{itemize}

\textbf{Adelie penguins:}   
    \begin{itemize}
        \item The distribution has one peak, meaning that it is unimodal, and is \textbf{symmetrical}.
        \item The widest section is centered at the \textbf{median (8.16 pounds)}, where most data points are concentrated, indicating that most Adelie penguins have body masses around this value.
        \item The density tapers off symmetrically on both sides, showing a relatively balanced distribution with fewer penguins at the lower and upper extremes.
    \end{itemize}

\textbf{Chinstrap penguins:}   
    \begin{itemize}
        \item The distribution also has one peak and is \textbf{symmetrical}. The peak near the median body mass is about 8.18 pounds.
        \item The peak is more distinct compared to Adelie penguins. The violin is wider, indicating that Chinstrap penguins' body masses are more tightly clustered around the median. The tails at both ends are skinnier than that of Adelie penguins. This suggests lower variability in body mass compared to Adelie penguins.
    \end{itemize}

\textbf{Gentoo penguins:}   
    \begin{itemize}
        \item The distribution is \textbf{bimodal}, meaning there are two distinct peaks in body mass. The first peak appears around \textbf{10.5 pounds}, and the second peak near \textbf{12 pounds}, forming two areas of high density. This suggests that Gentoo penguins exhibit greater variation in body mass, possibly due to differences in sex with males having distinctly higher body mass than females.
        \item The median body mass of \textbf{11.02 pounds} is \textbf{higher} than that of Adelie and Chinstrap penguins, and about as high as the heaviest penguins of the other species.
        \item The median is closer to the lower peak (10.5 pounds) rather than centered between the two peaks, indicating that more Gentoo penguins have body masses around the lower peak than the higher peak. This suggests an asymmetry in distribution, where the higher peak represents fewer but more widely spread heavier penguins.
        \item The violin also has a more elongated distribution compared to the other two species, showing Gentoo penguins have a larger body mass range.
    \end{itemize}

\subsubsection*{Step 3: Recap}

\begin{itemize}
    \item \textbf{Y-Axis}: Represents body mass in pounds, increasing from 5 to 14 (vertical scale).  
    \item \textbf{X-Axis}: Identifies penguin species (horizontal labels).  
    \item \textbf{Violin Shapes}: Show the distribution of body masses, with wider sections indicating higher density and horizontal lines marking the medians.
\end{itemize}

\clearpage
\section{Alt Text Corresponding to the Tactile Charts (Simple Alt Text)}
\label{sec:simple-alt-text}

In our evaluation study (\autoref{sec:evaluation-study}), we wrote the alt text for the simple dataset, which we refer to as \textit{simple alt texts}. These alt texts correspond to the tactile charts, the tactile chart exploration instructions, and the example chart referenced in the textual instructions. 

In this section, we present the simple alt texts for the clustered heatmap and the violin plot, which we used in our evaluation study. Alt texts for tactile charts of all four chart types are available on our accessible website: \companionweb.

\subsection{Simple Alt Text: Clustered Heatmap}
\label{sec:simple-alt-text-heatmap}

Below is the simple alt text of the clustered heatmap, which can also be accessed at \href{https://vdl.sci.utah.edu/tactile-charts/clustered-heatmap/alttext-simple}{\texttt{vdl\discretionary{}{.}{.}sci\discretionary{}{.}{.}utah\discretionary{}{.}{.}edu/tactile\discretionary{}{-}{-}charts\discretionary{/}{}{/}clustered\discretionary{}{-}{-}heatmap\discretionary{/}{}{/}alttext\discretionary{}{-}{-}simple}}.

\vspace{5mm}

This \textbf{clustered heatmap} visualizes how frequently actors appear in different movie genres. Similar patterns are grouped together. 

\begin{itemize}
    \item Julia Roberts and Tom Hanks show similar patterns, focusing primarily on Drama and Comedy 
    \item Jennifer Lawrence and Leonardo DiCaprio share a strong emphasis on Drama 
    \item Dwayne Johnson is  an outlier in this dataset, with a strong specialization in Action and no Romance films
\end{itemize}

The visualization combines a data table (heatmap) with tree-like diagrams (dendrograms) on its sides that show how similar rows and columns are. The heatmap displays \textbf{five actors} (rows):

\begin{itemize}
    \item Dwayne Johnson
    \item Julia Roberts
    \item Tom Hanks
    \item Jennifer Lawrence
    \item Leonardo DiCaprio
\end{itemize}

And \textbf{four movie genres} (columns):
\begin{itemize}
    \item Action
    \item Drama
    \item Comedy
    \item Romance
\end{itemize}

Each cell indicates the number of movies an actor has performed in for each genre using color saturation. Dark cells indicate many movies. The rows and columns are \textbf{clustered} based on similarity patterns. The most similar clusters of actors and genres are connected with the tree above and to the left of the matrix. 

\noindent \paragraph{Statistical Information}

\begin{itemize}
    \item \textbf{Highest Value}: \textbf{Dwayne Johnson}'s contributions to \textbf{Action} (45 movies).
    \item \textbf{Lowest Value}: \textbf{Dwayne Johnson}'s contributions to \textbf{Romance} (0 movies).
    \item \textbf{Most Represented Genre}: \textbf{Drama}, with a total of \textbf{145 movies} across all actors.
    \item \textbf{Least Represented Genre}: \textbf{Romance}, with a total of \textbf{49 movies} across all actors.
    \item \textbf{Person Who acted in the Most Films}: \textbf{Tom Hanks}, with a total of \textbf{102 movies} across all genres.
    \item \textbf{Person Who acted in the Fewest Films}: \textbf{Jennifer Lawrence}, with a total of \textbf{42 movies} across all genres.
\end{itemize}

\noindent \paragraph{Actor Clusters}

The actors are grouped into \textbf{two main clusters} and \textbf{one outlier} based on their genre preferences:

\textbf{Cluster 1:} \textbf{Julia Roberts} and \textbf{Tom Hanks}  
\begin{itemize}
    \item Both actors contributed a lot of movies in \textbf{Drama} and \textbf{Comedy}.
    \begin{itemize}
        \item Julia Roberts: 38 Drama, 29 Comedy
        \item Tom Hanks: 43 Drama, 37 Comedy
    \end{itemize}
    \item They have moderate contributions to \textbf{Romance}
    \begin{itemize}
        \item Julia Roberts: 22 Romance
        \item Tom Hanks: 14 Romance
    \end{itemize}
    \item They have low involvement in \textbf{Action}.
    \begin{itemize}
        \item Julia Roberts: 3 Action
        \item Tom Hanks: 8 Action
    \end{itemize}
\end{itemize}

\textbf{Cluster 2}: \textbf{Jennifer Lawrence} and \textbf{Leonardo DiCaprio}  
\begin{itemize}
    \item Both actors primarily contribute to \textbf{Drama}.  
    \item They show some involvement in \textbf{Action} but minimal contributions to \textbf{Comedy} and \textbf{Romance}.  
\end{itemize}

\textbf{Outlier: Dwayne Johnson}
\begin{itemize}
    \item Dwayne Johnson has lots of contributions in \textbf{Action} (45), while other actors contribute to Action only a little.  
    \begin{itemize}
        \item Julia Roberts: 3  
        \item Tom Hanks: 8  
        \item Jennifer Lawrence: 13  
        \item Leonardo DiCaprio: 6
    \end{itemize}
\end{itemize}

\noindent \paragraph{Movie Genre Clusters}

The movie genres are clustered incrementally in a step-by-step manner, starting with closely related pairs and gradually incorporating more distinct genres, emphasizing their hierarchical relationships. Generally, genres are fairly distinct. 

\textbf{Comedy and Romance:}
\begin{itemize}
    \item These genres are grouped together first  
    \item Julia Roberts and Tom Hanks are the most prominent contributors to these genres.
\end{itemize}

\textbf{Drama:}
\begin{itemize}
    \item Drama is added into the cluster in the next step.
\end{itemize}

\textbf{Action:}
\begin{itemize}
    \item This genre is the most distinct and isolated, and it is clustered last.
\end{itemize}

\clearpage
\subsection{Simple Alt Text: Violin Plot}
\label{sec:simple-alt-text-violin}

Below is the simple alt text of the violin plot, which can also be accessed at \href{https://vdl.sci.utah.edu/tactile-charts/violin-plot/alttext-simple}{\texttt{vdl\discretionary{}{.}{.}sci\discretionary{}{.}{.}utah\discretionary{}{.}{.}edu/tactile\discretionary{}{-}{-}charts\discretionary{/}{}{/}violin\discretionary{}{-}{-}plot\discretionary{/}{}{/}alttext\discretionary{}{-}{-}simple}}.

\vspace{5mm}

This is a violin plot showing the distribution of body mass (in pounds) across three species of penguins: Adelie, Chinstrap, and Gentoo. While \textbf{Adelie and Chinstrap penguins exhibit unimodal distributions with similar median values}, Gentoo penguins show a \textbf{bimodal distribution and a higher median.}

\begin{itemize}
    \item \textbf{Y-axis:} Represents body mass in pounds.   
    \item \textbf{X-axis:} Lists the three penguin species from left to right: \textbf{Adelie}, \textbf{Chinstrap}, and \textbf{Gentoo}.   
    \item \textbf{Three violins each shows the distribution of body masses for a species}
\end{itemize}

\noindent \paragraph{Distribution Properties}

\begin{itemize}
    \item \textbf{Adelie penguins:} The distribution is unimodal, and it has one peak in the middle. Most Adelie penguins’ body mass is concentrated close to the median (8.16 pounds).  
    \item \textbf{Chinstrap penguins:} The distribution is also unimodal, with a single peak in the middle. Most Chinstrap penguins' body masses are concentrated close to the median (8.18 pounds), and their weights are more tightly clustered near the middle compared to Adelie penguins.  
    \item \textbf{Gentoo penguins:} The median body mass of Gentoo penguins is 11.02 pounds, which is larger than Adelie and Chinstrap penguins. The distribution is bimodal. Gentoo penguins' body masses are concentrated around 10.5 pounds and 12 pounds, which form the two peaks. 
\end{itemize}

\noindent \paragraph{Statistical Information}

\textbf{Adelie penguins}:  
        \begin{itemize}
            \item Body mass ranges from 6.28 to 10.53 pounds.  
            \item The first quartile (Q1) is 7.39 pounds.  
            \item The median is 8.16 pounds.  
            \item The third quartile (Q3) is 8.82 pounds.  
        \end{itemize}
\textbf{Chinstrap penguins}:  
        \begin{itemize}
            \item Body mass ranges from 5.95 to 10.58 pounds.  
            \item The first quartile (Q1) is 7.76 pounds.  
            \item The median is 8.18 pounds.  
            \item The third quartile (Q3) is 8.74 pounds.  
        \end{itemize}
\textbf{Gentoo penguins}:  
        \begin{itemize}
            \item Body mass ranges from 8.71 to 13.89 pounds.  
            \item The first quartile (Q1) is 10.36 pounds.  
            \item The median is 11.02 pounds.  
            \item The third quartile (Q3) is 12.13 pounds.
            \end{itemize}

\clearpage
\section{Datasets for Complex Charts}
\label{sec:complex-datasets}

In our evaluation study (\autoref{sec:evaluation-study}), participants experienced two datasets for each chart type, which we refer to as the simple dataset and the complex dataset (see \autoref{fig:experiment-overview}).
In this section, we present the details of the preparation of the complex datasets.

\subsection{Clustered Heatmap}
For the complex clustered heatmap, we used a dataset from the European Values Study \cite{EVS} and selected countries based on Perin et al.'s work \cite{Perin:2014:Revisiting} on tabular visualizations. When creating the chart, we normalized each row based on the method described in the Perin et al.'s work \cite{Perin:2014:Revisiting}. We normalized the percentages independently for each row (country) to a scale of 0 to 1, to emphasize the relative distribution of responses across topics within a country.

\subsection{Violin Plot}
For the complex violin plot, we used a Human Development Index (HDI) dataset, which was downloaded from \href{https://data.humdata.org/dataset/human-development-data}{\texttt{data\discretionary{}{.}{.}humdata\discretionary{}{.}{.}org\discretionary{/}{}{/}dataset\discretionary{/}{}{/}human\discretionary{}{-}{-}development\discretionary{}{-}{-}data}}. We used the data of 2021.

\section{Python Generated Plots for Complex Charts}
\label{sec:complex-charts}
\begin{figure}[ht]
    \centering
        \includegraphics[width=1\columnwidth]{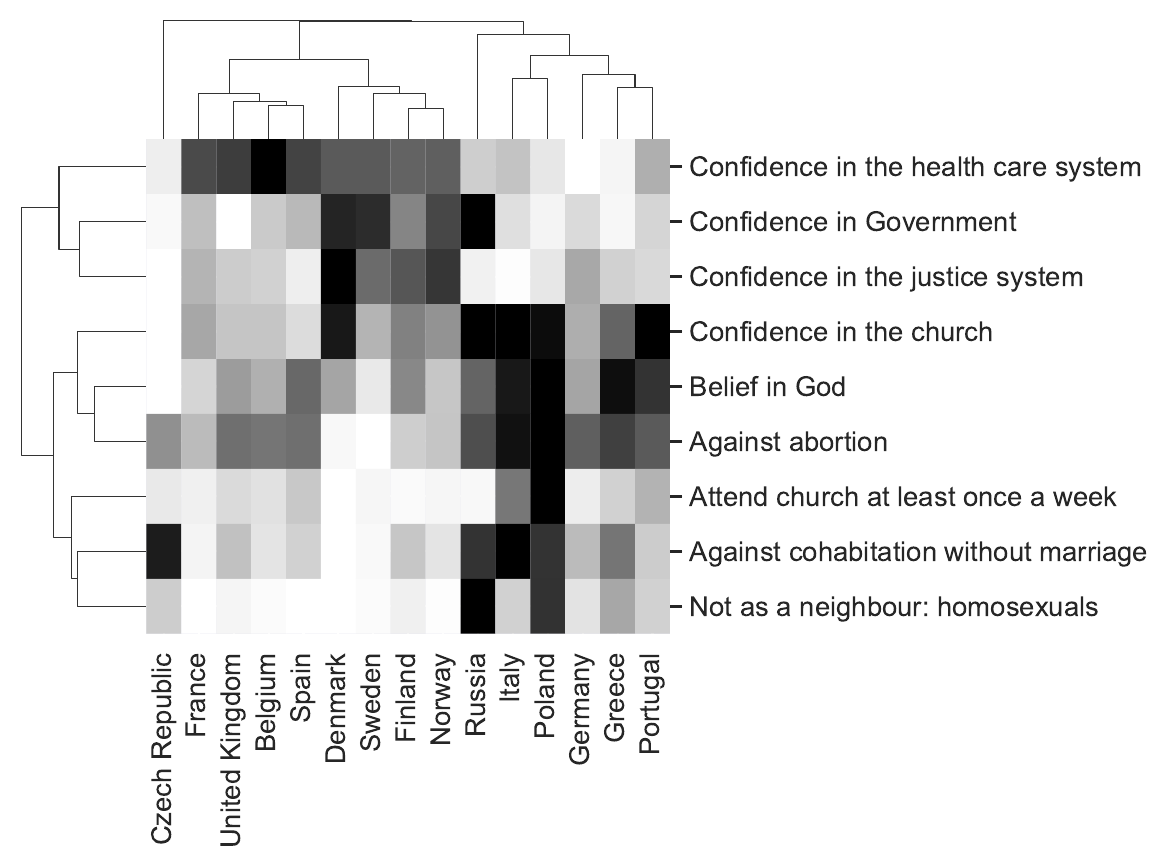}
    \caption{The clustered heatmap with the complex dataset.}
    \label{fig:heatmap_complex}
\end{figure}

\begin{figure}[ht]
    \centering
        \includegraphics[width=1\columnwidth]{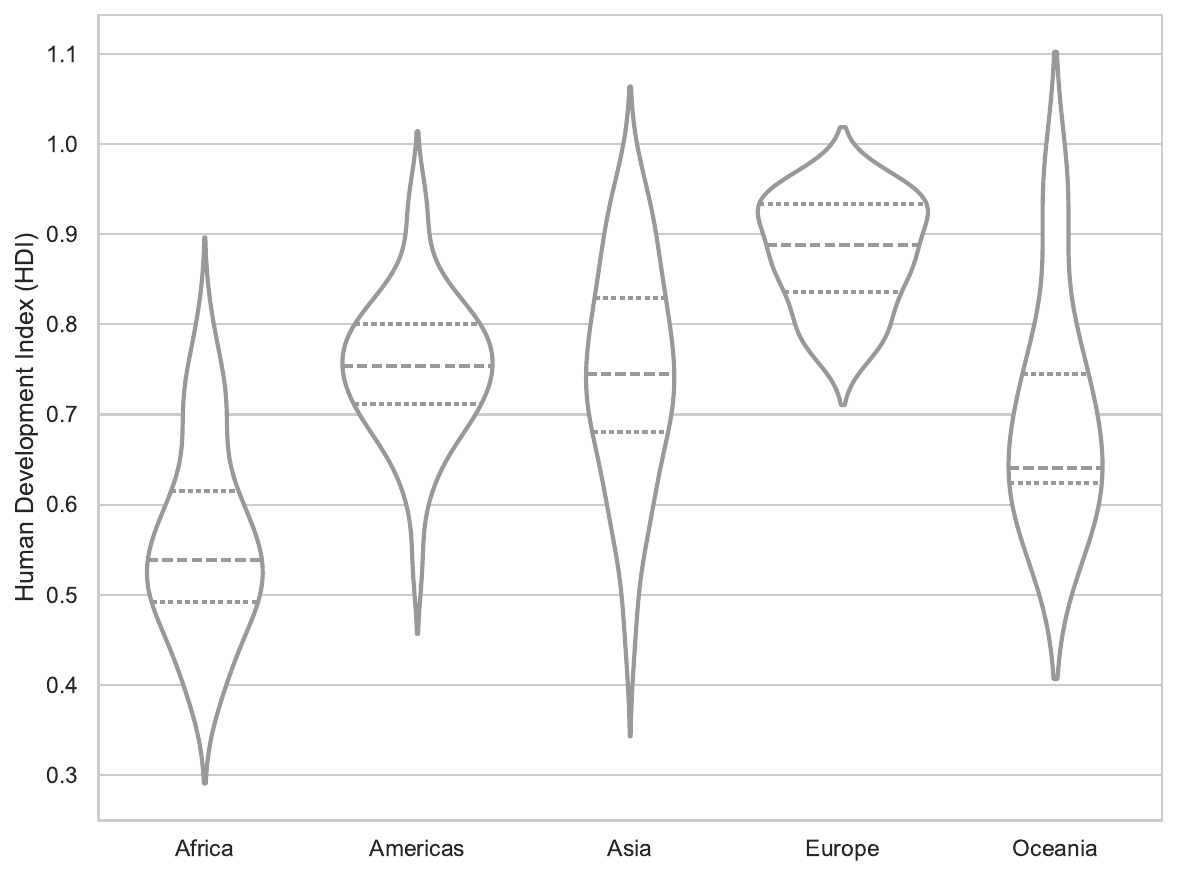}
    \caption{The violin plot with the complex dataset.}
    \label{fig:violin_complex}
\end{figure}

Based on our complex datasets, we created charts with Python (see \autoref{fig:experiment-overview} and present them in this section (\autoref{fig:heatmap_complex} and \autoref{fig:violin_complex}).

\clearpage
\section{Complex Alt Text}
\label{sec:complex-alt-texts}

In this section, we present the alt texts for the clustered heatmap and the violin plot based on complex datasets (see \autoref{sec:complex-charts}), which we refer to as \textit{complex alt texts} in our evaluation study (\autoref{sec:evaluation-study}).

\subsection{Complex Alt Text: Clustered Heatmap}
\label{sec:complex-alt-text-heatmap}

This visualization is a \textbf{clustered heatmap} comparing the relative proportion of importance citizens from 15 European countries assign to  sociocultural and religious topics, such as confidence in the government, belief in god, or cultural issues such as homosexuality. 

On a high level, the Nordic countries are the most confident in state institutions and most liberal. Religious values are strongest in Russia, Italy, Poland, Germany, Greece and Portugal. Poland, Italy, and Russia are most conservative on social issues. 

The \textbf{rows} represent countries, and the \textbf{columns} represent topics participants were asked about in a survey. The intensity of color in each \textbf{cell} indicates agreement levels, where \textbf{darker shades} represent higher agreement and \textbf{lighter shades} indicate lower agreement.

Both \textbf{countries} and \textbf{topics} are clustered based on similarity. The relationships among them are displayed in a \textbf{hierarchical tree}.

\noindent \paragraph{Topic Clusters}

Topics are grouped into three main clusters:

\textbf{Cluster 1: Confidence in the State Institutions, which includes}  
\begin{itemize}
    \item Confidence in the healthcare system  
    \item Confidence in the government  
    \item Confidence in the justice system  
\end{itemize}

\textbf{Cluster 2: Religious Issues, which includes}  
\begin{itemize}
    \item Confidence in the church  
    \item Belief in God  
    \item Opposition to abortion  
\end{itemize}

\textbf{Cluster 3: Conservative Social Issues, which includes}  
\begin{itemize}
    \item Attending church at least once a week  
    \item Opposition to cohabitation without marriage  
    \item Opposition to having homosexuals as neighbors
\end{itemize}

\noindent \paragraph{Country Clusters}

Countries are grouped into three clusters and one outlier:

\textbf{Cluster 1: Nordic Countries (Denmark, Sweden, Finland, Norway)}

\begin{itemize}
    \item High values on confidence in state institutions (healthcare, government, justice system).  
    \item Moderate values on religious issues, with exceptions:  
        \begin{itemize}
            \item Denmark shows higher confidence in the church.  
            \item Denmark and Sweden show lower opposition to abortion.  
        \end{itemize}
    \item Finland and Norway have the most similar response patterns among all countries.
\end{itemize}

\textbf{Cluster 2: Secular Western European Countries (France, United Kingdom, Belgium, Spain)}

\begin{itemize}
    \item High confidence in the healthcare system.  
    \item Moderate confidence in government and justice, with an exception:  The United 
    \item Moderate values on religious issues.  
    \item Low values on conservative social issues, but slightly higher than Nordic countries and lower than the next cluster (Religious European Countries).
\end{itemize}

\textbf{Cluster 3: Religious European Countries (Russia, Italy, Poland, Germany, Greece, Portugal)}

\begin{itemize}
    \item \textbf{Lower confidence} in state institutions, except \textbf{Russia, which has high confidence in the government.}
    \item \textbf{High values} on religious issues.
    \item \textbf{Mixed values} on conservative social issues:
    \begin{itemize}
        \item \textbf{Germany, Greece, and Portugal} form a \textbf{subcluster}, showing \textbf{moderate values} on all conservative social issues (attending church, opposing cohabitation without marriage, and opposing homosexual neighbors).
        \item \textbf{Italy and Poland} form another \textbf{subcluster}, with:  \textbf{Poland} showing \textbf{high values} on all conservative social issues.  \textbf{Italy} showing \textbf{high church attendance and opposition to cohabitation without marriage, but moderate opposition to homosexual neighbors.}
        \item \textbf{Russia} is an \textbf{outlier within this cluster}, showing \textbf{high opposition} to homosexual neighbors and cohabitation without marriage but \textbf{low church attendance.}
    \end{itemize}
\end{itemize}

\textbf{Outlier: Czech Republic. It has}

\begin{itemize}
    \item Low confidence in state institutions.  
    \item Low values on issues related to the church and belief in God, but only moderate opposition to abortion.  
    \item Moderate church attendance and opposition to homosexual neighbors, but high opposition to cohabitation without marriage.
\end{itemize}

\clearpage
\subsection{Complex Alt Text: Violin Plot}
\label{sec:complex-alt-text-violin}
This is a violin plot showing the distribution of the Human Development Index (HDI) across five continents, Africa, Americas, Asia, Europe, and Oceania, based on data from 2021\. The HDI aims to capture the overall well-being and development of countries based on health, education, and standard of living. Europe exhibits the highest and most consistent HDI values, while Asia has the widest variability. Africa and Oceania have non-normal  distributions, indicating groups of countries with different levels of development. HDI is measured from 0 (low) to 1 (high).

\noindent \paragraph{Distribution Properties}

\textbf{Africa:}   
\begin{itemize}
    \item The distribution is not a normal distribution but has a wider part at the upper end, in addition to a peak in the middle.   
    \item The majority of African countries have HDI values concentrated around the median of \textbf{0.54}, forming a primary peak.   
    \item A smaller wider section  is observed around \textbf{0.73}, indicating a group of countries with higher HDI.   
    \item The distribution is relatively spread out, spanning from \textbf{0.39 to 0.80}.  
\end{itemize}

\textbf{Americas:}   
\begin{itemize}
    \item The distribution is unimodal and symmetrical, peaking around the median \textbf{0.75}.   
    \item Most countries' HDI values are concentrated between \textbf{0.71 and 0.80}, with fewer extreme values, reflecting a less spread out level of human development across the region.  
\end{itemize}

\textbf{Asia:}   
\begin{itemize}
    \item The distribution is unimodal but highly dispersed, with HDI values spanning from \textbf{0.46 to 0.95}, the widest range among all continents.   
    \item The violin is relatively thin with no strong peak indicating that HDI values vary substantially across Asian countries.  
\end{itemize}

\textbf{Europe:}   
\begin{itemize}
    \item The distribution is unimodal but asymmetrical, with a clear peak above the median at \textbf{0.89}.   
    \item Europe has the highest median HDI among all continents and the smallest range \textbf{0.77 to 0.96}, suggesting high and consistent levels of human development.  
\end{itemize}

\textbf{Oceania:}   
\begin{itemize}
    \item The distribution is elongated and not normal bimodal, with a primary peak around the median of \textbf{0.64} and a smaller wider part near \textbf{0.94}.  
    \item The range is broad \textbf{0.56 to 0.95}, indicating variation in human development levels within the region.
\end{itemize}

\noindent \paragraph{Statistical Information}

\textbf{Africa}:  
\begin{itemize}
    \item HDI ranges from 0.39 to 0.80.  
    \item The first quartile (Q1) is 0.49.  
    \item The median is 0.54, the lowest median among all continents.   
    \item The third quartile (Q3) is 0.62.  
\end{itemize}

\textbf{Americas}:  
\begin{itemize}
    \item HDI ranges from 0.54 to 0.94.  
    \item The first quartile (Q1) is 0.71.  
    \item The median is 0.75.  
    \item The third quartile (Q3) is 0.80.  
\end{itemize}

\textbf{Asia}:  
\begin{itemize}
    \item HDI ranges from 0.46 to 0.95, the largest range among all continents.  
    \item The first quartile (Q1) is 0.68.  
    \item The median is 0.75.  
    \item The third quartile (Q3) is 0.83.  
\end{itemize}

\textbf{Europe}:  
\begin{itemize}
    \item HDI ranges from 0.77 to 0.96, the smallest range among all continents.  
    \item The first quartile (Q1) is 0.84.  
    \item The median is 0.89, the highest median among all continents.  
    \item The third quartile (Q3) is 0.93.  
\end{itemize}

\textbf{Oceania}:  
\begin{itemize}
    \item HDI ranges from 0.56 to 0.95.  
    \item The first quartile (Q1) is 0.62.  
    \item The median is 0.64.  
    \item The third quartile (Q3) is 0.75.
\end{itemize}

\clearpage
\section{Evaluation Study Scripts}
\label{sec:evaluation-study-scripts}

In this section, we present our full interview scripts used in the evaluation study (\autoref{sec:evaluation-study}).

\subsection*{Introduction}
Thank you for taking the time to speak with us today. My name is [Researcher A]. I am here with [Researcher B] and we are here to talk with you about tactile models and alternative text for data visualizations. 

Let's begin by going over the plan for today's interview. We are focusing on two specific types of data visualizations, which we will examine one at a time. For each visualization type, we will first introduce the chart type to you with a simple example, along with its corresponding alternative text. Then, we will provide an alternative text for a more complex example of the same visualization type. Throughout this process, we will discuss things from your perspective. We expect today to be 2 hours.

Any questions before we begin?

\subsection*{Audio Recording}

To help with the accuracy of our notes, we are requesting to record the audio and your screen from today's interview. The only people who will have access to this recording is us and our team. I am going to go ahead and start the recording now.

\vspace{3mm}

\noindent \textit{START SCREEN AND AUDIO RECORDING AFTER PARTICIPANT HAS CONSENTED}

\subsection*{Background Questions}

We will start by learning about you and your background. I will ask you questions, please answer them by speaking out loud. 

\vspace{3mm}
\noindent \textit{ONE EXPERIMENTER: INPUT ANSWERS IN REDCAP}

\subsubsection*{Demographics}

\begin{itemize}
    \item If you would like, please let us know your preferred pronouns.
    \item Year of birth
    \item Gender
    \item What is the highest level of education you have completed?
    \item What is your current status? Are you working or studying, and is it full-time or part-time? If so, what is your job or major?
    \item Vision loss level
    \item How would you describe your vision-loss level?  (Blind since birth / Lost vision suddenly / Lost vision gradually)
    \item How would you describe your vision level (\eg, completely blind, light perception, central vision loss, etc.)?
    \item (If the participant is not totally blind) If you know, what is your corrected visual acuity in either Snellen (e.g., 20/200) or LogMAR (\eg, 1.3)?
    \item If you are comfortable sharing, please indicate your visual pathology diagnosis. This information is optional and not required.
\end{itemize}

\subsubsection*{Screen Reader Experience}
\begin{itemize}
    \item Which screen reader do you use with your computer or other devices (e.g., NVDA, JAWS, VoiceOver, etc.)
    \item How long have you been using a screen reader?
    \item When using a screen reader, what is your preferred rate of speech?
    \item Do you use other accessibility devices or software in combination with a screen reader, such as screen magnification or a Braille display? If yes, please describe all accessibility devices or software you use in combination with a screen reader.
    \item Braille experience
    \item How long have you been reading Braille?
    \item How would your rate your skill reading Braille (1 - I can't read Braille / 5 - I'm very proficient at reading Braille)
    \item In what contexts did you read Braille (\eg, work, personal use)?
    \item Tactile chart experience
    \item How would you describe your familiarity with tactile charts or graphics? (1 - Not at all familiar / 5 - Extremely familiar)
    \item Can you describe in which context you have interacted with tactile charts or graphics?
\end{itemize}

\subsubsection*{Data Visualization Experience}
\begin{itemize}
    \item How many hours do you use a computer or smartphone each day?
    \item Would you consider your career to be data-intensive or numbers-driven (\eg, regularly work with large datasets, perform statistical analyses, or make decisions based on quantitative information)?
    \item How often do you interact with data visualizations, such as those for work, from news articles, in video games, etc.?
    \item In which context do you encounter data and data visualizations (\eg, work, news, leisure)?
    \item How do you typically view datasets? (\eg, download to excel, you don't, etc.)
\end{itemize}

\subsubsection*{Familiarity With the Two Chart Types}
In this study, we focus on two specific chart types. We want to know how familiar you are with each of them.

\begin{itemize}
    \item How familiar are you with \textbf{clustered heatmaps} - a visualization of tabular data that encodes values by color? (1 - Not at all familiar / 5 - Extremely familiar)
    \item Have you seen or touched \textbf{clustered heatmaps} through a tactile display or an embossed paper before? (Yes / No / I don't know what a clustered heatmap is)
    \item How familiar are you with \textbf{violin plots} - a visualization of a distribution of values? (1 - Not at all familiar / 5 - Extremely familiar)
    \item Have you seen or touched \textbf{violin plots} through a tactile display or an embossed paper before? (Yes / No / I don't know what a violin plot is)
    \item After you received our package, have you explored the model, or hear any of the instructions?
\end{itemize}

\begin{itemize}
    \item Researcher B, do you have any additional questions?
    \item To ensure that we are both on the same link during today's discussion, do you mind also sharing your screen?
\end{itemize}

\noindent \textit{IN ZOOM SETTINGS, MAKE SURE THAT ALL PARTICIPANTS CAN SHARE THEIR SCREEN}

\subsection*{Evaluation Part 1}

\subsubsection*{Training}

The first chart type we will explore is Clustered Heatmap / Violin Plot.

\paragraph{Chart Type Instructions}

I am going to send a link in the chat that will tell you more about this chart type. Please take your time to read and understand (and follow it to explore the physical model you received). 

After hearing the instructions, I am going to ask you several questions about this chart type.

\vspace{3mm}

\textit{DEPENDING ON THE CONDITIONS OF THE PARTICIPANT, SEND ONE OF THE FOLLOWING LINKS TO THEM:}
\begin{itemize}
    \item Clustered heatmap tactile chart exploration instructions (\autoref{sec:tactile-instructions-final-heatmap})
    \item Clustered heatmap textual instructions (\autoref{sec:textual-instructions-heatmap})
    \item Violin plot tactile chart exploration instructions (\autoref{sec:tactile-instructions-final-violin})
    \item Violin plot textual instructions (\autoref{sec:textual-instructions-violin})
\end{itemize}

\subparagraph{Questions for Both Chart Types}

\begin{itemize}
    \item What types of data are best suited for visualization using a clustered heatmap / violin plot?
    \item Can you provide an example scenario where you would use a clustered heatmap / violin plot?
    \item (Clustered Heatmap) We want to add another movie actor – Arnold Schwarzenegger – to this clustered heatmap. Could you describe how he would appear in the chart and how he relates to the others? You can make up data based on what you know about Arnold Schwarzenegger.
    \item (Violin Plot) Imagine a new species of very small penguins (little penguins) with its body mass distribution added to the example violin plot. Could you describe how this species would appear in the chart and how it relates to the others? You can make up data based on what you know about penguins.
    \item Researcher B, do you have any additional questions?
\end{itemize}

\paragraph{Simple Alt Text}

I am going to send you another link in the chart. This is an example of alt text, which describes the chart used in the instruction you just heard. You will hear alt text for another dataset of the same chart type later. After hearing the alt-text, I am going to ask you several questions based on it.

\vspace{3mm}

\textit{DEPENDING ON THE CONDITIONS OF THE PARTICIPANT, SEND ONE OF THE FOLLOWING LINKS TO THEM:}
\begin{itemize}
    \item Clustered heatmap simple alt text (\autoref{sec:simple-alt-text-heatmap})
    \item Violin plot simple alt text (\autoref{sec:simple-alt-text-violin})
\end{itemize}

\subparagraph{Questions for Clustered Heatmap}
\begin{itemize}
    \item What is the dataset about?
    \item Which actor is an outlier in this dataset, and why?
    \item Which two movie genres are most similar, and why?
\end{itemize}

\subparagraph{Questions for Violin Plot}
\begin{itemize}
    \item What is the dataset about?
    \item What is the key difference between the distributions of Gentoo penguins compared to the other two species?
    \item Which penguin species has the least variation of body mass values and why?
\end{itemize}

Researcher B, do you have any additional questions?

Do you have any questions about this chart type?

\subsubsection*{Testing}

If you have no more questions, I will send another link to the chat to another clustered heatmap / violin plot alt text. Please take your time to read and understand the text description. We will have some questions for you once you read it. 

\paragraph{Complex Alt Text}

\textit{DEPENDING ON THE CONDITIONS OF THE PARTICIPANT, SEND ONE OF THE FOLLOWING LINKS TO THEM:}
\begin{itemize}
    \item Clustered heatmap complex alt text (\autoref{sec:complex-alt-text-heatmap})
    \item Violin plot complex alt text (\autoref{sec:complex-alt-text-violin})
\end{itemize}

Now, I am going to ask you a series of questions about the alt text. 
We want to understand things from your perspective. It is important to highlight that this is not a test. So do not worry about the right or wrong to any of the questions, and I would like to ask you to be as honest as possible.

\subparagraph{Questions for Clustered Heatmap}
\begin{itemize}
    \item What is the dataset about?
    \item According to the description, how many clusters are the European countries divided into, and which country is an outlier?
    \item Name the three main topic clusters and one example from each.
    \item Which country cluster has the lowest values on social conservative issues?
    \item If a country strongly opposes having homosexual neighbors, which country cluster is it likely in?
    \item The Czech Republic is described as an outlier. Why is the Czech Republic an outlier relative to other countries?
    \item Describe what the corresponding tactile chart or visualization would be like for this dataset (e.g., What labels would be included, what would the height of the squares, etc.).
\end{itemize}

\subparagraph{Questions for Violin Plot}
\begin{itemize}
    \item What is the dataset about?
    \item Which continent has the most variability in HDI values?
    \item Which continents have non-normal distributions in their HDI values?
    \item Where is the peak for HDI in Africa’s distribution relative to the other countries?
    \item How does the HDI distribution of Asia differ from that of the Americas?
    \item Which continent has the highest levels of human development?
    \item Describe what the corresponding tactile chart or visualization would be like for this dataset (e.g., what would the violins look like for the countries, which would be the widest and skinniest plots, etc.)
\end{itemize}

\subparagraph{Questions for both chart types}

\begin{itemize}
    \item Do you feel like you have a good understanding of the dataset? Please rate your understanding on a 5-point scale, where 1 = I don't understand it at all and 5 = I completely understand it.
    \item What was difficult for you to understand about the dataset?
    \item What did you find interesting or surprising about the dataset?
    \item Do you have any additional comments on your experience with the clustered heatmap/violin plot?
\end{itemize}

\subsection*{Evaluation Part 2}

Now we will explore another chart type: violin plot / clustered heatmap. We will follow a similar procedure. 

\vspace{3mm}
\textit{SAME PROCEDURE AS EVALUATION PART 1}

\subsection*{Comparative Questions After Both Charts Were Explored}

\begin{itemize}
    \item You have learned about visualization types in two different ways: one by exploring a tactile model with instructions, and another by only hearing a textual explanation. Which training format do you prefer? Can you explain the reasons for your preference?
    \item How helpful is the tactile model used in the training for better understanding the chart type?
    \item After answering the question, could you please rate it on a scale of 1 to 5? (1 = Not helpful at all, 5 = Very helpful)
    \item Do you think the tactile model helped you understand the alt text of the dataset that was not shown in the chart? After answering the question, could you please rate it on a scale of 1 to 5? (1 = Did not help at all, 5 = Helped a lot)
    \item How easy was it to explore the tactile model?
    \item After answering the question, could you please rate it on a scale of 1 to 5? (1 = Not easy at all, 5 = Very easy)
    \item I believe that tactile models can teach me transferable knowledge about a chart type. After answering the question, could you please rate it on a scale of 1 to 5? (1 = I don't believe this at all, 5 = I strongly believe this)
    \item How useful is the training format (tactile model with exploration instructions) for BLV education in your opinion? After answering the question, could you please rate it on a scale of 1 to 5? (1 = Not useful at all, 5 = Very useful)
    \item How useful is the training format (only textual explanation) for BLV education in your opinion? After answering the question, could you please rate it on a scale of 1 to 5? (1 = Not useful at all, 5 = Very useful)
    \item If these tactile models were freely available, how beneficial would they be for blind or low-vision individuals? After answering the question, could you please rate it on a scale of 1 to 5? (1 = No benefit, 5 = Highly beneficial)
    \item Do you have any comments or feedback on the tactile models?
    \item Do you have any other feedback or comments that we didn't touch on today?
    \item Researcher B, do you have any follow-up questions?
\end{itemize}

\subsubsection*{Conclusion}
Thank you so much for sharing your insights today. Your feedback has been incredibly valuable and will play a crucial role in helping us improve text descriptions.

Do you have any final questions about the interview process or about our study that you would like to ask?

One final thing, to help us reach more individuals who may also be eligible for this study, would you be willing to provide contact information for people you know who might be interested and eligible to participate? Of course, sharing this information is entirely voluntary, and will be handled with care.

Alright, if anything comes to mind later, feel free to reach out. We will process your \$100 Amazon gift card and send it to your email within the next two weeks. Otherwise, thanks again for your time and have a wonderful day!

\vspace{3mm}
\textit{END SCREEN AND AUDIO RECORDING}

\clearpage

\section{\rev{Codebook}}
\label{sec:codebook}
\rev{In this section, we present the codebook from our thematic analysis, which includes the themes and corresponding codes. We discuss the themes in detail in \autoref{sec:themes}.}

\subsection*{\rev{Theme 1: Building Mental Models of Chart Types}}

\rev{This theme examines the ways in which tactile charts support BLV individuals in building mental models of different chart types.}

\begin{itemize}
    \item \rev{Supporting mental model creation}
    \begin{itemize}
        \item \rev{Chart layouts}
        \item \rev{Shapes of chart elements}
    \end{itemize}
    \item \rev{Learning by touch}
    \begin{itemize}
        \item \rev{Exploratory learning}
        \item \rev{Multisensory learning}
    \end{itemize}
\end{itemize}

\subsection*{\rev{Theme 2: Developing Transferable Knowledge}}

\rev{This theme examines how BLV individuals transfer and apply the knowledge of chart types acquired through tactile charts.}

\begin{itemize}
    \item \rev{Structural understanding of chart types}
    \item \rev{Understanding new alt text for the same chart type}
    \item \rev{Using template charts as references for comparison}
\end{itemize}

\subsection*{\rev{Theme 3: BLV Visualization Education}}

\rev{This theme examines the role of tactile charts in visualization education for BLV individuals.}

\begin{itemize}
    \item \rev{Challenges faced by BLV individuals in education}
    \item \rev{Interest in learning visualization}
    \item \rev{Empowering BLV education with tactile charts}
\end{itemize}

\subsection*{\rev{Theme 4: Other Modalities}}
\rev{This theme examines other modalities preferred by BLV individuals.}

\begin{itemize}
    \item \rev{Textual descriptions}
    \item \rev{Raw data and AI tools}
\end{itemize}

\subsection*{\rev{Theme 5: Design Better Tactile Charts For Chart Type Education}}

\rev{This theme examines design considerations for tactile charts intended for chart learning purposes.}

\begin{itemize}
    \item \rev{Topic familiarity}
    \item \rev{Clarity of chart elements}
    \item \rev{Use of Braille}
    \item \rev{Production methods}
    \item \rev{Supporting blind-sighted communication}
    \item \rev{Supporting independent learning}
\end{itemize}

\clearpage
\section{Template Chart Design Variations}
\label{sec:template-chart-design-variations}
\begin{figure}[ht]
    \centering
        \includegraphics[width=1\columnwidth]{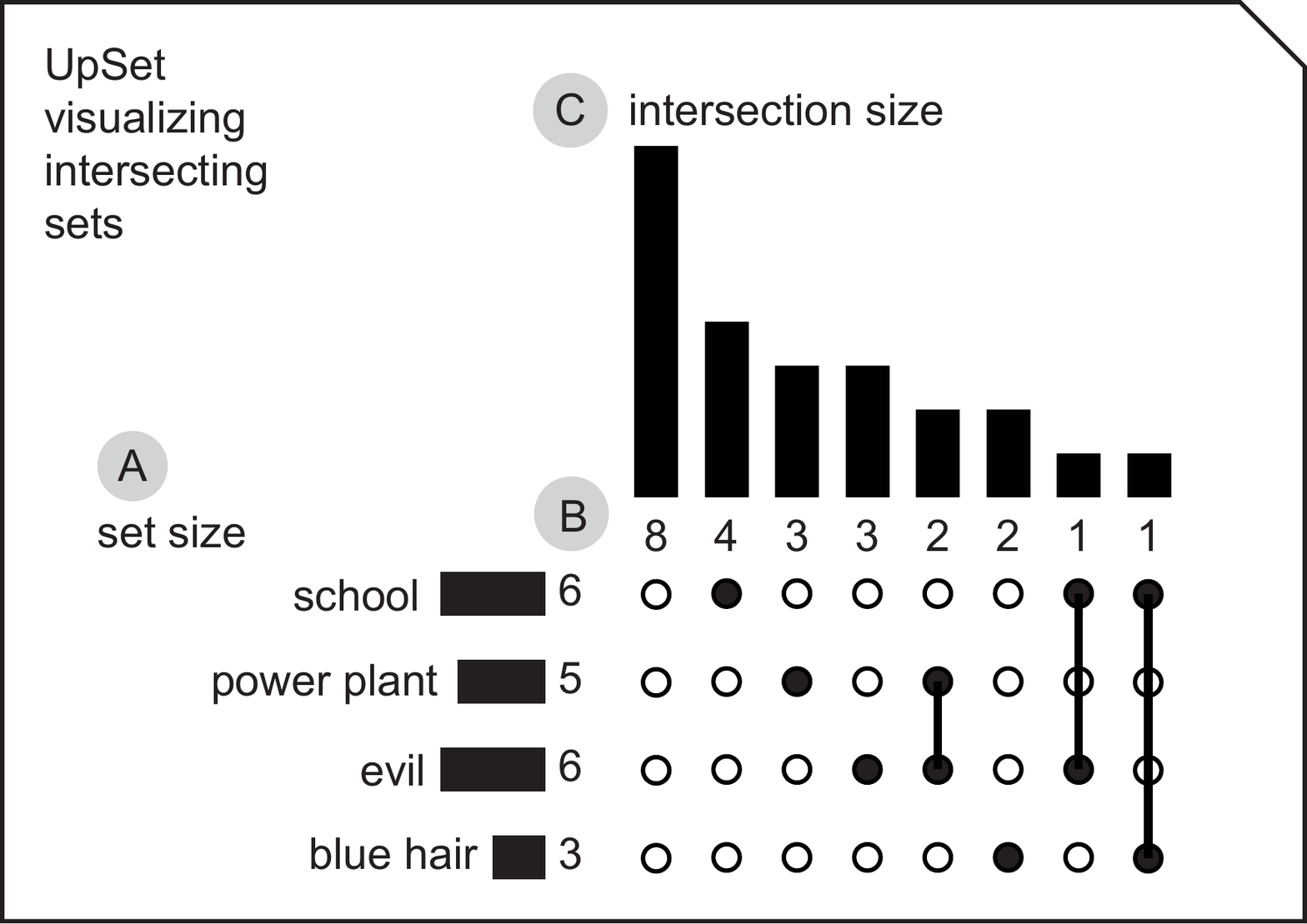}
    \caption{The tactile chart design for UpSet plot, Design 1, sighted version.}
    \label{fig:upset-design1-sighted}
\end{figure}

\begin{figure}[ht]
    \centering
        \includegraphics[width=1\columnwidth]{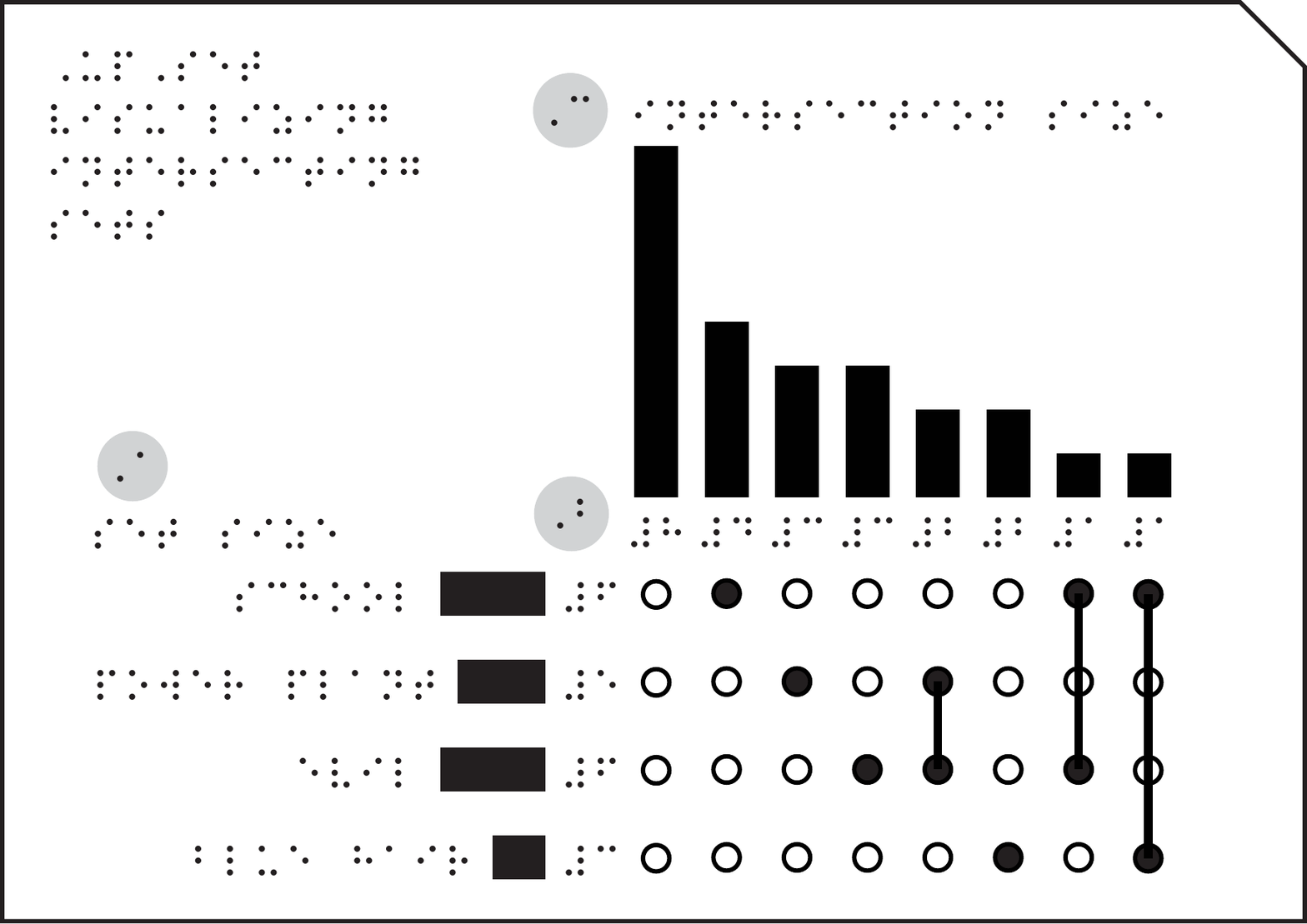}
    \caption{The tactile chart design for UpSet plot, Design 1, Braille version.}
    \label{fig:upset-design1-Braille}
\end{figure}

\begin{figure}[ht]
    \centering
        \includegraphics[width=1\columnwidth]{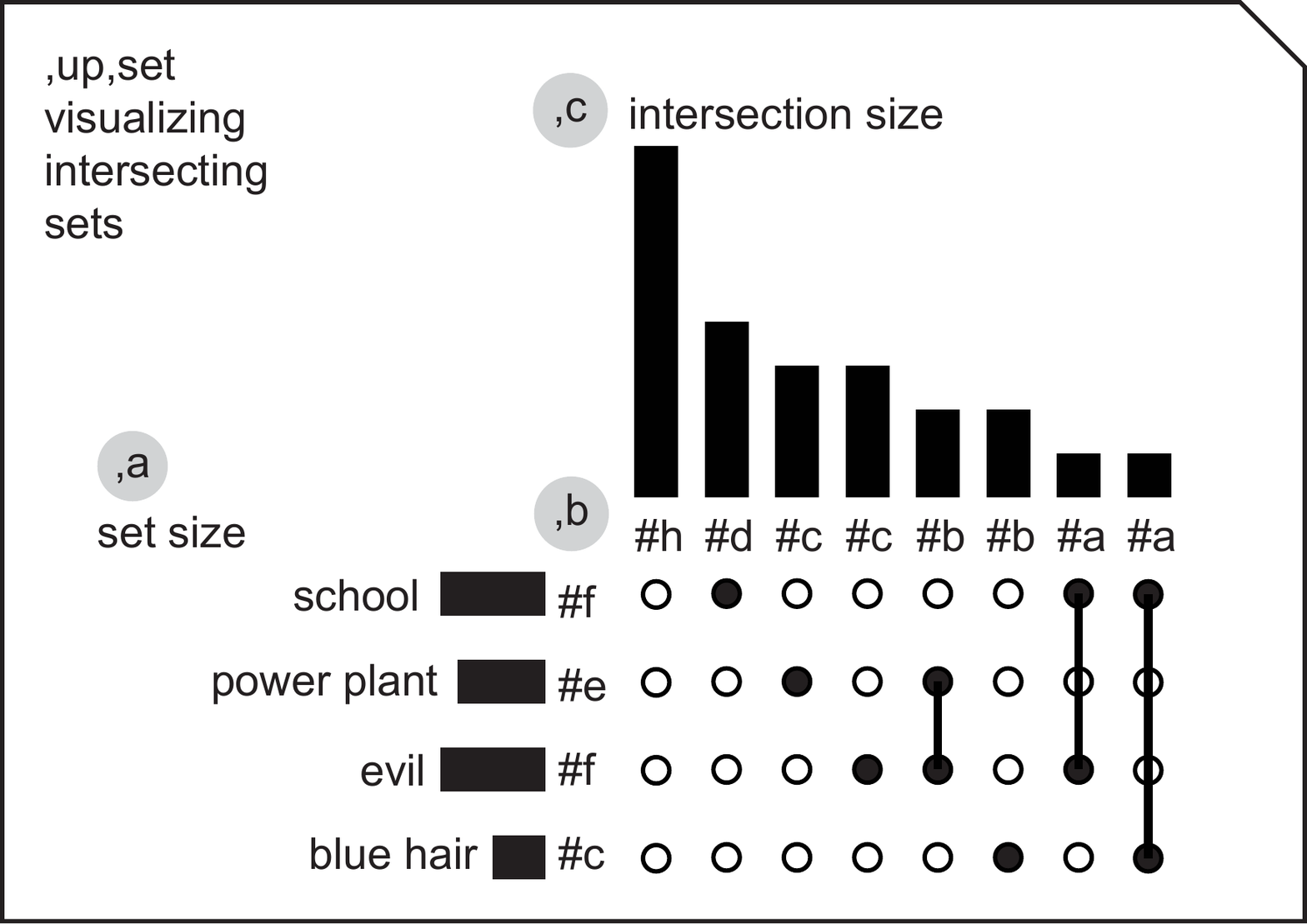}
    \caption{The tactile chart design for UpSet plot, Design 1, Braille-to-English letter-by-letter translation version.}
    \label{fig:upset-design1-characters}
\end{figure}

\begin{figure}[!t]
    \centering
        \includegraphics[width=1\columnwidth]{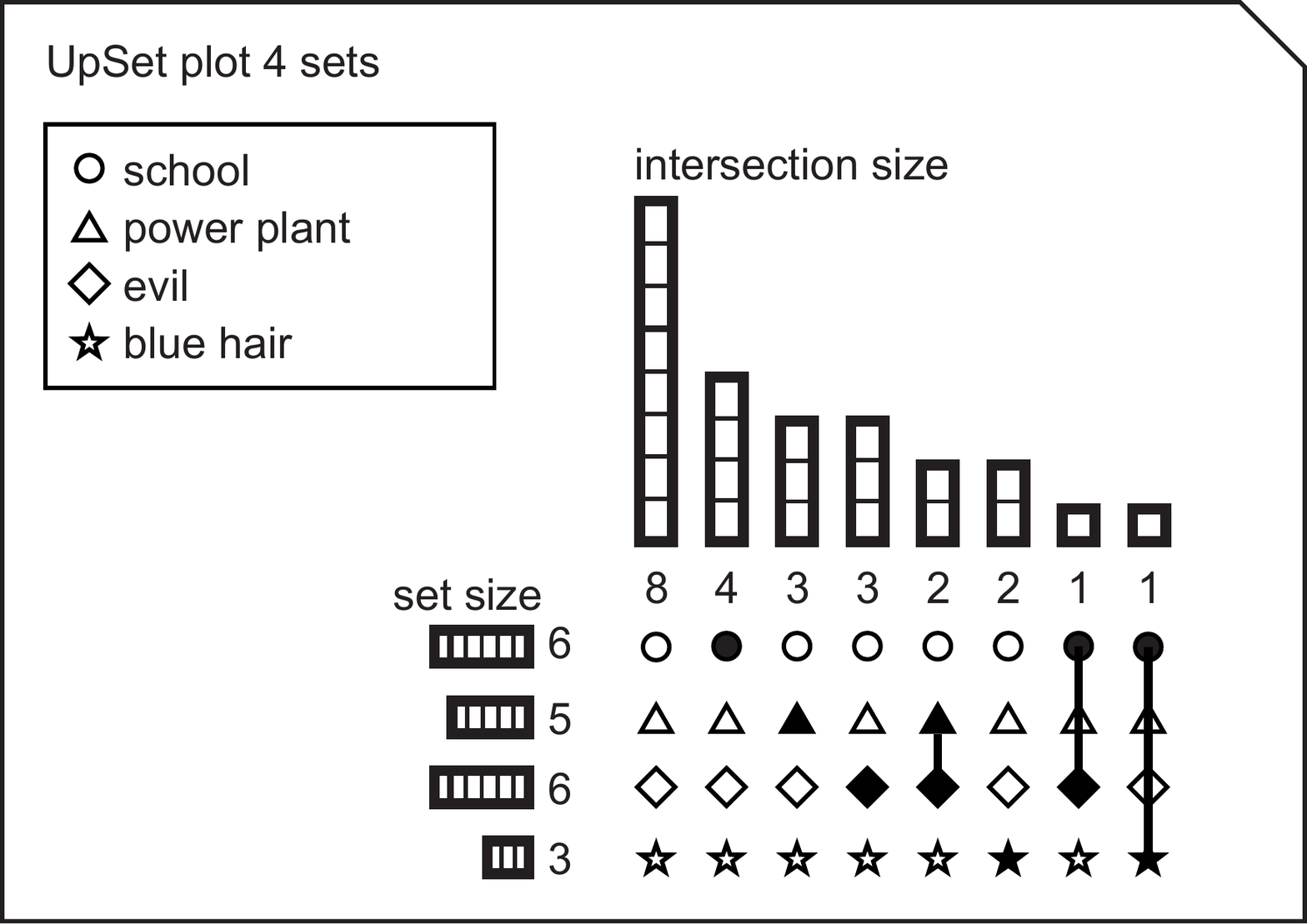}
    \caption{The tactile chart design for UpSet plot, Design 2, sighted version.}
    \label{fig:upset-design2-sighted}
\end{figure}

\begin{figure}[!t]
    \centering
        \includegraphics[width=1\columnwidth]{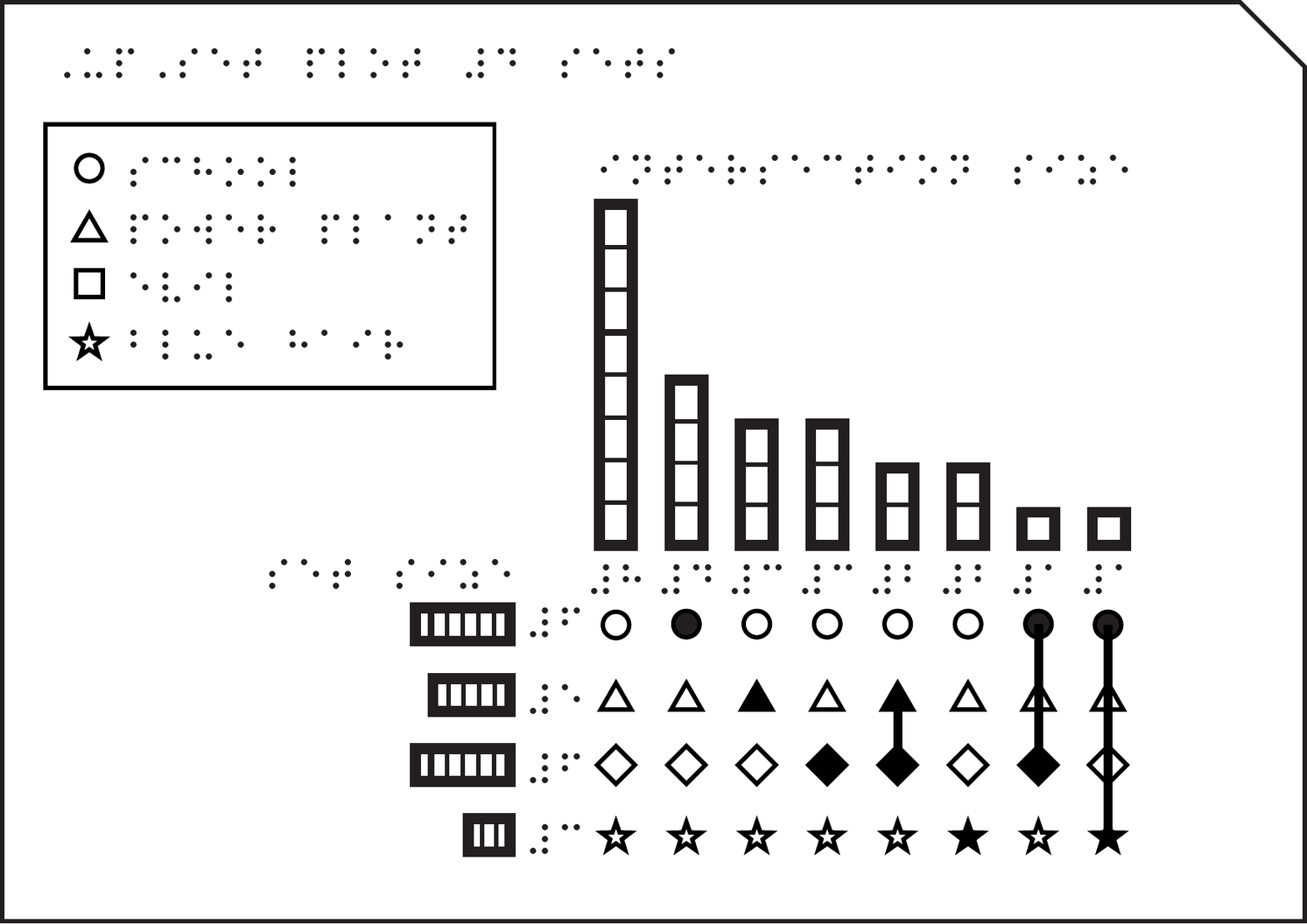}
    \caption{The tactile chart design for UpSet plot, Design 2, Braille version.}
    \label{fig:upset-design2-Braille}
\end{figure}

\begin{figure}[!t]
    \centering
        \includegraphics[width=1\columnwidth]{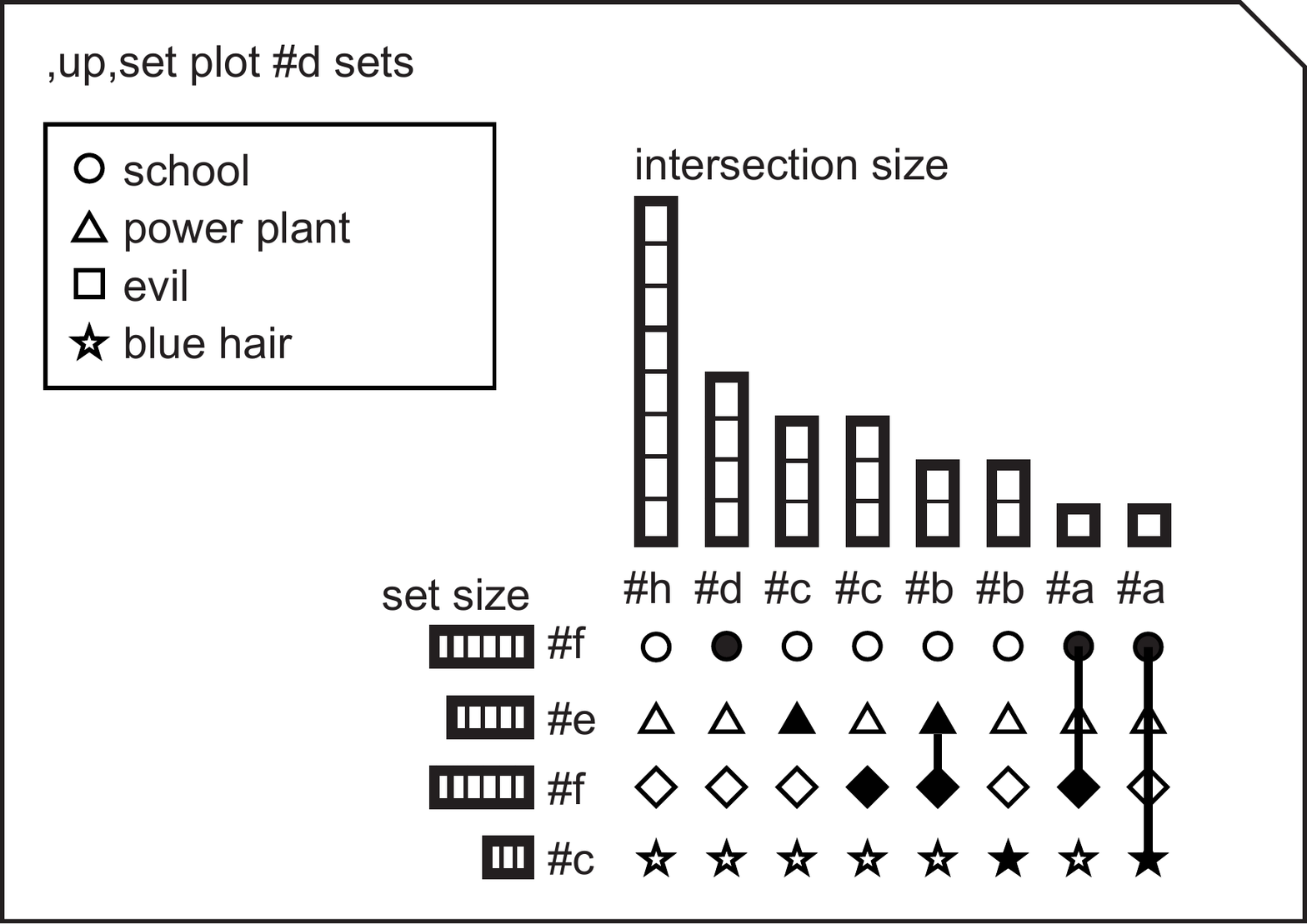}
    \caption{The tactile chart design for UpSet plot, Design 2, Braille-to-English letter-by-letter translation version.}
    \label{fig:upset-design2-characters}
\end{figure}

\begin{figure}[!t]
    \centering
        \includegraphics[width=1\columnwidth]{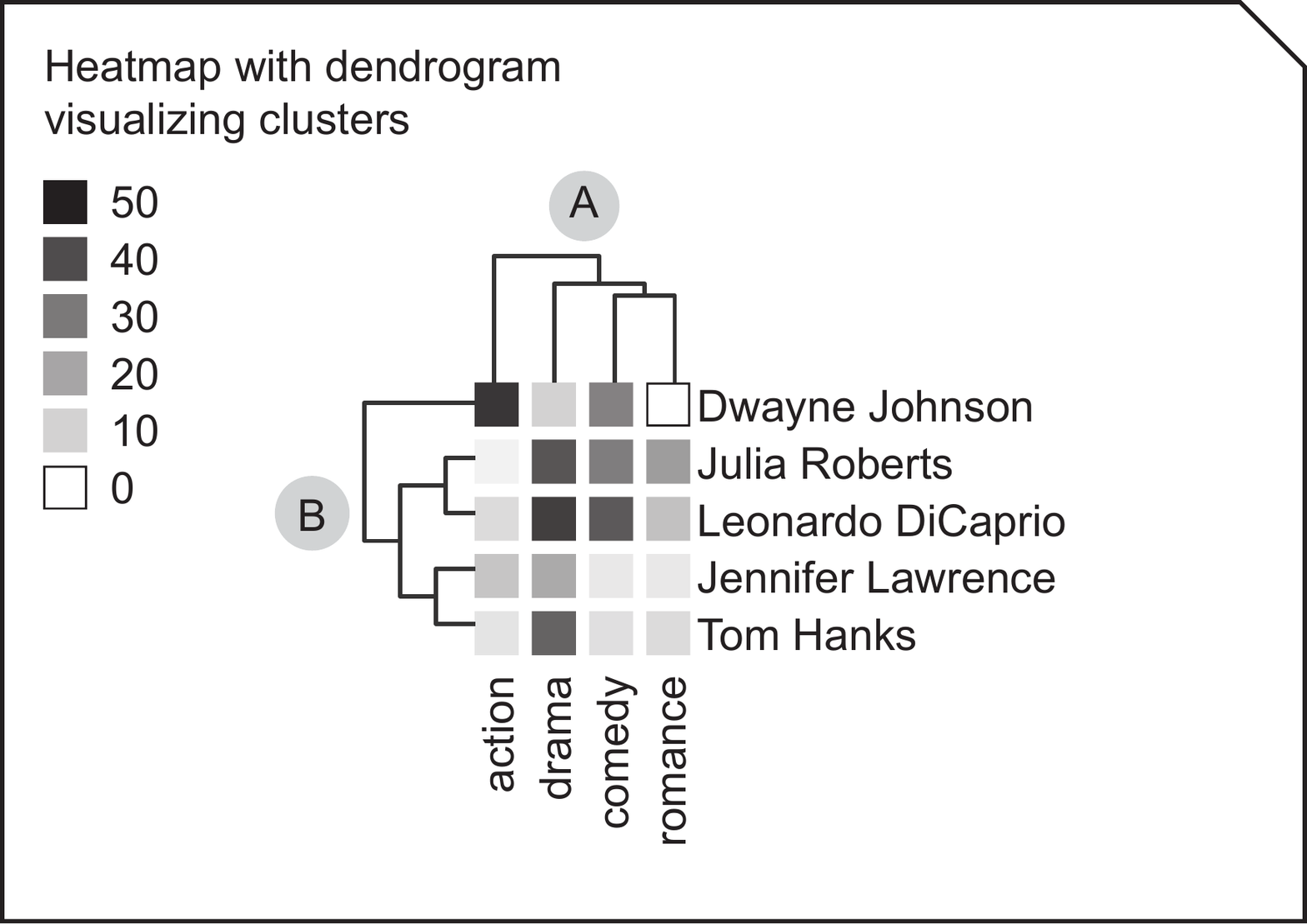}
    \caption{The tactile chart design for clustered heatmap, Design 1, sighted version.}
    \label{fig:heatmap-design1-sighted}
\end{figure}

\begin{figure}[!t]
    \centering
        \includegraphics[width=1\columnwidth]{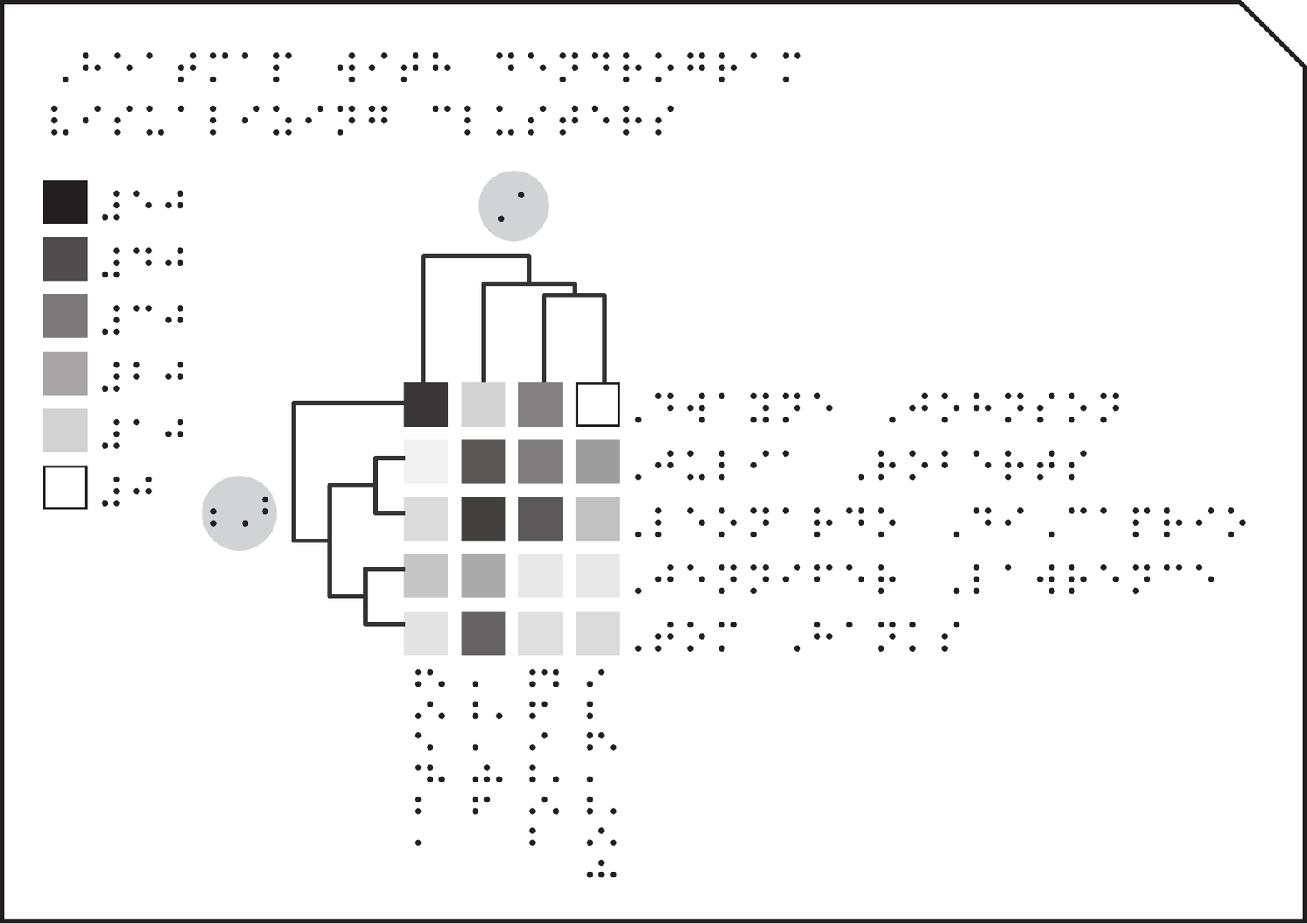}
    \caption{The tactile chart design for clustered heatmap, Design 1, Braille version.}
    \label{fig:heatmap-design1-Braille}
\end{figure}

\begin{figure}[!t]
    \centering
        \includegraphics[width=1\columnwidth]{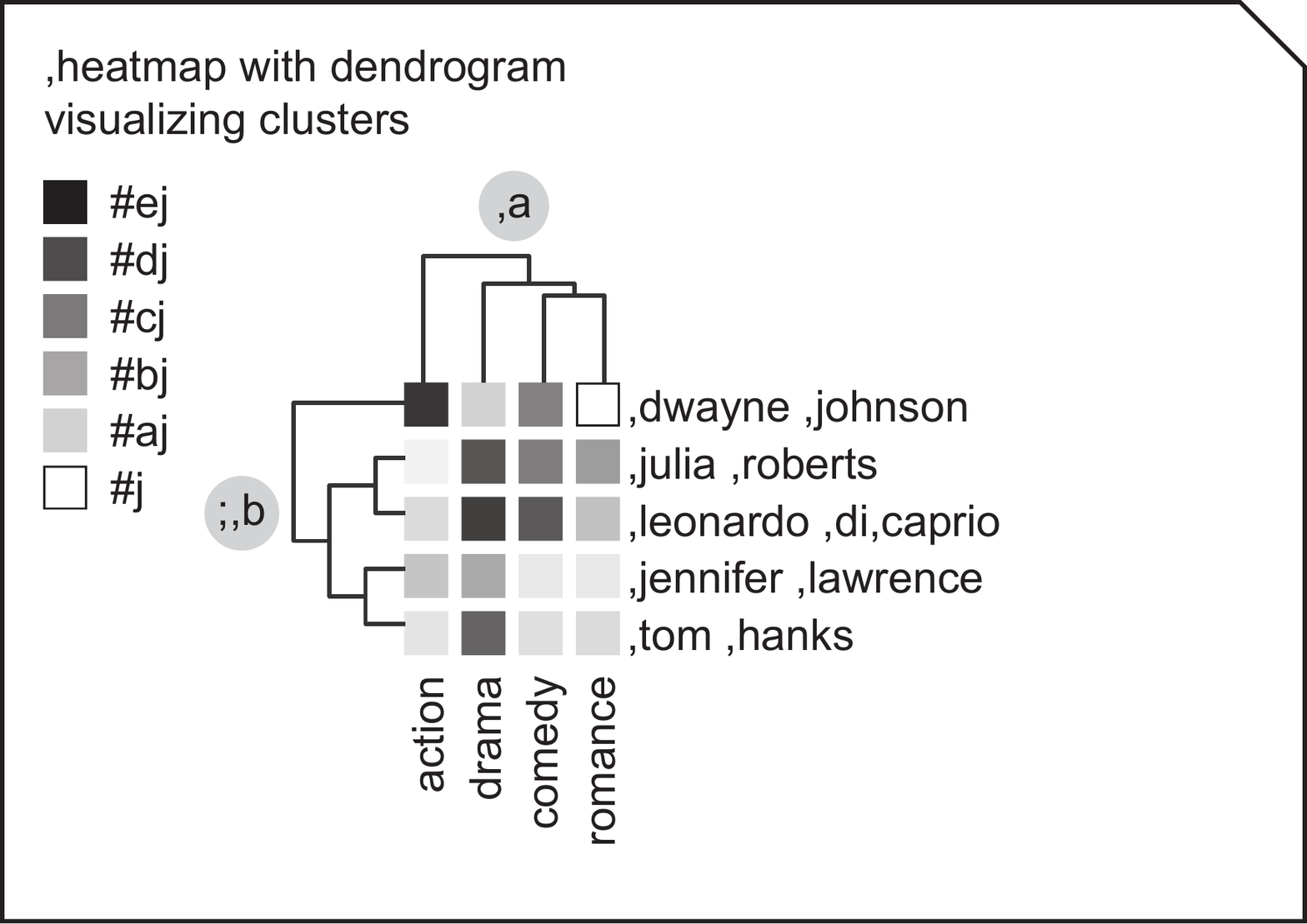}
    \caption{The tactile chart design for clustered heatmap, Design 1, Braille-to-English letter-by-letter translation version.}
    \label{fig:heatmap-design1-characters}
\end{figure}

\begin{figure}[!t]
    \centering
        \includegraphics[width=1\columnwidth]{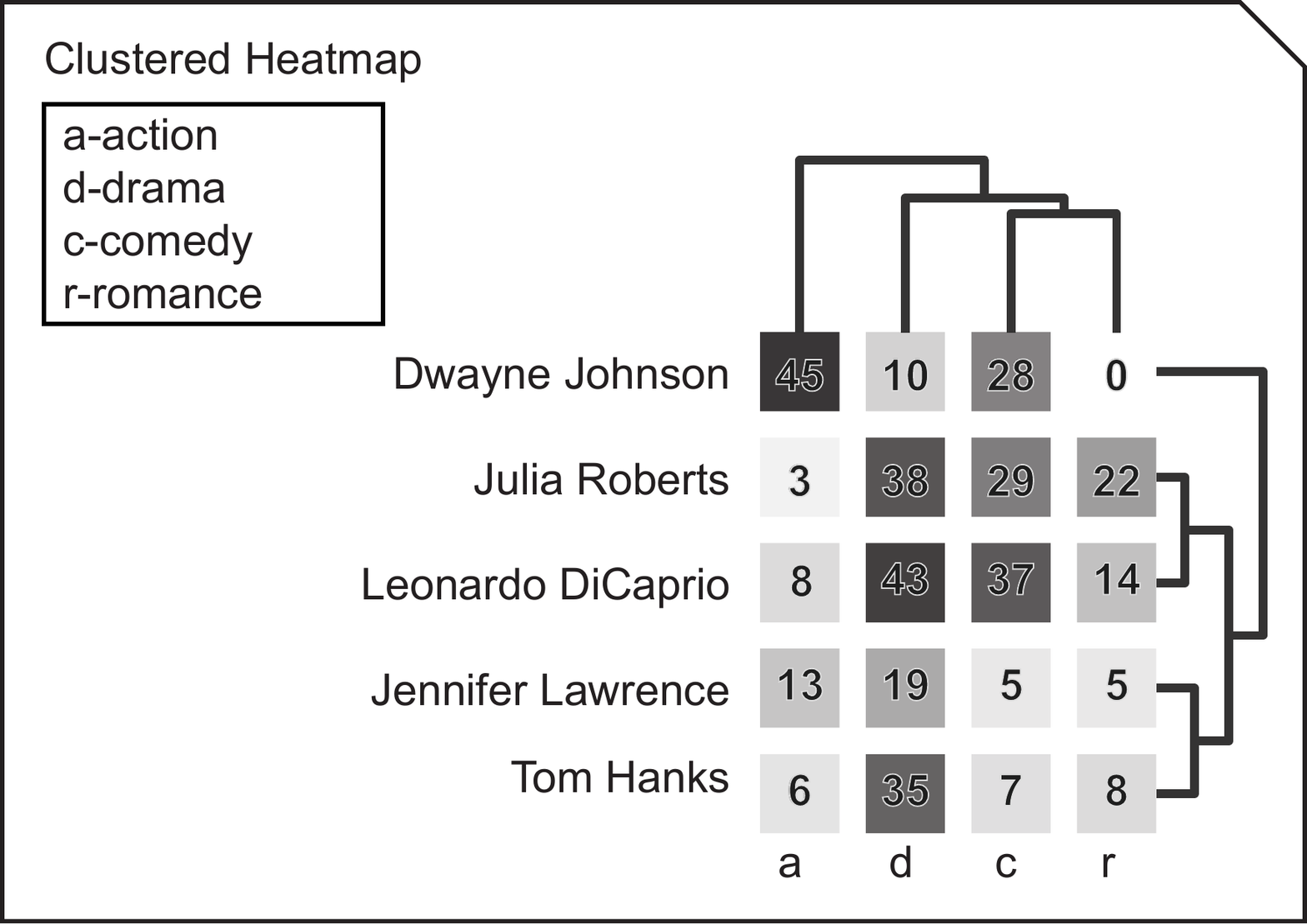}
    \caption{The tactile chart design for clustered heatmap, Design 2, sighted version.}
    \label{fig:heatmap-design2-sighted}
\end{figure}

\begin{figure}[!t]
    \centering
        \includegraphics[width=1\columnwidth]{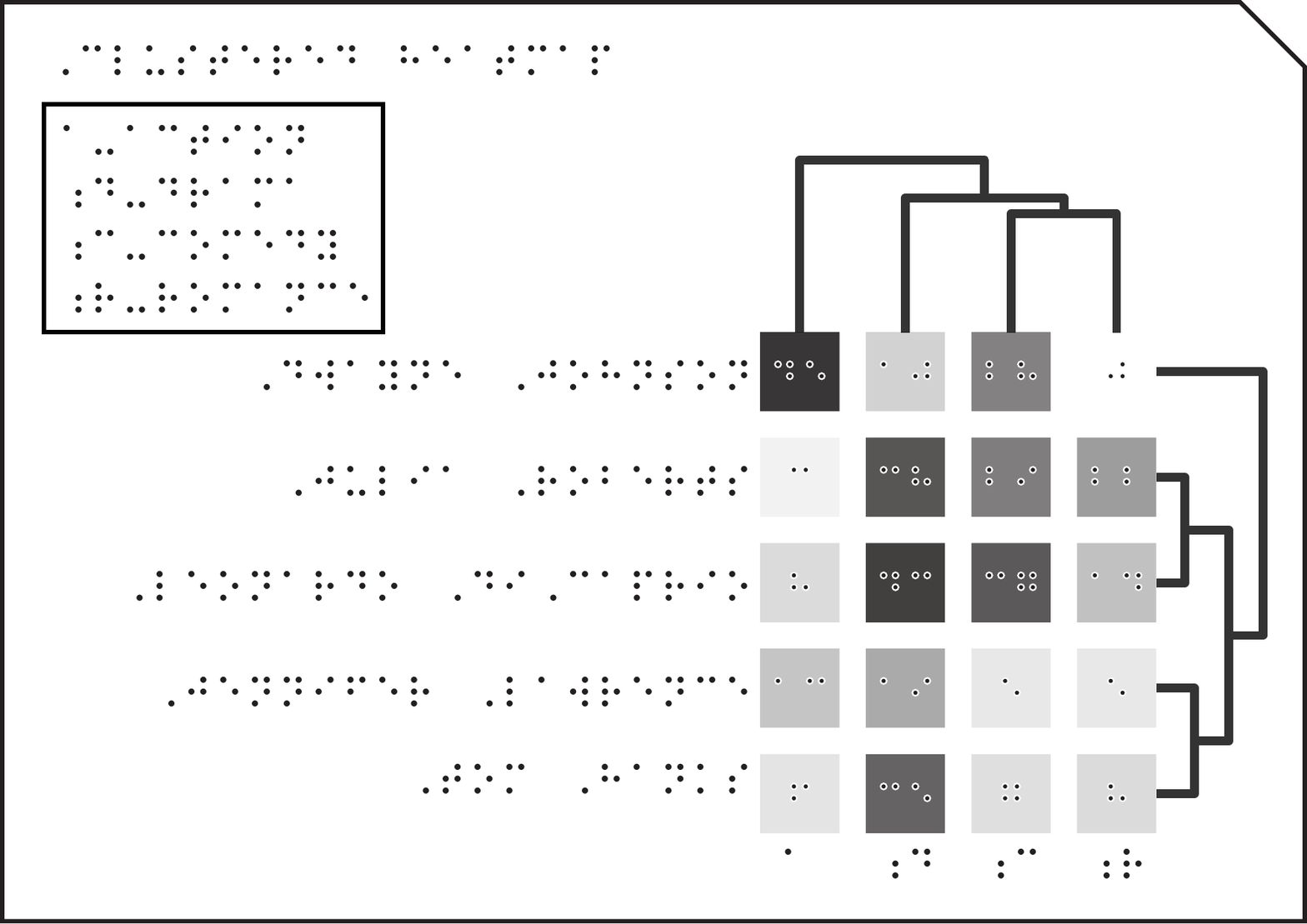}
    \caption{The tactile chart design for clustered heatmap, Design 2, Braille version.}
    \label{fig:heatmap-design2-Braille}
\end{figure}

\begin{figure}[!t]
    \centering
        \includegraphics[width=1\columnwidth]{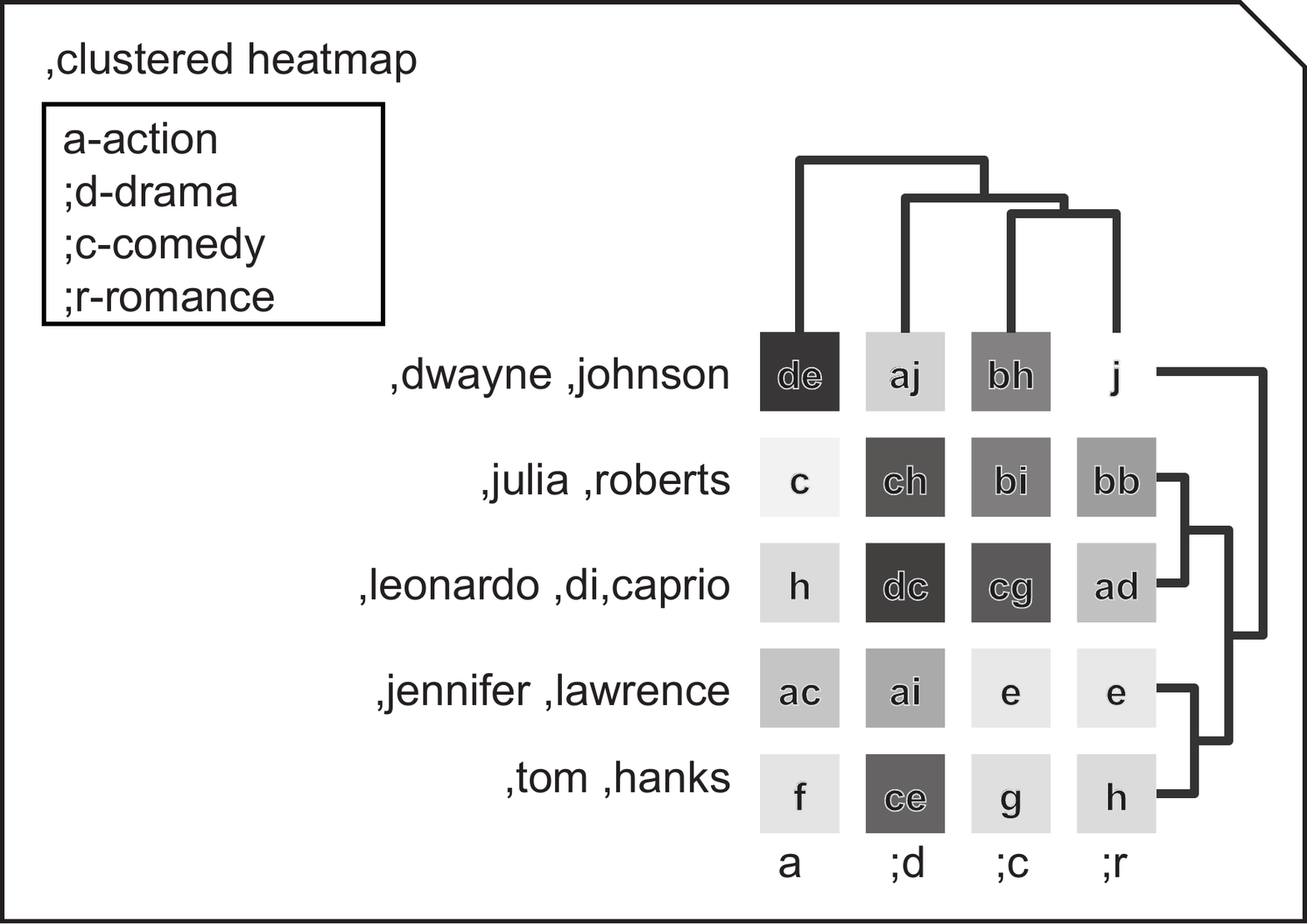}
    \caption{The tactile chart design for clustered heatmap, Design 2, Braille-to-English letter-by-letter translation version.}
    \label{fig:heatmap-design2-characters}
\end{figure}

\begin{figure}[!t]
    \centering
        \includegraphics[width=1\columnwidth]{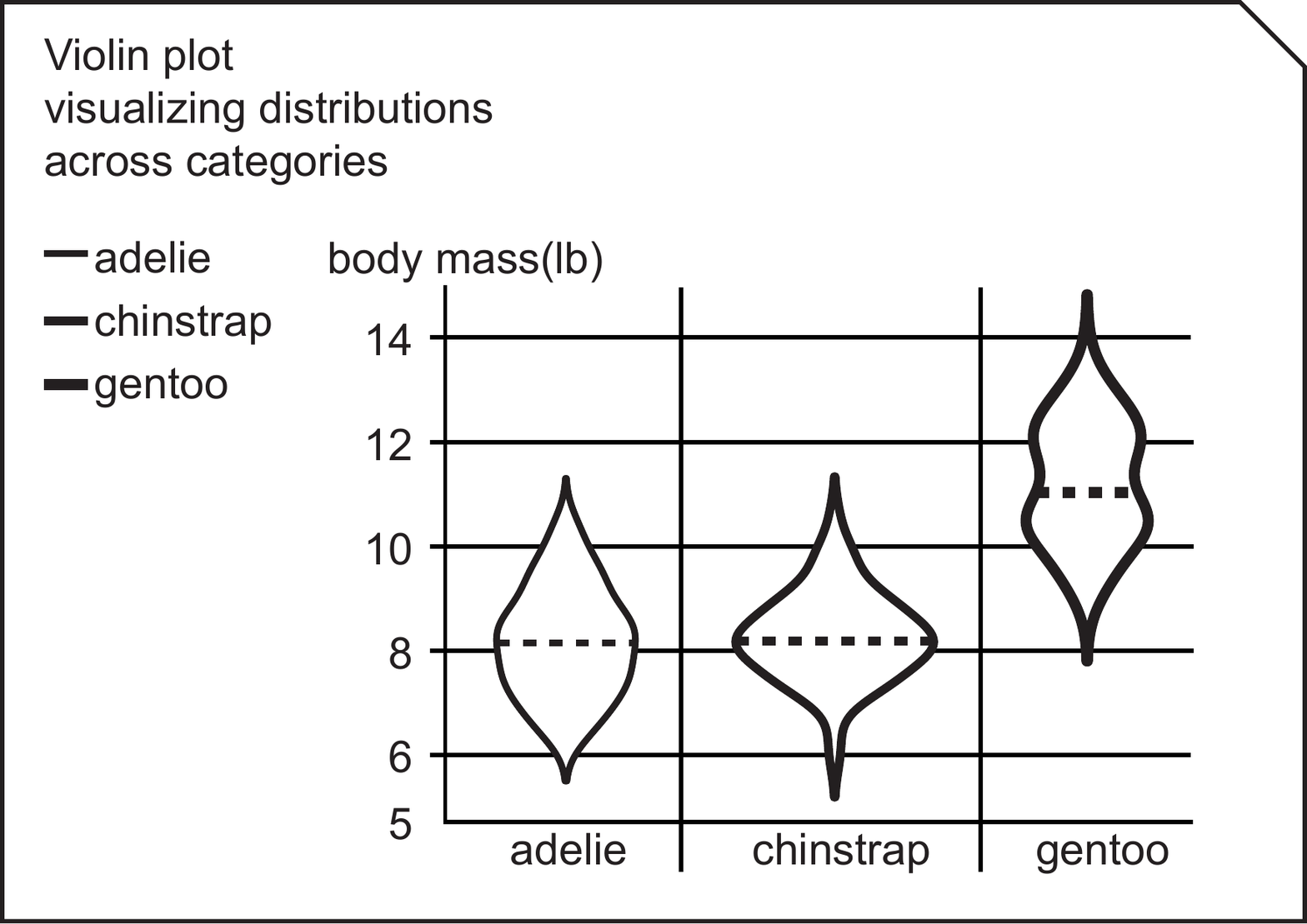}
    \caption{The tactile chart design for violin plot, Design 1, sighted version.}
    \label{fig:violin-design1-sighted}
\end{figure}

\begin{figure}[!t]
    \centering
        \includegraphics[width=1\columnwidth]{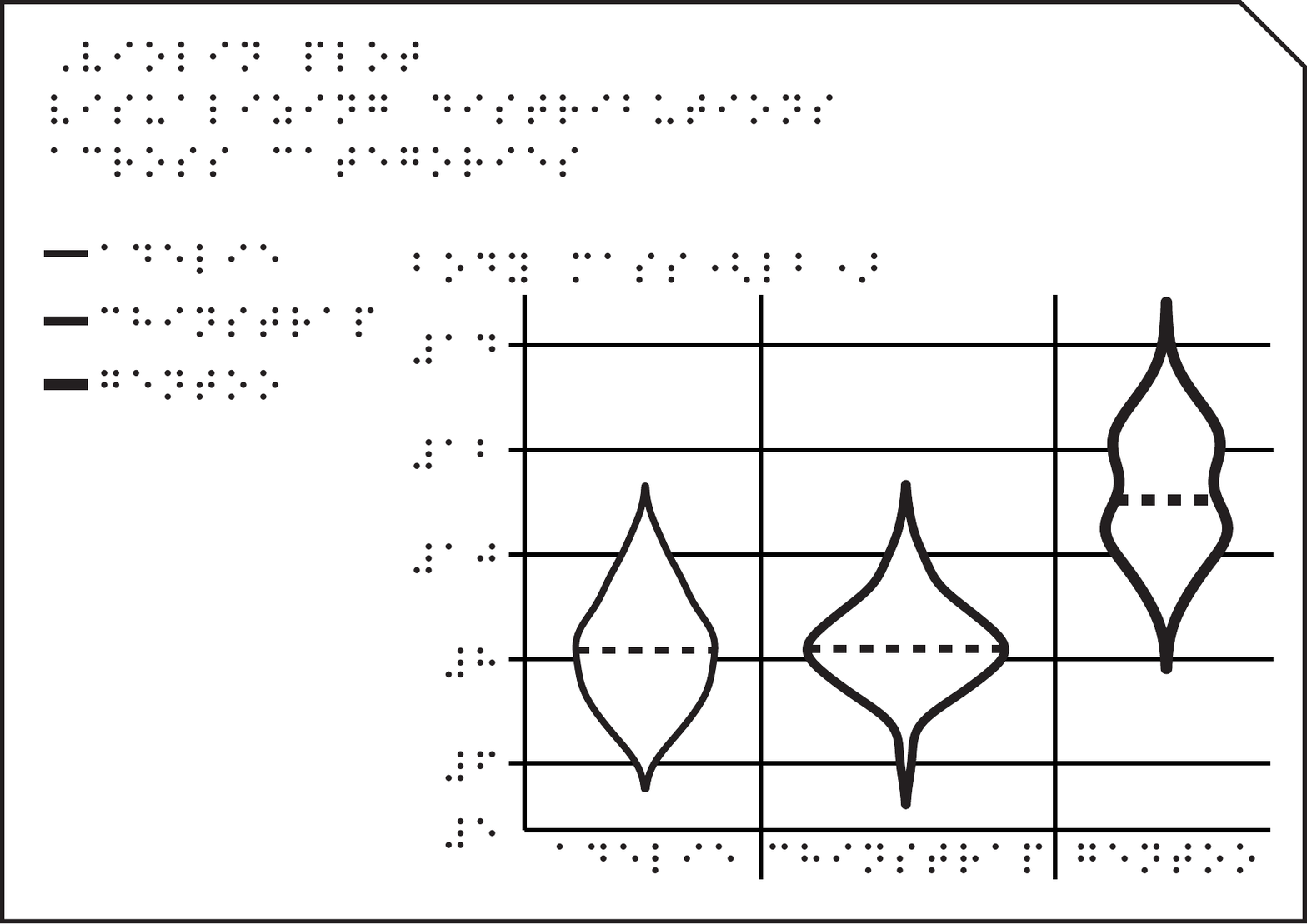}
    \caption{The tactile chart design for violin plot, Design 1, Braille version.}
    \label{fig:violin-design1-Braille}
\end{figure}

\begin{figure}[!t]
    \centering
        \includegraphics[width=1\columnwidth]{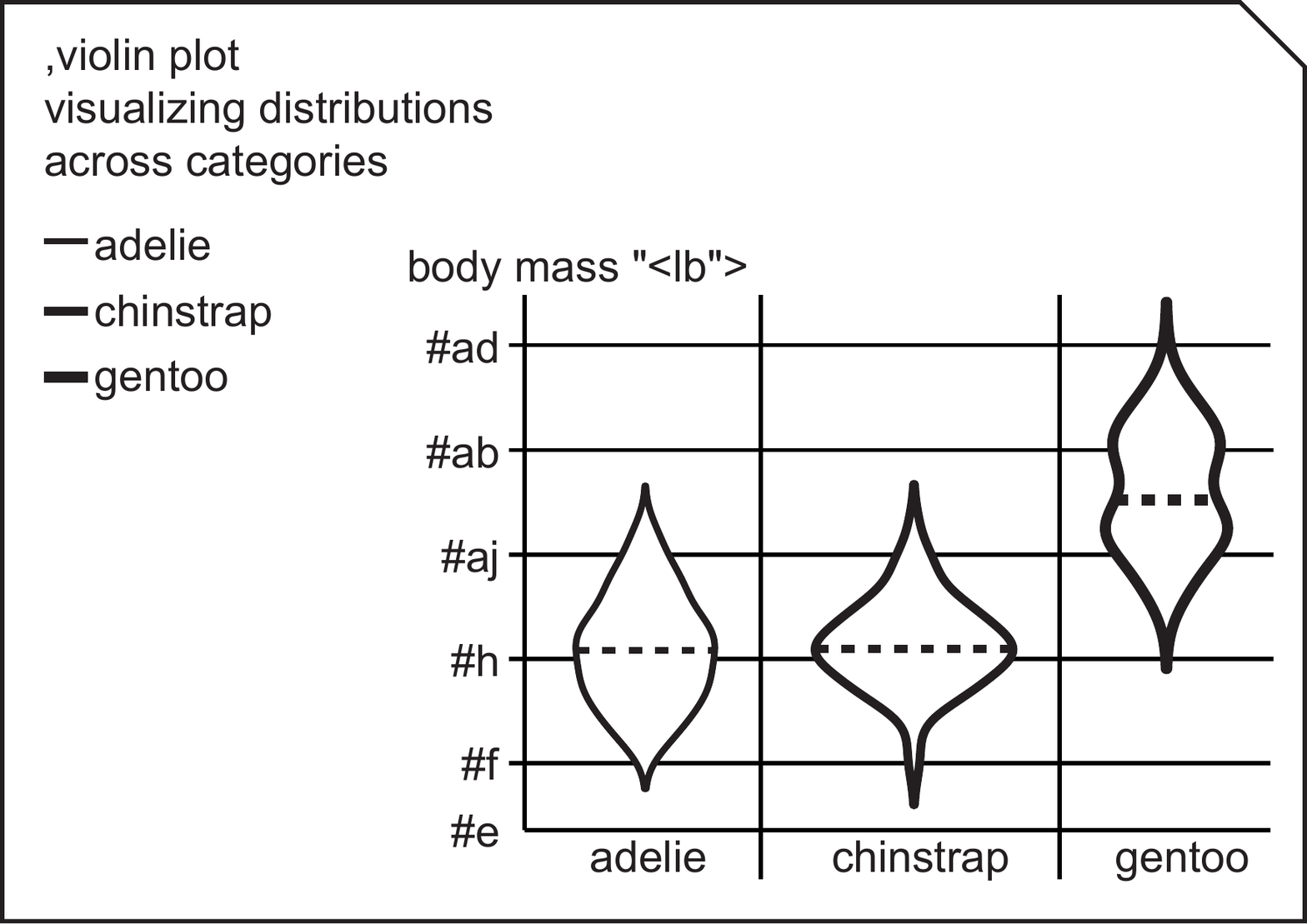}
    \caption{The tactile chart design for violin plot, Design 1, Braille-to-English letter-by-letter translation version.}
    \label{fig:violin-design1-characters}
\end{figure}

\begin{figure}[!t]
    \centering
        \includegraphics[width=1\columnwidth]{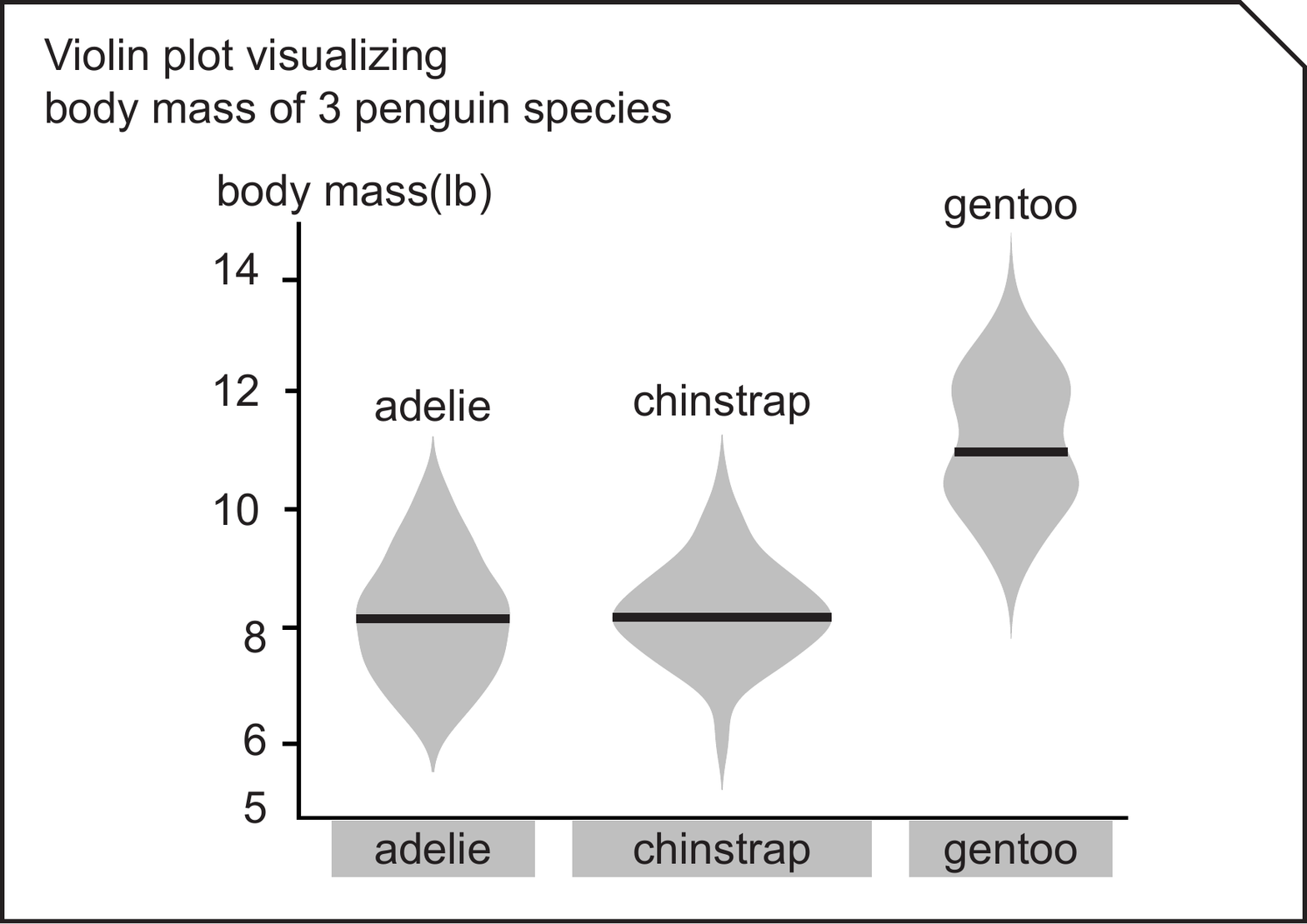}
    \caption{The tactile chart design for violin plot, Design 2, sighted version.}
    \label{fig:violin-design2-sighted}
\end{figure}

\begin{figure}[!t]
    \centering
        \includegraphics[width=1\columnwidth]{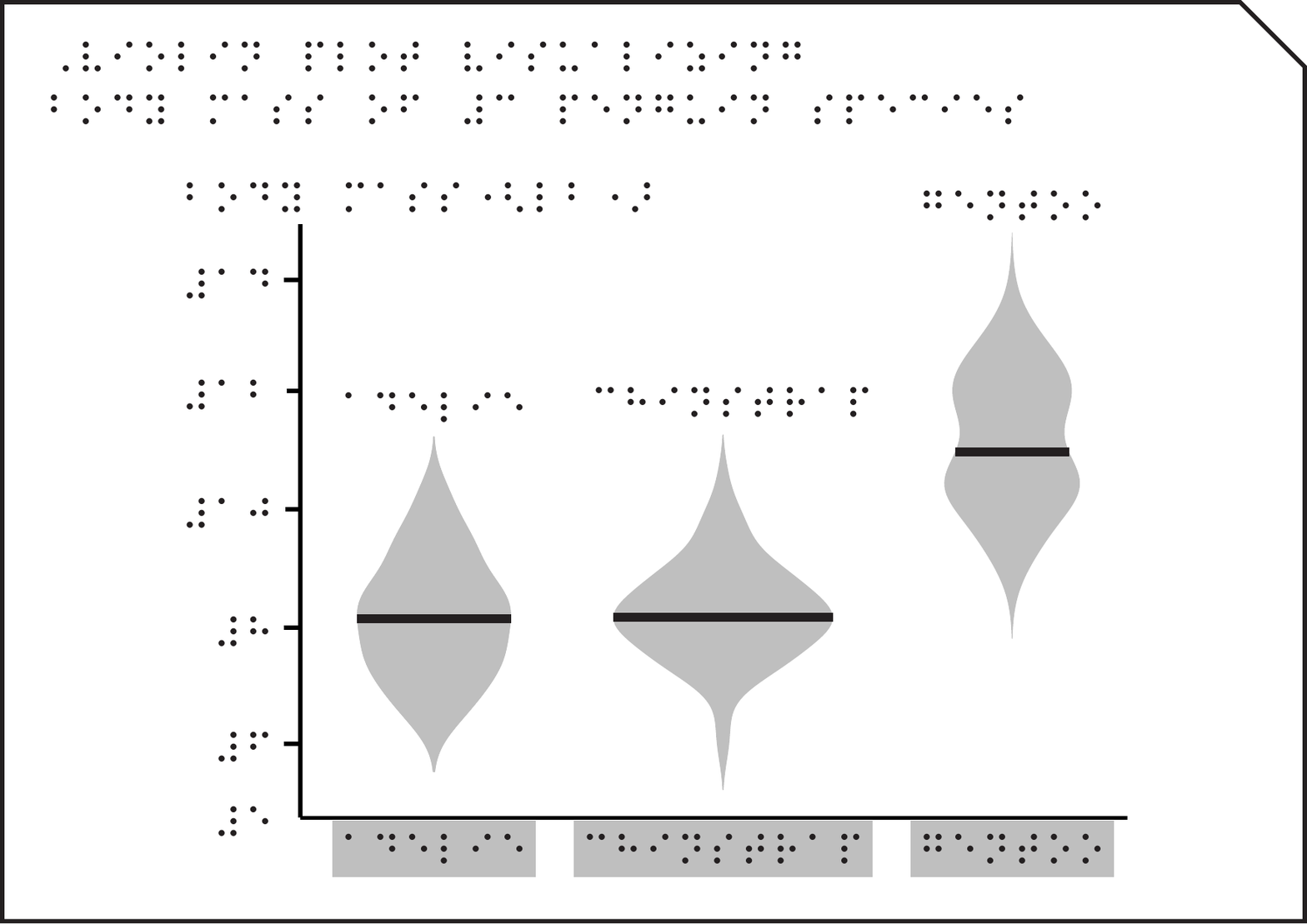}
    \caption{The tactile chart design for violin plot, Design 2, Braille version.}
    \label{fig:violin-design2-Braille}
\end{figure}

\begin{figure}[!t]
    \centering
        \includegraphics[width=1\columnwidth]{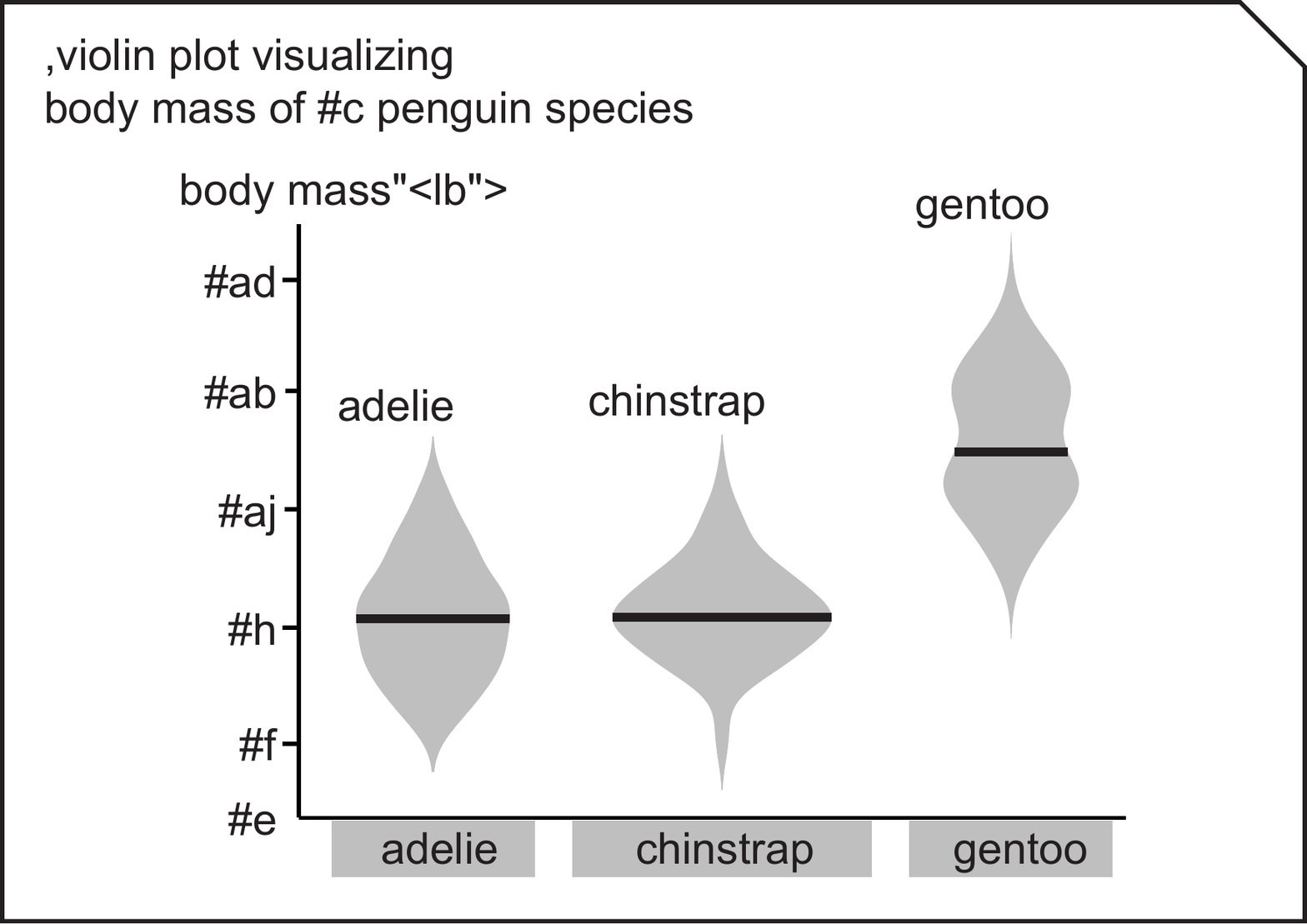}
    \caption{The tactile chart design for violin plot, Design 2, Braille-to-English letter-by-letter translation version.}
    \label{fig:violin-design2-characters}
\end{figure}

\begin{figure}[!t]
    \centering
        \includegraphics[width=1\columnwidth]{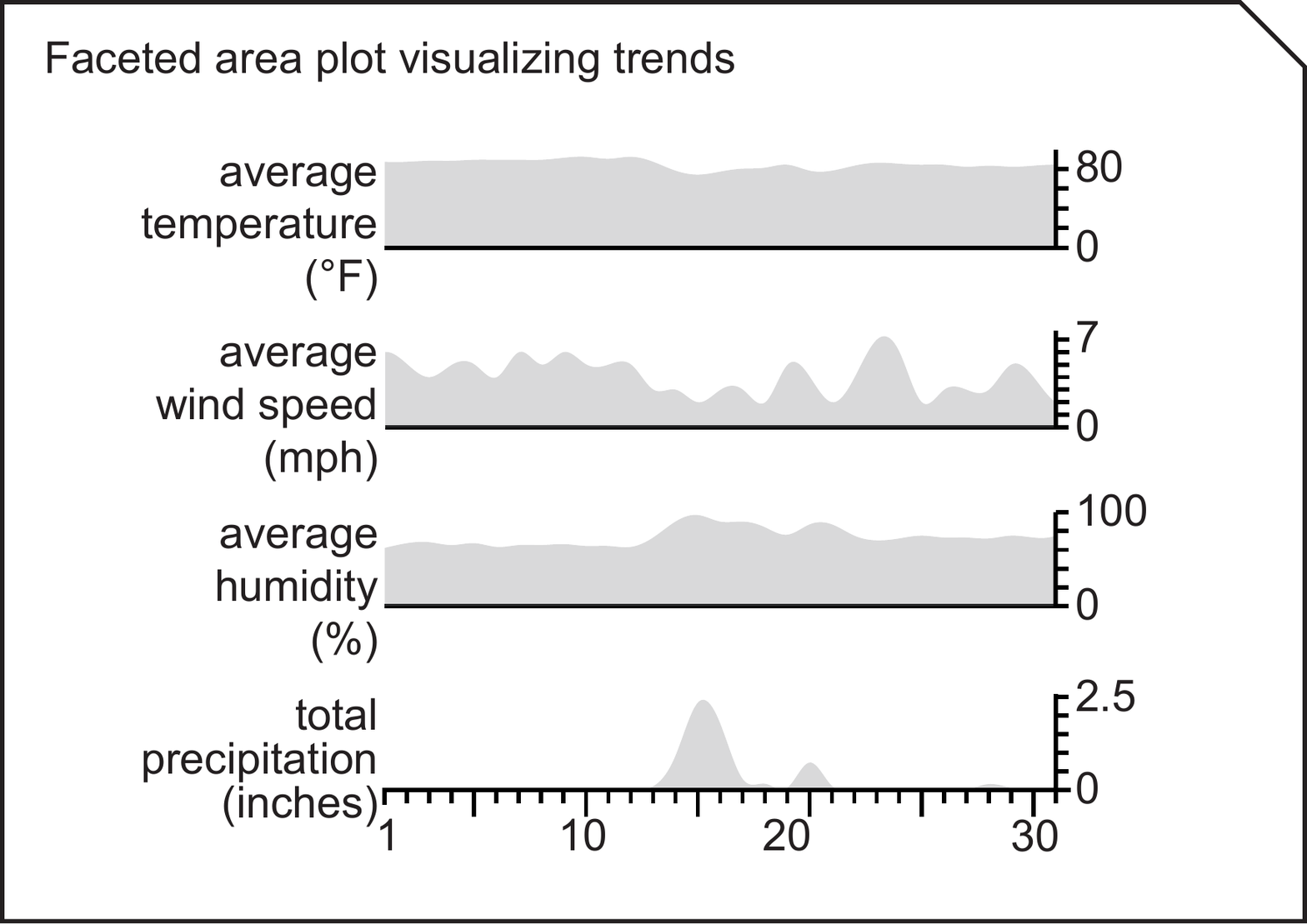}
    \caption{The tactile chart design for faceted plot, Design 1, sighted version.}
    \label{fig:faceted-design1-sighted}
\end{figure}

\begin{figure}[!t]
    \centering
        \includegraphics[width=1\columnwidth]{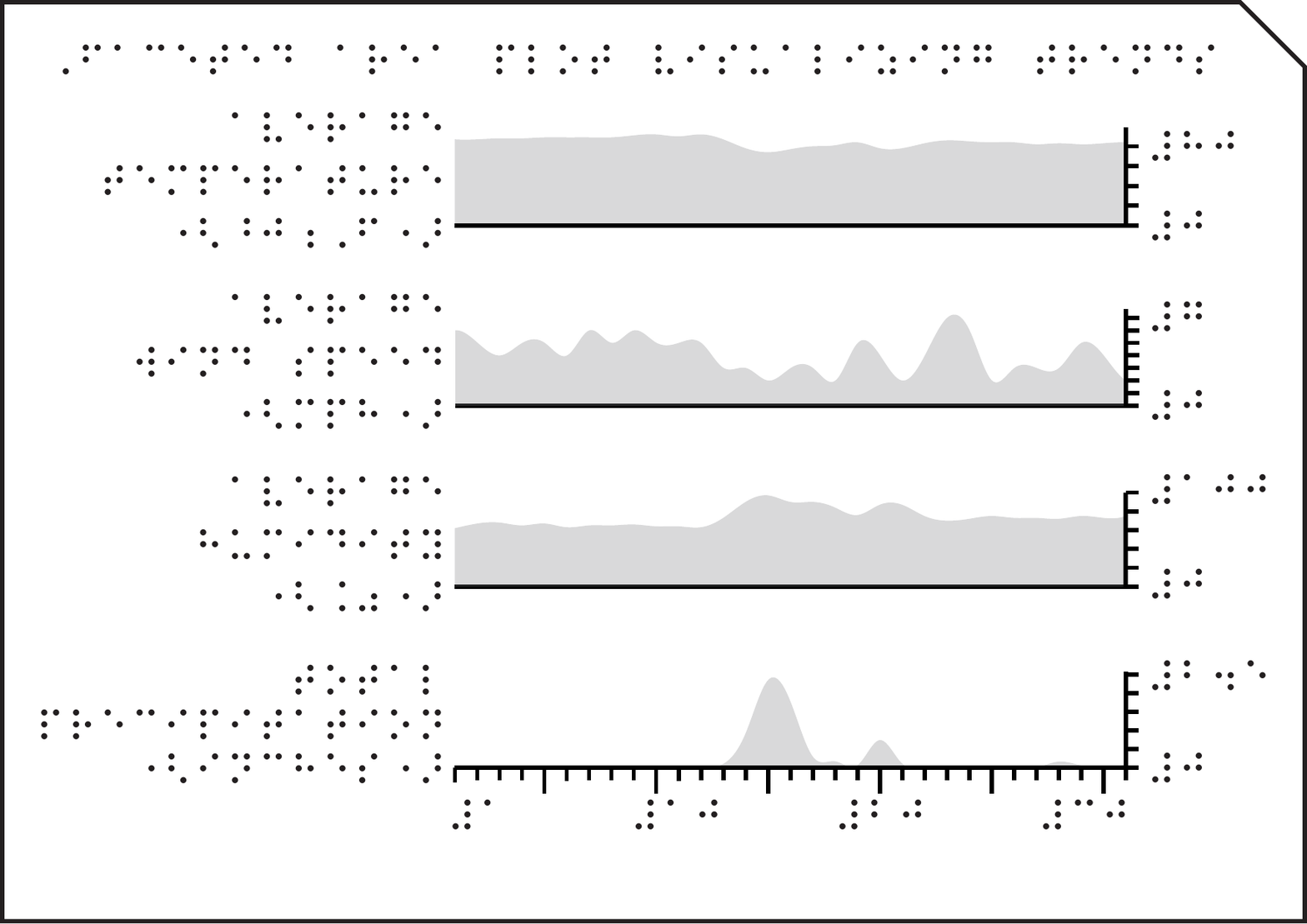}
    \caption{The tactile chart design for faceted plot, Design 1, Braille version.}
    \label{fig:faceted-design1-Braille}
\end{figure}

\begin{figure}[!t]
    \centering
        \includegraphics[width=1\columnwidth]{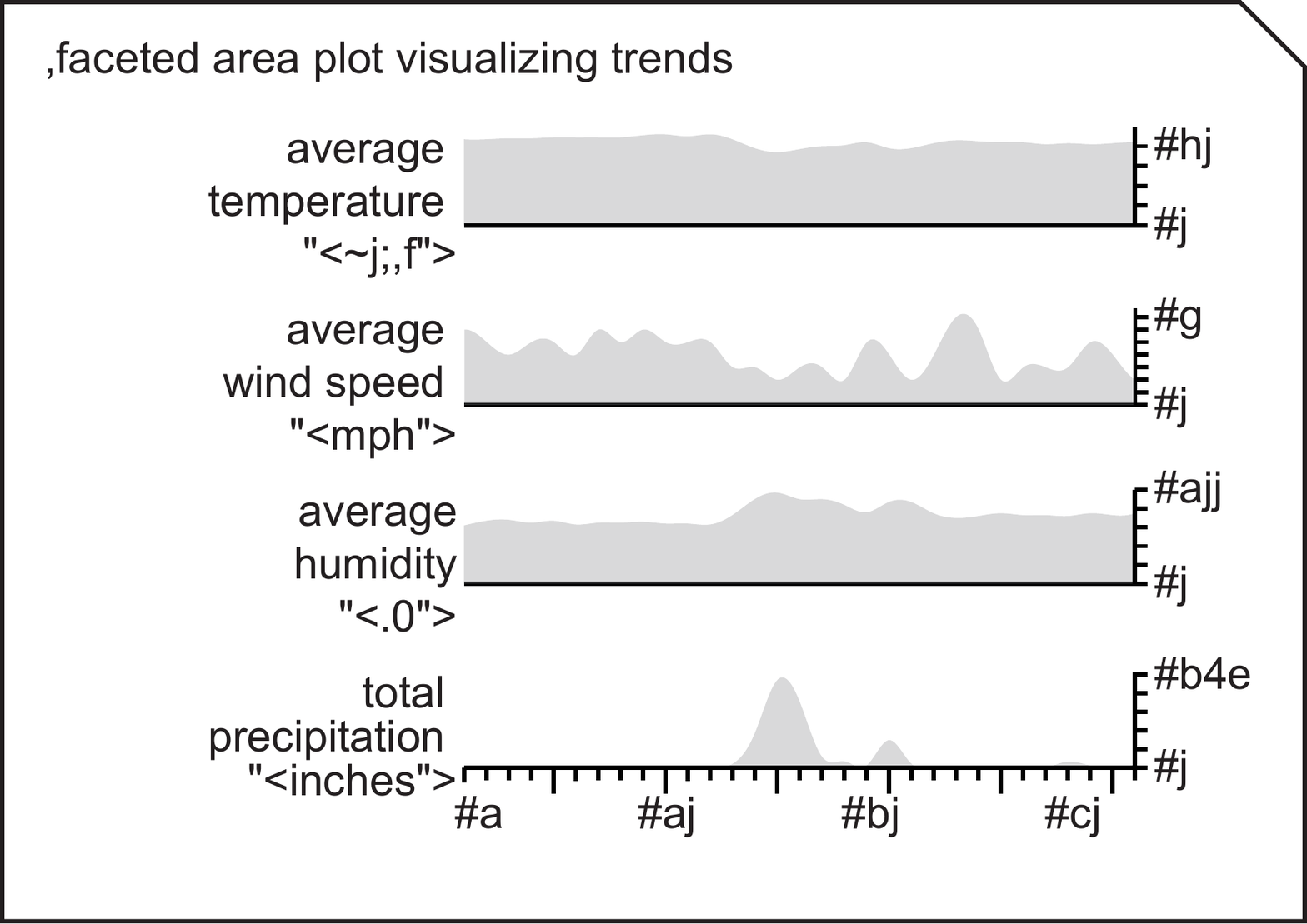}
    \caption{The tactile chart design for faceted plot, Design 1, Braille-to-English letter-by-letter translation version.}
    \label{fig:faceted-design1-characters}
\end{figure}

\begin{figure}[!t]
    \centering
        \includegraphics[width=1\columnwidth]{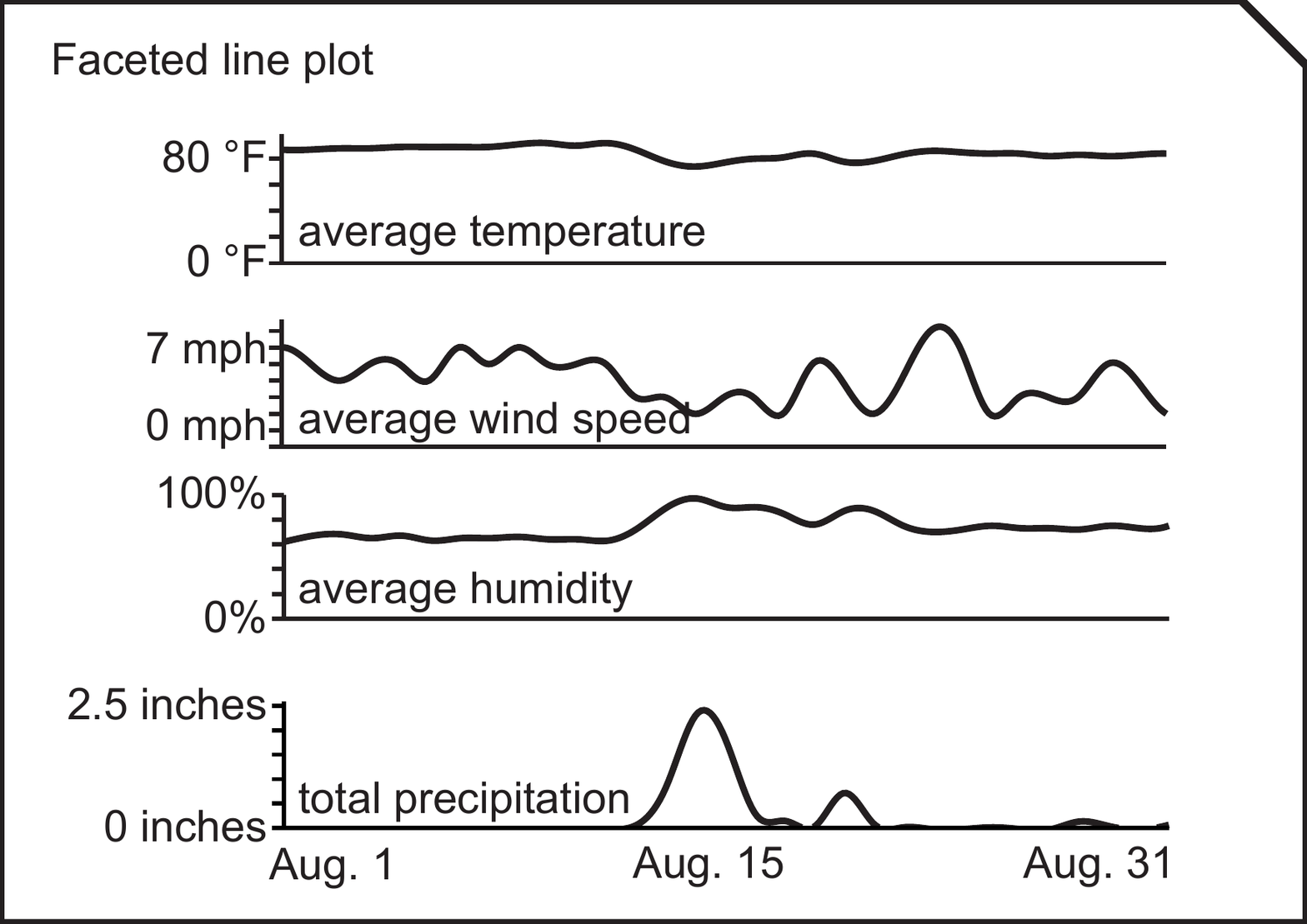}
    \caption{The tactile chart design for faceted plot, Design 2, sighted version.}
    \label{fig:faceted-design2-sighted}
\end{figure}

\begin{figure}[!t]
    \centering
        \includegraphics[width=1\columnwidth]{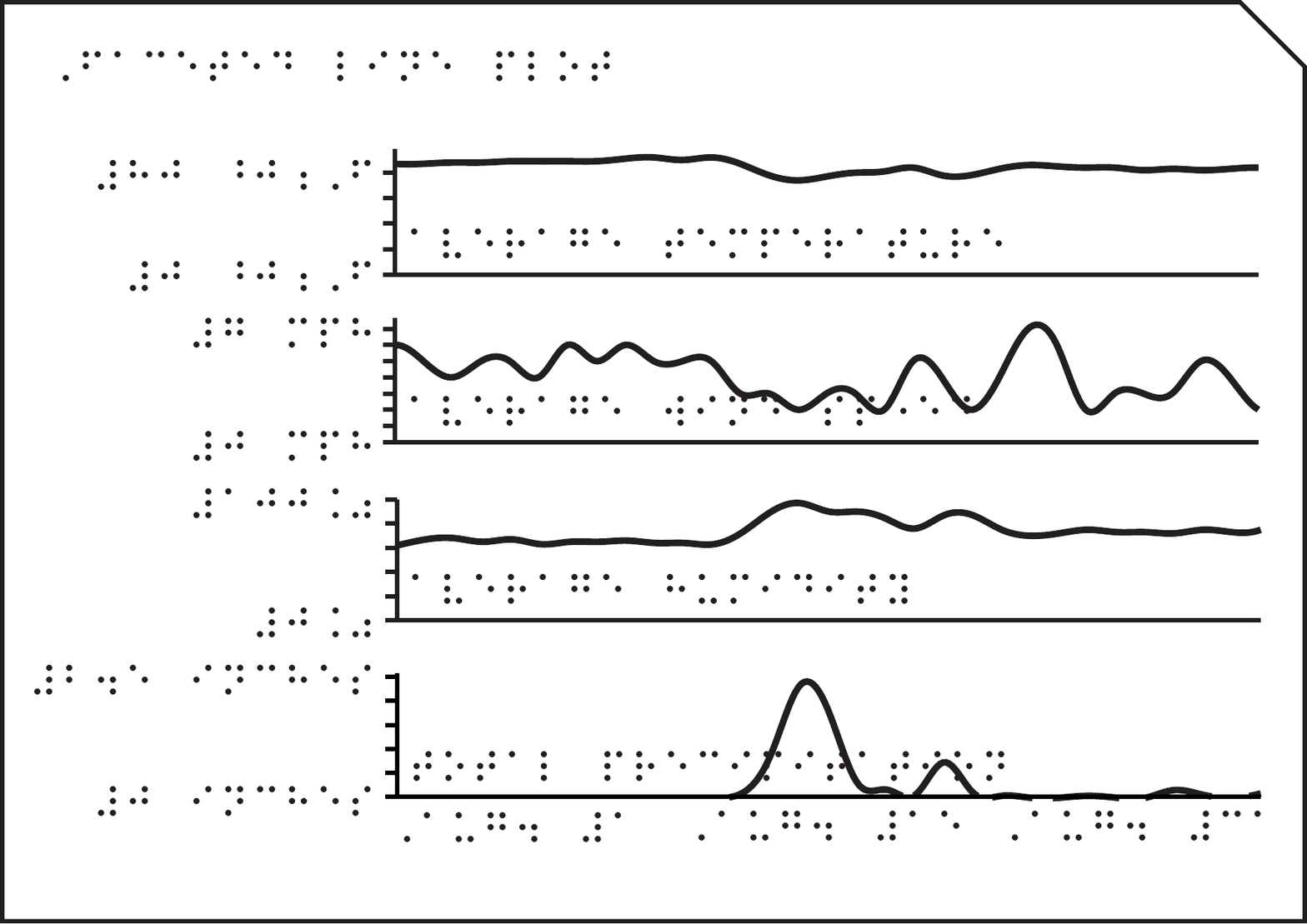}
    \caption{The tactile chart design for faceted plot, Design 2, Braille version.}
    \label{fig:faceted-design2-Braille}
\end{figure}

\begin{figure}[!t]
    \centering
        \includegraphics[width=1\columnwidth]{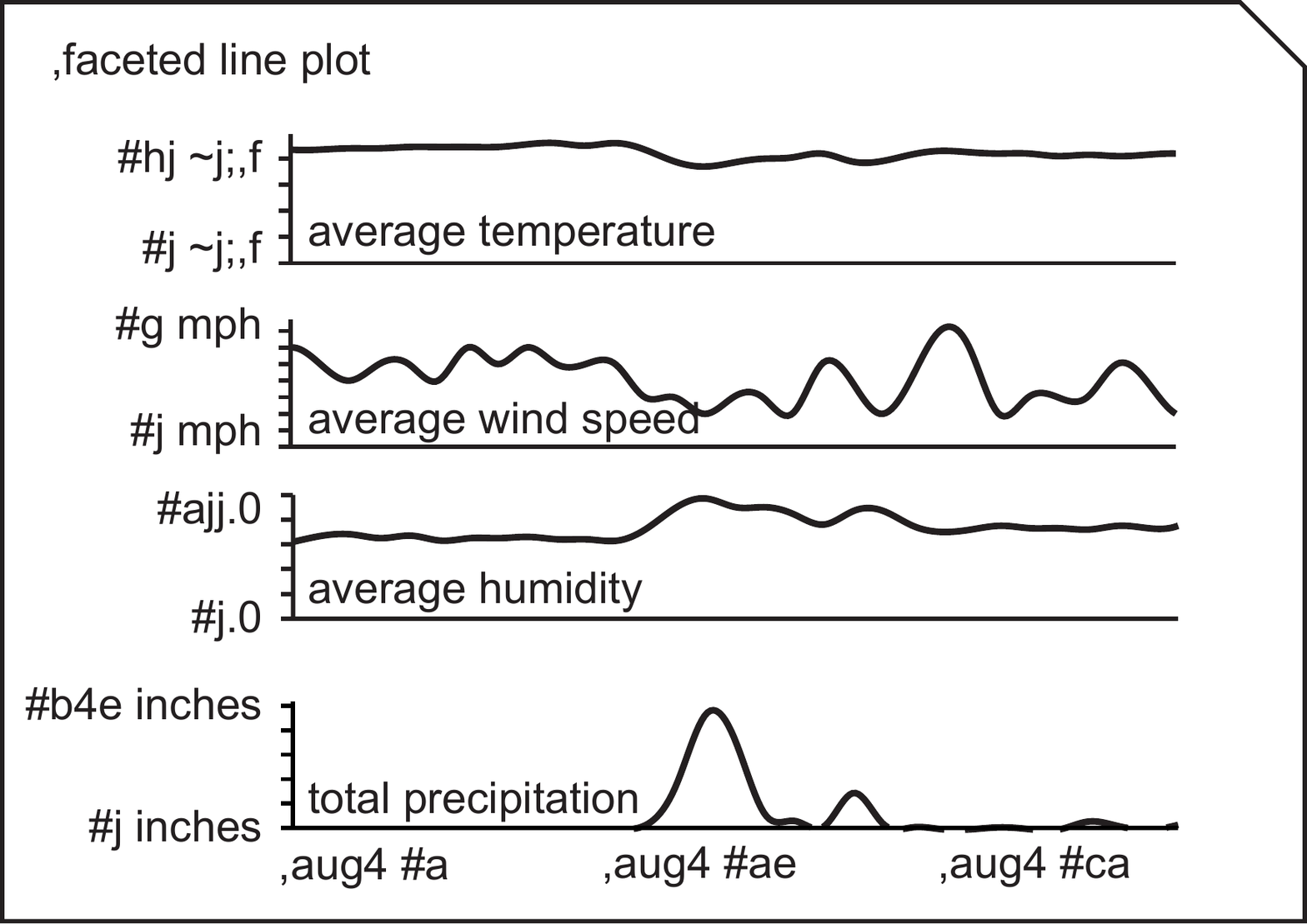}
    \caption{The tactile chart design for faceted plot, Design 2, Braille-to-English letter-by-letter translation version.}
    \label{fig:faceted-design2-characters}
\end{figure}

Based on the tactile graphics guidelines, we transcribed the Python generated charts to the 2D designs for tactile charts. 
In this section, we present the 2D tactile chart designs for the four chart types.
We provide two design variations for each chart type. Each chart has three versions: sighted version, Braille version, and a Braille-to-English letter-by-letter translation version (\autoref{fig:upset-design1-sighted}--\ref{fig:faceted-design2-characters}).

% UpSet plot - Design 1

% UpSet plot - Design 2

% Clustered heatmap - Design 1

% Clustered heatmap - Design 2

% Violin plot - Design 1

% Violin plot - Design 2

% Faceted plot - Design 1

% Faceted plot - Design 2

% \clearpage
\section{Model Photos of Template Chart Design Variations}
\label{sec:template-chart-model-photos-design-variations}
\begin{figure}[!t]
    \centering
        \includegraphics[width=1\columnwidth]{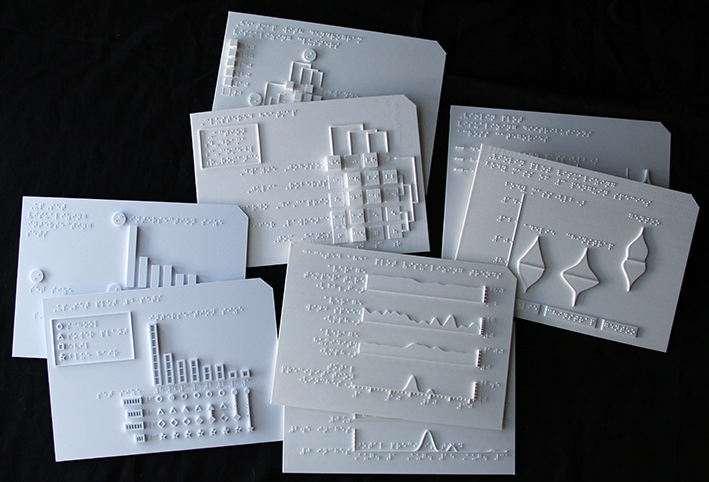}
   \caption{All eight 3D-printed tactile charts created in our design process, including two design variations for each chart type: UpSet plot, clustered heatmap, violin plot, and faceted line chart.}
    \label{fig:two-variations-mess}
\end{figure}

\begin{figure}[!t]
    \centering
        \includegraphics[width=1\columnwidth]{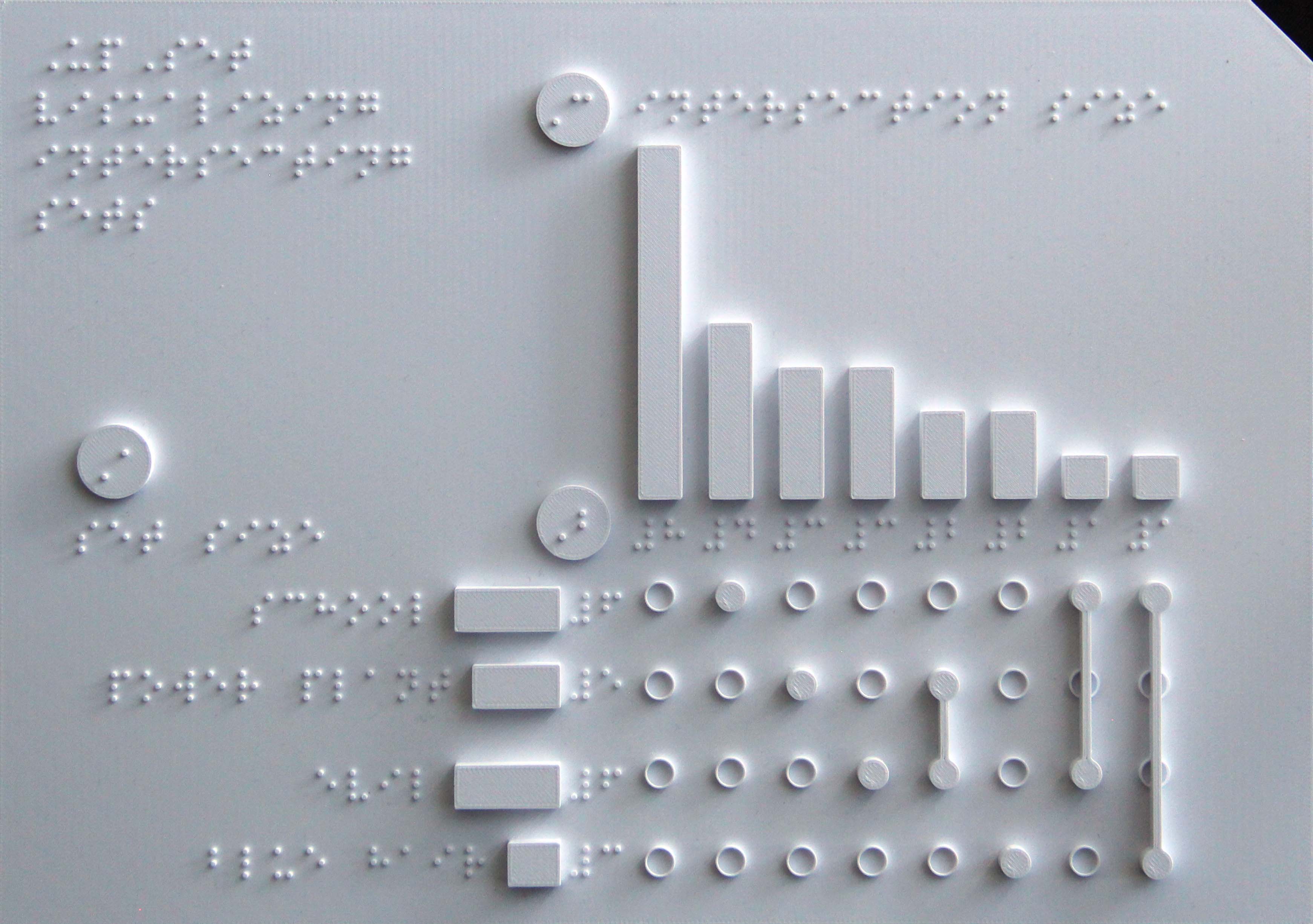}
    \caption{The 3D printed tactile chart for UpSet plot, Design 1, front view.}
    \label{fig:upset-design1-front}
\end{figure}

\begin{figure}[!t]
    \centering
        \includegraphics[width=1\columnwidth]{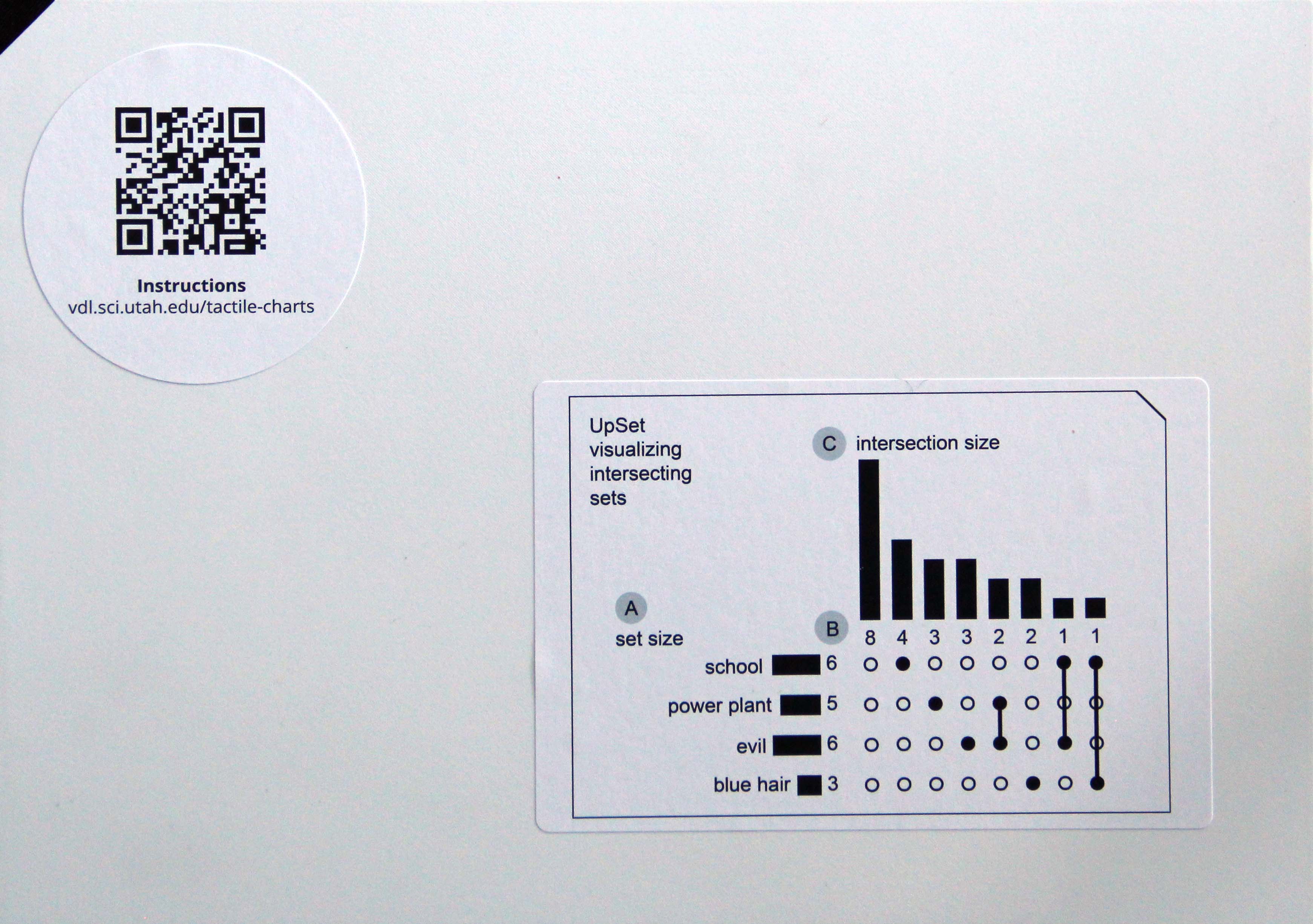}
    \caption{The 3D printed tactile chart for UpSet plot, Design 1, back view.}
    \label{fig:upset-design1-back}
\end{figure}

\begin{figure}[!t]
    \centering
        \includegraphics[width=1\columnwidth]{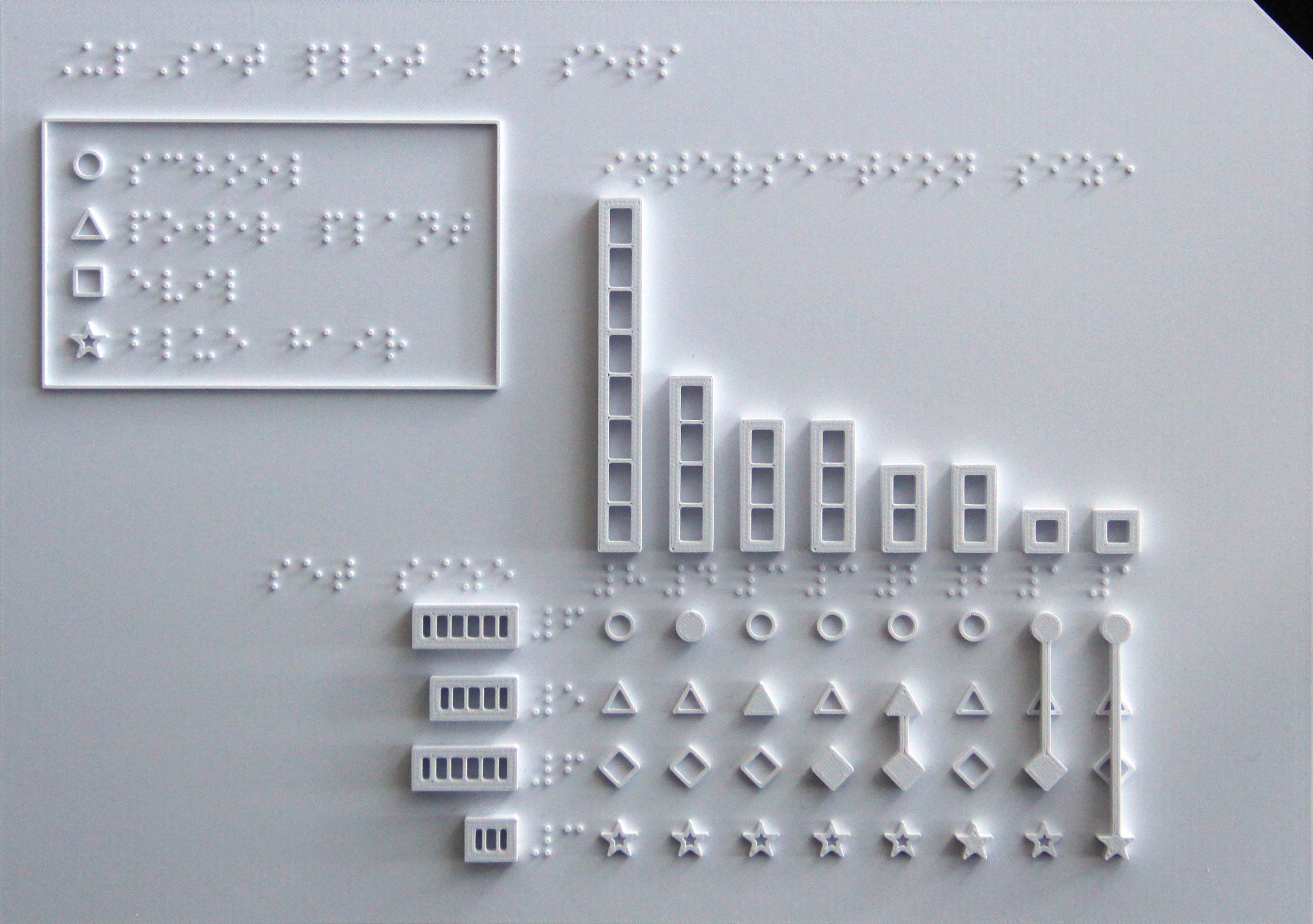}
    \caption{The 3D printed tactile chart for UpSet plot, Design 2, front view.}
    \label{fig:upset-design2-front}
\end{figure}

\begin{figure}[!t]
    \centering
        \includegraphics[width=1\columnwidth]{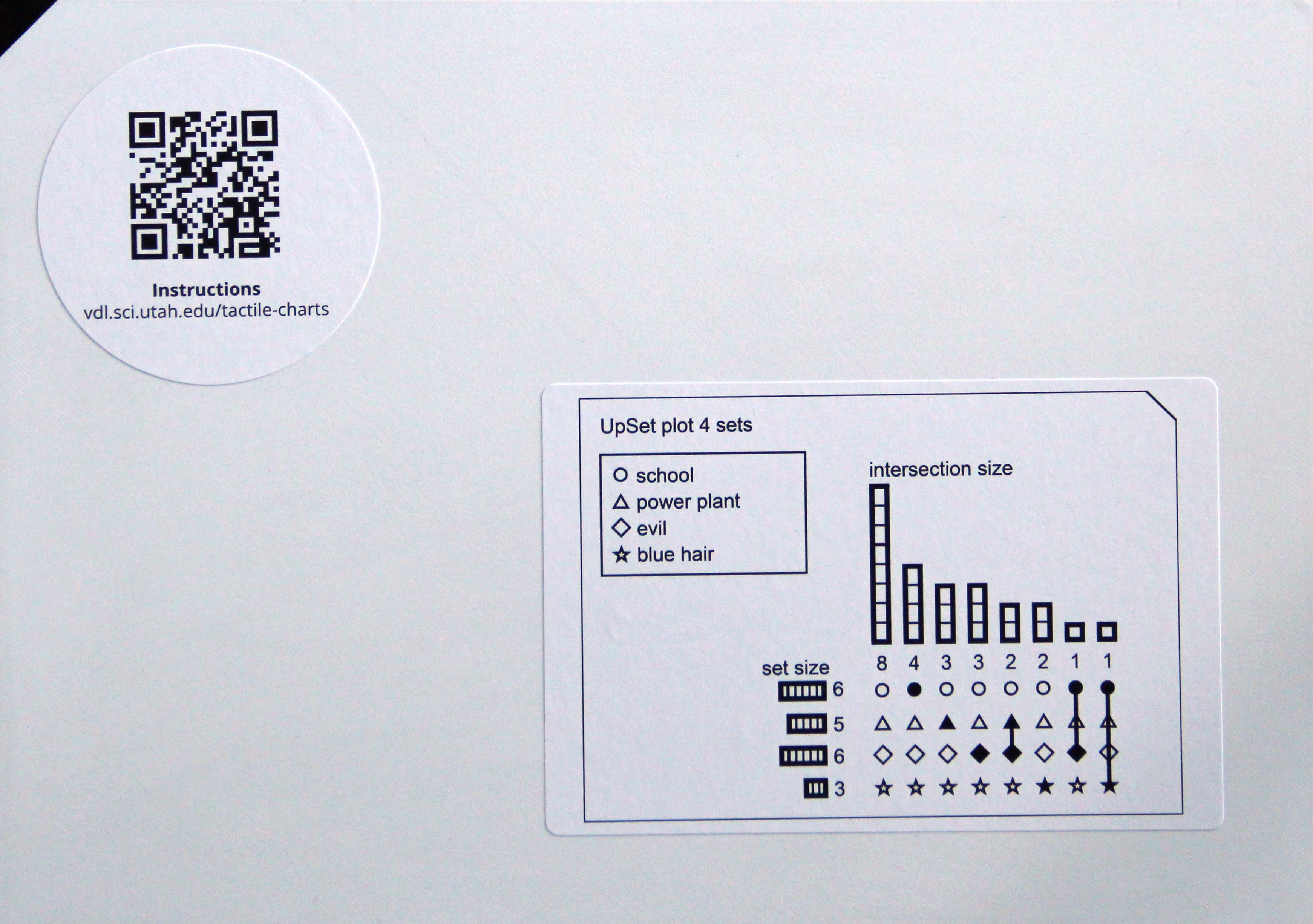}
    \caption{The 3D printed tactile chart for UpSet plot, Design 2, back view.}
    \label{fig:upset-design2-back}
\end{figure}

\begin{figure}[!t]
    \centering
        \includegraphics[width=1\columnwidth]{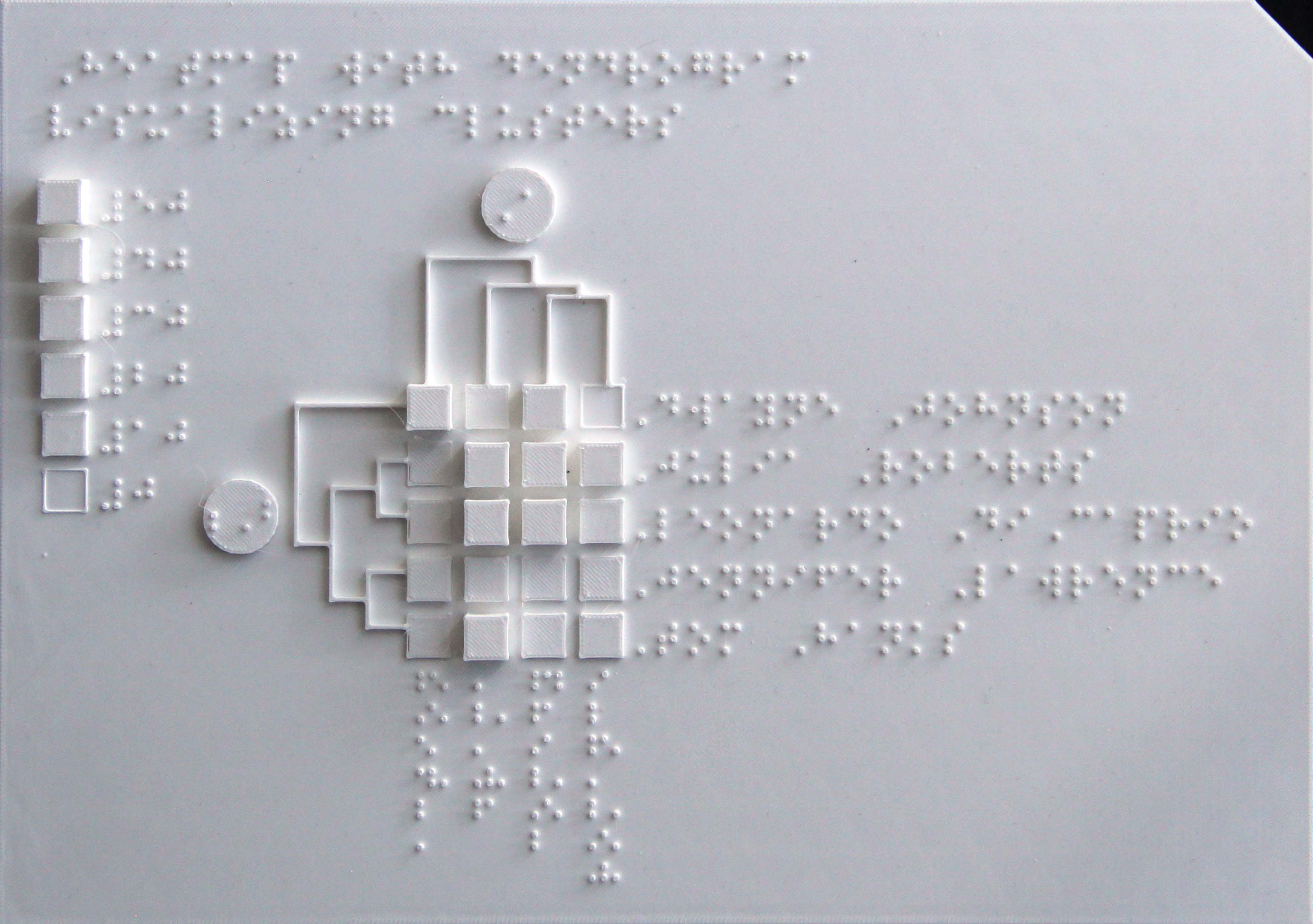}
    \caption{The 3D printed tactile chart for clustered heatmap, Design 1, front view.}
    \label{fig:heatmap-design1-front}
\end{figure}

\begin{figure}[!t]
    \centering
        \includegraphics[width=1\columnwidth]{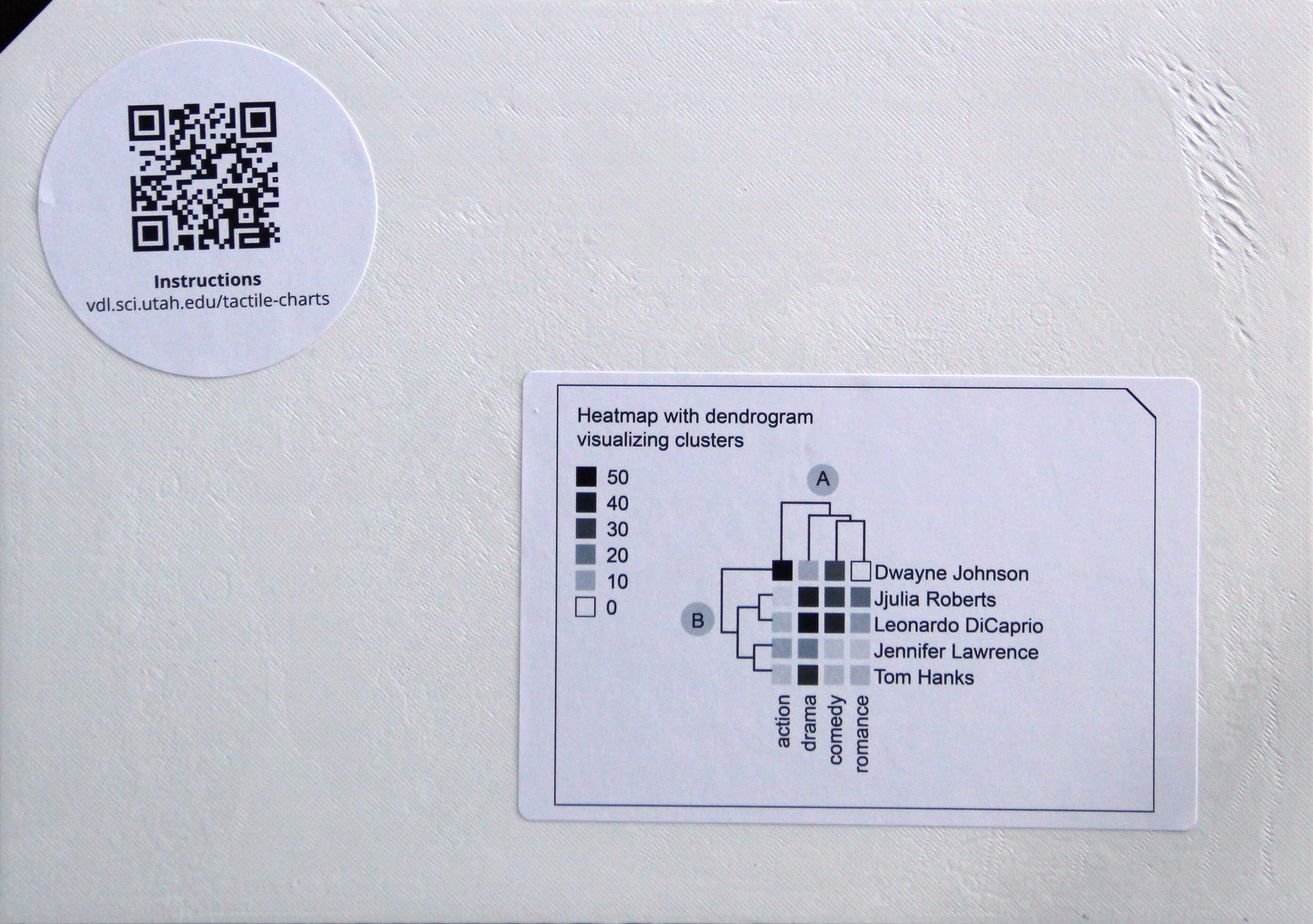}
    \caption{The 3D printed tactile chart for clustered heatmap, Design 1, back view.}
    \label{fig:heatmap-design1-back}
\end{figure}

\begin{figure}[!t]
    \centering
        \includegraphics[width=1\columnwidth]{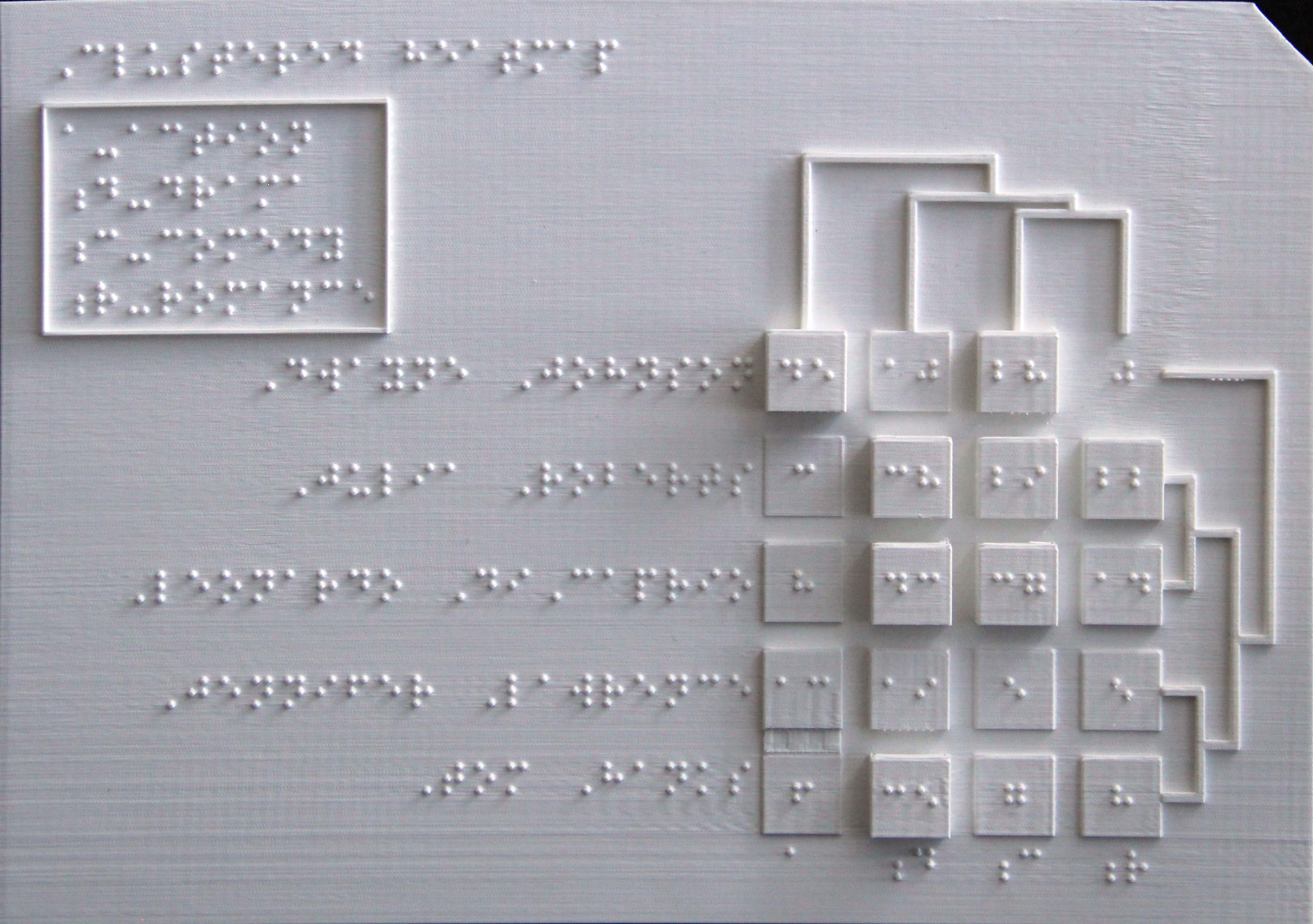}
    \caption{The 3D printed tactile chart for clustered heatmap, Design 2, front view.}
    \label{fig:heatmap-design2-front}
\end{figure}

\begin{figure}[!t]
    \centering
        \includegraphics[width=1\columnwidth]{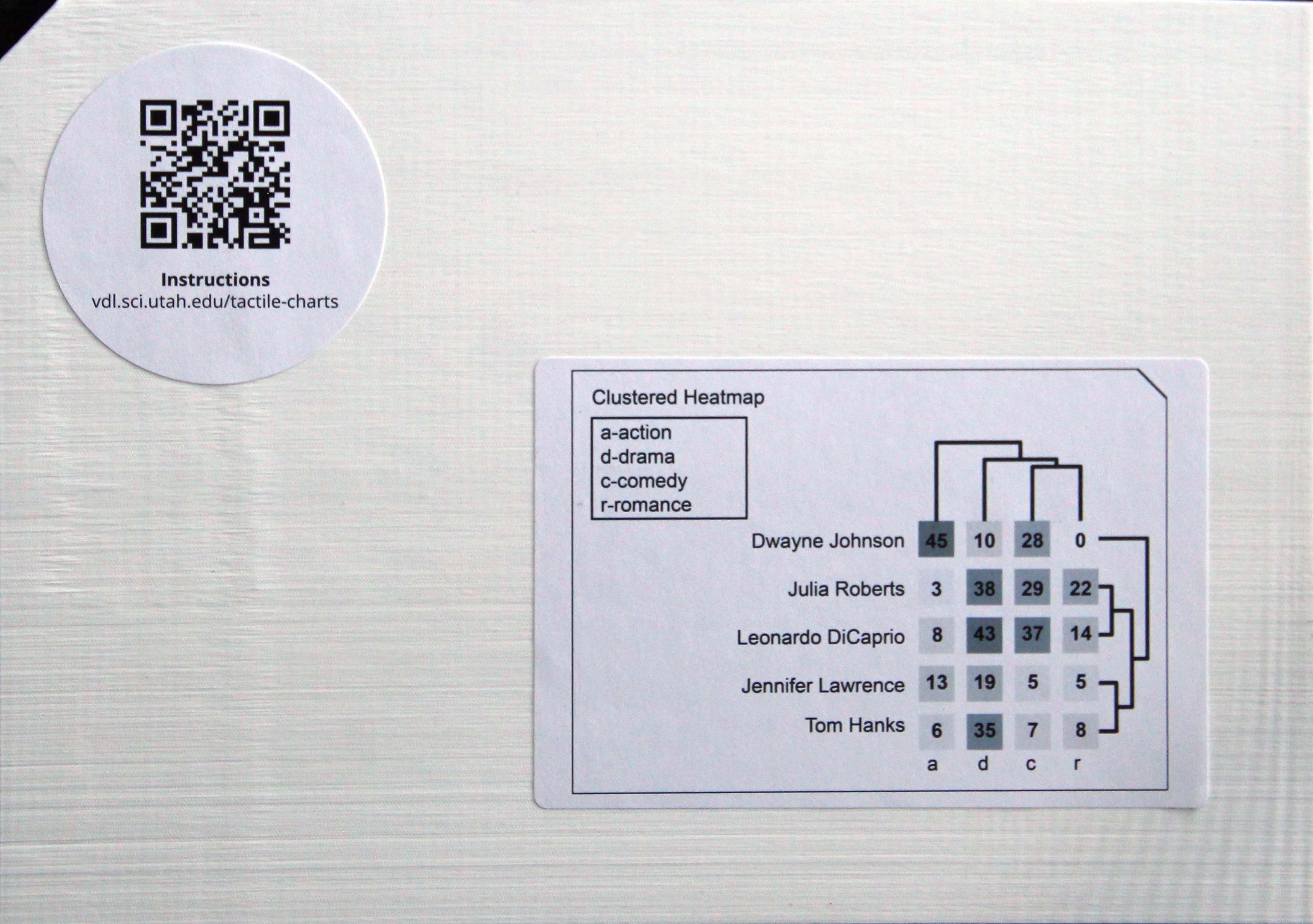}
    \caption{The 3D printed tactile chart for clustered heatmap, Design 2, back view.}
    \label{fig:heatmap-design2-back}
\end{figure}

\begin{figure}[!t]
    \centering
        \includegraphics[width=1\columnwidth]{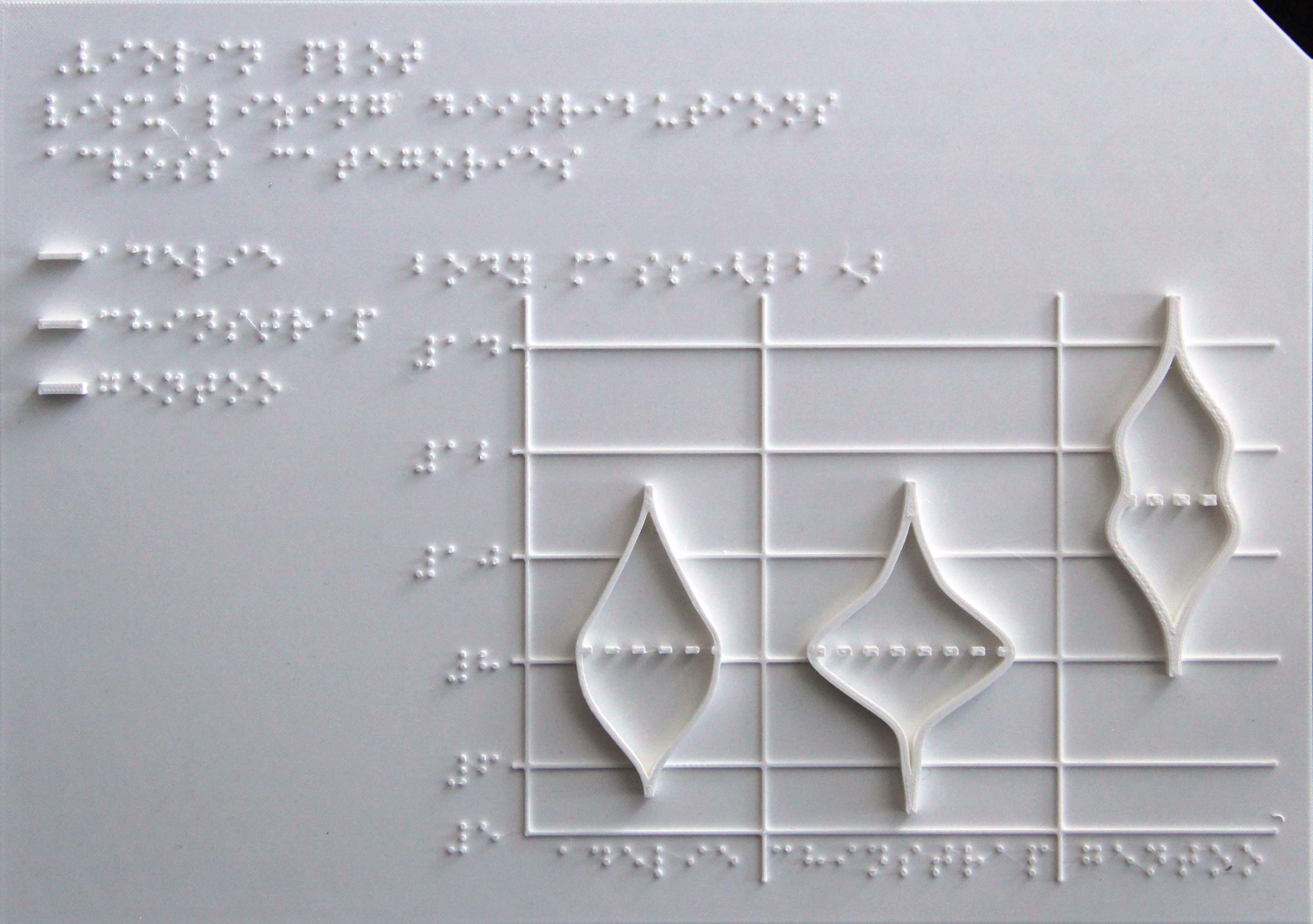}
    \caption{The 3D printed tactile chart for violin plot, Design 1, front view.}
    \label{fig:violin-design1-front}
\end{figure}

\begin{figure}[!t]
    \centering
        \includegraphics[width=1\columnwidth]{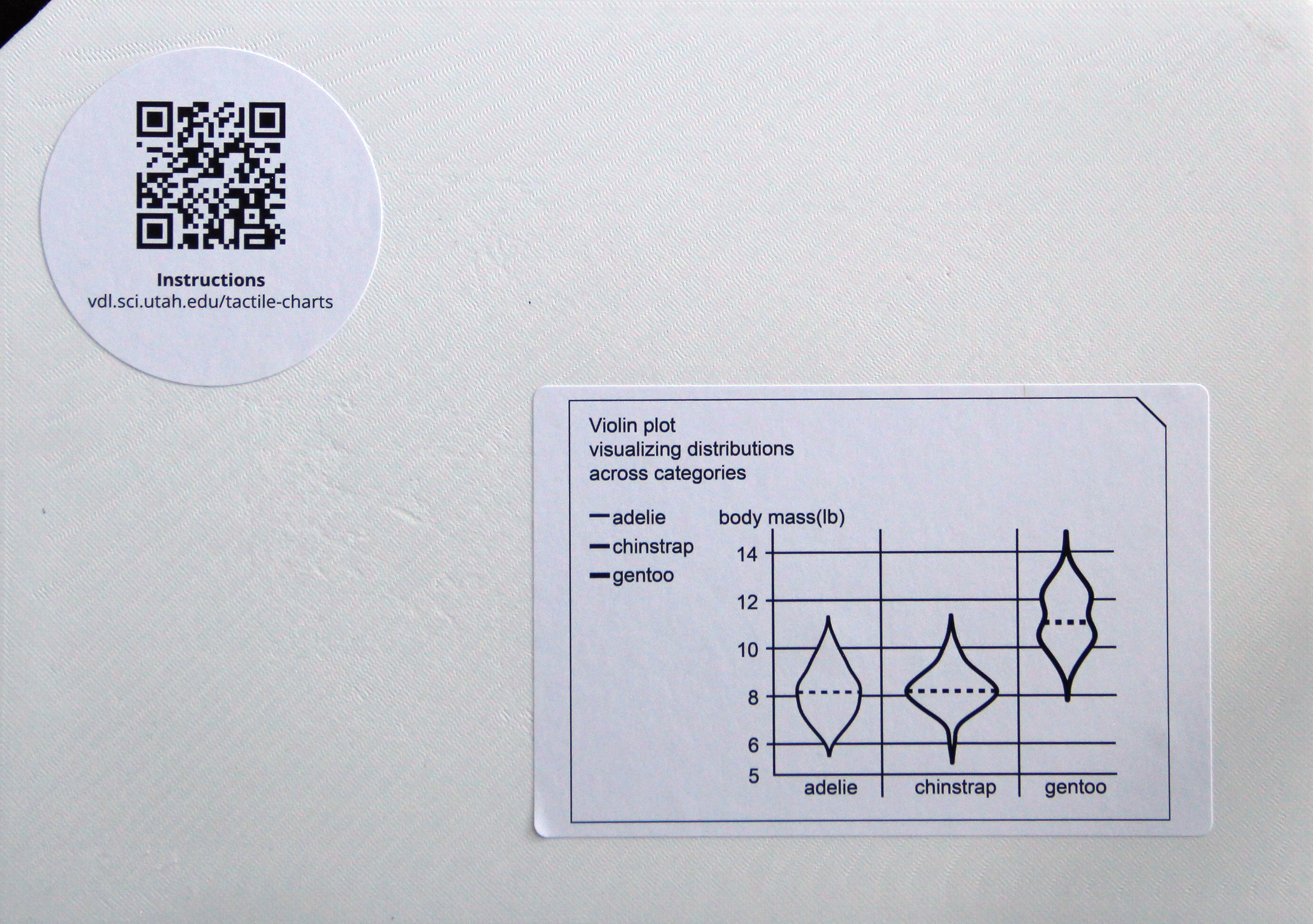}
    \caption{The 3D printed tactile chart for violin plot, Design 1, back view.}
    \label{fig:violin-design1-back}
\end{figure}

\begin{figure}[!t]
    \centering
        \includegraphics[width=1\columnwidth]{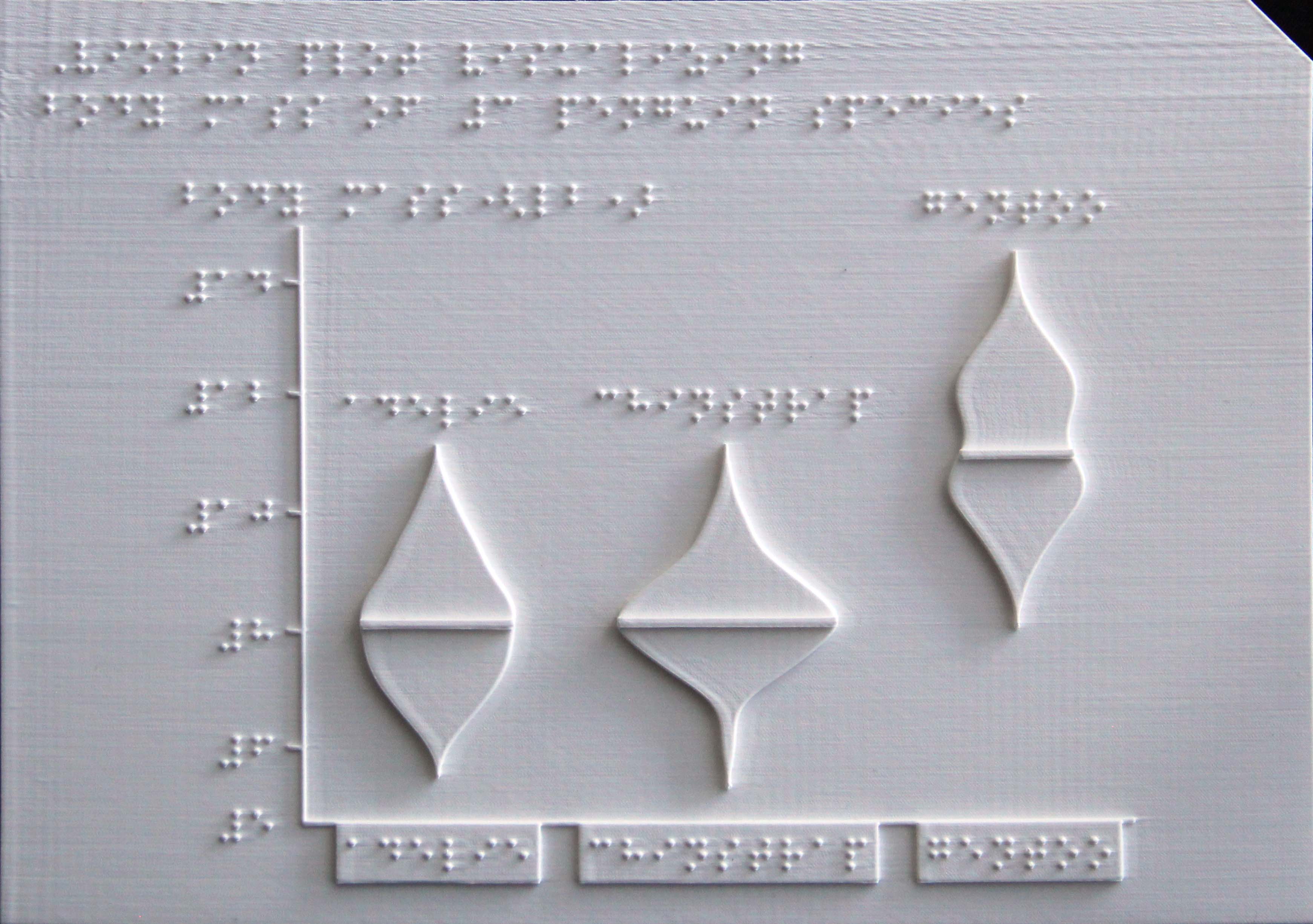}
    \caption{The 3D printed tactile chart for violin plot, Design 2, front view.}
    \label{fig:violin-design2-front}
\end{figure}

\begin{figure}[!t]
    \centering
        \includegraphics[width=1\columnwidth]{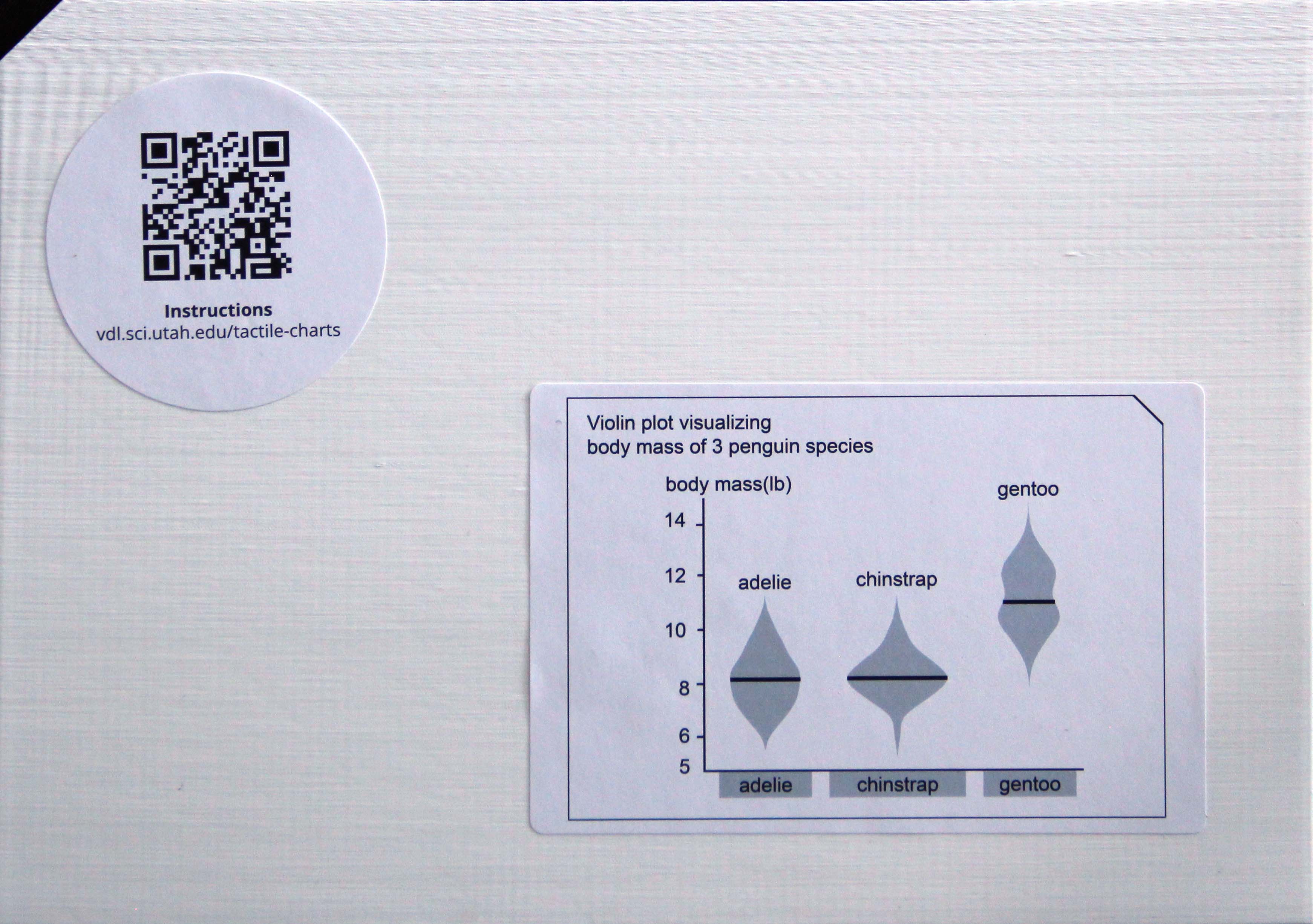}
    \caption{The 3D printed tactile chart for violin plot, Design 2, back view.}
    \label{fig:violin-design2-back}
\end{figure}

\begin{figure}[!t]
    \centering
        \includegraphics[width=1\columnwidth]{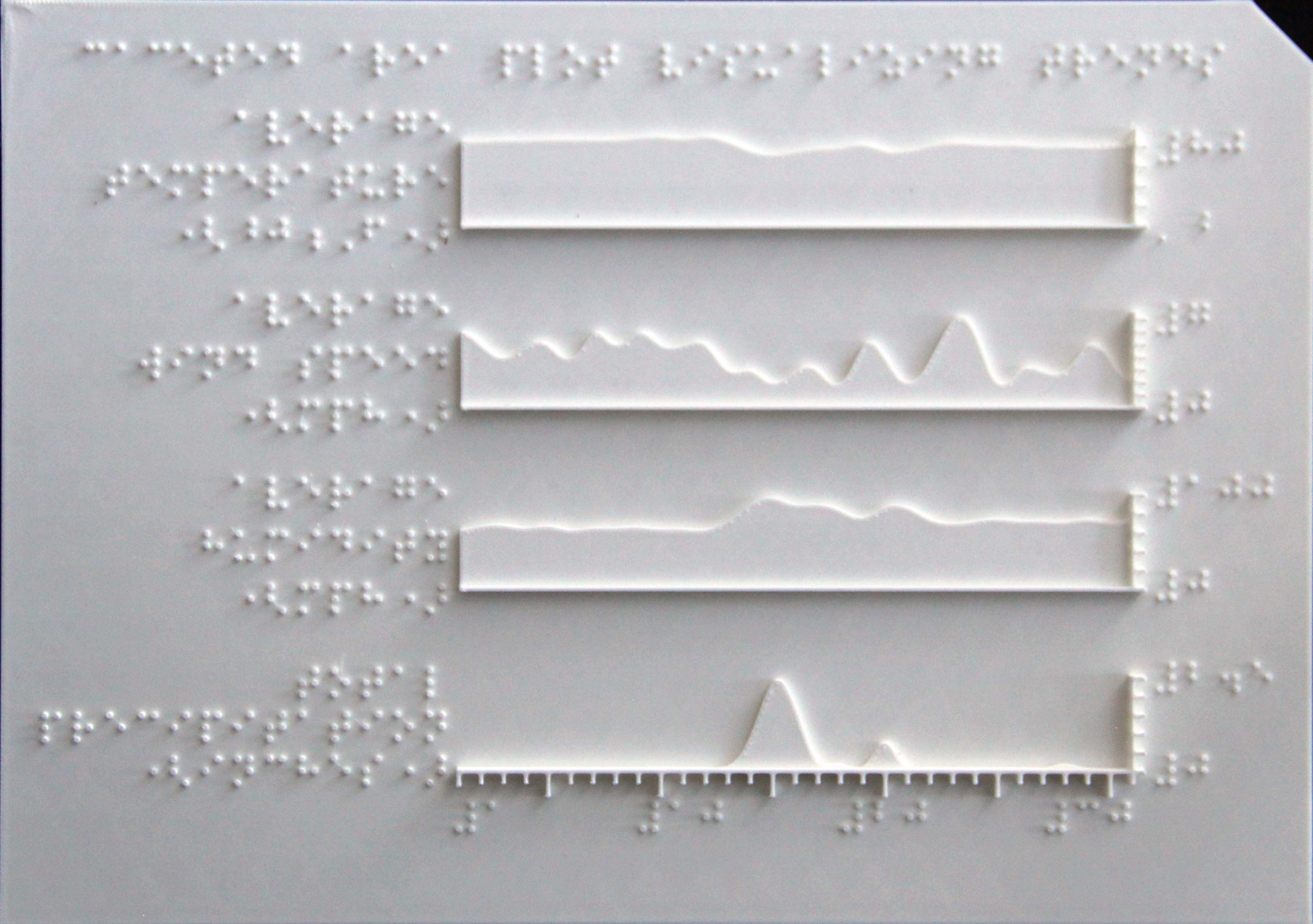}
    \caption{The 3D printed tactile chart for faceted plot, Design 1, front view.}
    \label{fig:faceted-design1-front}
\end{figure}

\begin{figure}[!t]
    \centering
        \includegraphics[width=1\columnwidth]{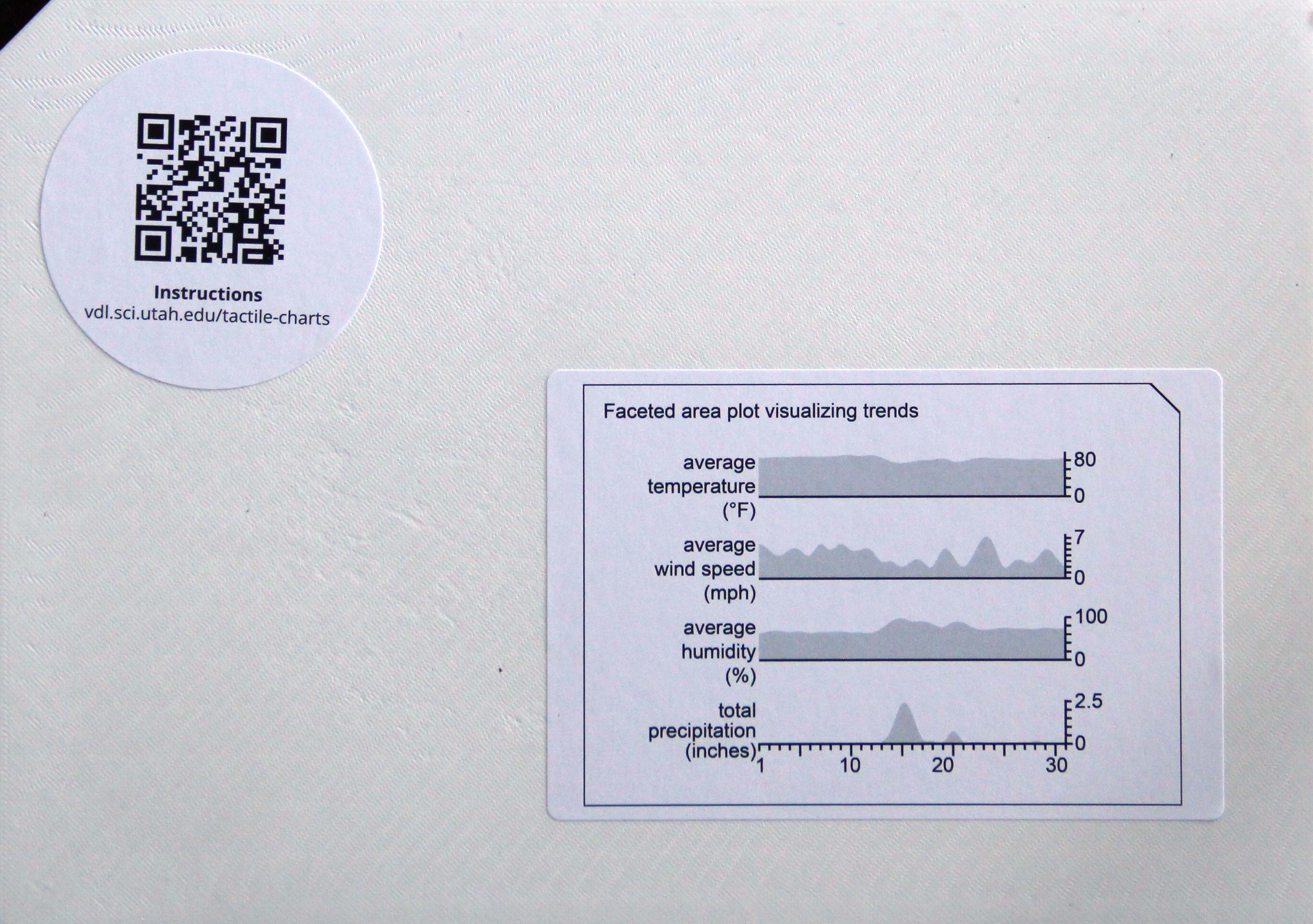}
    \caption{The 3D printed tactile chart for faceted plot, Design 1, back view.}
    \label{fig:faceted-design1-back}
\end{figure}

\begin{figure}[!t]
    \centering
        \includegraphics[width=1\columnwidth]{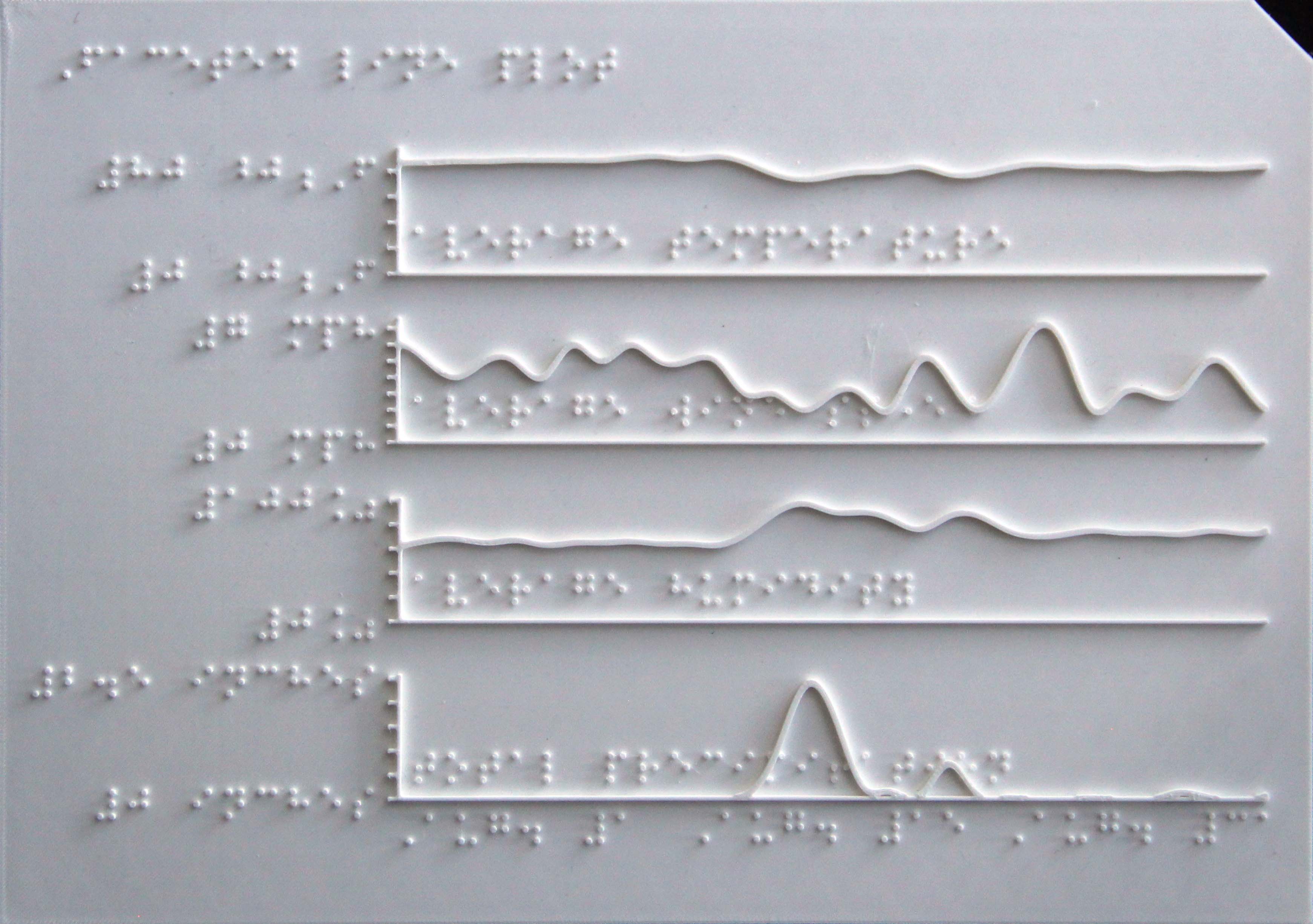}
    \caption{The 3D printed tactile chart for faceted plot, Design 2, front view.}
    \label{fig:faceted-design2-front}
\end{figure}

\begin{figure}[!t]
    \centering
        \includegraphics[width=1\columnwidth]{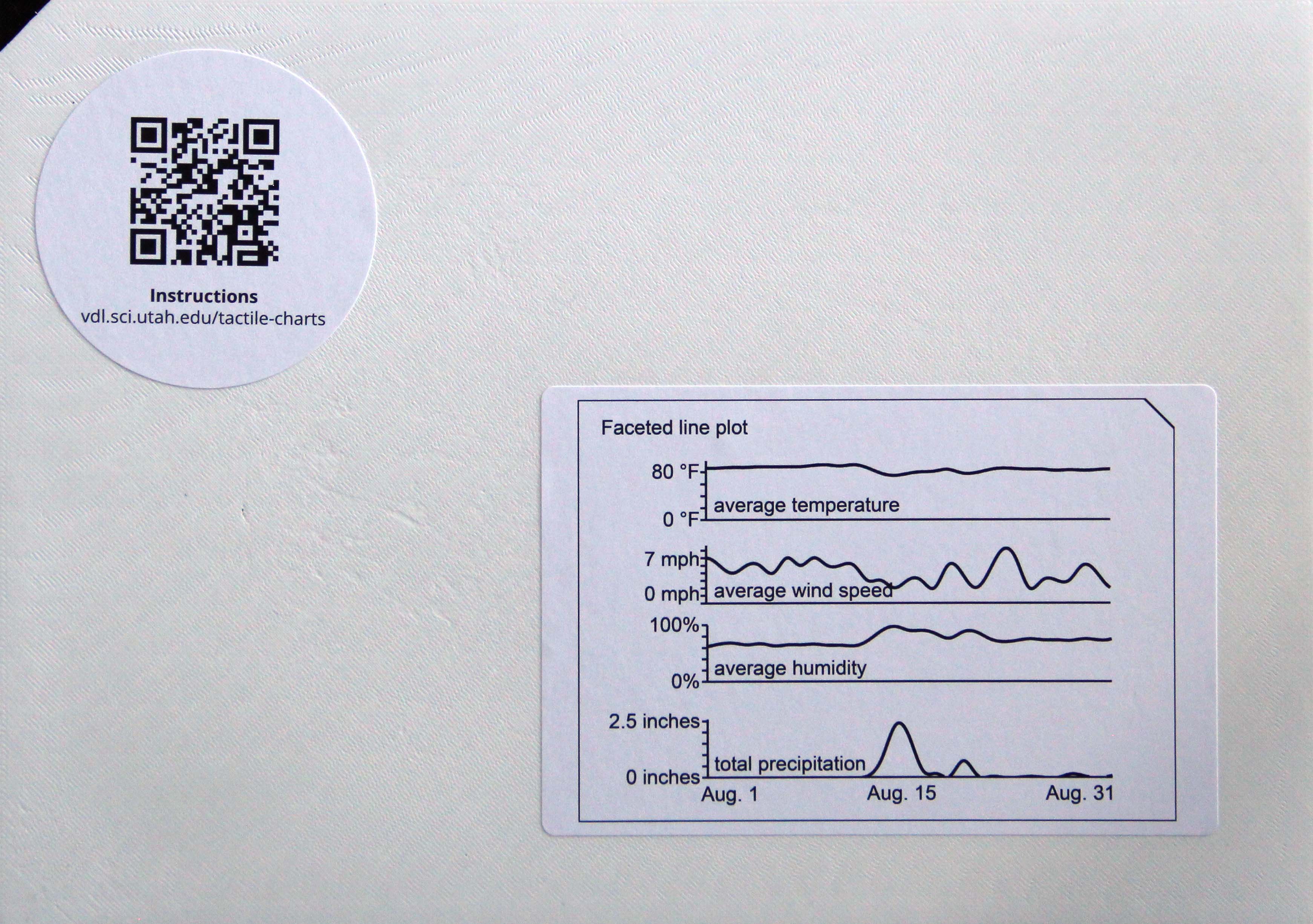}
    \caption{The 3D printed tactile chart for faceted plot, Design 2, back view.}
    \label{fig:faceted-design2-back}
\end{figure}

In this section, we present the photos of the 3D printed tactile charts for the four chart types in \autoref{fig:two-variations-mess}--\ref{fig:faceted-design2-back}.
Each chart type has two design variations. We show two views of each model: the front view and the back view.

% UpSet plot - Design 1

% UpSet plot - Design 2

% Clustered heatmap - Design 1

% Clustered heatmap - Design 2

% Violin plot - Design 1

% Violin plot - Design 2

% Faceted plot - Design 1

% Faceted plot - Design 2

\clearpage
\section{Template Chart Final Designs}
\label{sec:template-chart-final-design}
\begin{figure}[ht]
    \centering
        \includegraphics[width=1\columnwidth]{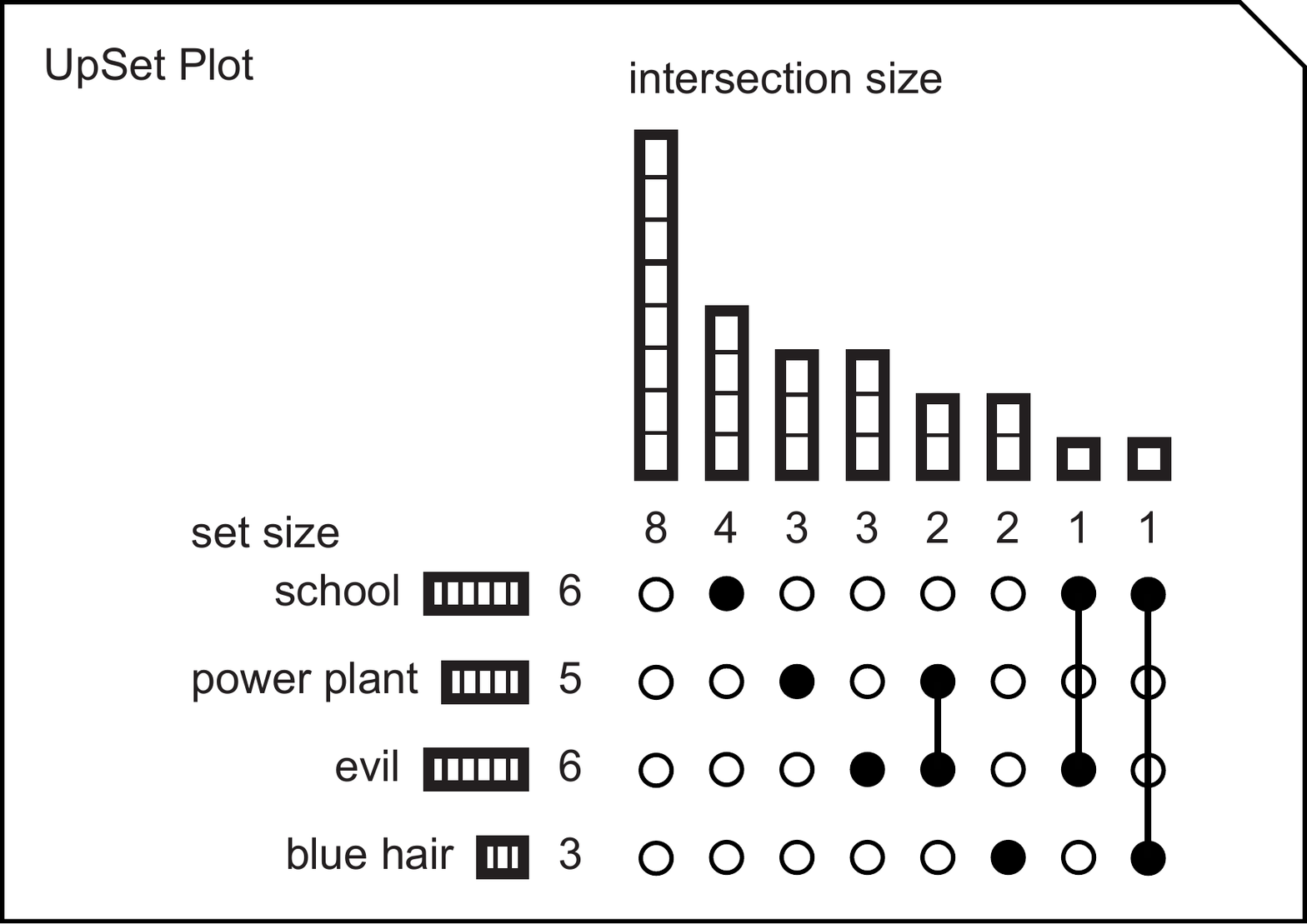}
    \caption{The final tactile chart design for UpSet plot, sighted version.}
    \label{fig:upset-final-sighted}
\end{figure}

\begin{figure}[ht]
    \centering
        \includegraphics[width=1\columnwidth]{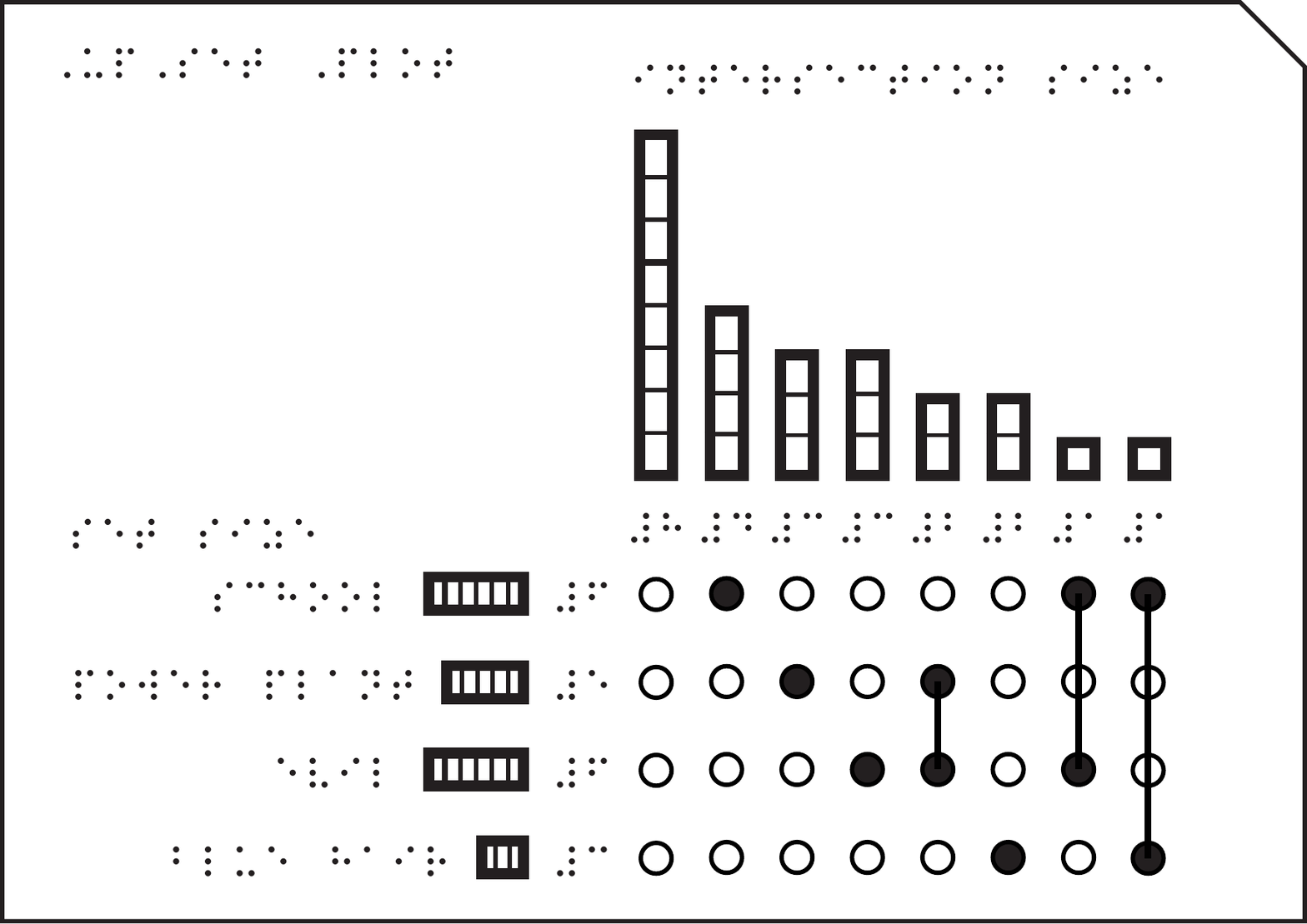}
    \caption{The final tactile chart design for UpSet plot, Braille version.}
    \label{fig:upset-final-Braille}
\end{figure}

\begin{figure}[ht]
    \centering
        \includegraphics[width=1\columnwidth]{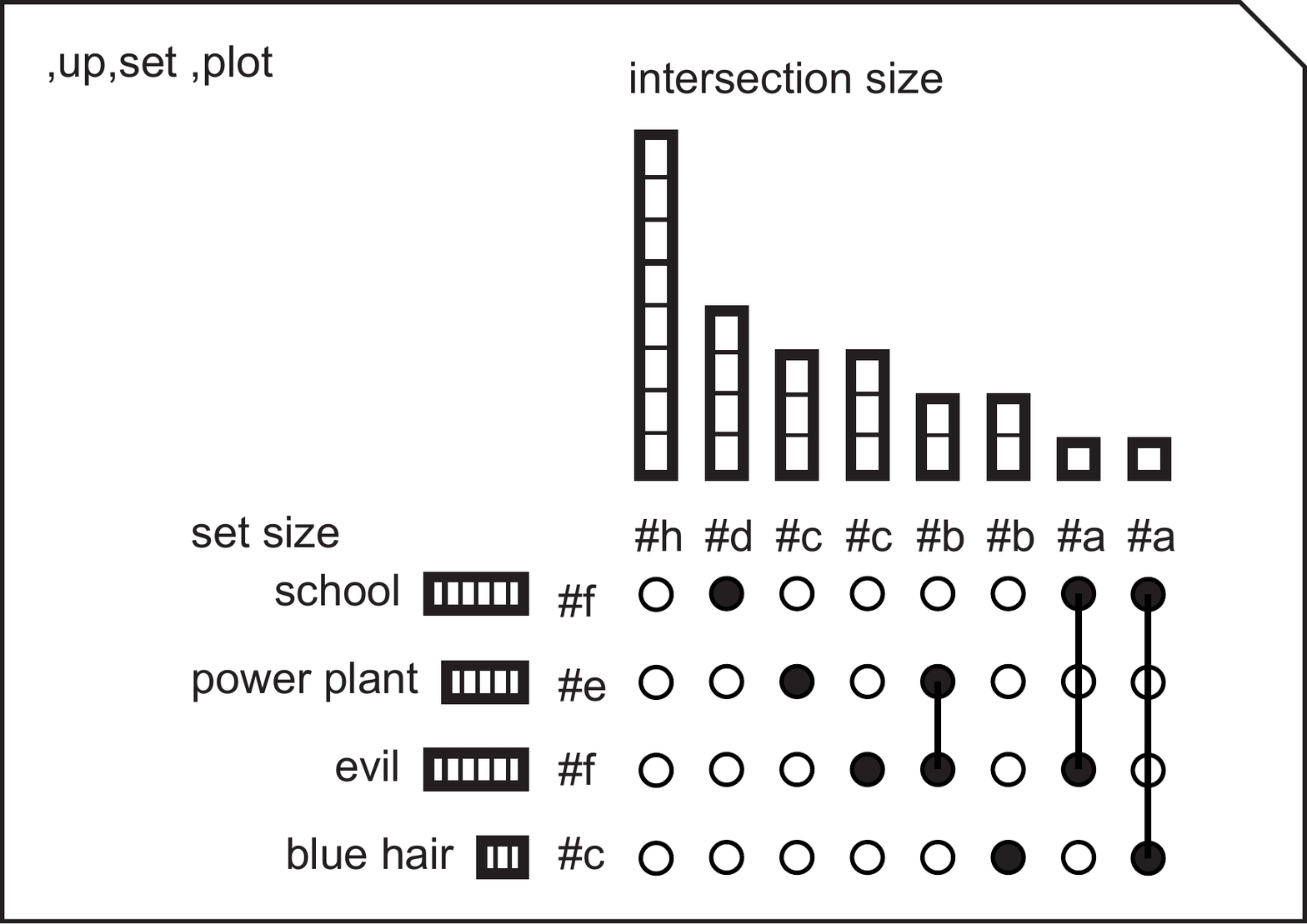}
    \caption{The final tactile chart design for UpSet plot, Braille-to-English letter-by-letter translation version.}
    \label{fig:upset-final-characters}
\end{figure}

\begin{figure}[ht]
    \centering
        \includegraphics[width=1\columnwidth]{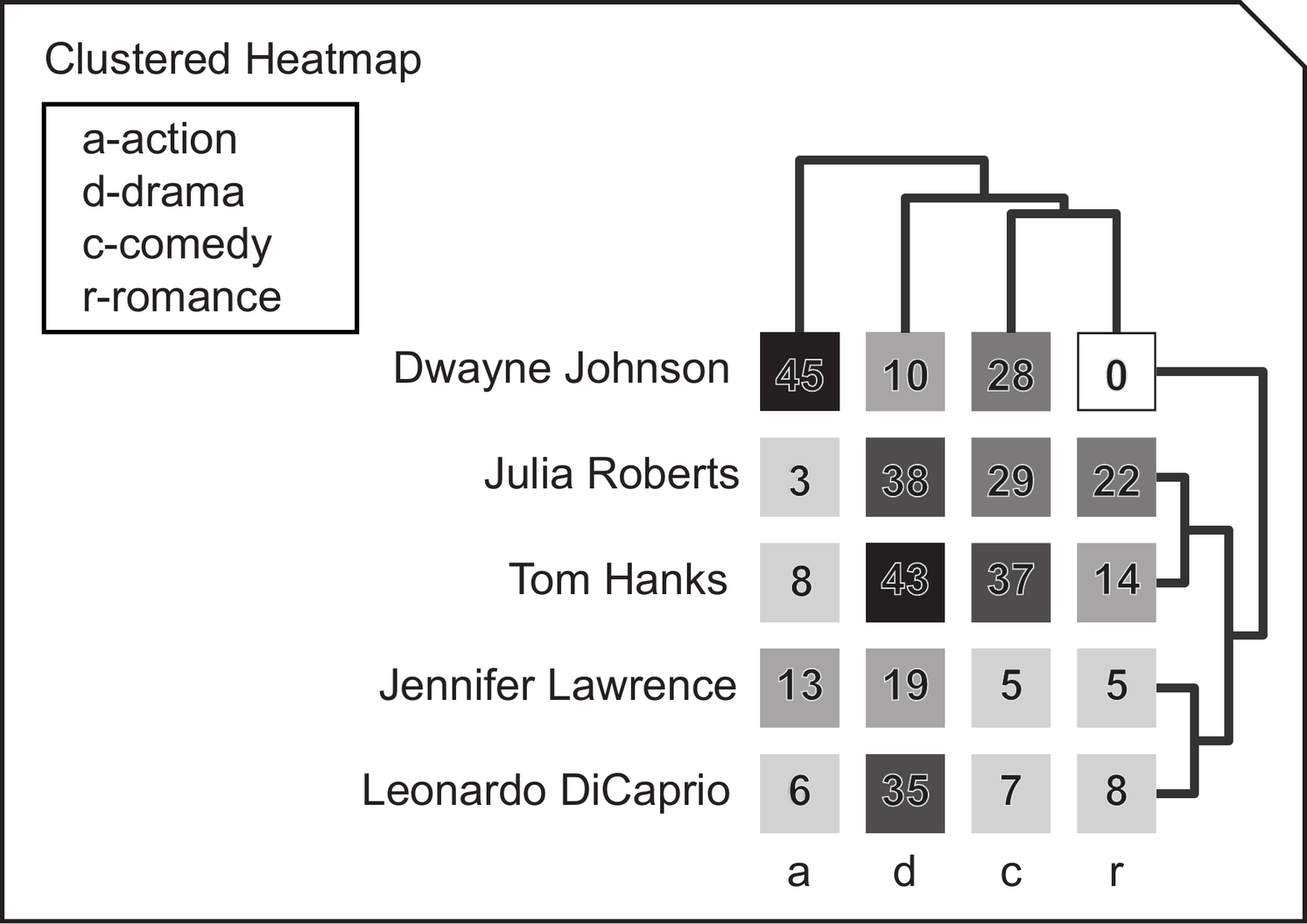}
    \caption{The final tactile chart design for clustered heatmap, sighted version.}
    \label{fig:heatmap-final-sighted}
\end{figure}

\begin{figure}[ht]
    \centering
        \includegraphics[width=1\columnwidth]{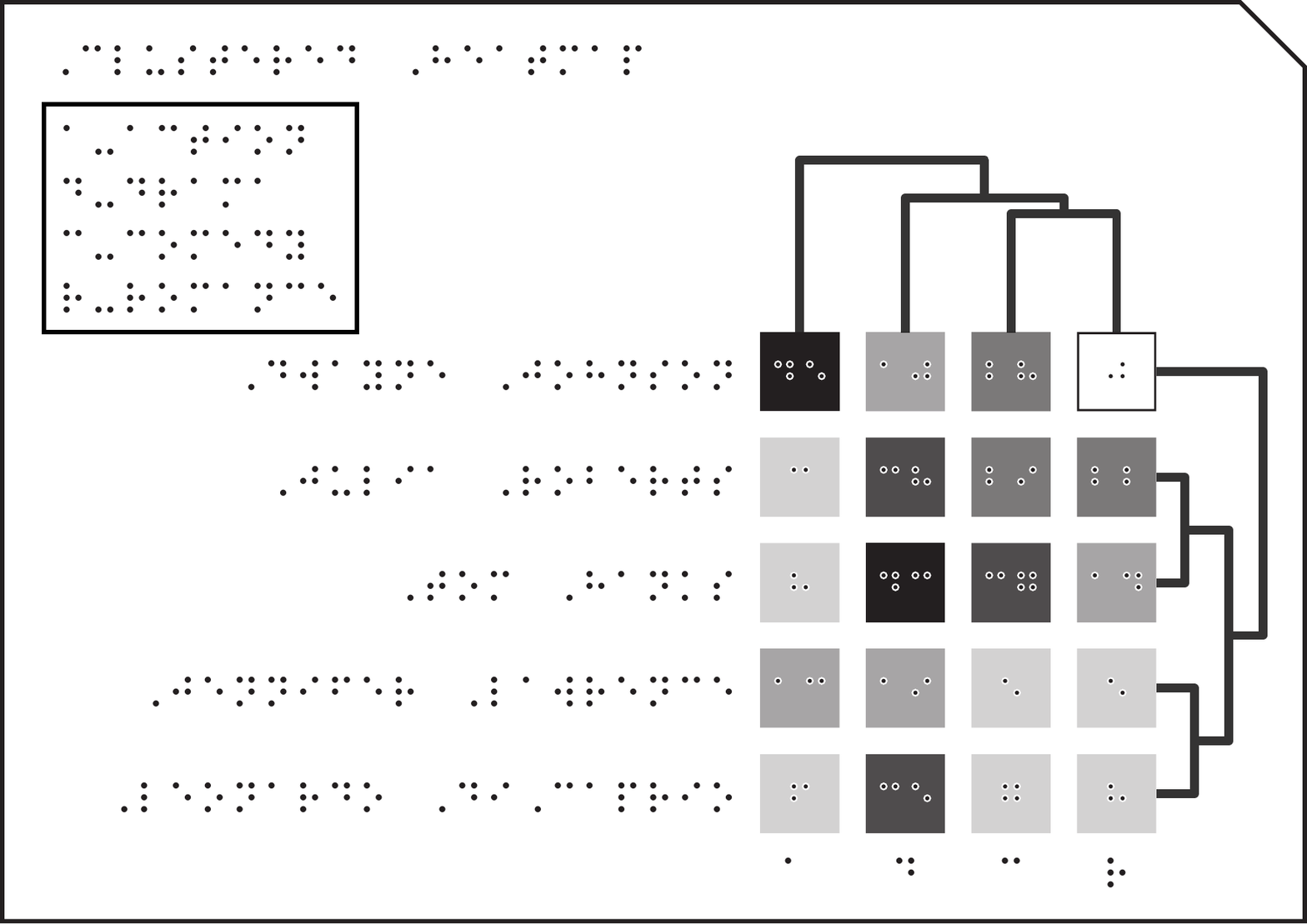}
    \caption{The final tactile chart design for clustered heatmap, Braille version.}
    \label{fig:heatmap-final-Braille}
\end{figure}

\begin{figure}[ht]
    \centering
        \includegraphics[width=1\columnwidth]{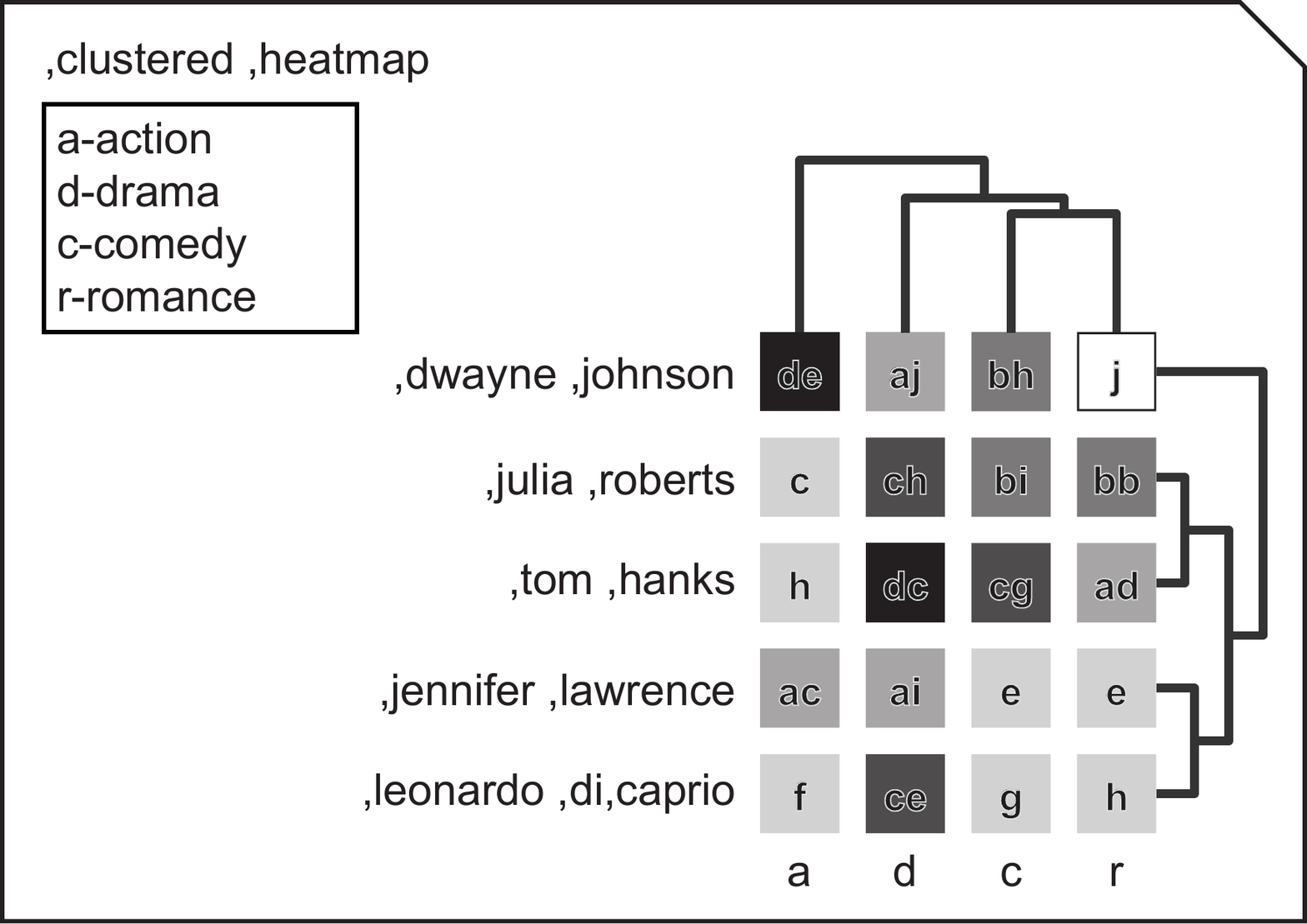}
    \caption{The final tactile chart design for clustered heatmap, Braille-to-English letter-by-letter translation version.}
    \label{fig:heatmap-final-characters}
\end{figure}

\begin{figure}[ht]
    \centering
        \includegraphics[width=1\columnwidth]{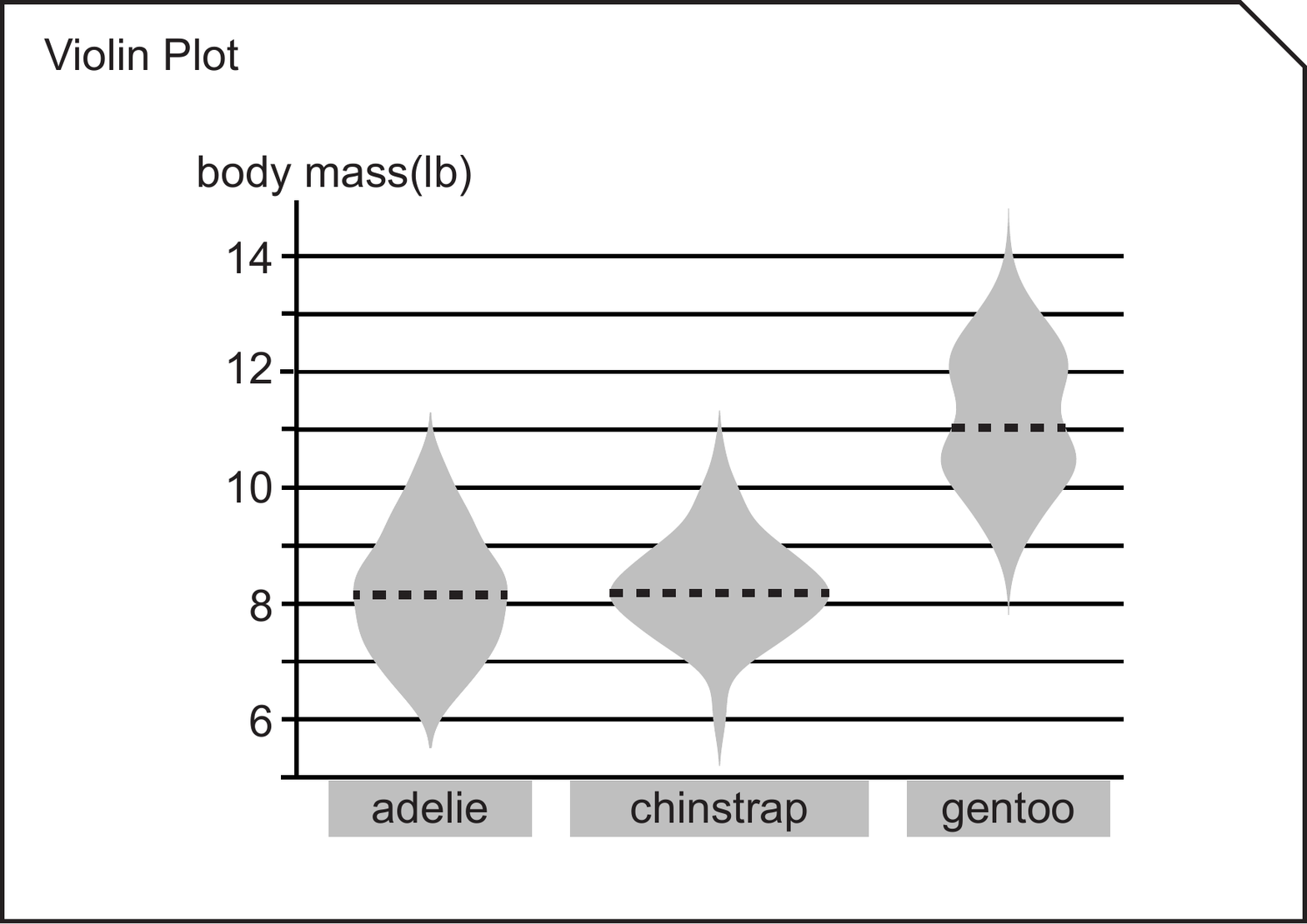}
    \caption{The final tactile chart design for violin plot, sighted version.}
    \label{fig:violin-final-sighted}
\end{figure}

\begin{figure}[ht]
    \centering
        \includegraphics[width=1\columnwidth]{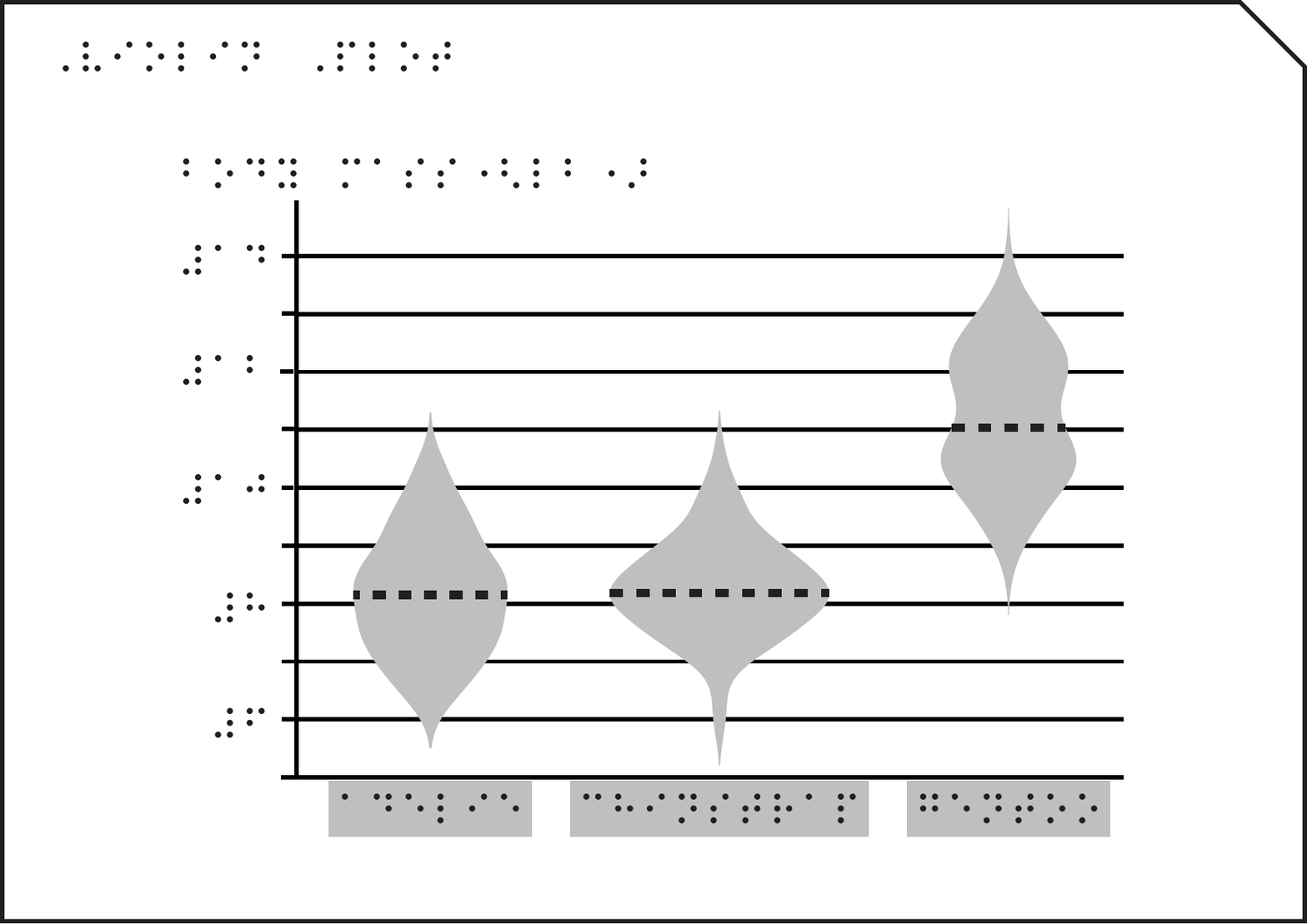}
    \caption{The final tactile chart design for violin plot, Braille version.}
    \label{fig:violin-final-Braille}
\end{figure}

\begin{figure}[!t]
    \centering
        \includegraphics[width=1\columnwidth]{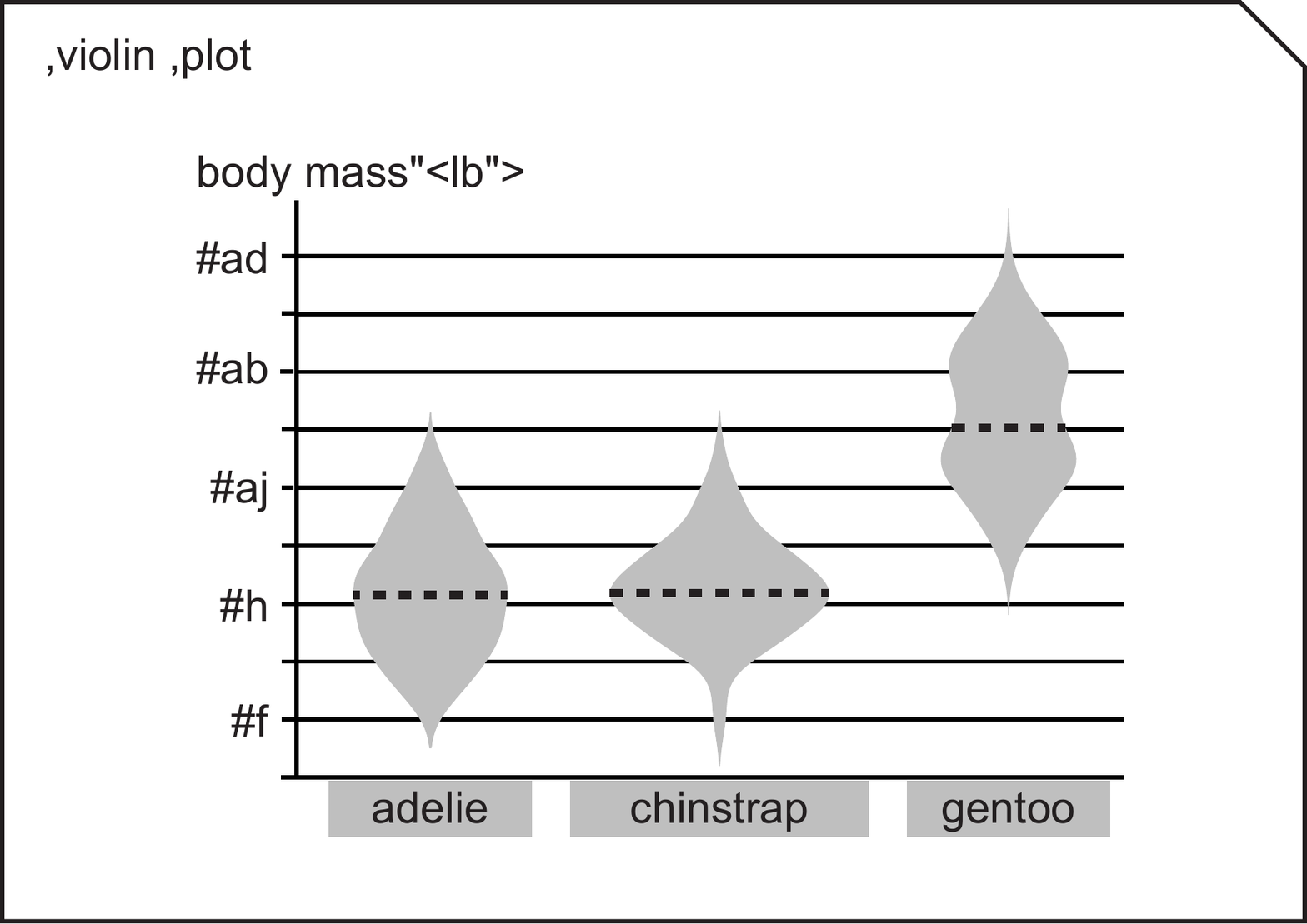}
    \caption{The final tactile chart design for violin plot, Braille-to-English letter-by-letter translation version.}
    \label{fig:violin-final-characters}
\end{figure}

\begin{figure}[!t]
    \centering
        \includegraphics[width=1\columnwidth]{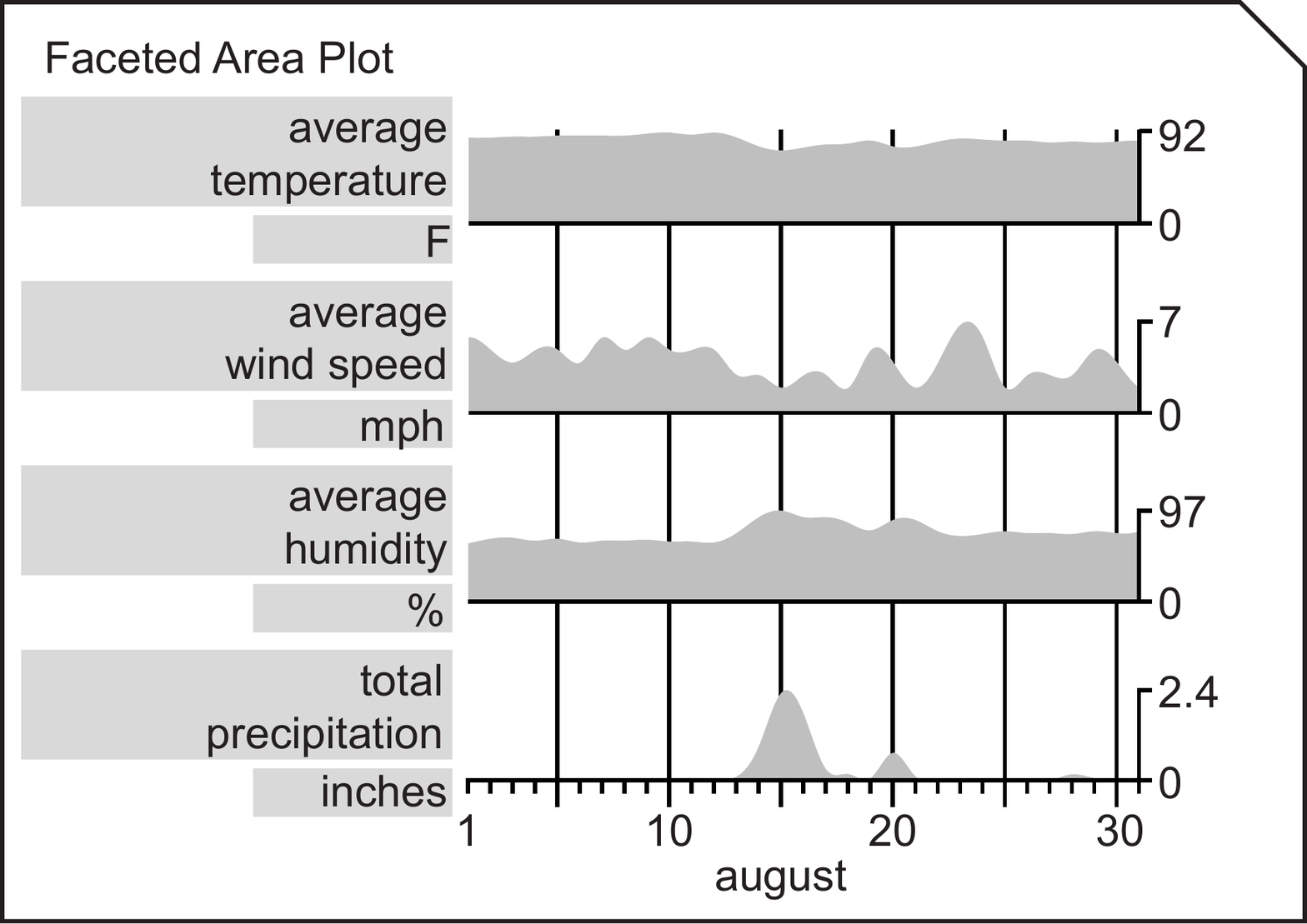}
    \caption{The final tactile chart design for faceted plot, sighted version.}
    \label{fig:faceted-final-sighted}
\end{figure}

\begin{figure}[!t]
    \centering
        \includegraphics[width=1\columnwidth]{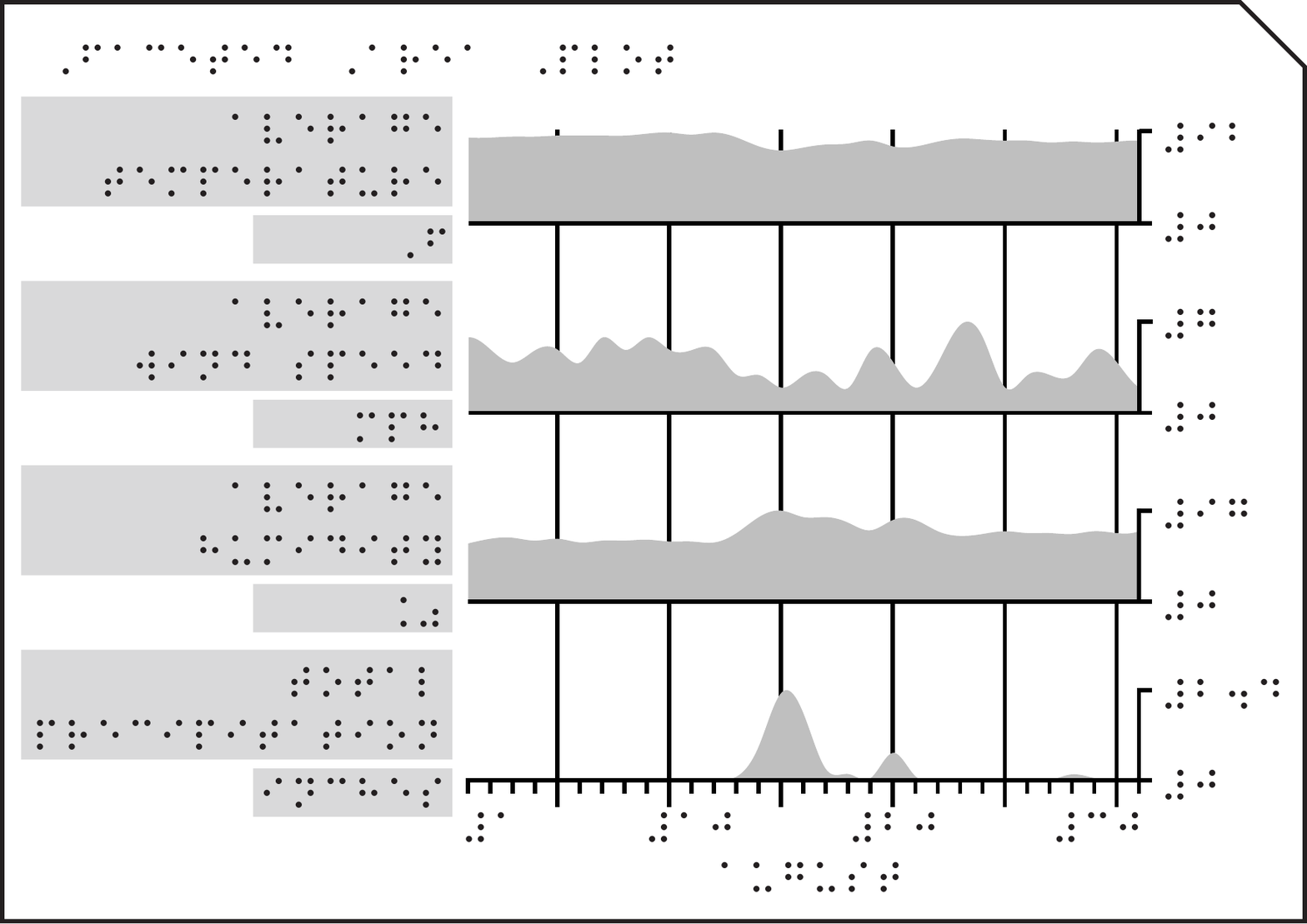}
    \caption{The final tactile chart design for faceted plot, Braille version.}
    \label{fig:faceted-final-Braille}
\end{figure}

\begin{figure}[!t]
    \centering
        \includegraphics[width=1\columnwidth]{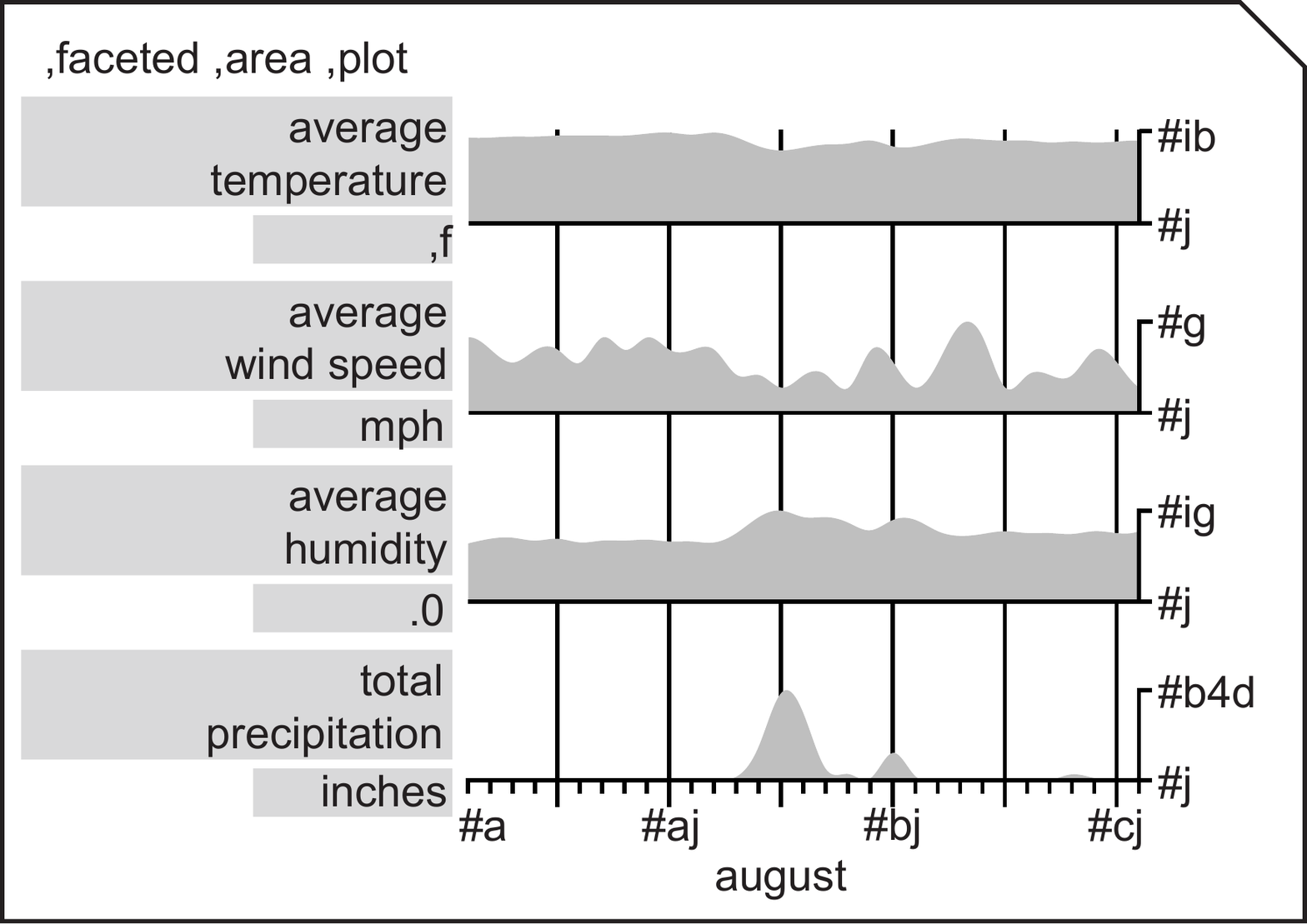}
    \caption{The final tactile chart design for faceted plot, Braille-to-English letter-by-letter translation version.}
    \label{fig:faceted-final-characters}
\end{figure}

Based on the feedback from the consultation sessions with the two blind participants, we made the final design for our four chart types and present them in this section (\autoref{fig:upset-final-sighted}--\ref{fig:faceted-final-characters}).
For each chart, we show three versions: sighted version, Braille version, and a Braille-to-English letter-by-letter translation version.

% UpSet plot - Final design

% Clustered heatmap - Final design
% \clearpage

% Violin plot - Final design

% \clearpage
% Faceted plot - Final design

% \clearpage
\section{Model Photos of Template Chart Final Designs}
\label{sec:template-chart-model-photos-final-design}
\begin{figure}[!t]
    \centering
        \includegraphics[width=1\columnwidth]{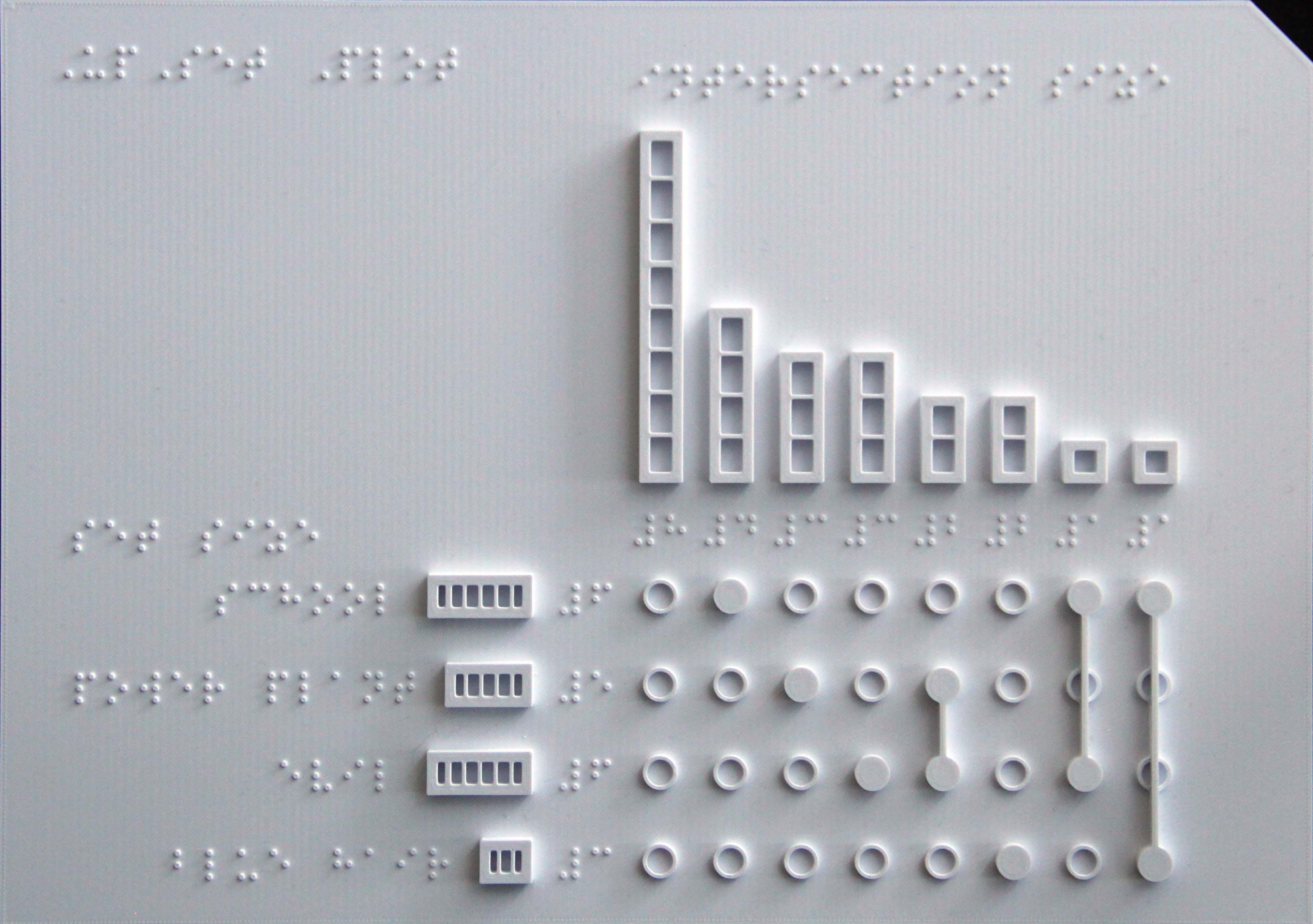}
    \caption{The 3D printed tactile chart for UpSet plot, final design, front view.}
    \label{fig:upset-final-front}
\end{figure}

\begin{figure}[!t]
    \centering
        \includegraphics[width=1\columnwidth]{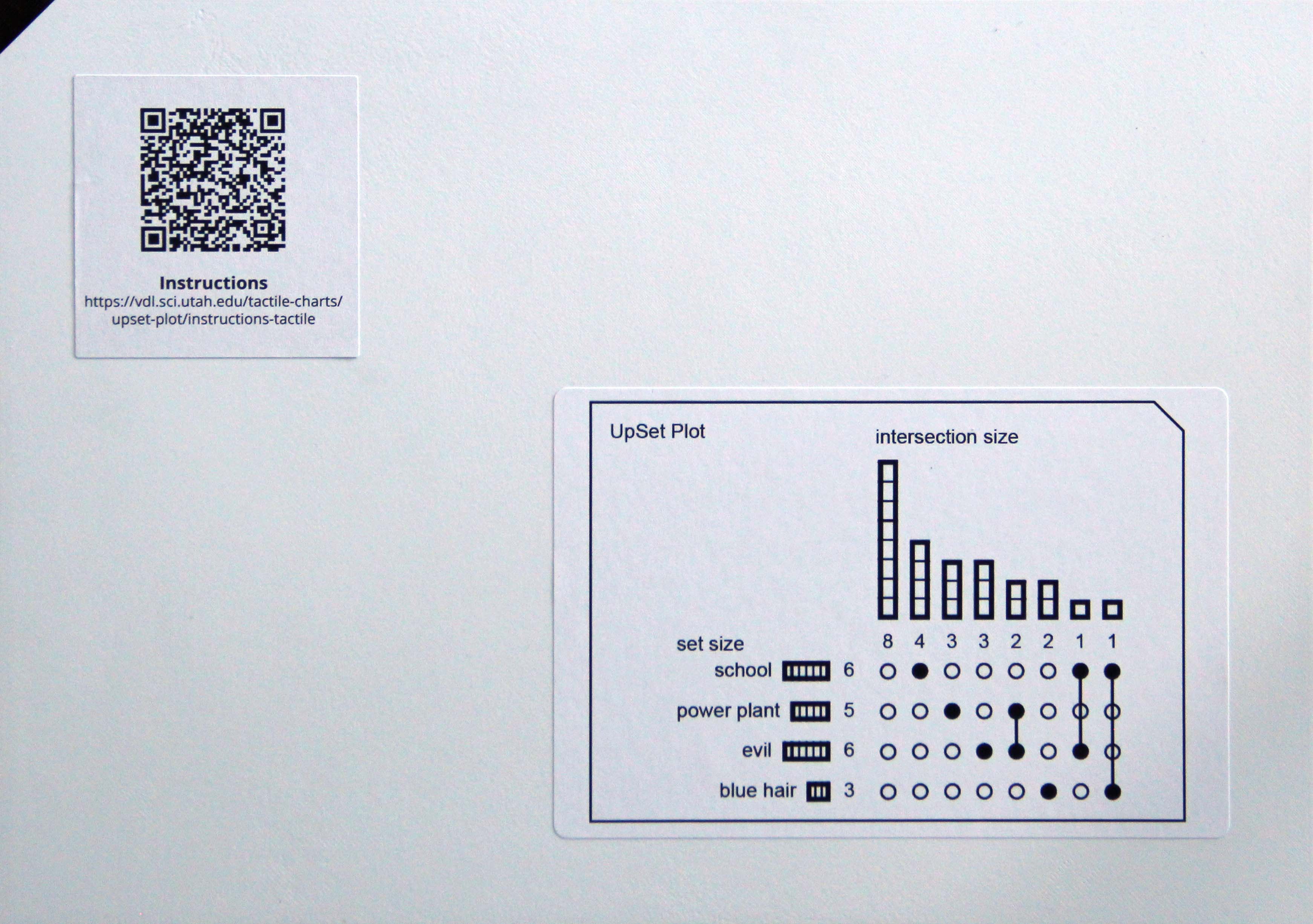}
    \caption{The 3D printed tactile chart for UpSet plot, final design, back view.}
    \label{fig:upset-final-back}
\end{figure}

\begin{figure}[!t]
    \centering
        \includegraphics[width=1\columnwidth]{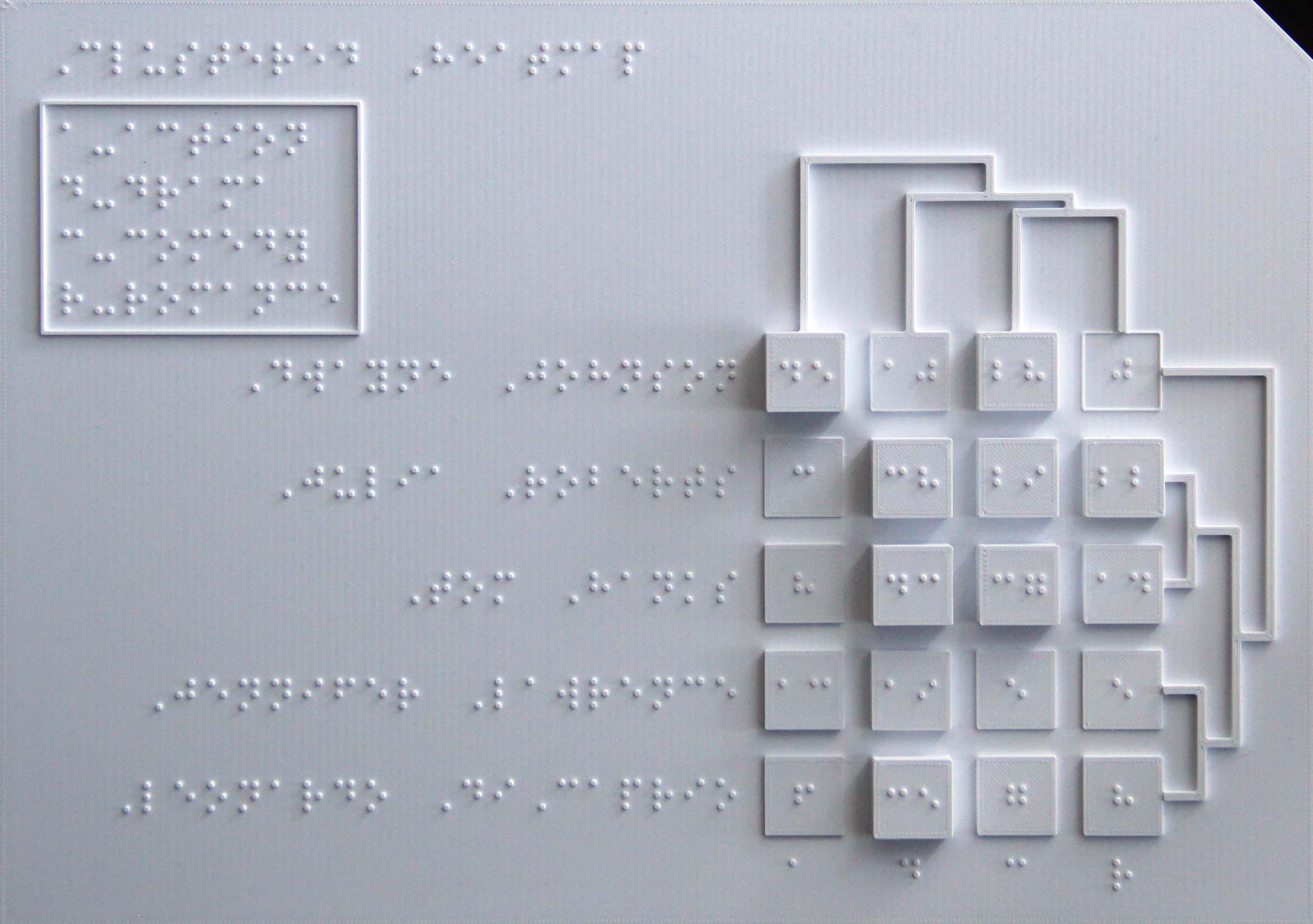}
    \caption{The 3D printed tactile chart for clustered heatmap, final design, front view.}
    \label{fig:heatmap-final-front}
\end{figure}

\begin{figure}[!t]
    \centering
        \includegraphics[width=1\columnwidth]{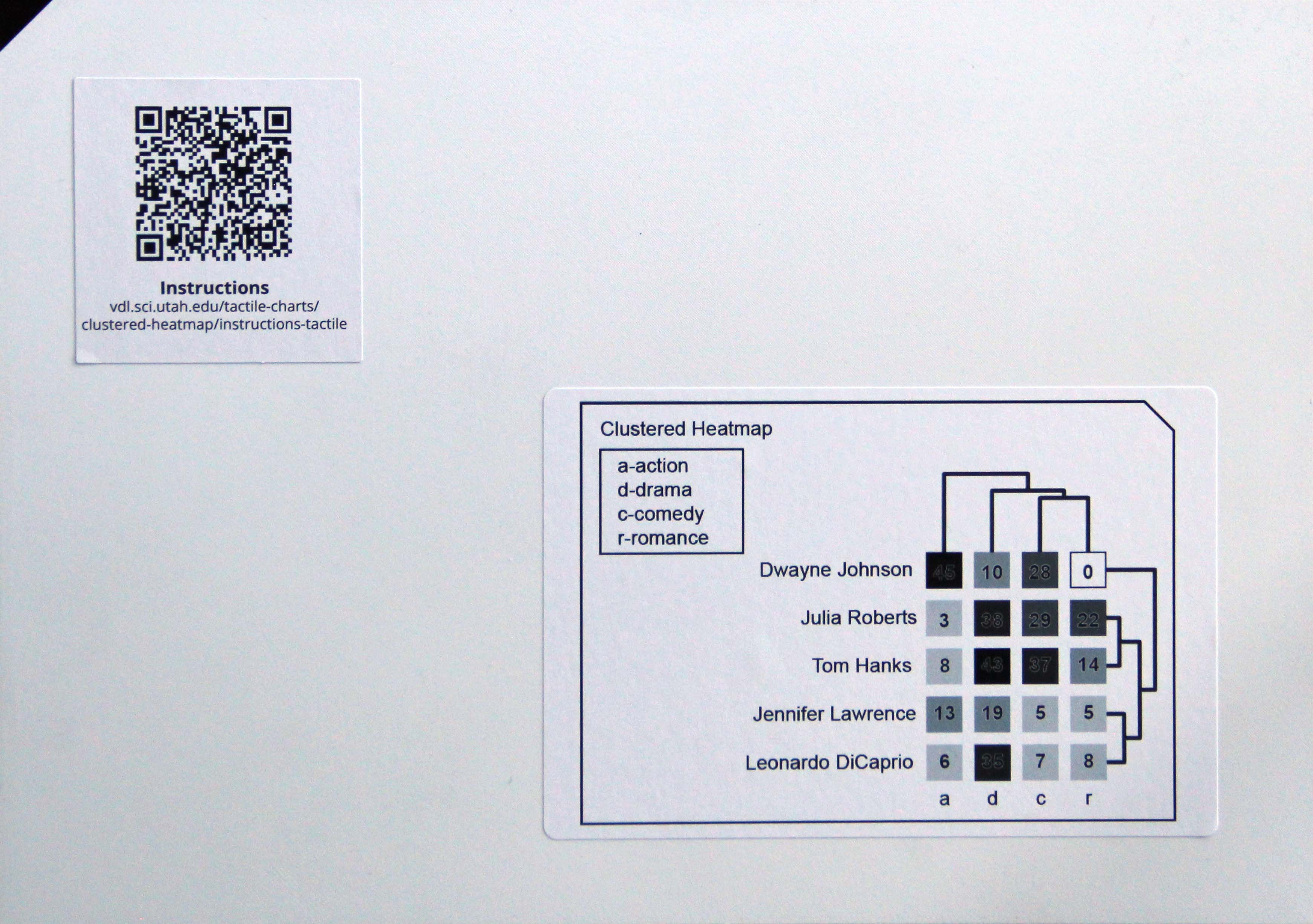}
    \caption{The 3D printed tactile chart for clustered heatmap, final design, back view.}
    \label{fig:heatmap-final-back}
\end{figure}

\begin{figure}[!t]
    \centering
        \includegraphics[width=1\columnwidth]{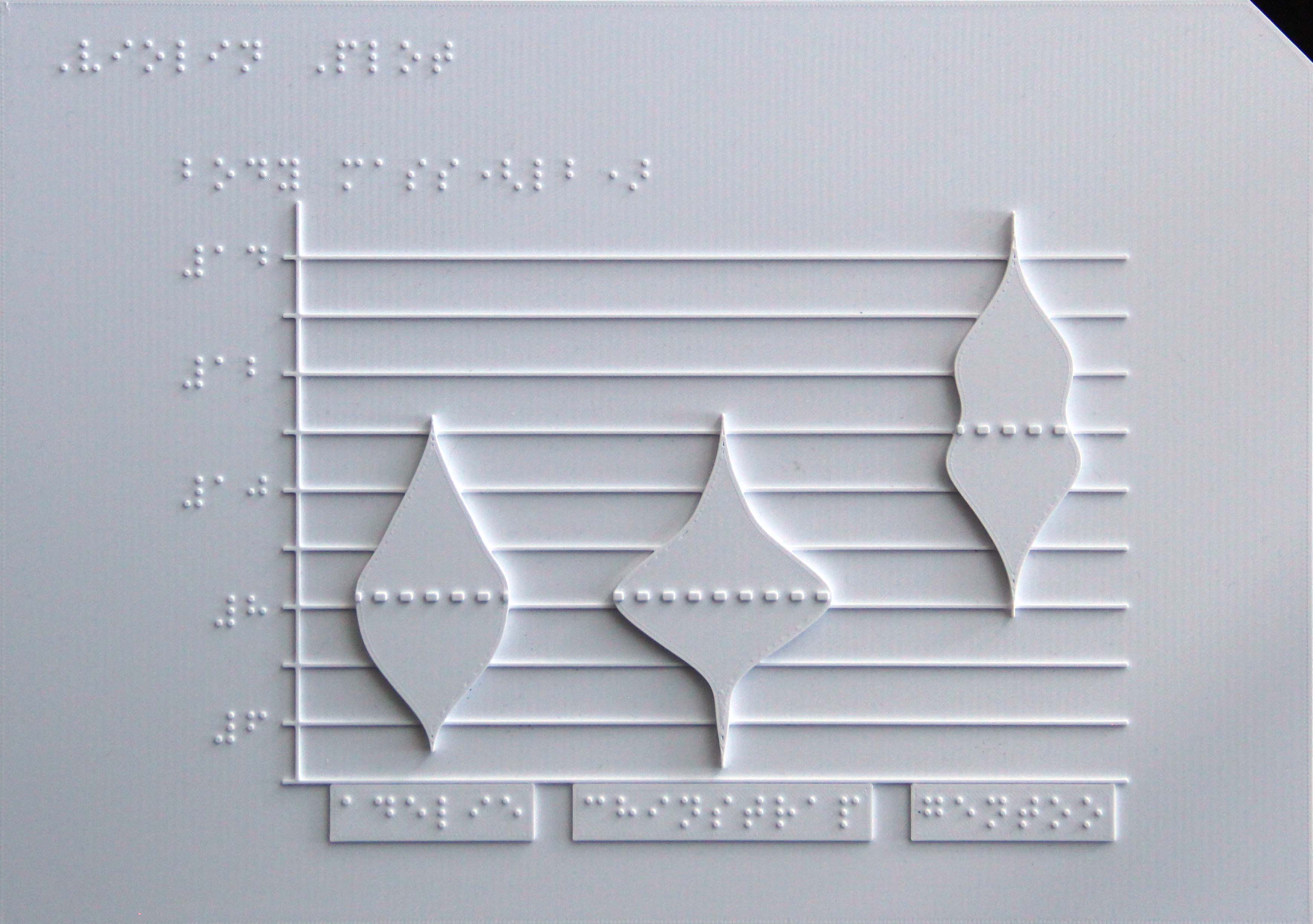}
    \caption{The 3D printed tactile chart for violin plot, final design, front view.}
    \label{fig:violin-final-front}
\end{figure}

\begin{figure}[!t]
    \centering
        \includegraphics[width=1\columnwidth]{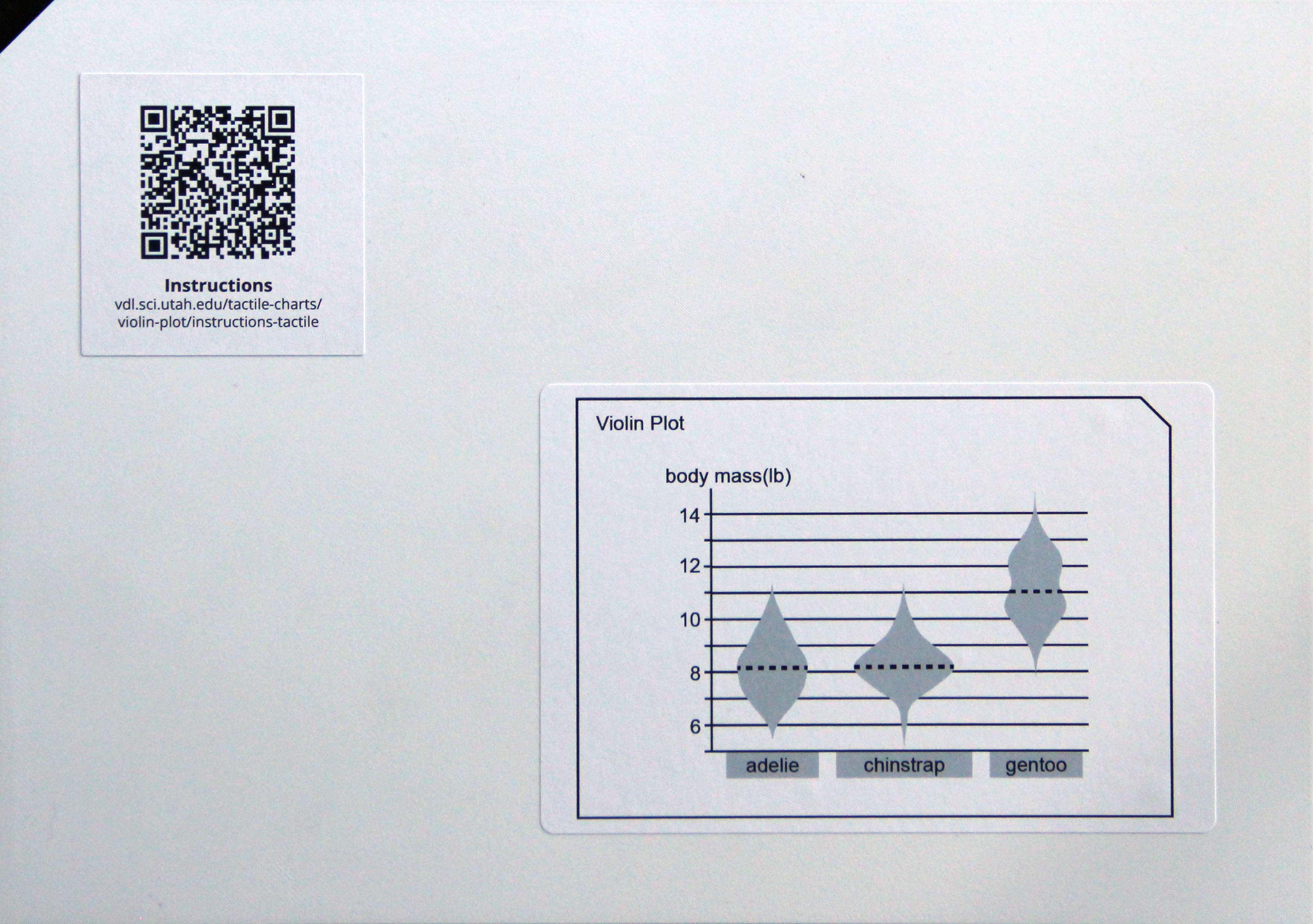}
    \caption{The 3D printed tactile chart for violin plot, final design, back view.}
    \label{fig:violin-final-back}
\end{figure}

\begin{figure}[!t]
    \centering
        \includegraphics[width=1\columnwidth]{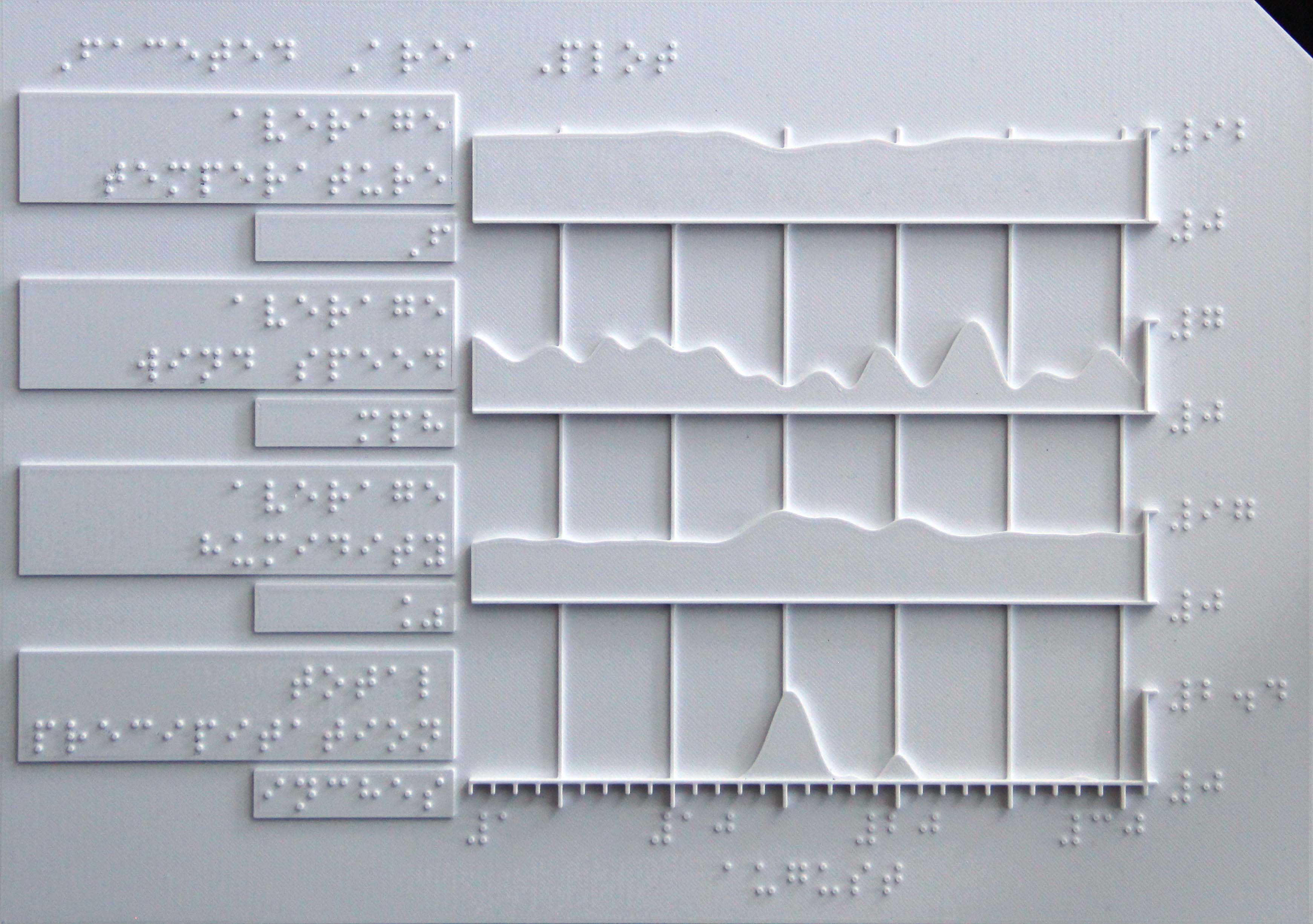}
    \caption{The 3D printed tactile chart for faceted plot, final design, front view.}
    \label{fig:faceted-final-front}
\end{figure}

\begin{figure}[!t]  
    \centering
        \includegraphics[width=1\columnwidth]{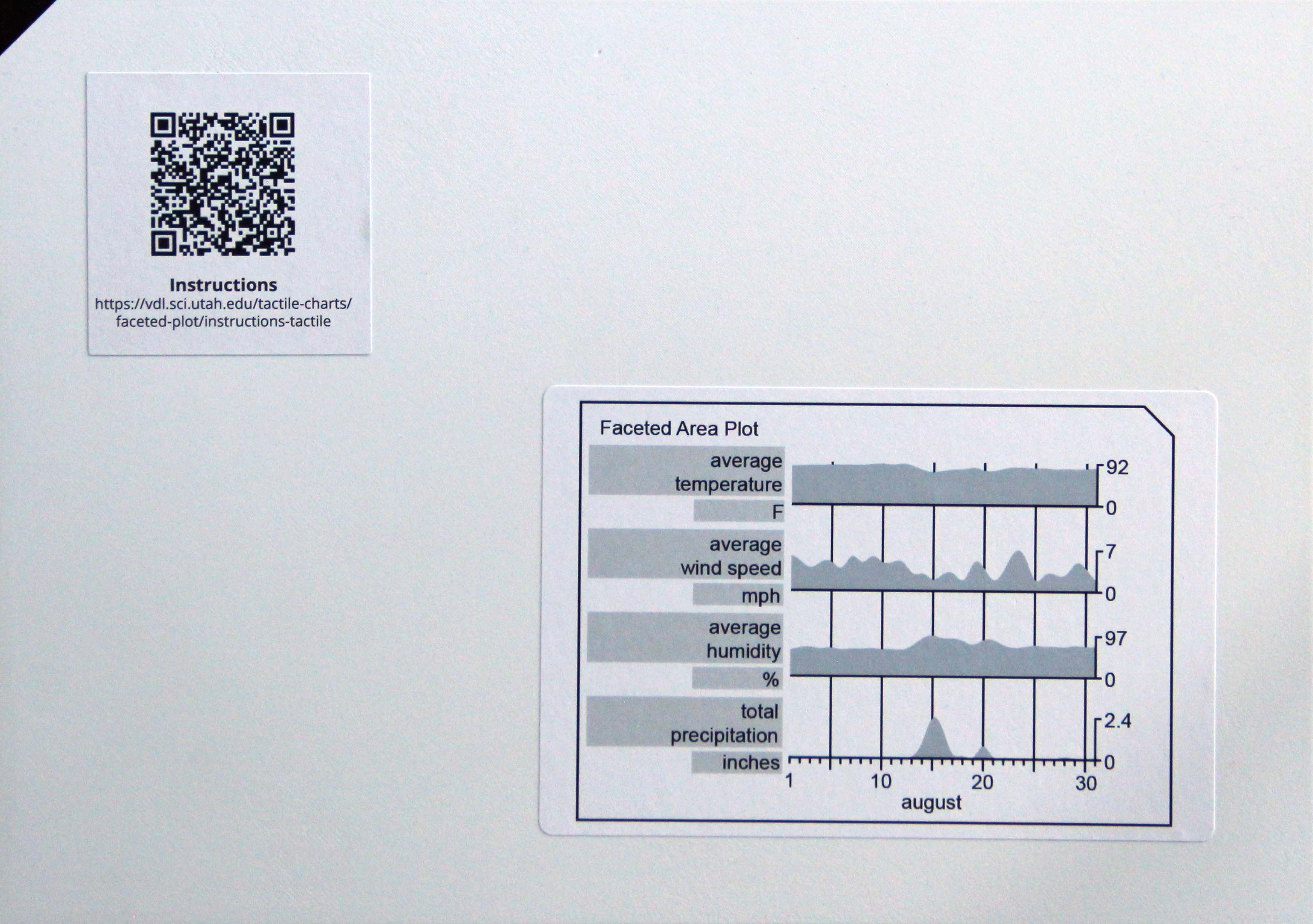}
    \caption{The 3D printed tactile chart for faceted plot, final design, back view.}
    \label{fig:faceted-final-back}
\end{figure}

In this section, we present the 3D printed models for the final tactile charts (\autoref{fig:upset-final-front}--\ref{fig:faceted-final-back}). We show two views of each model: front view and back view.

\begin{figure}[!t]
    \centering
        \includegraphics[width=1\columnwidth]{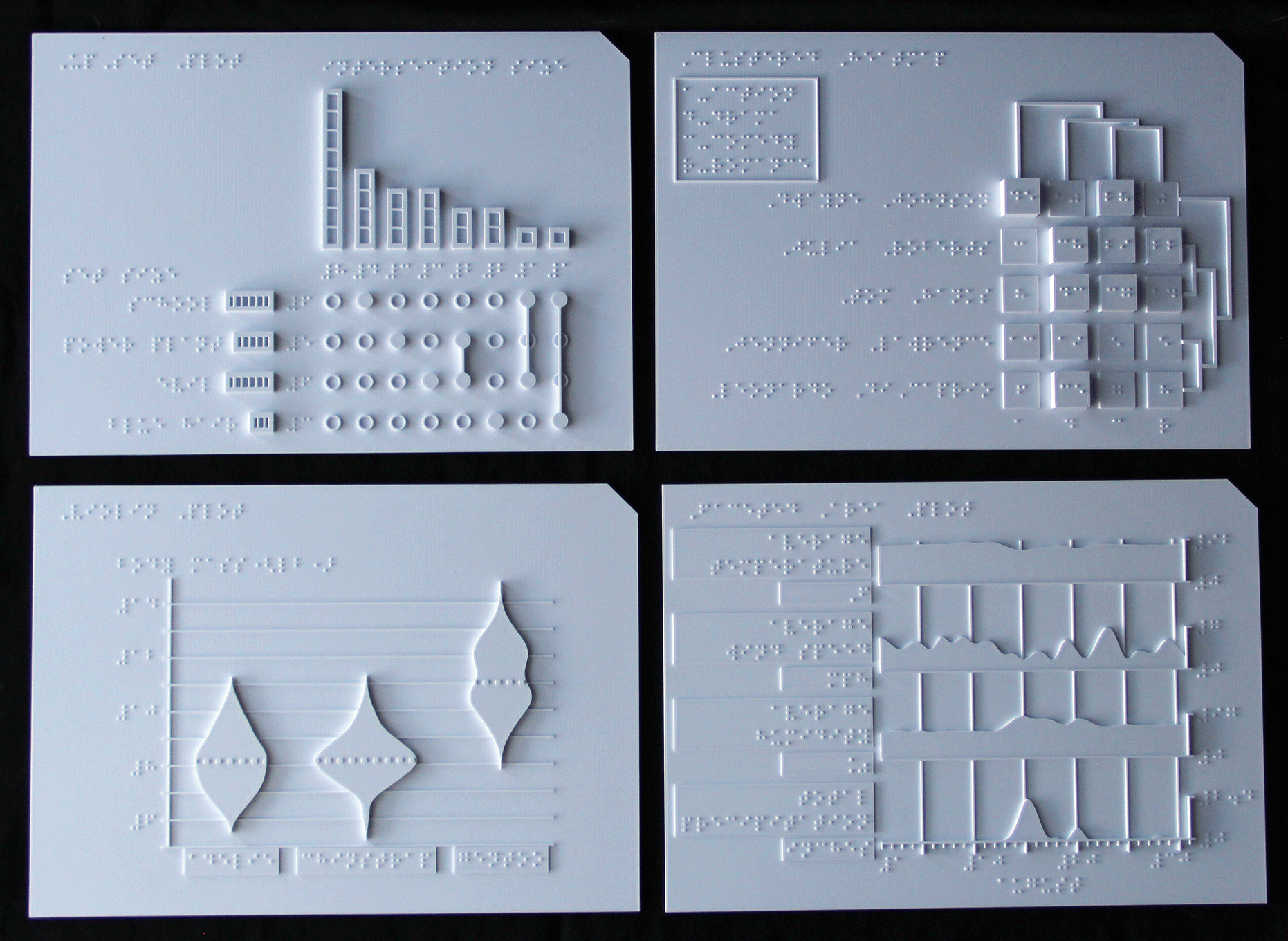}
   \caption{Four final 3D-printed tactile charts, one for each chart type: UpSet plot, clustered heatmap, violin plot, and faceted line chart.}
    \label{fig:final-models}
\end{figure}

% UpSet plot - Final design

% Clustered heatmap - Final design

% Violin plot - Final design

% Faceted plot - Final design

\end{document}